\documentclass[a4paper]{article}
\usepackage{geometry}
\usepackage{graphicx}
\usepackage{subcaption}
\usepackage{xcolor}
\usepackage{multirow}
\usepackage{booktabs}
\usepackage{makecell}
\usepackage{adjustbox}
\usepackage{changepage}
\usepackage{amsmath}
\usepackage{amssymb}
\usepackage{authblk}
\usepackage{cite}

\usepackage{url}
\usepackage[colorlinks = true,
            linkcolor = blue,
            urlcolor  = blue,
            citecolor = blue,
            anchorcolor = blue]{hyperref}

\usepackage{txfonts}

\newcommand\snowmass{\begin{center}\rule[-0.2in]{\hsize}{0.01in}\\\rule{\hsize}{0.01in}\\
\vskip 0.1in Submitted to the  Proceedings of the US Community Study\\ 
on the Future of Particle Physics (Snowmass 2021)\\ 
\rule{\hsize}{0.01in}\\\rule[+0.2in]{\hsize}{0.01in} \end{center}}

\usepackage{amsmath}
\begin{document}

   \title{Higgs Self Couplings Measurements at \\Future proton-proton Colliders: a Snowmass White Paper}

    \author[1]{Angela Taliercio}
    \author[1]{Paola Mastrapasqua}
    \author[1]{Claudio Caputo}
    \author[1]{Pietro Vischia}
    \author[2]{Nicola de Filippis}
    \author[3]{Pushpa Bhat}

   \affil[1]{Centre for Cosmology, Particle Physics and Phenomenology (CP3), Universit\'e catholique de Louvain, Belgium}
   \affil[2]{Politecnico and INFN, Bari (Italy)}
   \affil[3]{Fermilab, Batavia (US)}
   
      \maketitle

   \snowmass

\begin{center}
    \textbf{Abstract}
\end{center}
The Higgs boson trilinear and quartic self-couplings are directly related to the shape of the Higgs potential; measuring them with precision is extremely important, as they provide invaluable information on the electroweak symmetry breaking and the electroweak phase transition. 
\\In this paper, we perform a detailed analysis of double Higgs boson production, through the gluon-gluon fusion process, in the most promising decay channels $b\bar{b} \gamma\gamma$, $b\bar{b} \tau\tau$, and $b\bar{b}b\bar{b}$ for several future colliders: the HL-LHC at 14 TeV and the FCC-hh at 100 TeV, assuming respectively 3 $ab^{-1}$ and 30 $ab^{-1}$ of integrated luminosity.\\
In the HL-LHC scenario, we expect an upper limit on the di-Higgs cross section production of 0.76 at 95\% confidence level, corresponding to a significance of 2.8 $\sigma$.
In the FCC-hh scenario, depending on the assumed detector performance and systematic uncertainties, we expect that the Higgs self-coupling will be measured with a precision in the range 4.8-8.5\% at 95\% confidence level.\\

\newpage
\section*{Executive Summary}

The study of the Higgs boson pair production (HH) is one of the main goals of the scientific program at future colliders. It offers a direct experimental access to the Higgs boson trilinear self coupling and hence to the structure of the scalar potential itself, allowing an unprecedented insight in the electroweak symmetry breaking mechanism.
\\In the Standard Model (SM), the Higgs boson potential is described in terms of the Higgs boson trilinear ($\lambda_{3H}$) and quartic ($\lambda_{4H}$) couplings as \cite{yell_rep}:
\begin{equation}
   \scalebox{1.1}{$V(h) = \frac{m_{H}^{2}}{2} h^{2} + \lambda_{3H} v h^{3} + \lambda_{4H} vh^{4}$}
\end{equation}
\noindent where $h$ is the Higgs field, $v$ = 246 GeV and for Higgs mass $m_{H}$ = 125 GeV, $\lambda_{SM} = \lambda_{3H} = \lambda_{4H} = m_{H}^{2}/2 v^{2} \approx 0.13$.\\

\noindent The anomalous self-coupling $\kappa_{\lambda}$, defined as the ratio $\lambda/\lambda_{SM}$ of the experimental measurement of the coupling to its SM predicted value, is used to parameterize any deviation from the SM expectations.
The measurement of the Higgs self-couplings would represent an important test bench of the SM. At the same time, it could help in probing theories beyond the Standard Model (BSM): if large deviations from the SM predictions occur, these could be interpreted  as an exciting sign of New Physics.\\
The di-Higgs phenomenology is dominated by the very tiny cross section of 37~fb in SM at NNLO at $\sqrt{s} $= 14 TeV, a result of the destructive interference of the box and triangle diagrams \cite{hh_theory}. 

\begin{figure}[h!]
    \centering
     \includegraphics[width=0.6\textwidth]{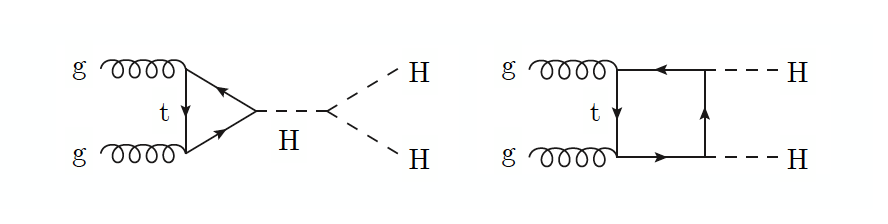}
   \caption{The principal Feynman diagrams of the pair production of Higgs bosons via gluon-gluon fusion. Left: Higgs self-coupling diagram. Right: top box diagram.}
   \label{pic:box_tri}
\end{figure}

\begin{figure}[h!]
    \centering
     \includegraphics[width=0.45\textwidth]{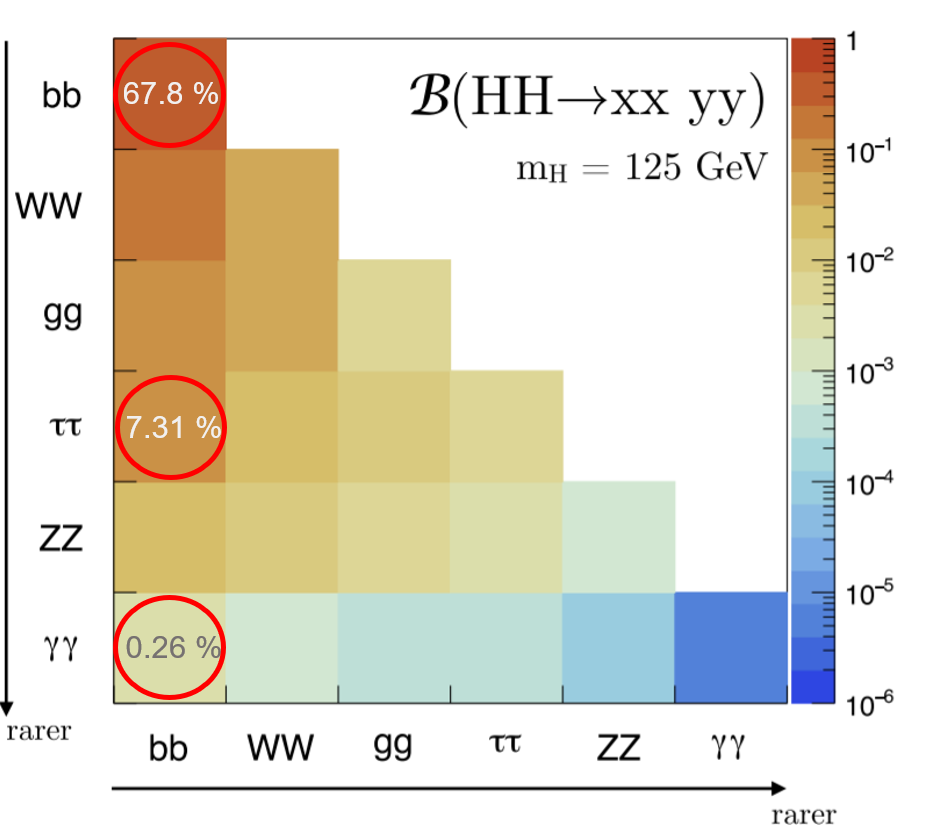}
   \caption{Branching ratios of the main HH decays assuming the SM Higgs boson. The BR for the channels considered in this paper are reported.}
   \label{pic:BR}
\end{figure}

\noindent For that reason, in order to access experimentally the HH phase space it is essential to find a trade-off between keeping the branching ratio high enough and enhancing the signal purity by selecting and combining different Higgs boson decays.
The branching ratios for the main combinations of Higgs decay channels are specified in Figure \ref{pic:BR}. To keep the branching ratio as high as possible, the majority of di-Higgs searches are forced to have one Higgs boson decaying into two b quarks.\\
The most sensitive channels are $b\bar{b} \gamma\gamma$, $b\bar{b}\tau\tau$ and $b\bar{b}b\bar{b}$, as shown in Figure~\ref{pic:BR}:
\begin{itemize}
    \item[$\blacksquare$] $b\bar{b} \gamma\gamma$ is the final state with the highest purity, has the benefit that all objects can be reconstructed but suffers from a very low branching ratio
    \item[$\blacksquare$] $b\bar{b} \tau\tau$ has the second highest branching ratio, is easy to trigger on due to the presence of leptons, and has a relatively low background
    \item[$\blacksquare$] $b\bar{b}b\bar{b}$ is the final state with the highest branching ratio but it does not include any particular object to trigger on and suffers from high QCD- and tt-induced background. 
\end{itemize}

\noindent In this paper, we perform a detailed analysis of HH production, via the gluon-gluon fusion production process (ggF), in the most sensitive decay channels $b\bar{b} \gamma\gamma$, $b\bar{b} \tau\tau$, and $b\bar{b}b\bar{b}$ for several future collider options: the HL-LHC at 14 TeV and the FCC-hh at 100 TeV, assuming respectively 3 $ab^{-1}$ and 30 $ab^{-1}$ of integrated luminosity.\\
\\ \noindent In the HL-LHC scenario, the combined significance is expected to be 2.8 standard deviations, considering both statistical and systematic uncertainties. As in this case the significance of the process is not enough to claim its observation, the study is used to derive an upper limit on the production rate of the process, which we estimate to be 0.76 times the SM prediction, at the 95\% CL. Prospects for the measurement of the trilinear coupling are also studied, leading to a constraint on $\kappa_{\lambda}$ of $[-0.02,3.05]$ at the 95\% CL.\\
\\In the FCC-hh scenario, the significance for a HH signal is expected to lead to an observation. Depending on the assumed detector performance and systematic uncertainties, the Higgs boson trilinear self-coupling and the signal strength will be measured with a precision in the range $4.8-8.5$\% at 95\% CL ($2.4-3.9$ at 68\% CL) and $4-8$\% at 95\% CL ($2-3.6$\% at 68\% CL), respectively. 
\newpage
\tableofcontents
\newpage
\section{Monte Carlo event simulation}
\label{sec:powheg}

The signal and background processes in proton-proton (pp) collisions at $\sqrt{s} = 14$ and 100 TeV are modelled using Monte Carlo (MC) event generators, which simulate the hard process; the hadronisation and fragmentation effects are handled by using the PYTHIA8~\cite{pythia} program.

\noindent Signal processes from gluon-gluon fusion (ggF) HH production are simulated at next-to-leading order (NLO) accuracy in quantum chromodynamics (QCD) with POWHEG 2.0~\cite{powheg_1, powheg_2, powheg_3} for $\kappa_{\lambda}$ values of 1, 2.45, and 5. The distributions of expected ggF signal events are scaled by functions of $\kappa_{\lambda}$ defined according to the known dependence of the ggF HH cross section \cite{hh_scaling} and added together. The total prediction is normalized to the corresponding next-to-NLO (NNLO) cross section~\cite{hh_nnlo} to model the signal for an arbitrary $k_{\lambda}$ value. 

\section{Detector simulation and physics object reconstruction}

All the simulated samples are processed with the DELPHES~\cite{delphes} fast simulation program to  model the detector response and performance; these were chosen to mimik the parameterized behaviour of the LHC experiments in the HL-LHC scenario. The simulation accounts also for pileup contributions by overlaying an average of 200 (1000) minimum bias interaction events simulated with PYTHIA8 at center-of-mass energies of 14(100) TeV. 

\noindent The performance of the reconstruction and the identification algorithms are modeled for electrons, muons, tau leptons decaying to hadrons ($\tau_{h}$) and a neutrino, photons, jets, including those containing heavy flavour particles, and the missing transverse momentum vector. In particular, the physics object resolutions, the energy and momentum scales, the efficiencies, and misidentification rates for the various objects, as well as the reconstruction algorithms, follow the studies made both for the CMS Phase II Technical Design Reports~\cite{CERN-LHCC-2017-009}, and for the FCC-hh Technical Design Reports~\cite{FCC-tdr}.

\noindent A brief description of the modeling of the relevant physics objects in DELPHES is reported below:
\begin{itemize}
    \item[$\blacksquare$] photons are built from neutral energy excess in a simplified version of the electromagnetic calorimeter. Photon objects do not have a MVA score for photon ID, nor a reliable isolation value. Indeed, they are just categorized in Tight/Loose categories using a parameterized formula in $p_{T}$ and $\eta$ for efficiency and fake rate, miming the performance of the full simulation for PhaseII CMS detector;

    \item[$\blacksquare$] jet objects are built through a particle-flow algorithm starting from the smeared tracks, in order to account for the tracker resolution, and the energy deposits in the simplified versions of ECAL, HCAL, and HGCal. The jet collection used for the analysis is PUPPI jet, which are jets partially cleaned of the pileup by an algorithm emulating the PUPPI algorithm \cite{puppi}. The b and $\tau$ tagging jet is performed via parameterized efficiencies and mistag rates that take into account the presence of the 'MIP-timing detector';

    \item[$\blacksquare$] electrons are seeded by generator-level electrons and are identified with an efficiency that is parameterized  as a function of the energy and the pseudorapidity. The energy resolution of reconstructed electrons is a function of the ECAL tracker resolution;

    \item[$\blacksquare$] muons are seeded by generator-level muons and identified with a parameterized efficiency given by the muon chambers. The momentum of the reconstructed muons is obtained by smearing the generator-level muon 4-momenta, with a resolution parameterized as a fuction of the transverse momentum and the pseudorapidity;

    \item[$\blacksquare$] missing transverse momentum is calculated for each event by using the particle flow objects information and corrected for the pile up (PUPPI algorithm \cite{puppi}.)
    
\end{itemize}

\section{Data analysis framework}

\noindent The data analysis for the three aforementioned double Higg decay channels has been done by using the Bamboo framework~\cite{Bamboo}; this program automatically constructs lightweight python wrappers based on the structure of the ROOT TTrees,
which allow to construct physics oriented expressions with high-level code. By constructing
an object representation of the expression, a few powerful operations can be used to compose complex expressions.
The mechanics of loading data samples, processing them locally or on a batch
system, combining the outputs for different samples in an overview, designing selection criteria and defining plots is very similar over a broad range of use cases; therefore is can profit from a common implementation, some user-defined instructions and a configuration file with a list of samples, and instructions how to display them.

\clearpage
\section{$\boldsymbol{HH\to b\bar{b} \gamma\gamma}$ analysis}
\label{sec:bbgg}

The most sensitive channel for HH production is undoubtedly the $bb\gamma\gamma$ one. On one hand it suffers from a low branching ratio of only 0.26\%, on the other it benefits from the high photon resolution and the possibility of reconstruct fully and unambiguously the decay products of both Higgs bosons. \\Using the $bb\gamma\gamma$ final state, the current limit at 95\% for the trilinear coupling modifier is $-3.3<\kappa_{\lambda}< 8.5$, while a cross section greater than 7.7 times the SM prediction is excluded, based on data collected by CMS during its Run-2 at $\sqrt{s} = 13$ TeV and with 138 $fb^{-1}$ of integrated luminosity~\cite{bbgg_run2}. The LHC Run-2 target integrated luminosity is larger by a factor larger than 10 with respect to the Run-2 (target of 250 $fb^{-1}$ per year, with the goal of 3000 $fb^{-1}$ in the 12 years after the upgrade \cite{HL-LHC}). Results for HH production are therefore expected to be significantly improved, as we will prove in the following.
\\The analysis strategy followed here consist of kinematic selections to identify the Higgs boson candidates, implementation of multivariate classifiers to improve the signal-to-background ratio, event categorization, and extraction of results from the diphoton invariant mass fit. 

\subsection{Simulated samples}
Signal samples for SM and BSM hypotheses with different values (1, 2.45, and 5) of the trilinear Higgs boson coupling are simulated in order to study the expected constraint on the $k_\lambda$.
The backgrounds considered for the study can be divided in two groups: resonant processes, where a single Higgs boson decays to photons, and nonresonant processes, where no physical Higgs bosons are produced. The main resonant background sources are the single Higgs processes produced via ggF and in associated production with top pairs. 
The nonresonant background sources contain two isolated, energetic photons and can be divided into QCD- and tt-induced events. Table~\ref{tab:ev} reports the simulated samples with their relative cross sections.

 \begin{table}[h!]
	\centering
    \begin{tabular}{c|c|c} 
    \hline
    
    &\textbf{Process}  & \makecell[c]{\textbf{Cross section} (fb)} \\
    
    \hline
    
	\multirow{3}{*}{Signal} & \makecell[l]{$(gg)HH \rightarrow b\Bar{b} \gamma\gamma$ ($\kappa_{\lambda} =1$)} & $9.70 \times 10^{-2}$ \\
	& \makecell[l]{$(gg)HH \rightarrow b\Bar{b} \gamma\gamma$  ($\kappa_{\lambda} =2.45$)} & $4.09 \times 10^{-2}$ \\
	& \makecell[l]{$(gg)HH \rightarrow b\Bar{b} \gamma\gamma$  ($\kappa_{\lambda} =5$) }& $2.96 \times 10^{-1}$ \\
	
	\hline
	
     \multirow{4}{*}{\makecell[c]{Single \\ Higgs }}& $(gg)H \rightarrow \gamma\gamma$ &  $1.24 \times 10^{2}$\\
    & $qqH \rightarrow \gamma\gamma$  &  $ 9.71$\\
    & $VH \rightarrow \gamma\gamma$  & $ 5.67$\\
    & $ttH \rightarrow \gamma\gamma$ &  $ 1.39$\\
    
	\hline
	
	 \multirow{3}{*}{\makecell[c]{QCD- \\ induced }}& $pp\rightarrow \gamma\gamma +jets$ & $9.46 \times 10^{4}$\\
    & $pp\rightarrow \gamma +jets$ &  $1.04 \times 10^{6}$\\
    & $pp\rightarrow jets$ &  $ 1.41 \times 10^{8}$\\
    
	\hline
	
    \multirow{5}{*}{\makecell[c]{tt- \\ induced }}& $pp\rightarrow t\bar{t} \gamma\gamma$ &  $1.86 \times 10^{1}$\\
    & $pp \rightarrow t\bar{t}\gamma \;had$  &  $ 7.92 \times 10^{2}$\\
    & $pp \rightarrow t\bar{t}\gamma \;semi\,lep$ &  $ 7.71 \times 10^{2}$\\
    & $pp \rightarrow t\bar{t}\gamma \; fully\,lep$ & $ 6.23 \times 10^{2}$\\
    & $pp \rightarrow t\bar{t} \; inclusive$ & $ 8.64 \times 10^{5}$\\
    
	\hline
	\end{tabular}
	\caption{List of simulated samples for $bb\gamma\gamma$ channel.}. 
	\label{tab:ev}
\end{table}

\subsection{Event selection}
Signal events are characterized by four exclusive objects, namely two photons and two jets. Several kinematic requirements are imposed on jet and photon objects to select only events with signal-like topology.
We select loose photons and require them to satisfy the tight isolation criteria~\cite{delphes}. The pseudorapity is required to be less than 2.5, with the exclusion of the transition region between endcap and barrel stations. Photons are ordered according to the transverse momentum and the (sub)leading photon is requested to have a $p_{T}$ greater then 30 (20) GeV. Leading and subleading photons are used to build one of the two Higgs boson candidates, whose invariant mass is required to be in the range $[100,180]$ GeV around the Higgs mass nominal value of 125 GeV. Table~\ref{tab:ph_jet_sel} left summarizes the photon kinematic selections.\\
Jets satisfying the tight identification and the loose b tagging criteria are selected~\cite{delphes}. To avoid overlay with photon objects, the $\Delta R = \sqrt{\Delta\phi^{2}+\Delta\eta^{2}}$ between selected jets and photons is required to be greater than 0.4. The pseudorapidity is requested to be smaller than 2.5, and the $p_{T}$ greater than 30 GeV. The two highest-$p_{T}$ jets are used to build the second Higgs boson candidate, whose mass is restricted to be inside the range $[80,200]$ GeV. Table~\ref{tab:ph_jet_sel} right summarizes the jet kinematic selections.

 \begin{table}[h!]
	\centering
	\begin{minipage}{0.5\linewidth}
	\centering
    \begin{tabular}{c|c } 
    \hline
    \textbf{Variable} & \textbf{Requirement} \\
    \hline
    ID & loose \\
    ISO & tight \\
    $|\eta|$ & $ < $ 1.44 or in $[1.57,2.5]$ \\
    $p_{T}$ (sub)lead  & $>$ 30 (20) GeV \\
    $p_{T}/m_{\gamma\gamma}$ (sub)lead & $>$ 1/3 (1/4) \\
    $m_{\gamma\gamma}$ & in $[100,180]$ GeV \\
	\hline
	\end{tabular}
	\end{minipage}%
	\begin{minipage}{0.5\linewidth}
	\centering
    \begin{tabular}{c|c } 
    \hline
    \textbf{Variable} & \textbf{Requirement} \\
    \hline
    ID & tight \\
    b-tag & loose \\
    $|\eta|$ & $<$ 2.5 \\
    $p_{T}$   &  $>$ 30 GeV \\
    $m_{jj}$ & in $[80,200]$ GeV \\
    &\\
	\hline
	\end{tabular}
	\end{minipage}
	
	\caption{Photon (Left) and Jet (Right) kinematic selections}. 
	\label{tab:ph_jet_sel}
\end{table}

\begin{figure}[h!]
    \centering
    \begin{subfigure}[t]{0.45\textwidth}
    	\centering
        \includegraphics[width=0.99\textwidth]{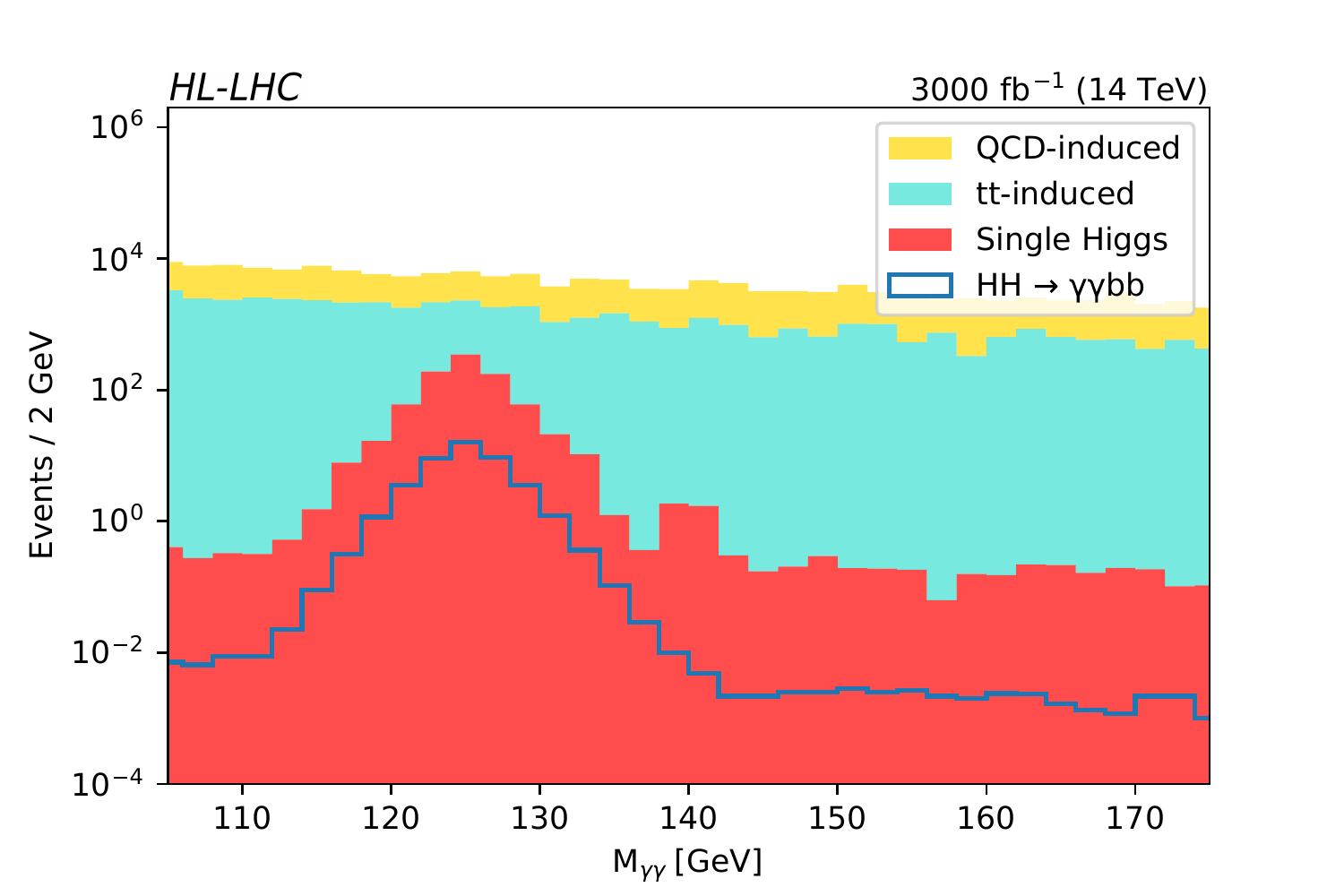}
    \end{subfigure}
    \begin{subfigure}[t]{0.45\textwidth}
       \centering
       \includegraphics[width=0.99\textwidth]{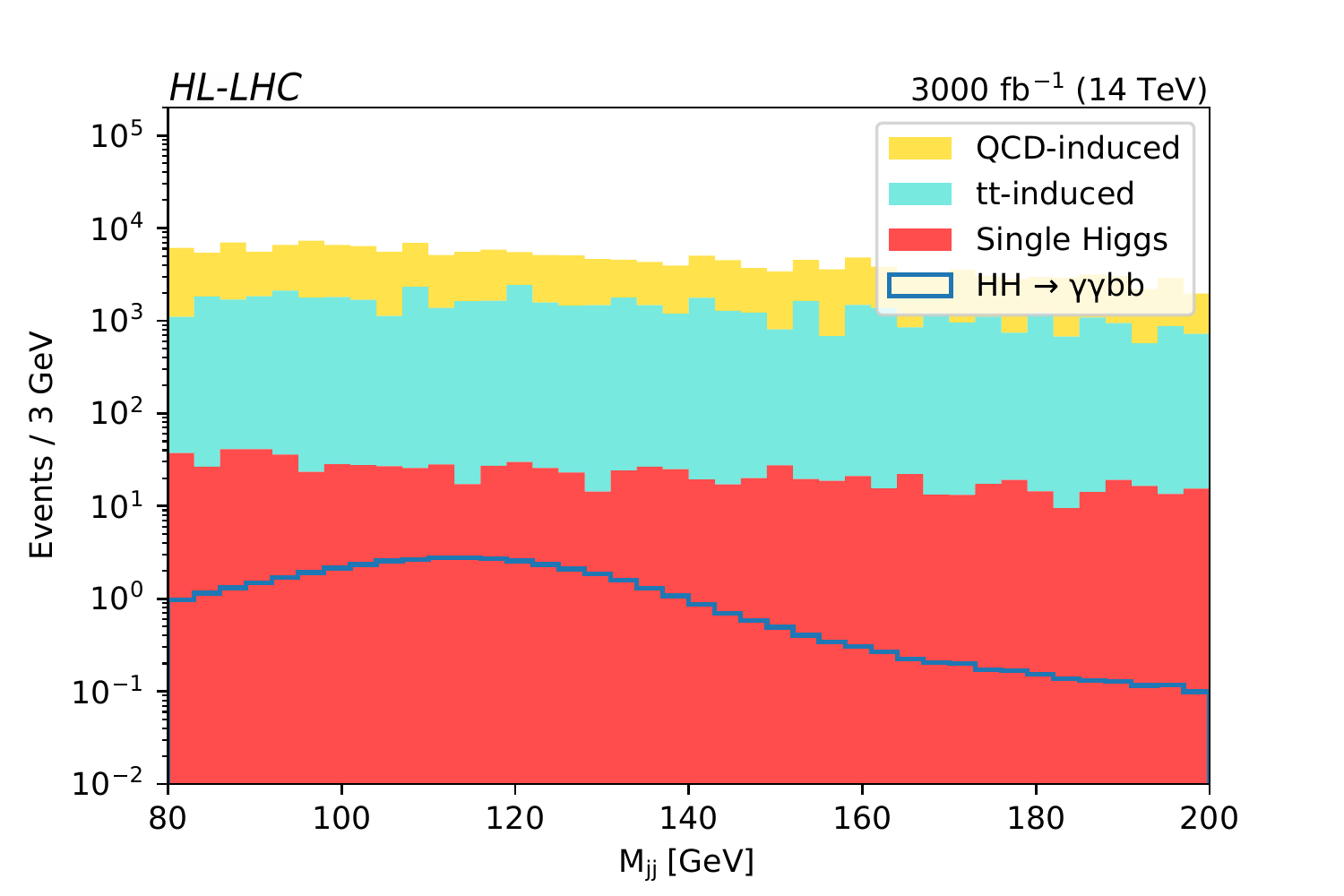}
   \end{subfigure}
   \caption{(Left) Diphoton and (Right) dijets invariant mass after kinematic selections, for signal and background processes. Histograms are scaled to cross section and luminosity.}
   \label{pic:inv_mass}
\end{figure}

 \noindent Table~\ref{tab:yields} reports the yields for all the simulated processes after the kinematic selections. It can be noticed that the selection efficiency for the ttH process is similar to the signal one, as it mimics well the signal topology. A dedicated tagger to discriminate against the ttH events is therefore needed, and will be described in detail in the next section.
\\Figure \ref{pic:inv_mass} shows the invariant mass distributions after the selections of diphotons and dijets candidates for the signal process and the backgrounds (divided in resonant and non-resonant ones).

\begin{table}[h!]
	\centering
    \begin{tabular}{c|c} 
    \hline
    \textbf{Process} & \textbf{Yields} \\
    \hline
	 \makecell[l]{$(gg)HH \rightarrow b\Bar{b} \gamma\gamma$  $\kappa_{\lambda} = 1$} & 45 $\pm$ 1  \\
	 \makecell[l]{$(gg)HH \rightarrow b\Bar{b} \gamma\gamma$  $\kappa_{\lambda} = 2.45$} & 18 $\pm$ 2 \\
	 \makecell[l]{$(gg)HH \rightarrow b\Bar{b} \gamma\gamma$  $\kappa_{\lambda} = 5$} & 97 $\pm$ 2 \\
	\hline
      $(gg)H \rightarrow \gamma\gamma$ & 275 $\pm$ 32 \\
     $qqH \rightarrow \gamma\gamma$  & 40 $\pm$ 3 \\
     $VH \rightarrow \gamma\gamma$ & 110 $\pm$ 3 \\
     $ttH \rightarrow \gamma\gamma$ & 476 $\pm$ 12\\
	\hline
	 $pp\rightarrow \gamma\gamma +jets$ & 85997 $\pm$ 2286  \\
     $pp\rightarrow \gamma +jets$ & 41270 $\pm$ 2737  \\
     $pp\rightarrow jets$ & 0.0 \\
	\hline
     $pp\rightarrow t\bar{t} \gamma\gamma$ & 562 $\pm$ 17 \\
     $pp \rightarrow t\bar{t}\gamma$ &  3939 $\pm$ 133  \\
     $pp \rightarrow t\bar{t} \; $ & 49060 $\pm$ 2020  \\
	\hline
	\end{tabular}
	\caption{Yields and selection efficiencies for all the simulated processes.}
	\label{tab:yields}
\end{table}

\subsection{DNN-based ttH tagger}
\label{sec:ttH}
\begin{figure}[h!]
    \centering
    \begin{subfigure}[t]{0.45\textwidth}
    	\centering
        \includegraphics[width=0.99\textwidth]{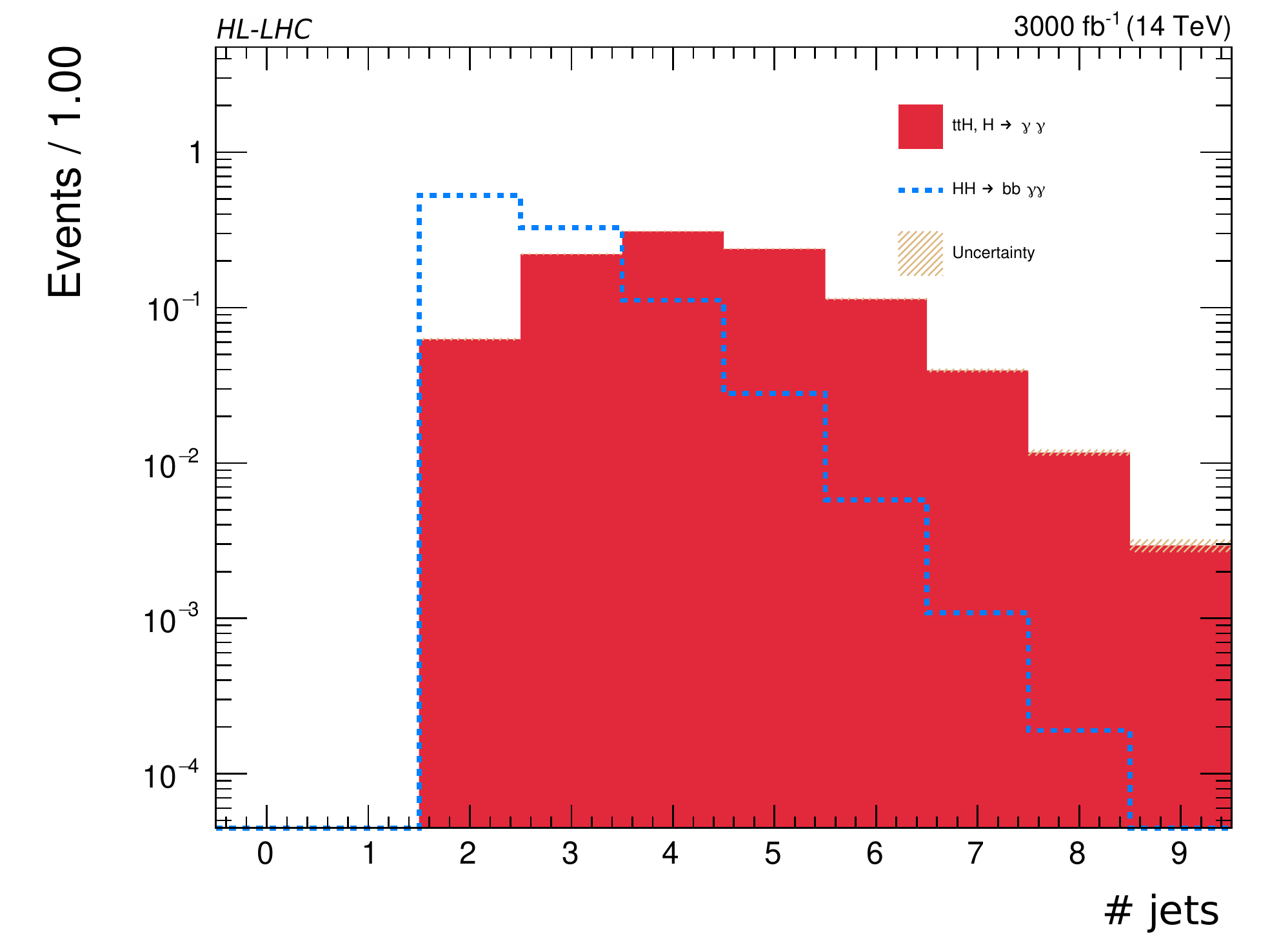}
    \end{subfigure}
    \begin{subfigure}[t]{0.45\textwidth}
       \centering
       \includegraphics[width=0.99\textwidth]{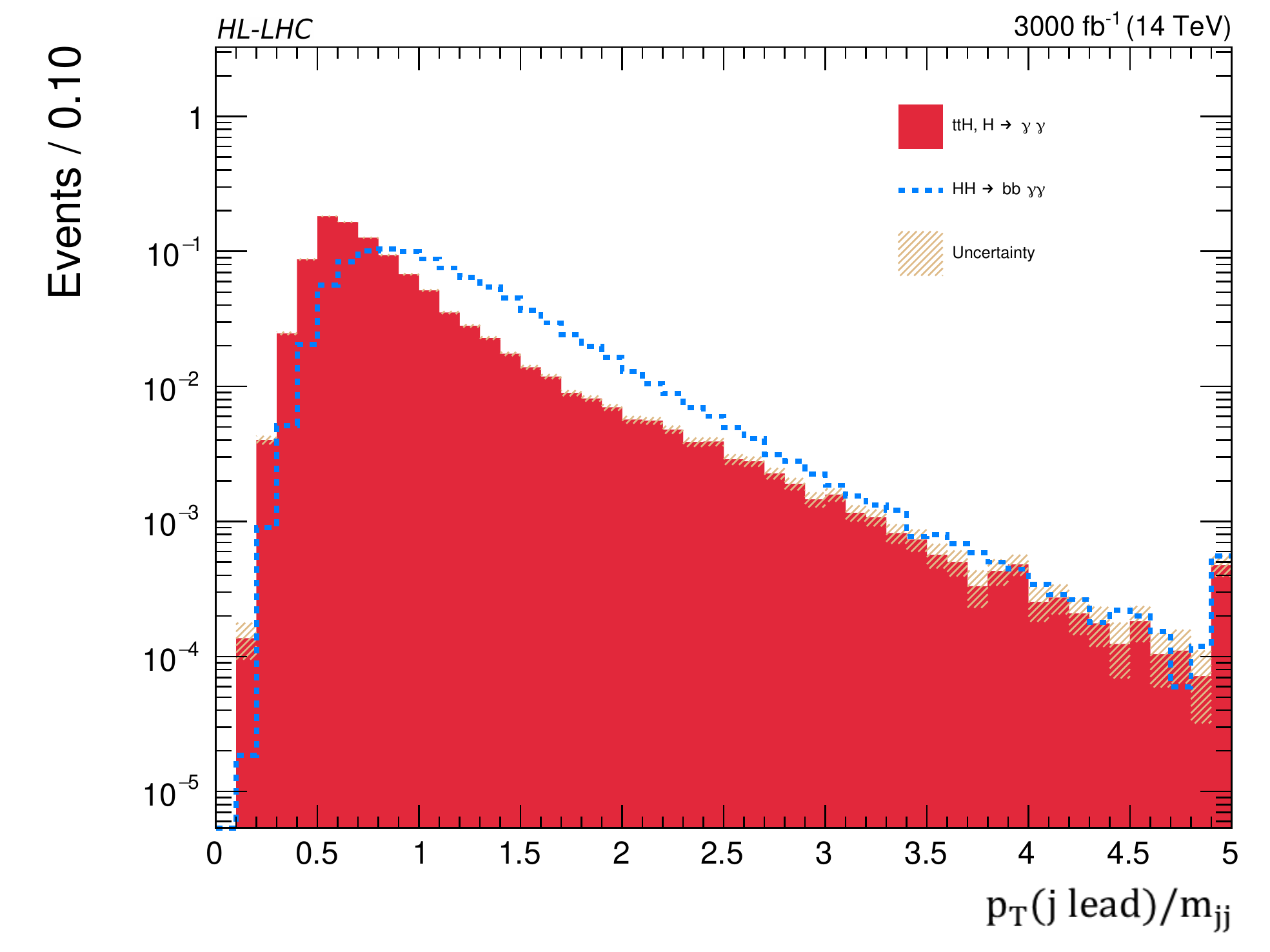}
   \end{subfigure}
   \caption{(Left) Number of jets per event and (Right) transverse momentum of the leading jet divided by the invariant mass of the dijet object, for signal and ttH background. Histograms are scaled to unity to compare the shapes. }
   \label{pic:jet}
\end{figure}

\begin{figure}[h!]
    \centering
    \begin{subfigure}[t]{0.45\textwidth}
    	\centering
        \includegraphics[width=0.99\textwidth]{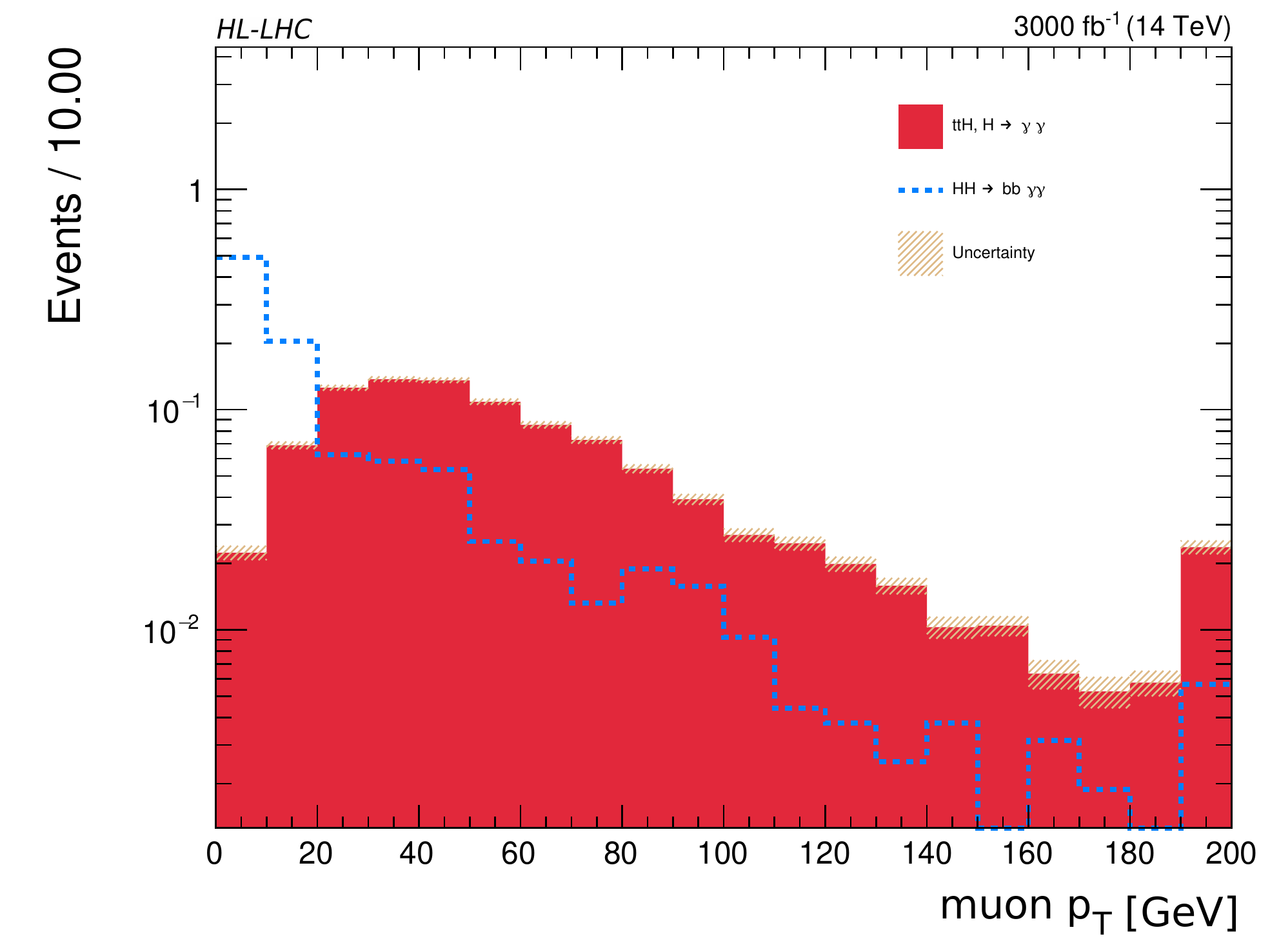}
    \end{subfigure}
    \begin{subfigure}[t]{0.45\textwidth}
       \centering
       \includegraphics[width=0.99\textwidth]{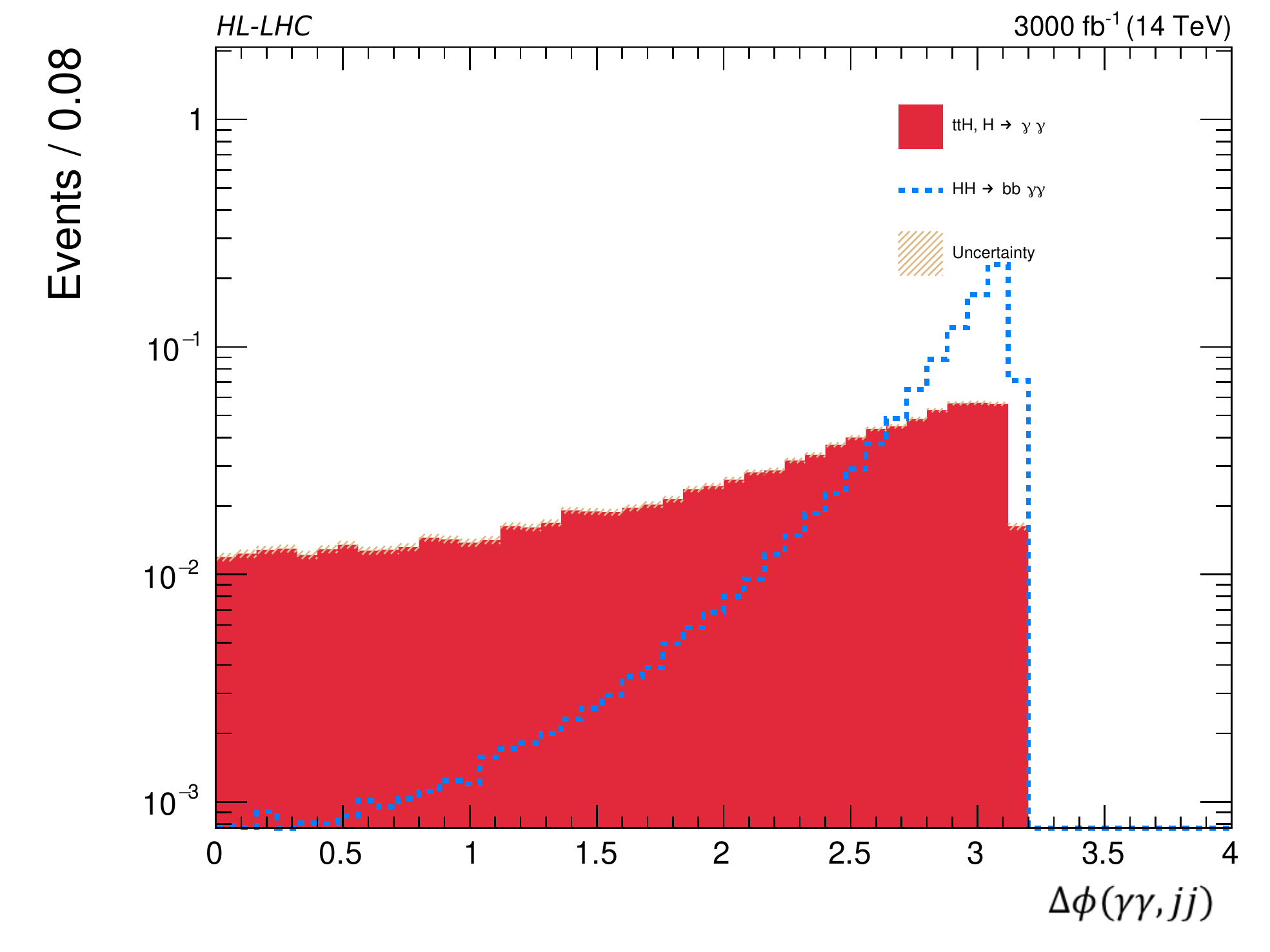}
   \end{subfigure}
   \caption{(Left) Transverse momentum of the selected muons and (Right) $\Delta\phi$ between the diphoton and the dijet objetcs for signal and ttH background. Histograms are scaled to unity to compare the shapes.}
   \label{pic:muon}
\end{figure}

In order to discriminate the signal against the ttH background, several variables are combined together into a single powerful tagger through a Deep Neural Network (DNN) algorithm implemented using a Keras frontend~\cite{keras} with a tensorflow backend~\cite{tensorflow}. The following features are exploited:
\begin{itemize}
    \item[$\blacksquare$] The number of jets (with no b tag requirement), shown in Figure~\ref{pic:jet} (left);
    \item[$\blacksquare$] The b tag of the leading and subleading jet;
    \item[$\blacksquare$] $p_{T}(j)/m(jj)$ of the leading and subleading jet, shown in Figure~\ref{pic:jet} (right);
    \item[$\blacksquare$] $p_{T}(jj)/m(jj)$ of the dijet object;
    \item[$\blacksquare$] $p_{T}(\gamma)/m(\gamma\gamma)$ of the leading and subleading photon;
    \item[$\blacksquare$] $p_{T}(\gamma\gamma)/m(\gamma\gamma)$ of the diphoton object;
    \item[$\blacksquare$] The scalar sum of the jet $p_{T}$;
    \item[$\blacksquare$] The $\Delta R$ between the closest photon-jet pair;
    \item[$\blacksquare$] The $\Delta R$ between the other photon-jet pair;
    \item[$\blacksquare$] The $\Delta \phi$ and $\Delta \eta$ between the leading and subleading photon;
    \item[$\blacksquare$] The $\Delta \phi$ and $\Delta \eta$ between the leading and subleading jet;
    \item[$\blacksquare$] The $\Delta \phi$ and $\Delta \eta$ between the diphoton and the dijet object, illustrated in Figure~\ref{pic:muon} (right);
    \item[$\blacksquare$] The angle between the diphoton object and the beam axis in the dijet rest frame;
    \item[$\blacksquare$] The angle between the leading jet and the beam axis in the dijet rest frame;
    \item[$\blacksquare$] The angle between the leading photon and the beam axis in the diphoton rest frame;
    \item[$\blacksquare$] Number of leptons, i.e. muons and electrons identified by the cuts in Table~\ref{tab:lep_sel};
    \item[$\blacksquare$] $p_{T}$ of muons and electrons (Figure \ref{pic:muon} left).
\end{itemize}

 \begin{table}[h!]
	\centering
    \begin{tabular}{c|c|c } 
    \hline
    & \textbf{Variable} & \textbf{Requirement} \\
    \hline
    \multirow{4}{*}{Electrons}
    &ID & tight \\
    &ISO & tight \\
    &$|\eta|$ & $<$ 1.44 or in $[1.57,2.5]$ \\
    &$p_{T}$  &  $>$ 10 GeV \\
	\hline
	\multirow{4}{*}{Muons}
    &ID & tight \\
    &ISO & tight \\
    &$|\eta|$ & $<$ 2.5 \\
    &$p_{T}$  &  $>$ 10 GeV \\
    \hline
	\end{tabular}
	\caption{Lepton kinematic selections}. 
	\label{tab:lep_sel}
\end{table}


\begin{figure}[h!]
    \centering
    \begin{subfigure}[t]{0.45\textwidth}
    	\centering
        \includegraphics[width=0.99\textwidth]{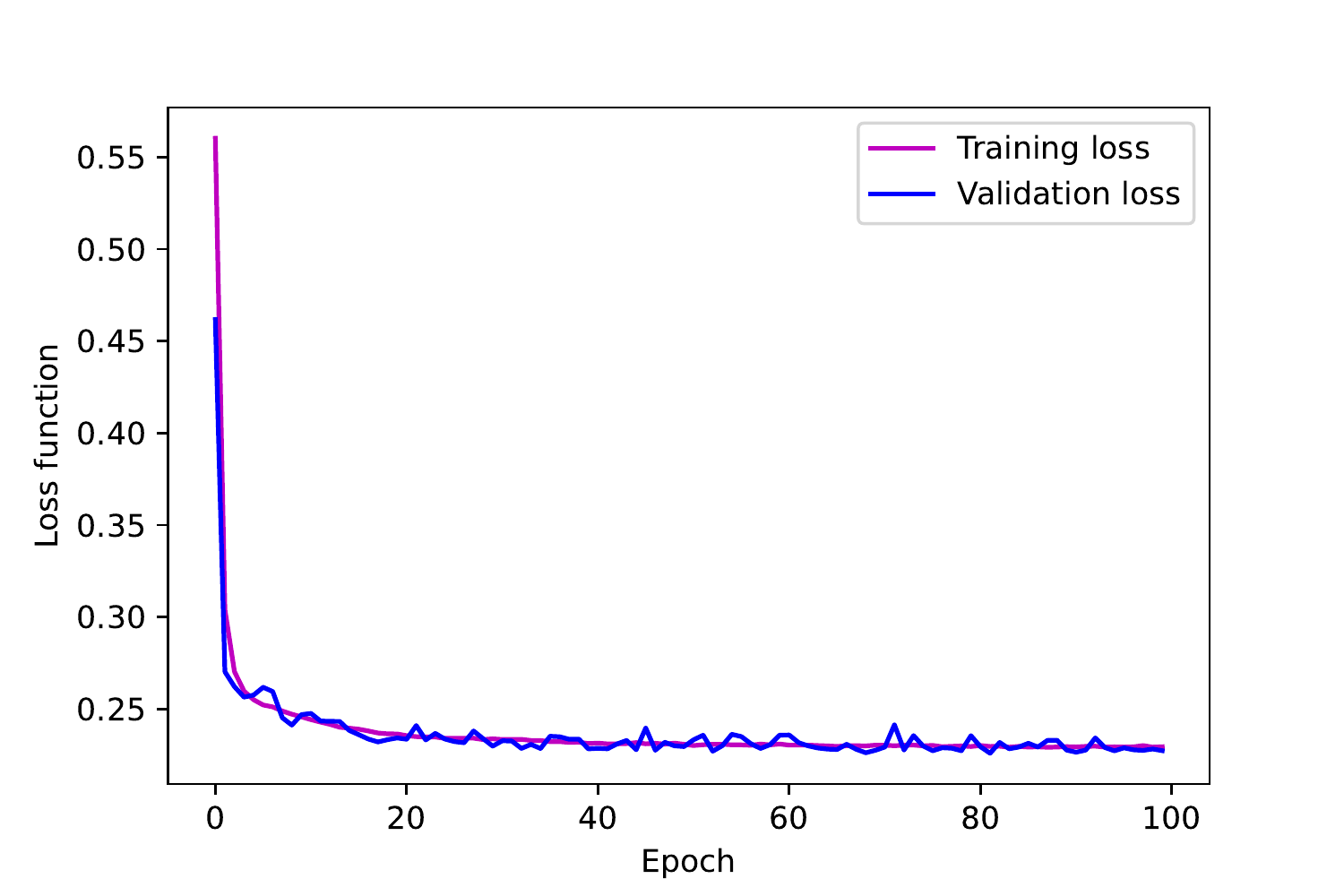}
    \end{subfigure}
    \begin{subfigure}[t]{0.45\textwidth}
       \centering
       \includegraphics[width=0.99\textwidth]{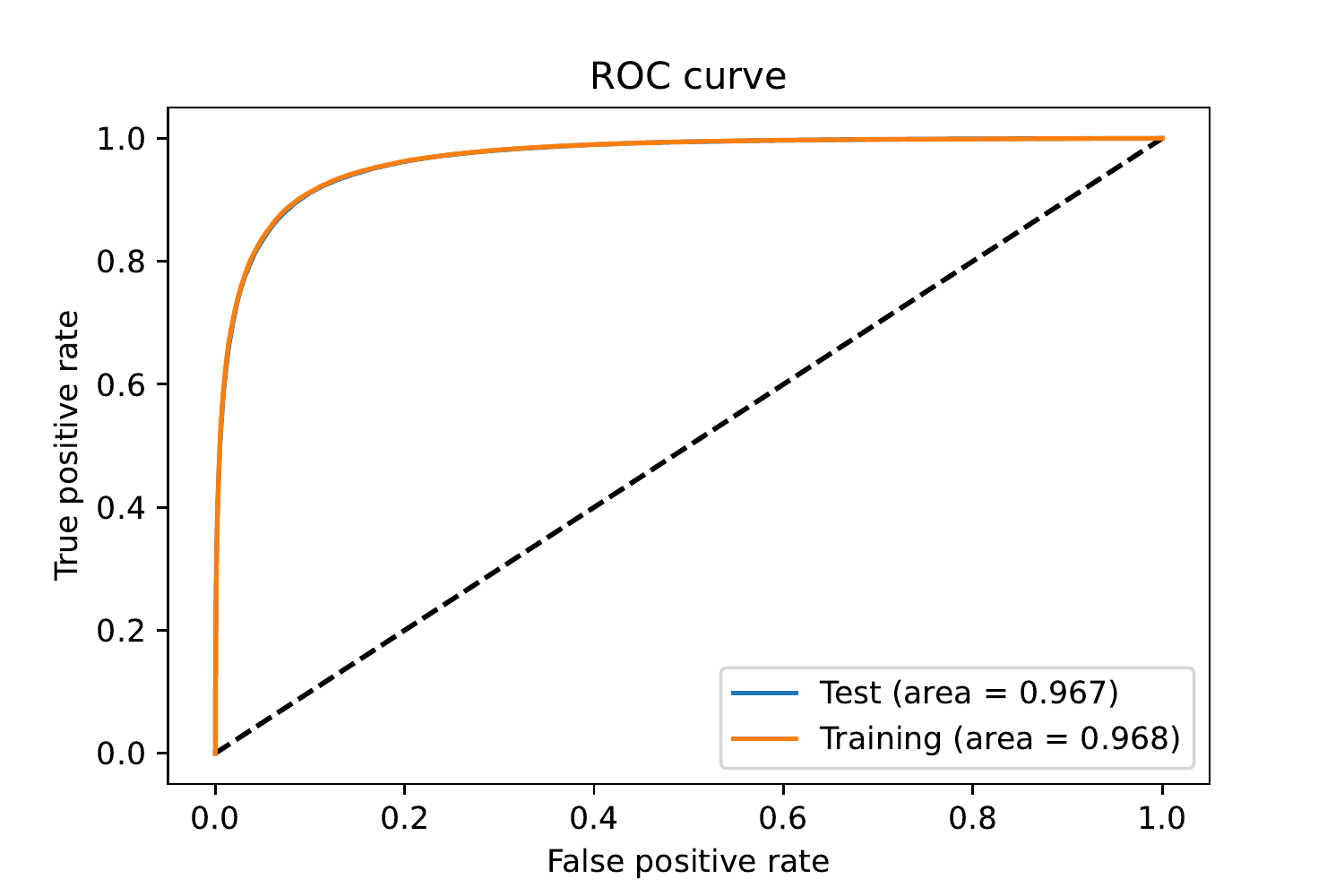}
   \end{subfigure}
   \caption{(Left) Loss function plotted against epochs for training and validation samples. (Right) ROC curves for testing and training samples with relative AUC value.}
   \label{pic:perf_ttH}
\end{figure}

\noindent As evident from the examples in Figures~\ref{pic:jet} and~\ref{pic:muon}, ttH events are characterized by a larger number of reconstructed jets, which are on average less energetic. The transverse momentum of the reconstructed muons, on the contrary, is in general higher for ttH events (where muons come directly from the W bosons) with respect to signal ones (where muons can come only from heavy hadron decay). Finally, a back-to-back production of the diphoton and dijet objects is typical of the signal, while this behaviour is less pronounced for background events. 
\\The DNN is trained with half of the selected events and tested with the other half. Figure~\ref{pic:perf_ttH} (left) reports the learning curve (i.e the loss function over the number of training epochs) for the training dataset, to give an idea of how well the model is learning, and for the validation dataset  to understand how well the model is generalizing. The learning algorithm shows a good fit (no hints of overtraining or undertraining), identified by a training and validation loss that decreases to a point of stability with a minimal gap between the two curves. The performance of the DNN are represented in Figure~\ref{pic:perf_ttH} (right) through the Receiver Operating Characteristic (ROC) curve, which plots the true positive rate versus the false positive one of the classification model at all possible classification thresholds. The Area Under the Curve (AUC) is also reported for all ROC curves, to provide an aggregate measure of performance across all possible classification thresholds. 

\begin{figure}[h!]
    \centering
    \begin{subfigure}[t]{0.45\textwidth}
    	\centering
        \includegraphics[width=0.99\textwidth]{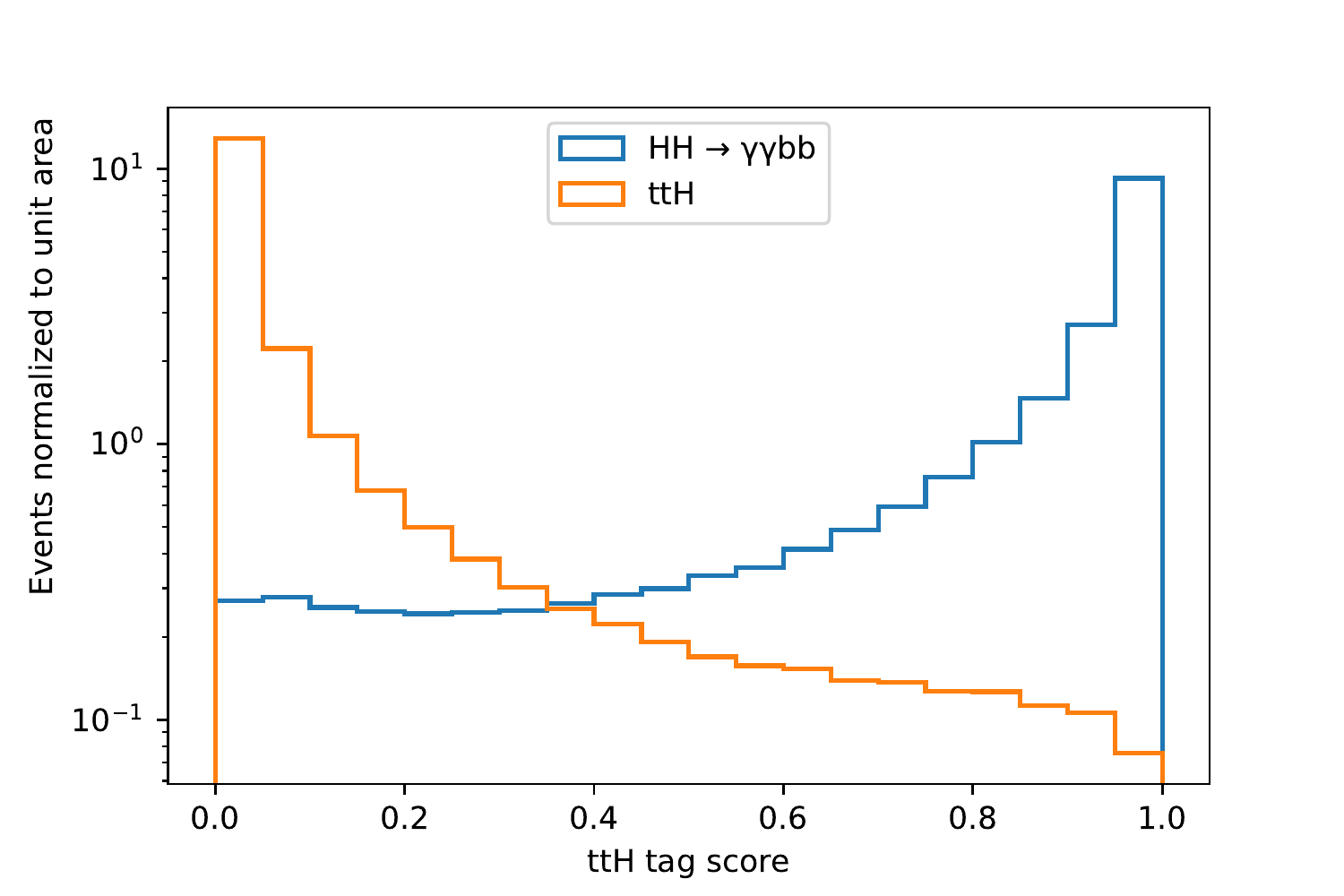}
    \end{subfigure}
    \begin{subfigure}[t]{0.45\textwidth}
       \centering
       \includegraphics[width=0.99\textwidth]{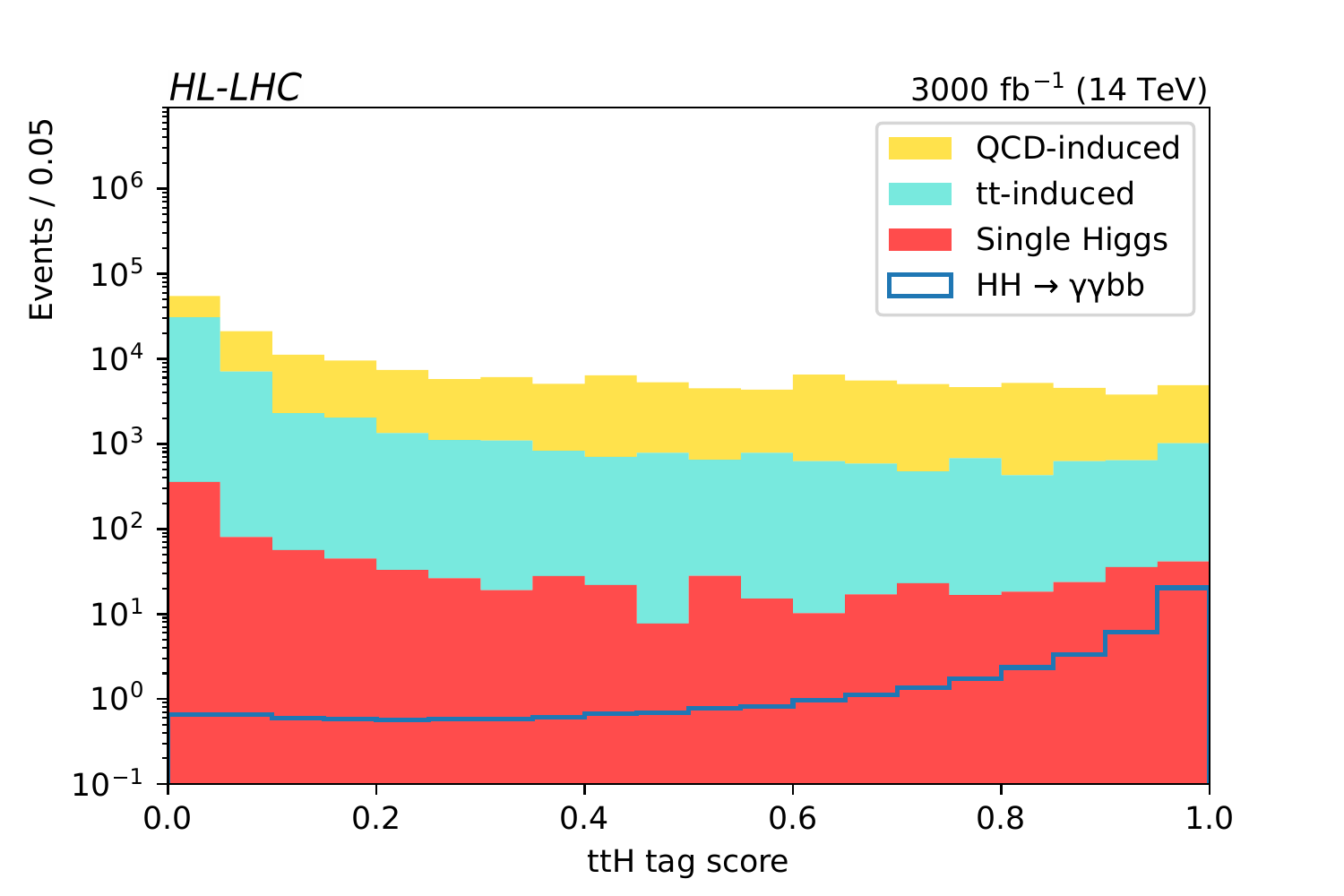}
   \end{subfigure}
   \caption{(Left) ttH tagger score for HH signal and ttH background normalized to unit area. (Right) ttH tagger score for HH signal and stacked background scaled to cross section and luminosity.}
   \label{pic:ttH_score}
\end{figure}

\noindent The ROC curve together with the DNN score for signal and ttH background reported in Figure~\ref{pic:ttH_score} (left), demonstrate a good separation between HH events and ttH ones. Figure~\ref{pic:ttH_score} (right) shows the tag score for all the background processes. It is interesting to notice that the tagger, which is trained only on ttH events, is equally efficient in rejecting also the nonresonant tt-induced background. For example, at 95\% signal efficiency, the contamination of ttH and tt-induced processes is around 15\%, while the tagger is less effective on QCD-induced processes, where the contamination reaches $50--60$\%, and on the other single Higgs processes, whose contamination is around $60--70$\%.  
\\In the analysis, a loose cut on the tag score is imposed to reject background-like events and the exact value of the working point is optimized together with the event categorization explained in Section~\ref{sec:evt_cat}. 

\subsection{Event categorization}
\label{sec:evt_cat}
\begin{figure}[h!]
    	\centering
        \includegraphics[width=0.60\textwidth]{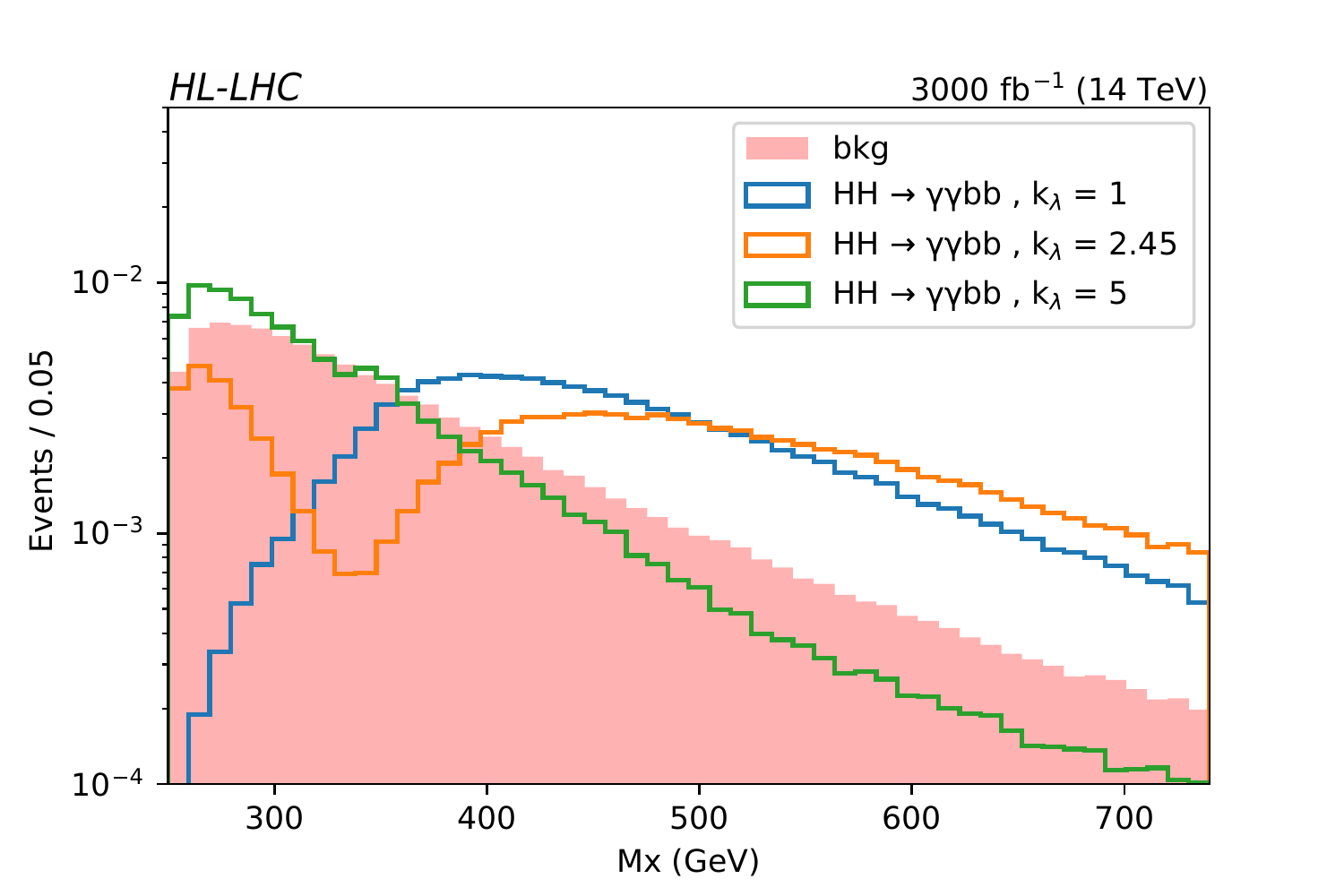}
   \caption{Distribution of the reconstructed Mx for different anomalous coupling hypothesis and the background.}
   \label{pic:Mx}
\end{figure}

\begin{figure}[h!]
    \centering
    \begin{subfigure}[t]{0.45\textwidth}
    	\centering
        \includegraphics[width=0.99\textwidth]{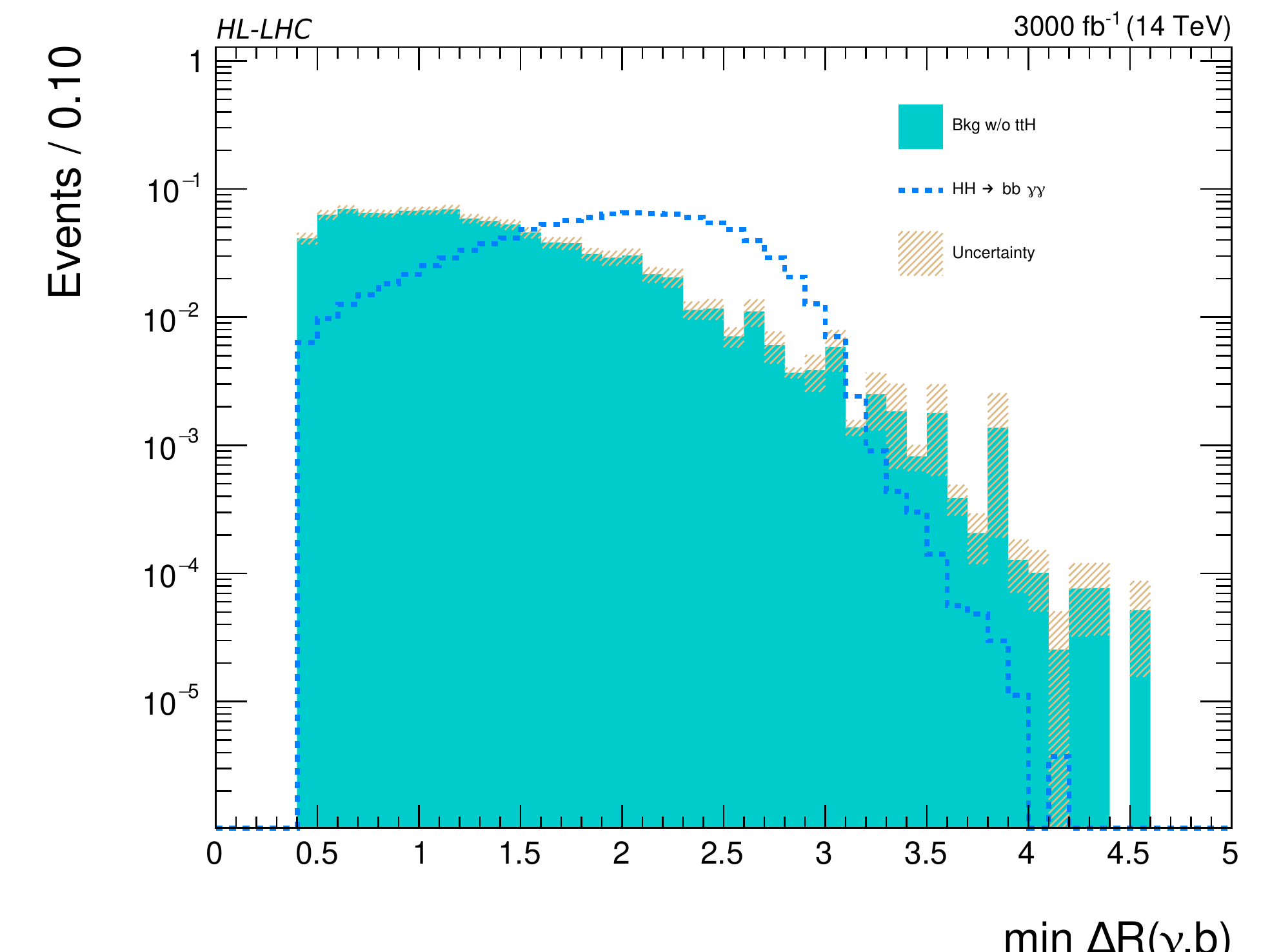}
    \end{subfigure}
    \begin{subfigure}[t]{0.45\textwidth}
       \centering
       \includegraphics[width=0.99\textwidth]{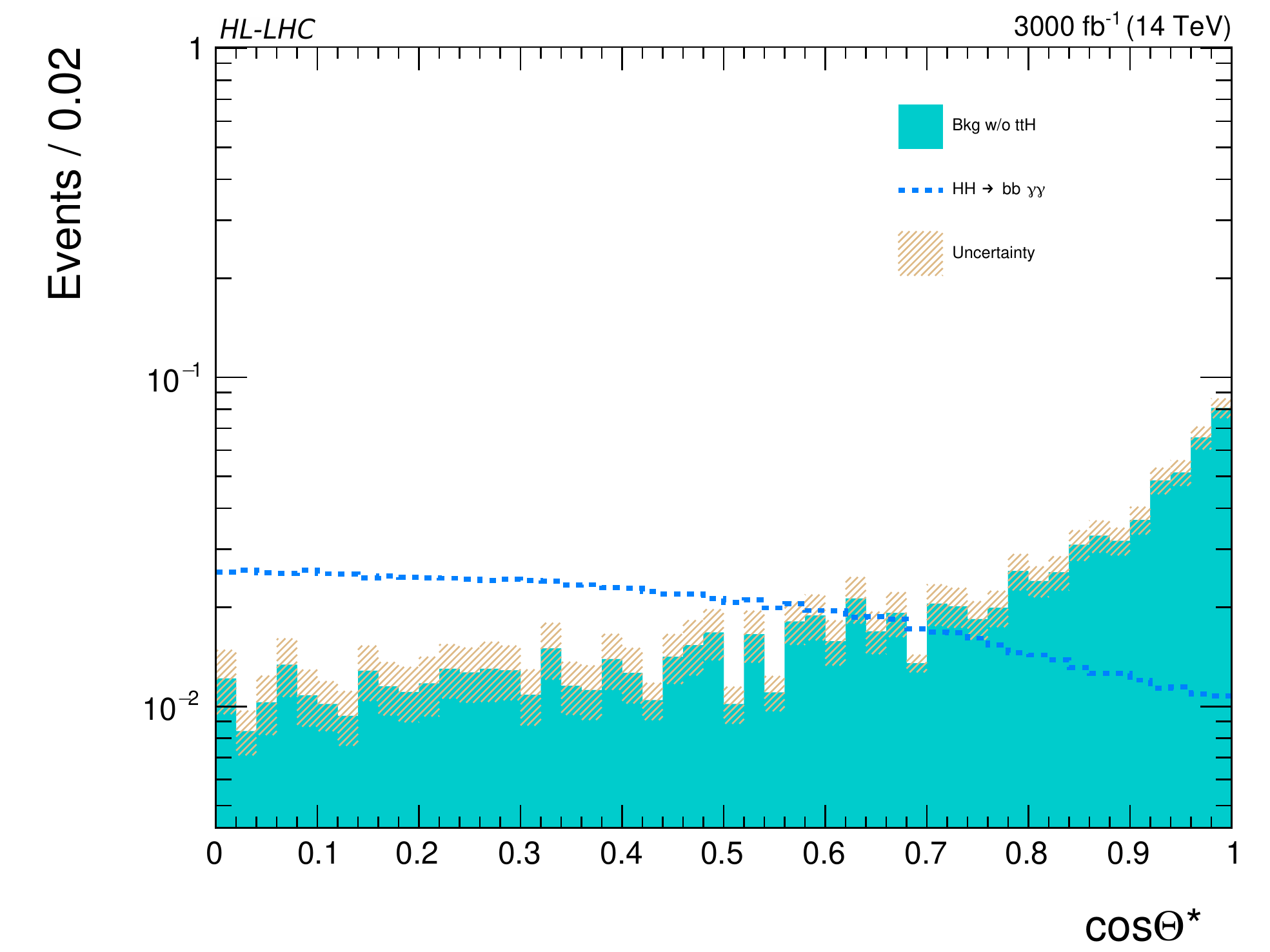}
   \end{subfigure}
   \caption{(Left) Minimum $\Delta R$ between the selected photons and jets and (Right) polar angle of the diphoton system in the diphoton-dijet rest frame, for the HH signal and all background processes combined except ttH. Histograms are normalized to unit area to compare the shapes.  }
   \label{pic:dnn_features}
\end{figure}

A preliminary event categorization is imposed according to the $Mx$ variable, defined as $Mx = m_{\gamma\gamma jj}-m_{\gamma\gamma}-m_{jj}+250 \; GeV$, that is, in a first approximation, the invariant mass of the diphoton and dijet system cleaned out of the jet and photon energy resolution dependence. Figure~\ref{pic:Mx} shows the distribution of this variable for the SM HH signal, two different scenarios with $\kappa_{\lambda}$ values of 2.45, and 5 and for the background. In the SM scenario, the $Mx <350$ GeV region is characterized by a very small S/B ratio, while in BSM scenarios that region becomes much more populated by the signal. For this reason, the value of 350 GeV in Mx is a natural boundary for the event categorization.
Two separate DNN (one per each region of Mx) have been trained to discriminate the signal against all other background processes except ttH. The use of two different discriminators is needed because of the change in kinematic between the two Mx regions. The same variables described in Section \ref{sec:ttH} are exploited, with the addition of the number of loose, medium and tight b-tagged jets.
\\Two examples of variable distribution are reported in Figure \ref{pic:dnn_features}: on the left, the minimum $\Delta R$ separation between a selected photon and a selected jet shows that the objects are on average more well resolved in signal events with respect to background ones, while, on the right, the polar angle of the diphoton system in the diphoton-dijet rest frame is reported.

\begin{figure}[h!]
    \centering
    \begin{subfigure}[t]{0.45\textwidth}
    	\centering
        \includegraphics[width=0.99\textwidth]{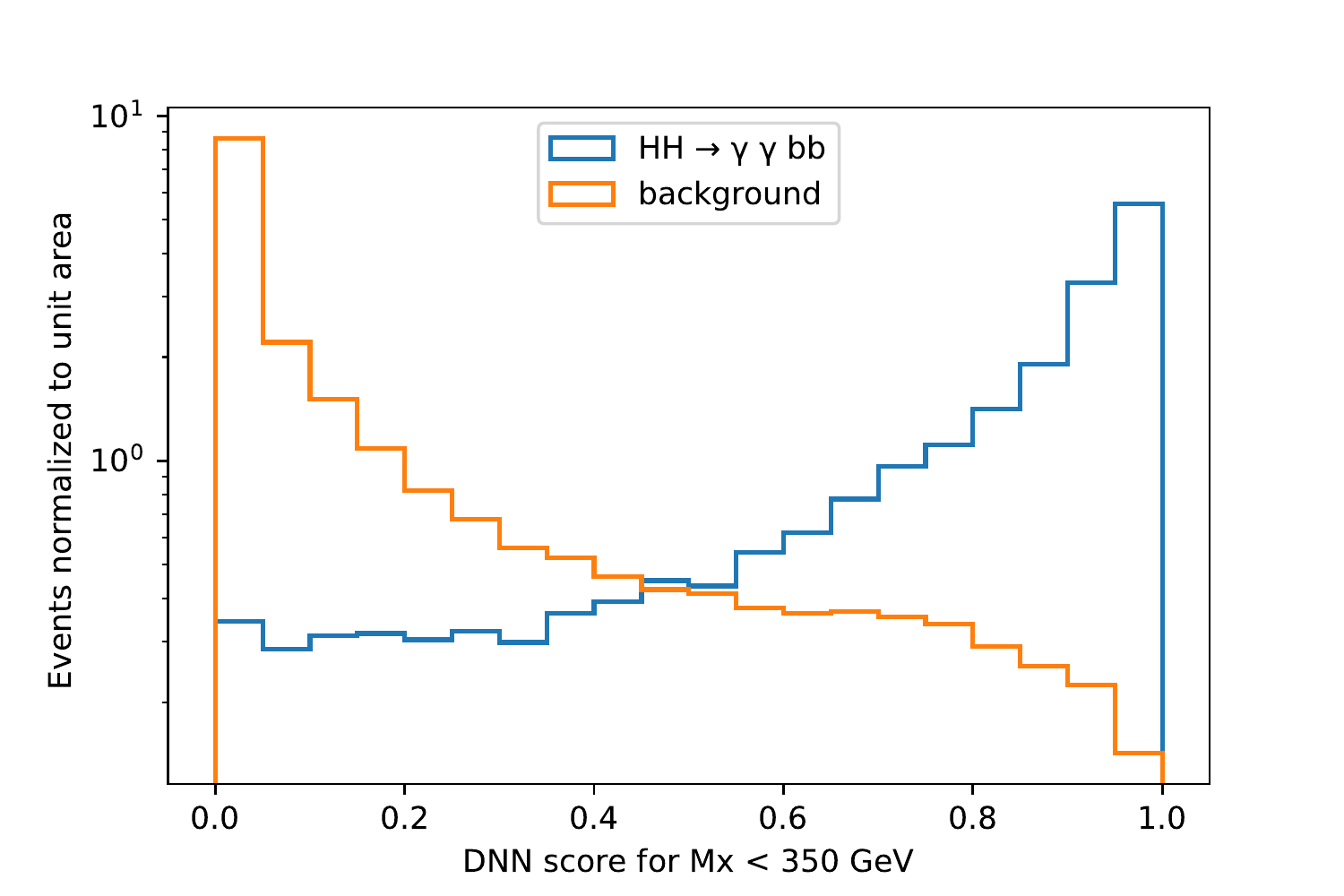}
    \end{subfigure}
    \begin{subfigure}[t]{0.45\textwidth}
       \centering
       \includegraphics[width=0.99\textwidth]{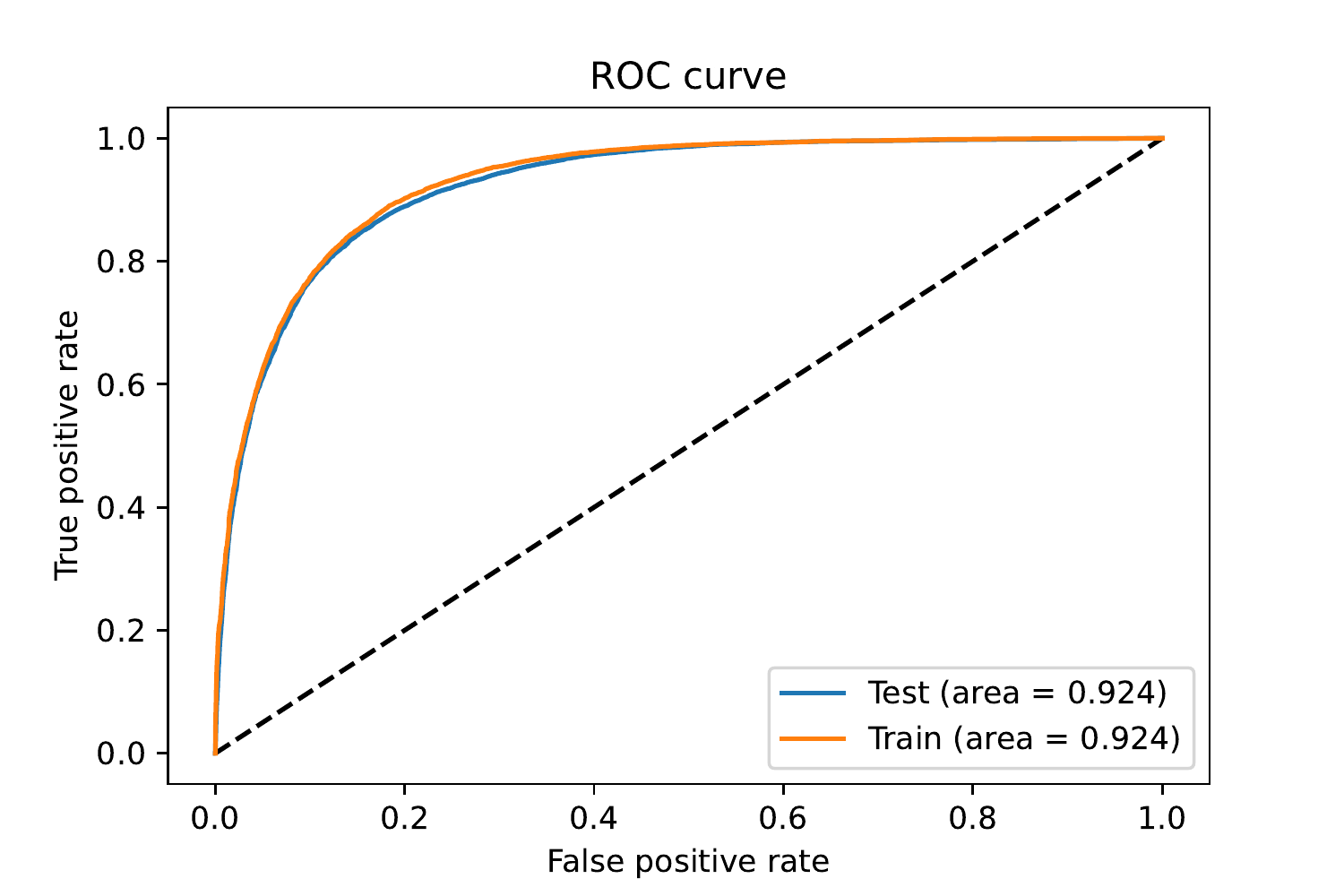}
   \end{subfigure}
   \caption{(Left) DNN score for signal and backgrounds (except ttH) in the Mx  <350 GeV region. (Right) ROC curve for the training and testing samples.}
   \label{pic:dnn_plot_small}
\end{figure}

\begin{figure}[h!]
    \centering
    \begin{subfigure}[t]{0.45\textwidth}
    	\centering
        \includegraphics[width=0.99\textwidth]{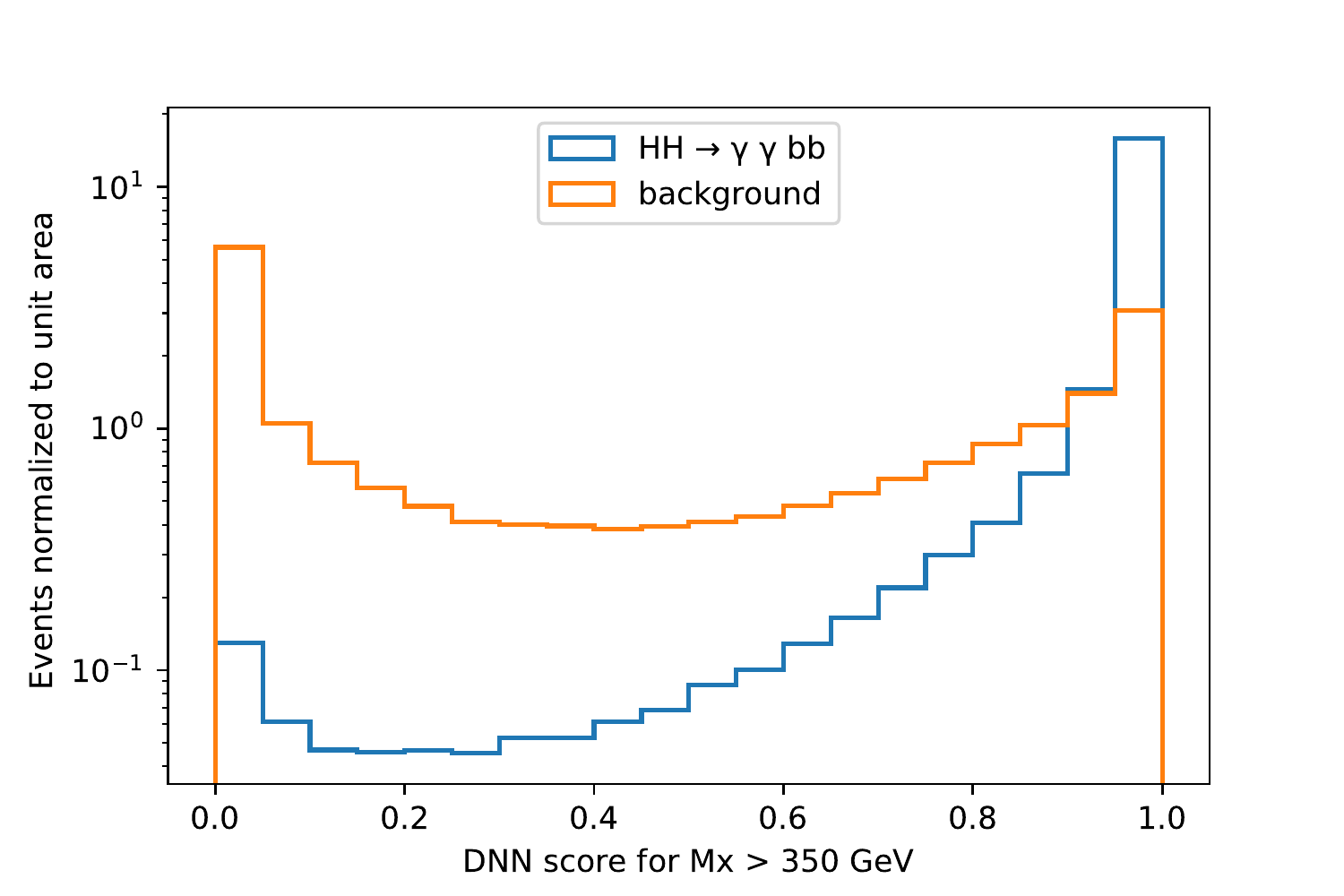}
    \end{subfigure}
    \begin{subfigure}[t]{0.45\textwidth}
       \centering
       \includegraphics[width=0.99\textwidth]{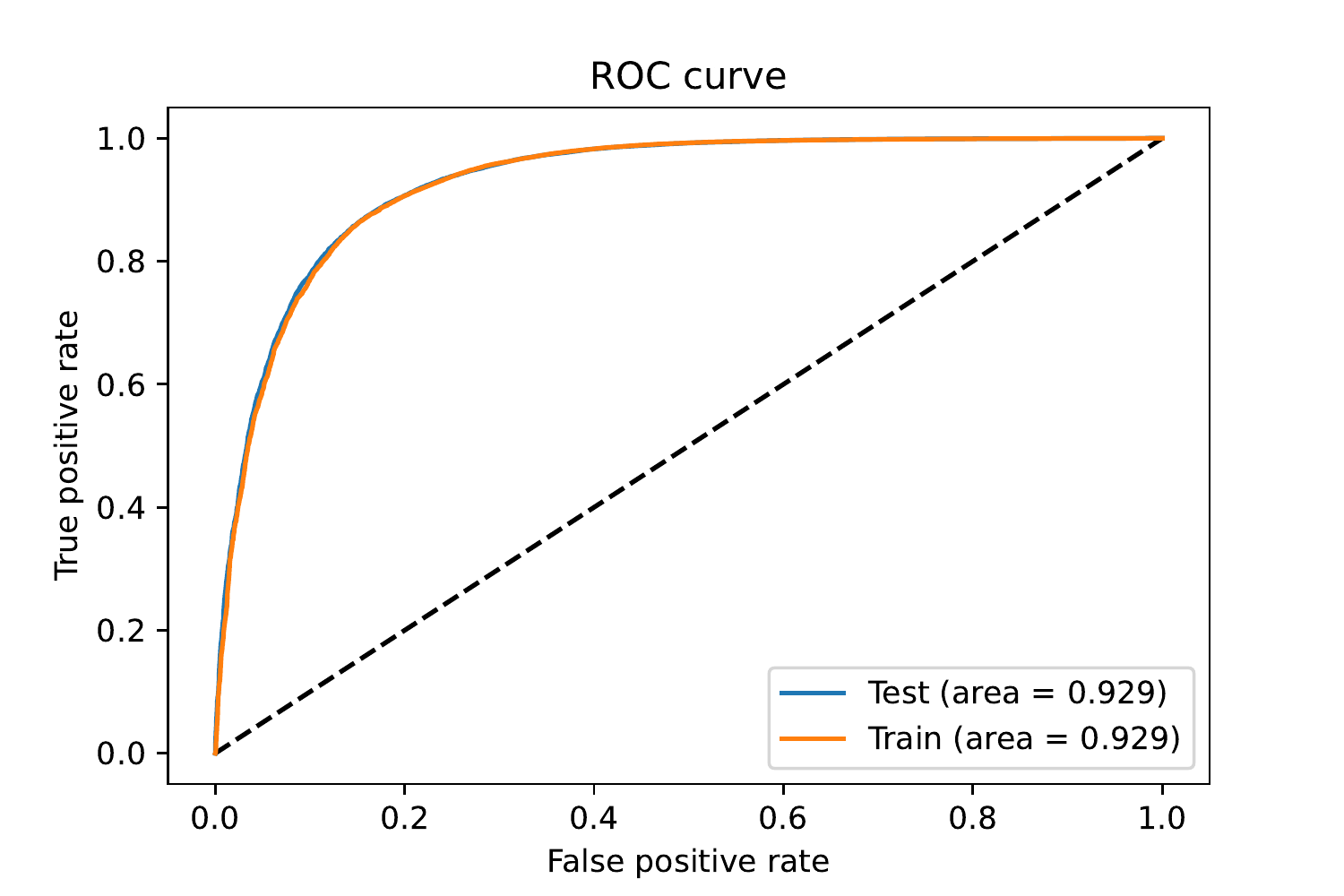}
   \end{subfigure}
   \caption{(Left) DNN score for signal and backgrounds (except ttH) in the Mx >350 GeV region. (Right) ROC curve for the training and testing samples.}
   \label{pic:dnn_plot_great}
\end{figure}

\begin{figure}[h!]
    \centering
    \begin{subfigure}[t]{0.45\textwidth}
    	\centering
        \includegraphics[width=0.99\textwidth]{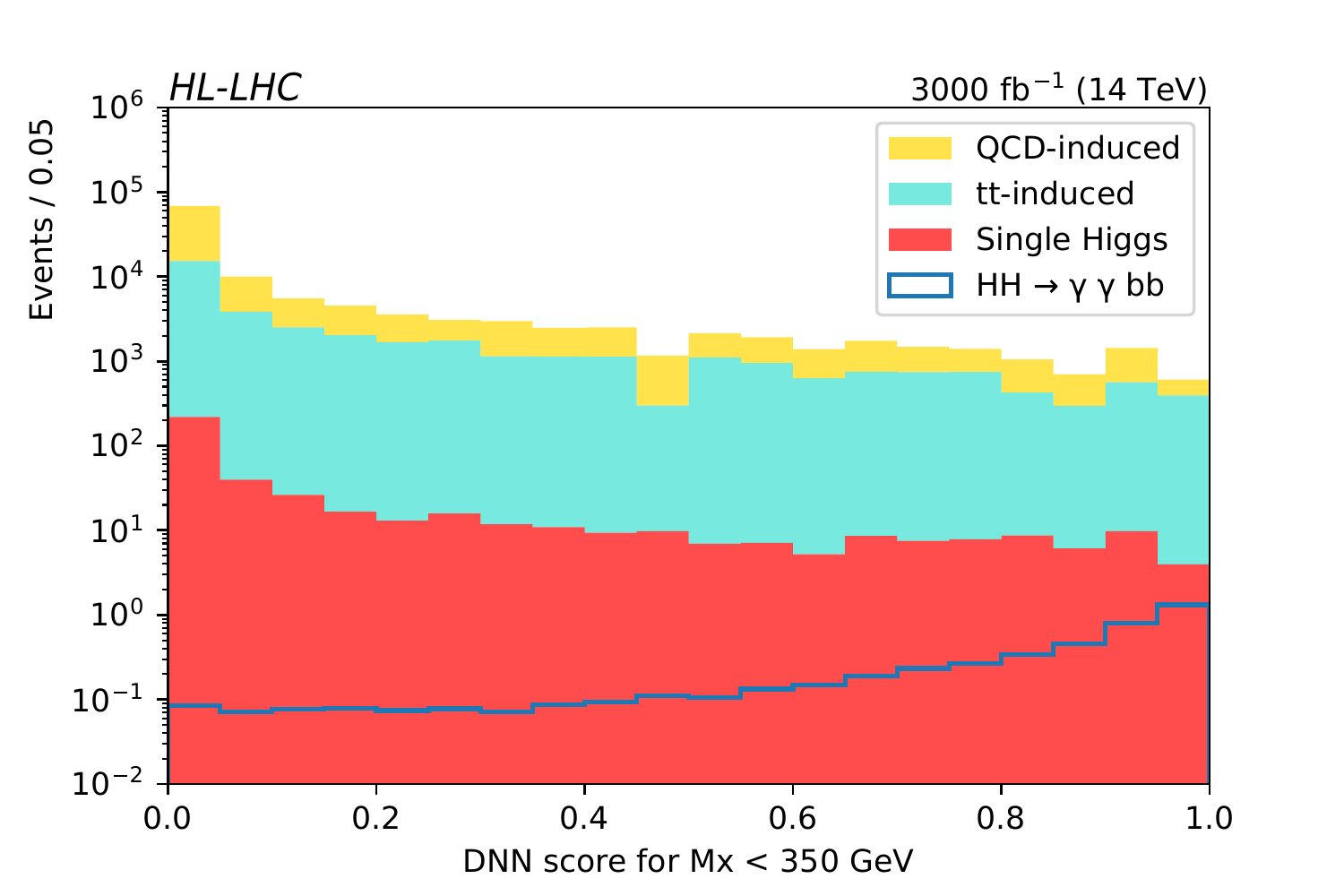}
    \end{subfigure}
    \begin{subfigure}[t]{0.45\textwidth}
       \centering
       \includegraphics[width=0.99\textwidth]{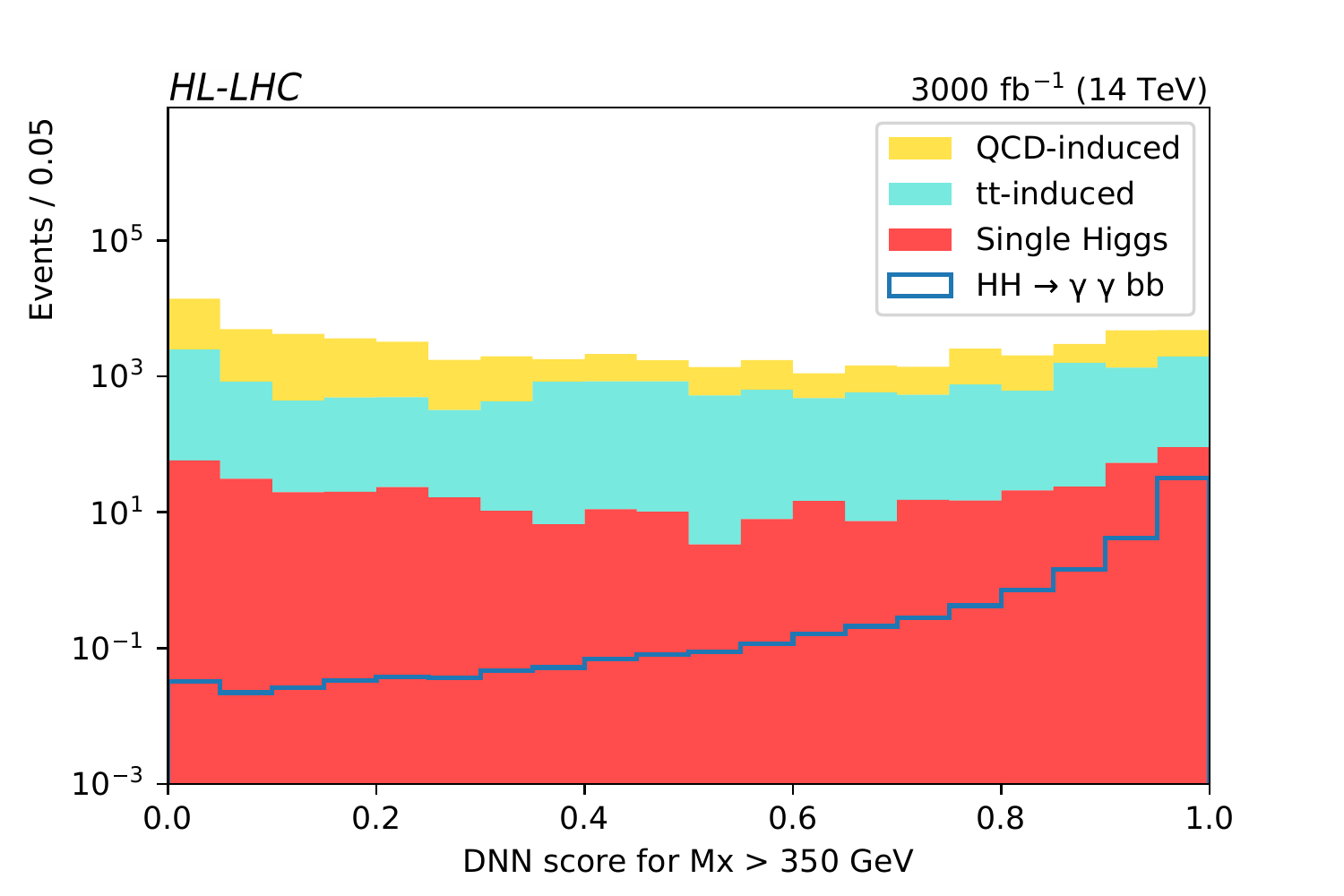}
   \end{subfigure}
   \caption{(Left) DNN score for the M< 350 GeV region.(Right) DNN score for the M> 350 GeV region. Histograms are scaled to cross section and luminosity. The background processes are stacked. }
   \label{pic:dnn}
\end{figure}

\noindent Figures \ref{pic:dnn_plot_small} and \ref{pic:dnn_plot_great} show the DNN score for signal and backgrounds in the two Mx regions, together with their ROC curves. Both networks show good performance in separating the signal topology from the background one. The distributions scaled to cross section and luminosity and considering all backgrounds stacked are reported in Figure \ref{pic:dnn}.

\begin{figure}[h!]
    \centering
    \begin{subfigure}[t]{0.45\textwidth}
    	\centering
        \includegraphics[width=0.99\textwidth]{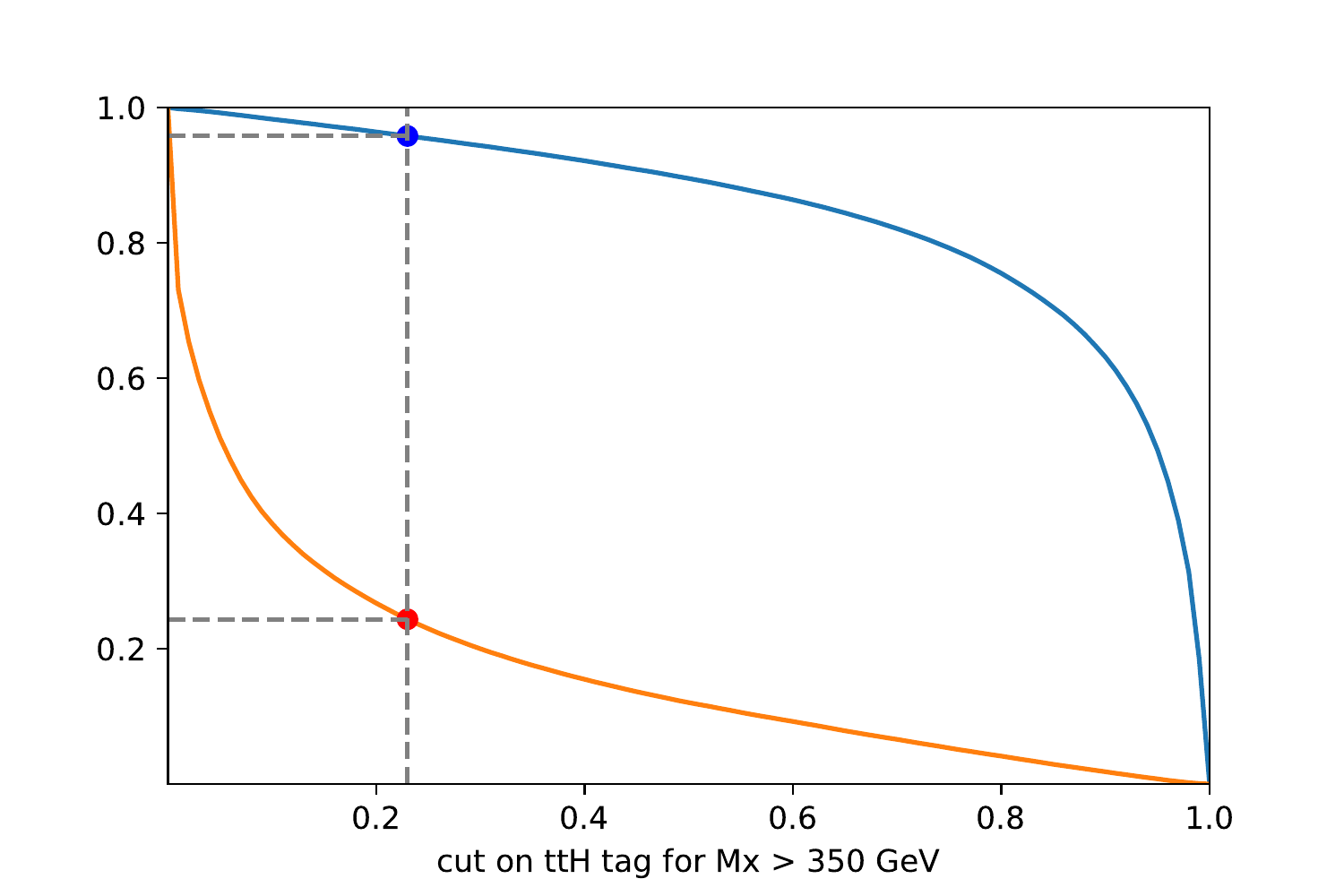}
    \end{subfigure}
    \begin{subfigure}[t]{0.45\textwidth}
       \centering
       \includegraphics[width=0.99\textwidth]{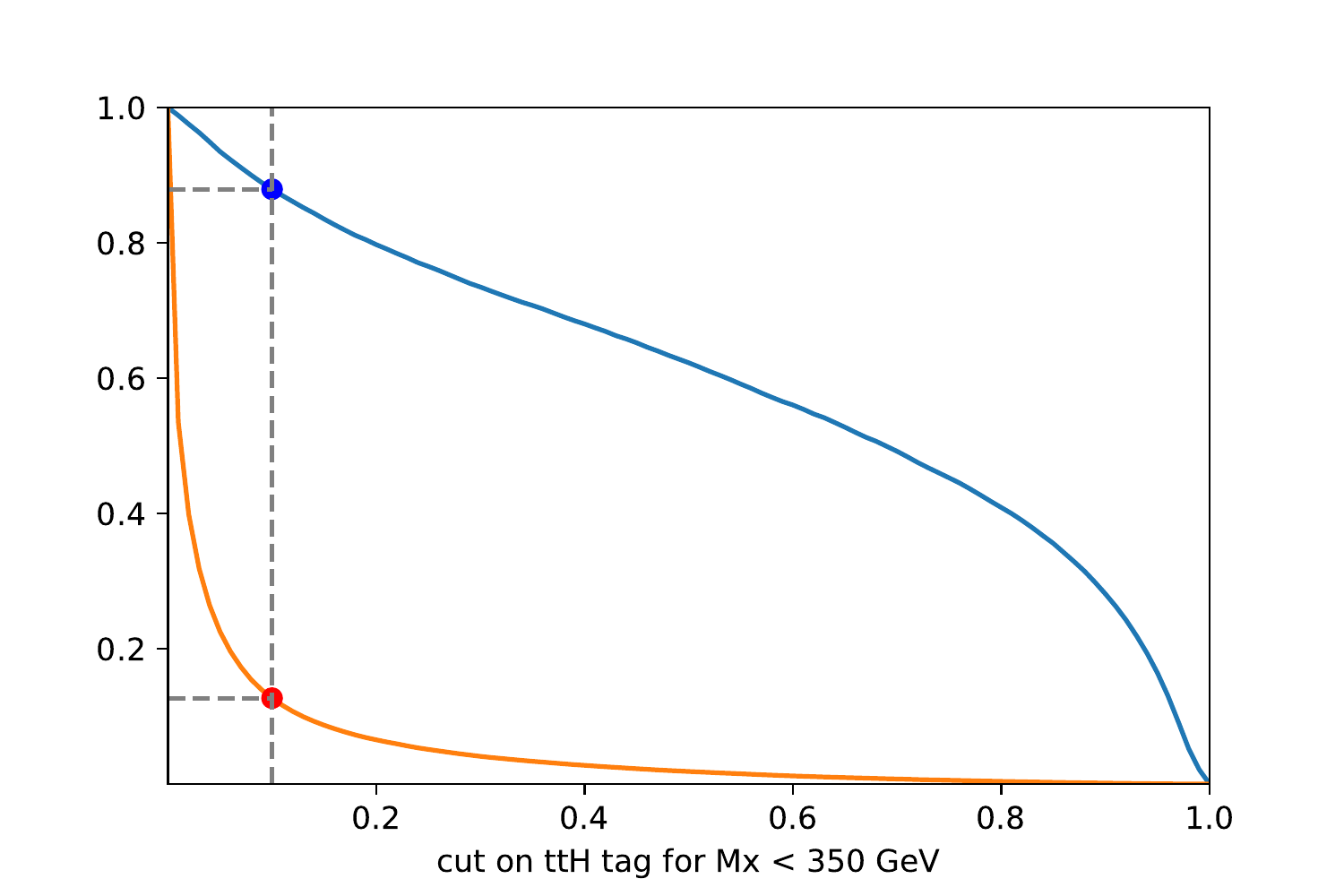}
   \end{subfigure}
   \caption{Signal efficiency in blue and background contamination in orange as a function of the cut on the ttH tagger score for the high mass region (Left) and low mass region (Right).}
   \label{pic:eff_vs_cut}
\end{figure}

\noindent The events are further categorized according to the DNN score, defining two different purity region: a medium purity region with score within $[x_{1}, x_{2}]$ and a high purity region with score greater then $x_{2}$. The values of $x_{1}$ and $x_{2}$ are chosen independently for the two different mass category and are optimized together with the cut on the ttH tagger score, with the aim of increasing the signal strength. Table \ref{tab:opt_cut} reports the employed values, while Figure \ref{pic:eff_vs_cut} reports the efficiency curves for signal and ttH background, underlying the working points chosen for the tagger. These working points corresponds to 95\% and 90\% signal efficiency for high mass and low mass region, respectively. 

\begin{table}[h!]
	\centering
    \begin{tabular}{c|c |c|c } 
    \hline
    &&&\\
    &  \textbf{ttH tag cut} & $ \boldsymbol{x_{1}}$ & $ \boldsymbol{x_{2}}$  \\
    &&&\\
    \hline
    &&&\\
    \textbf{Mx < 350 GeV} & 0.10 & 0.70 & 0.90  \\
    \textbf{Mx > 350 GeV} & 0.23 & 0.05 & 0.95  \\
    &&&\\
    \hline
	\end{tabular}
	\caption{Chosen values for the ttH tagger cut and $x_{1}$ and $x_{2}$ which define the medium and high purity region.}. 
	\label{tab:opt_cut}
\end{table}

\begin{figure}[t]{}
    	\centering
        \includegraphics[width=0.5\textwidth]{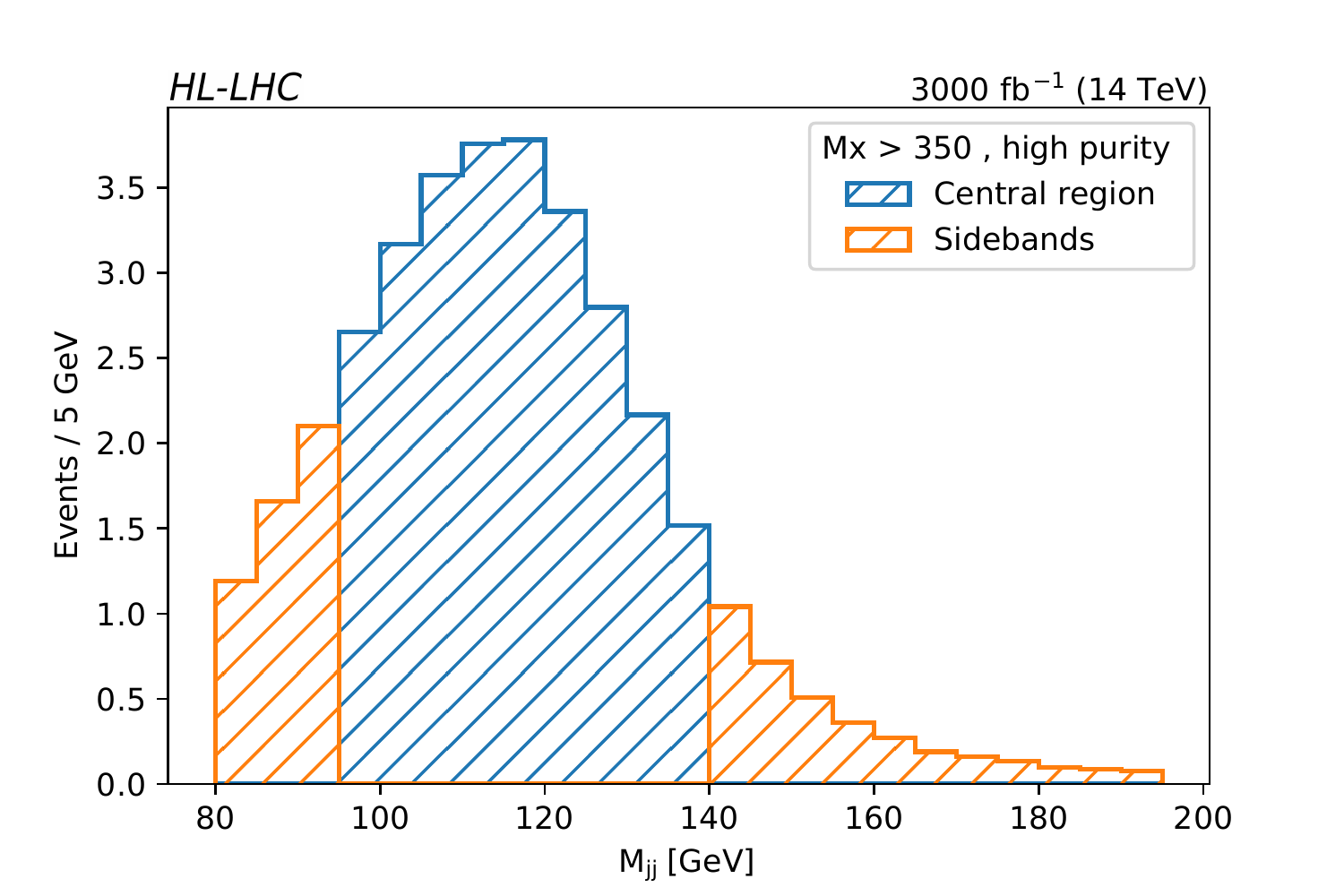}
        \caption{dijet invariant mass distribution for the high mass, high purity category, showing the division in central region and sidebands. The histogram is scaled to cross section and luminosity.}
   \label{pic:m_jj_split}
    \end{figure}

\noindent A final split is made according to the dijet invariant mass: a central region and sidebands are defined as in Figure \ref{pic:m_jj_split}. The exact thresholds are chosen such that maximize the signal to noise ratio.  

\subsection{Statistical analysis and results}
\begin{figure}[h!]
    \centering
    \begin{subfigure}[t]{0.45\textwidth}
    	\centering
        \includegraphics[width=0.99\textwidth]{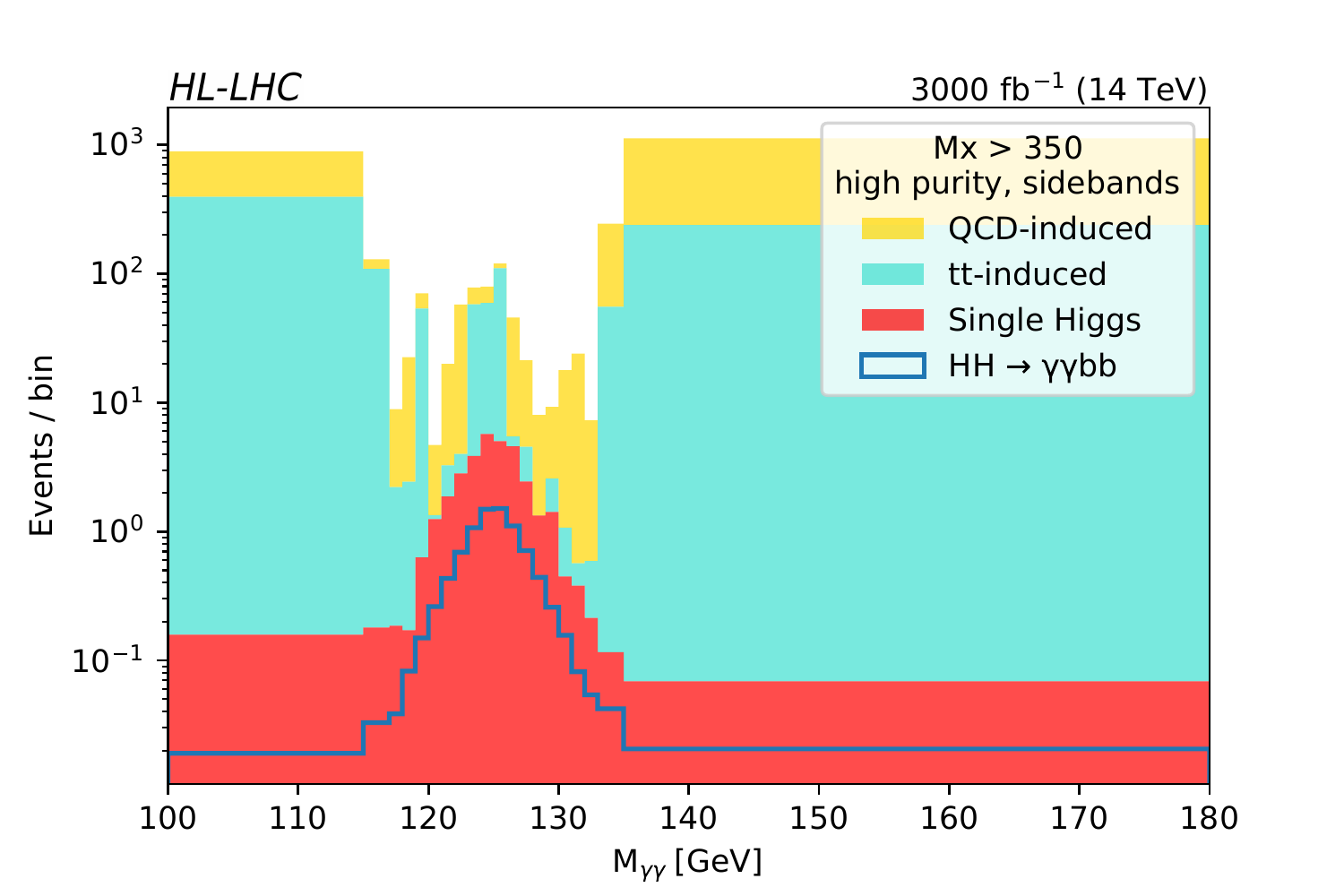}
    \end{subfigure}
    \begin{subfigure}[t]{0.45\textwidth}
       \centering
       \includegraphics[width=0.99\textwidth]{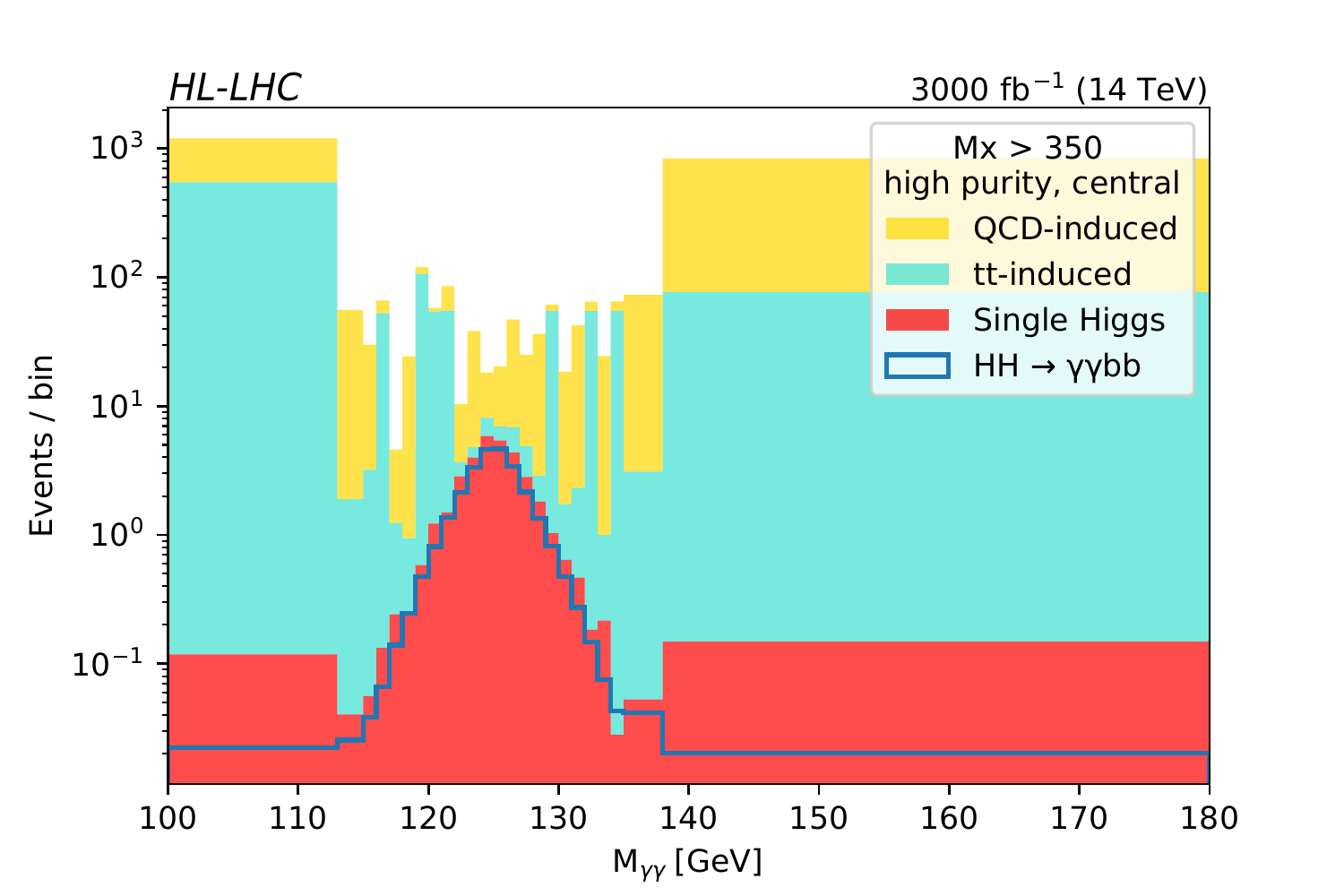}
   \end{subfigure}
   \caption{diphoton invariant mass for high mass, high purity, sidebands (Left) central (Right) categories.}
   \label{pic:inv_mass_binned}
\end{figure}

The purpose of the statistical inference in this context is to determine what is the likelihood that the analysed sample contains signal or whether it can be described by the background-only hypothesis.
If the significance of the signal process is not enough to claim its existence, then its study can be used to place an upper limit on the rate of production of this process. That's indeed the case expected for the double Higgs production at HL-LHC.  
\\In this projection study, the expected discovery significance and cross section upper limits at 95\% confidence level are determined by considering the diphoton invariant mass distribution for each of the eight defined category. Histograms (examples in Figure \ref{pic:inv_mass_binned}) are built with variable bin width in order to better capture the shape of the distributions but, at the same time, not causing the statistical uncertainties in each bin to become too large (the relative uncertainty in each bin for the signal is required to be less then 10\%).

\begin{figure}[h!]
    	\centering
        \includegraphics[width=0.5\textwidth]{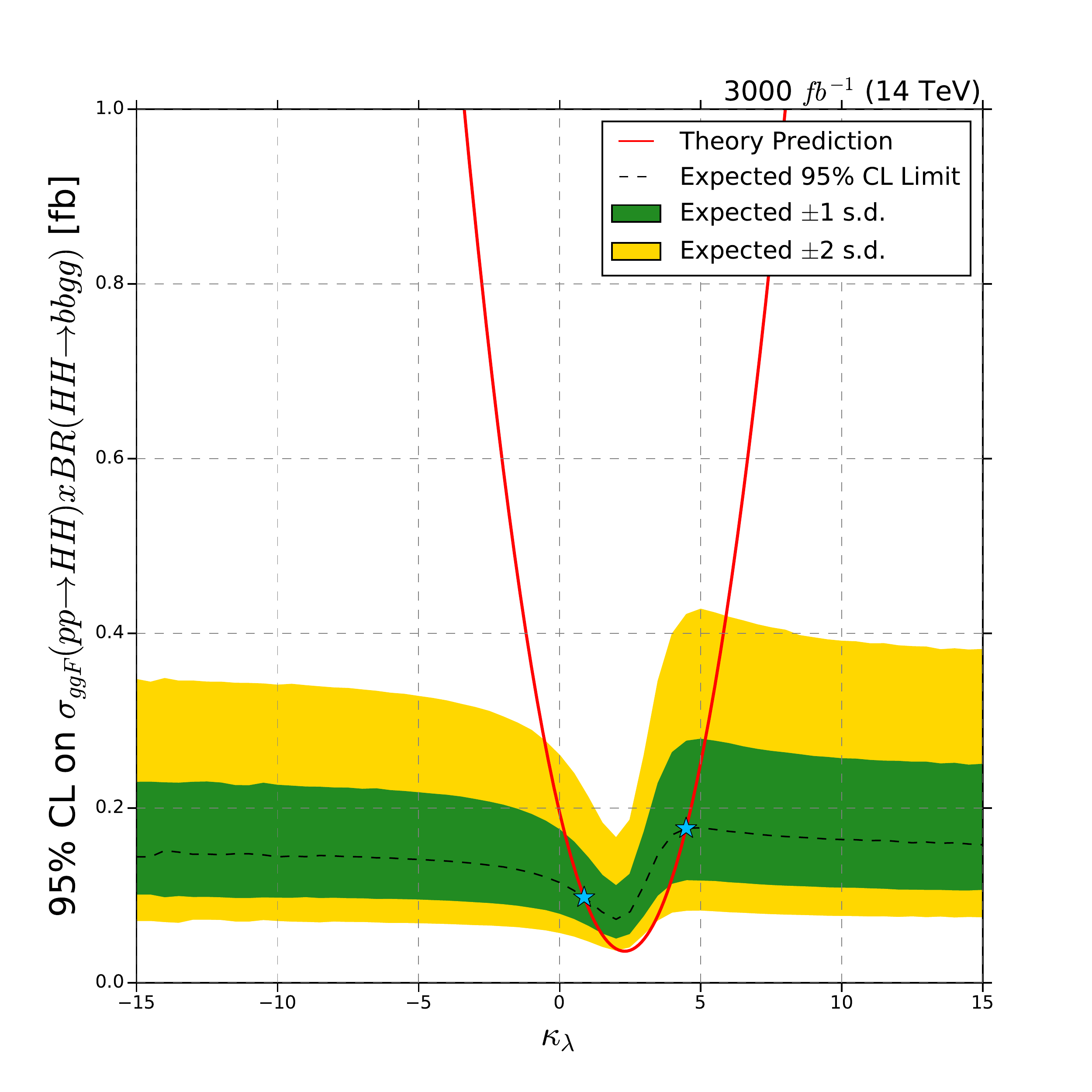}
        \caption{Expected upper limit at the 95\% CL on the HH production cross section as a function of $\kappa_{\lambda}$ with $1\sigma$ and $2\sigma$ bands. The red curve indicates the theoretical prediction.}
   \label{pic:scan_kl_bbgg}
    \end{figure}

\noindent A simultaneous fit is performed on the expected event distributions for the eight considered categories. The systematic uncertainties taken into account are reported in Table \ref{tab:sys_bbgg} and are drawn from Ref. \cite{sys}. 
\\Expected results in terms of 95\% CL upper limits and HH signal significance are reported in Table \ref{tab:results_bbgg} with and without systematic uncertainties. In the worst considered scenario, an upper limit on the HH cross section times branching fraction of 1.09 times the SM prediction is obtained, corresponding to a significance of 1.94.
\\Prospects for the measurement of the trilinear coupling are also studied. Under the assumption that no HH signal exists, 95\% CL upper limits on the SM HH production cross section are derived as a function of $\kappa_{\lambda}$ as visible in Figure \ref{pic:scan_kl_bbgg}. A variation of the excluded cross section, directly related to changes in the HH kinematic properties, can be observed as a function of $\kappa_{\lambda}$. The intersection between the expected curve and the theory prediction is used to estimate the constraint on the anomalous coupling.

\begin{table}[h!]
	\centering
    \begin{tabular}{c|c } 
    \hline
     \textbf{Systematic uncertainty source} & \textbf{Impact on yields} \\
    \hline
    Luminosity & $\pm$ 1.0 \% \\
    $m_{\gamma\gamma}$ scale & $\pm$ 0.5 \% \\
    Photon energy scale & $\pm$ 2.0 \% \\
    Diphoton trigger  & $\pm$ 2.0\% \\
    Photon ID efficiency  & $\pm$ 1.0 \%\\
    Jet Energy Scale  & $\pm$ 1.0\% \\
    B-tag efficiency & $\pm$ 1.0\% \\
	\hline
	\multirow{5}{*}{QCD scale}
    & +4.6\% / -6.7\% (ggH) \\
    & +0.4\% / -0.7\% (VH)\\
    & +0.5\% / -0.3\% (VBFH)\\
    & +6.0\% / -9.2\% (ttH)\\
    & +2.4\% / -3.6\% (tt)\\
    \hline
    \multirow{5}{*}{Pdf scale}
    & $\pm$3.2\% (ggH) \\
    & $\pm$1.8\% (VH)\\
    & $\pm$2.1\% (VBFH)\\
    & $\pm$3.5\% (ttH) \\
    & $\pm$4.2\% (tt) \\
    \hline
    \multirow{3}{*}{{\makecell[c]{Signal theoretical \\ uncertainties}}}
    & +2.1\% / -4.9\% (QCD scale)\\
    & $\pm$3.0\% (pdf scale)\\
    & +4.0\% / -18.0\% (top mass)\\
    \hline
	\end{tabular}
	\caption{Systematic uncertainties for $bb\gamma\gamma$ channel.}. 
	\label{tab:sys_bbgg}
\end{table}

 \begin{table}[h!]
	\centering
    \begin{tabular}{c|c|c|c } 
    \hline
     
     \textbf{Condition} & \textbf{Significance (in $\sigma$)} & \makecell[c]{\textbf{Upper limit on}\\ \textbf{ $\mu$ at 95\% CL}} & \textbf{$\kappa_{\lambda}$ constraint}\\
    
    \hline
    
    stat only & 1.99 & 0.99 & [1.01,4.36] \\
    
    stat + sys & 1.94 & 1.09 & [0.87, 4.48] \\
     
    \hline
	\end{tabular}
	\caption{Results for $bb\gamma\gamma$ channel.}. 
	\label{tab:results_bbgg}
\end{table}

\clearpage
\section{$\boldsymbol{HH \rightarrow b\bar{b}\tau\tau}$ analysis}

\label{sec:bbtautau}

The second most sensitive channel for the double Higgs production is $HH \rightarrow b\bar{b} \tau\tau$.\\
This channel takes the advantage of a very high branching ratio from the $\tau \tau$ decay (around 6\% second highest) furthermore the presence of light lepton (muons and electrons), coming from the $\tau$ decay, in the final state helps a lot in suppressing the background mainly coming from QCD sources.\\
\noindent Top-quark pair production, Drell-Yan and single Higgs processes with their vastly greater production cross-section, are the background that most contaminate the signal phase space making the $HH \rightarrow b\bar{b} \tau \tau$ search challenging. The use of machine learning will be therefore crucial to improve the statistical power of this analysis.

\subsection{Simulated samples}

Signal samples for SM and BSM hypotheses with different value of the trilinear Higgs boson couplings are generated with different values of $k_\lambda$ (Table~\ref{tab:ev_bbtautau}).
The sources of background contamination considered for this analysis are: tt, Drell-Yan to di-lepton plus jets, production of single top plus a vector boson, di-vector boson production and W+jets. Single Higgs production in association with a vector boson or with $t\bar{t}$ pair is one of the irreducible background for this analysis.\\
Even if the contamination of QCD background is expected to be very small, this process has a relative large production cross section and it can also constitute a non-negligible background; estimating this contamination is not trivial and is left for a future study.\\
The list of background samples with their cross section is reported in Table~\ref{tab:ev_bbtautau}.

 \begin{table}[h!]
	\centering
    \addtolength{\leftskip} {-2.5cm}
    \addtolength{\rightskip}{-3.0cm}
    \begin{tabular}{c|c|c} 
    \hline
    
    &\textbf{Process}  & \makecell[c]{\textbf{Cross section}  (fb)} \\
    
    \hline
    
	\multirow{3}{*}{Signal} & \makecell[l]{$(gg)HH \rightarrow b\Bar{b} \tau\tau$  ($\kappa_{\lambda} =1$)} & $2.68$ \\
	& \makecell[l]{$(gg)HH \rightarrow b\Bar{b} \tau\tau$  ($\kappa_{\lambda} =2.45$)} &  $1.13$ \\
	& \makecell[l]{$(gg)HH \rightarrow b\Bar{b} \tau\tau$  ($\kappa_{\lambda} =5$)} &  $8.18$ \\
	
	\hline
	
     \multirow{11}{*}{\makecell[c]{Single \\ Higgs }}& $(gg)H \rightarrow b\bar{b}$ & $3.18 \times 10^{4}$\\
    & $(gg)H \rightarrow \tau\tau$  &  $ 3.43  \times 10^{3}$\\
    & $ttH \rightarrow b\bar{b}$  &  $ 3.574  \times 10^{2}$\\
    & $ttH \not\rightarrow b\bar{b}$  & $ 2.563 \times 10^{2}$\\
    & $ZH, Z \rightarrow q\bar{q}, H \rightarrow b\bar{b}$  & $ 4.02 \times 10^{2}$\\
    & $ZH, Z \rightarrow ll, H \rightarrow b\bar{b}$  & $ 1.94  \times 10^{1}$\\
    & $W^{+}H, W \rightarrow q\bar{q'}, H \rightarrow b\bar{b}$  & $ 3.62 \times 10^{2}$\\
    & $W^{+}H, W \rightarrow ll, H \rightarrow b\bar{b}$   & $ 6.03\times 10^{1}$\\
    & $W^{-}H, W \rightarrow q\bar{q'}, H \rightarrow b\bar{b}$  &  $ 2.32 \times 10^{2}$\\
    & $W^{-}H, W \rightarrow ll, H \rightarrow b\bar{b}$   & $3.87  \times 10^{1}$\\
    & $VH, H \not\rightarrow b\bar{b}$  & $ 1.46 \times 10^{3}$\\
    
	\hline
	
	\multirow{4}{*}{\makecell[c]{Single \\ Boson}}& $tW$ & $4.506 \times 10^{4}$\\
	& $\bar{t}W$  & $4.502 \times 10^{4}$\\
    & $tZq, Z \rightarrow ll$ & $8.5 \times 10^{1}$ \\
    & $W \rightarrow l\nu +jets$ &  $ 6.052 \times 10^{7}$\\
    
    \hline
	
    \multirow{2}{*}{\makecell[c]{Double \\ Boson}}& $WW$ & $1.31 \times 10^{5}$ \\
    & $ZZ \rightarrow llq\bar{q}$ & $3.721 \times 10^{3}$\\
   
	\hline
	
	 \multirow{7}{*}{Drell-Yan} & \makecell[c]{$DY \rightarrow ll+jets$  HT 100 to 200} &  $1.5 \times 10^{5}$\\
	 & \makecell[c]{$DY \rightarrow ll+jets$  HT 200 to 400} & $3.295 \times 10^{4}$\\
	 & \makecell[c]{$DY \rightarrow ll+jets$  HT 400 to 600} &  $3.911 \times 10^{3}$\\
	 & \makecell[c]{$DY \rightarrow ll+jets$  HT 600 to 800} & $8.301 \times 10^{2}$\\
	 & \makecell[c]{$DY \rightarrow ll+jets$  HT 800 to 1200} &  $3.852 \times 10^{2}$\\
	 & \makecell[c]{$DY \rightarrow ll+jets$  HT 1200 to 2500} &  $8.874 \times 10^{1}$\\
	 & \makecell[c]{$DY \rightarrow ll+jets$  HT 2500 to Inf} &  $1.755$\\

    \hline
    
    tt &  $t\bar{t}$ inclusive& $8.644\times 10^{5}$\\
    
	\hline
	\end{tabular}
	\caption{List of simulated samples for $bb\tau\tau$ channel.}. 
	\label{tab:ev_bbtautau}
\end{table}

\subsection{Event selection}

According to the branching ratio of $H \rightarrow \tau \tau$, six possible scenarios are possible: $\tau_h \mu$, $\tau_h e$, $\tau_h \tau_h$, $\mu e$, $ee$ and $\mu \mu$, where $\tau_h$ indicates the a hadronically decaying tau lepton (a tau jet). Among these six final states, we considered only the final states that involves at least one $\tau_h$.\\
Following the lepton and $\tau_h$ requirements defined in Table \ref{tab:tt_sel}, events are exclusively accepted into the following three categories:

\begin{table}[h!]
	\centering
    \begin{tabular}{cccc } 
    \hline
    Lepton             & Min $p_T$ & Max $\eta$ & Max iso \\
    \hline
    Primary muon       & 23        & 2.1        & 0.15 \\
    Primary electron   & 27        & 2.1        & 0.1 \\
    Veto muon/electron & 10        & 2.4        & 0.3 \\
    \hline
    Hadronic $\tau$    &           &            & \\
    \hline
    $lep$ $\tau_h$     & 20        & 2.3        &  \\
    $\tau_h$ $\tau_h$     & 45        & 2.1       & \\
    \hline
	\end{tabular}
	\caption{Kinematic requirements of leptons and hadronic taus}. 
	\label{tab:tt_sel}
\end{table}

\begin{itemize}
    \item[$\blacksquare$] $\mu \tau_h$ ($e \tau_h$): exactly one primary muon (electron) and at least one $\tau_h$ with opposite charge to the selected muon (electron). If more then one $\mu \tau_h$ ($e \tau_h$) couple pass the selection, the couple with the highest isolation is selected.
    Exactly 0 veto electrons (muons) are requested;
    \item[$\blacksquare$] $\tau_h \tau_h$: exactly zero veto muons or electrons and at least two hadronic taus of opposite charge to one another. In case of multiple choices of hadronic tau, the highest $p_T$ one(s) is/are selected.
\end{itemize}

\noindent To consider the $H \rightarrow b\bar{b}$ decay, events that are accepted into one of the three categories, are required to have at least 2 b-jets at medium working point with $p_T > 30$ GeV and $|\eta| < 2.4$. The jets are also required to be isolated from the leptons: $\Delta R(lept, jet) > 0.5$. The expected yields and efficiency for each process and each category after the final selection are written in Table \ref{tab:ev_cat}.

 \begin{table}[h!]
	\centering
    \begin{tabular}{ c|c|c|c}

    \hline
    
     \textbf{Process} & $\boldsymbol{\tau_{\mu}\tau_{h}}$ & $\boldsymbol{\tau_{e}\tau_{h}}$& $\boldsymbol{\tau_{h}\tau_{h}}$ \\
   
    \hline
    
     \makecell[l]{$HH \rightarrow b\bar{b} \tau\tau$  $\kappa_{\lambda} = 1$} &          $101 \pm 3$ &      $68 \pm 2$ &    $58 \pm 2$ \\
     \makecell[l]{$HH \rightarrow b\bar{b} \tau\tau$  $\kappa_{\lambda} = 2.45$} &          $45 \pm 2$ &    $43 \pm 2$ &    $61 \pm 3$  \\
     \makecell[l]{$HH \rightarrow b\bar{b} \tau\tau$  $\kappa_{\lambda} = 5$} &        $371 \pm 19$ &  $298 \pm 17$ &  $374 \pm 20$ \\
    
     \hline
     
     $ggH, H \rightarrow b\bar{b}$ &        $899 \pm 202$ &      $90 \pm 64$ &     $0.0 \pm 0.0$\\

     $ggH, H \rightarrow \tau \tau$ &         $312 \pm 63$ &    $262 \pm 58$ &     $125 \pm 40$ \\
     
     $ttH, H \rightarrow b\bar{b}$ &       $6499 \pm 168$ &   $3420 \pm 91$ & $365 \pm 134$ \\

     $ttH, H \not\rightarrow b\bar{b}$ &       $4725 \pm 122$ &     $2840 \pm 74$ & $869 \pm 25$  \\
     
     $ZH, H\rightarrow b\bar{b}$ &         $523 \pm 18$ &      $188 \pm 8$ &     $103 \pm 6$  \\
     
     $WH, H\rightarrow b\bar{b}$ &         $642 \pm 27$ &     $170 \pm 11$ &       $14 \pm 3$ \\
     $VH, H \not\rightarrow b\bar{b}$ &         $378 \pm 66$ &     $229 \pm 52$ &      $97 \pm 40$ \\
     
     \hline
     
     $tW$ &    $133035 \pm 3944$ &   $80708 \pm 2614$ &   $4633 \pm 418$  \\
     
     $tZq, Z \rightarrow ll$ &      $626 \pm 24$ &     $409 \pm 18$ &       $116 \pm 8$  \\
     
     $W \rightarrow l\nu + jets$ &            $0.0 \pm 0.0$ &        $0.0 \pm 0.0$ &        $0.0 \pm 0.0$ \\
    
     \hline
     
     $WW$ &       $2374 \pm 323$ &     $1145 \pm 222$ &      $42 \pm 42$  \\
     $ZZ \rightarrow ll q \bar{q}$ &      $1347 \pm 84$ &      $877 \pm 64$ &     $427 \pm 44$\\
     
     \hline
    
     $DY \rightarrow ll + jets$  &     $38341 \pm 1553$ &  $24055 \pm 1141$ &   $11822 \pm 717$ \\
     
     \hline
     
     $t\bar{t}$ &  $3261832 \pm 82589$ &  $1952133 \pm 49842$ &  $91079 \pm 3156.9$ \\
     
    \hline
\end{tabular}
\caption{Yields and selection efficiencies for $bb\tau\tau$ channel.}. 
	\label{tab:ev_cat}
\end{table}

\clearpage

\begin{figure}[h!]
    \centering
    \begin{subfigure}[t]{0.45\textwidth}
    	\centering
        \includegraphics[width=0.99\textwidth]{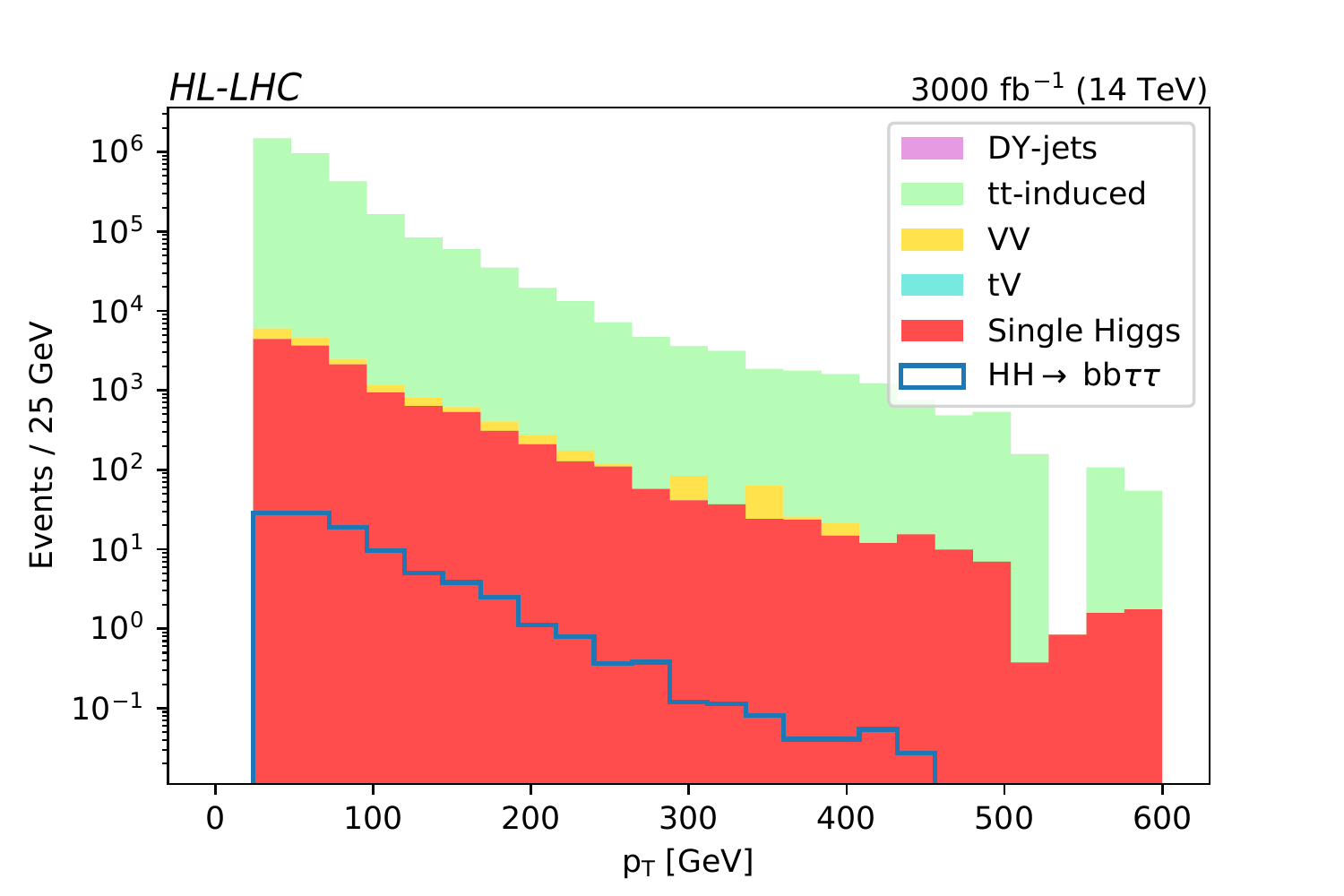}
    \end{subfigure}
    \begin{subfigure}[t]{0.45\textwidth}
    	\centering
        \includegraphics[width=0.99\textwidth]{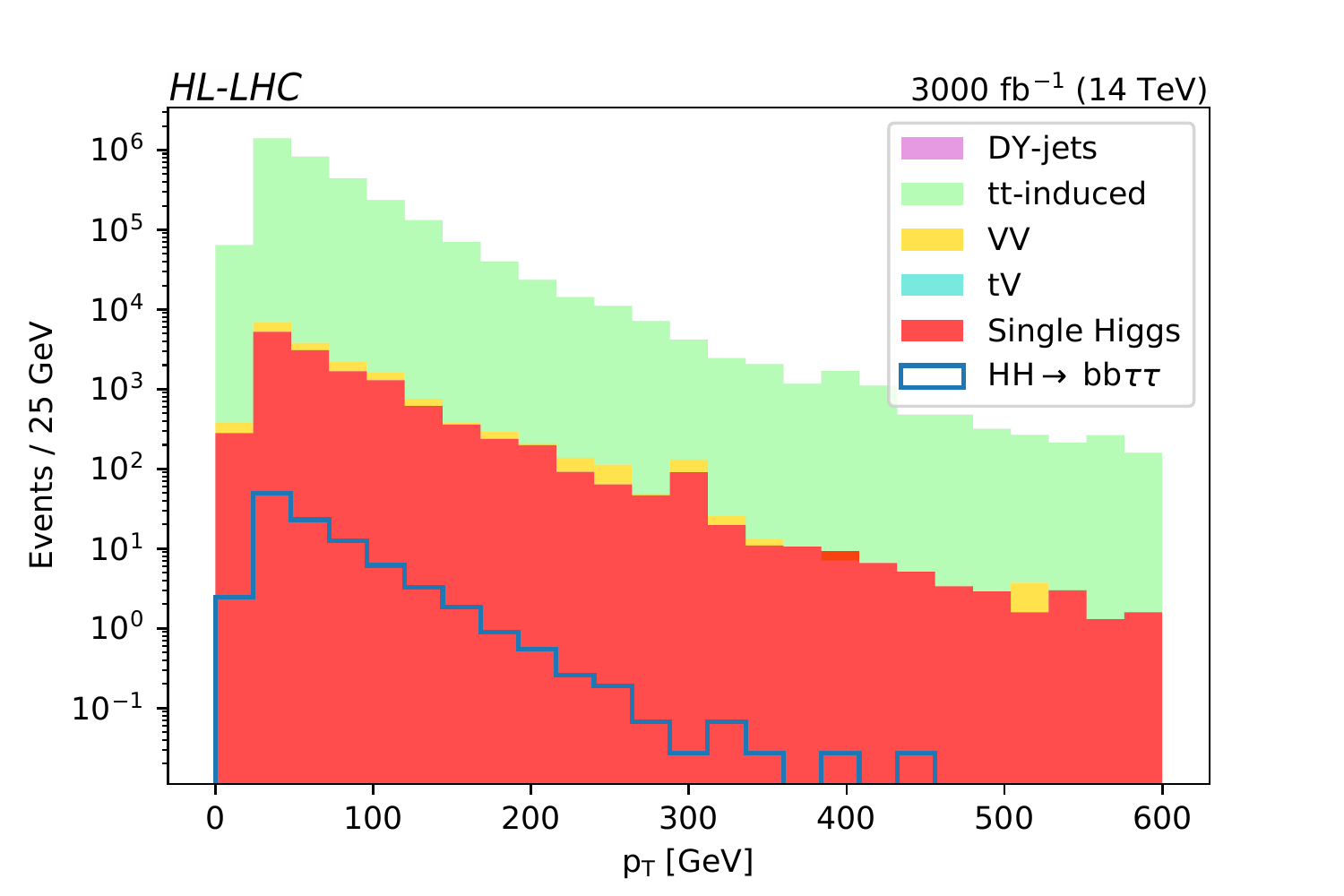}
    \end{subfigure}
    \begin{subfigure}[t]{0.45\textwidth}
    	\centering
        \includegraphics[width=0.99\textwidth]{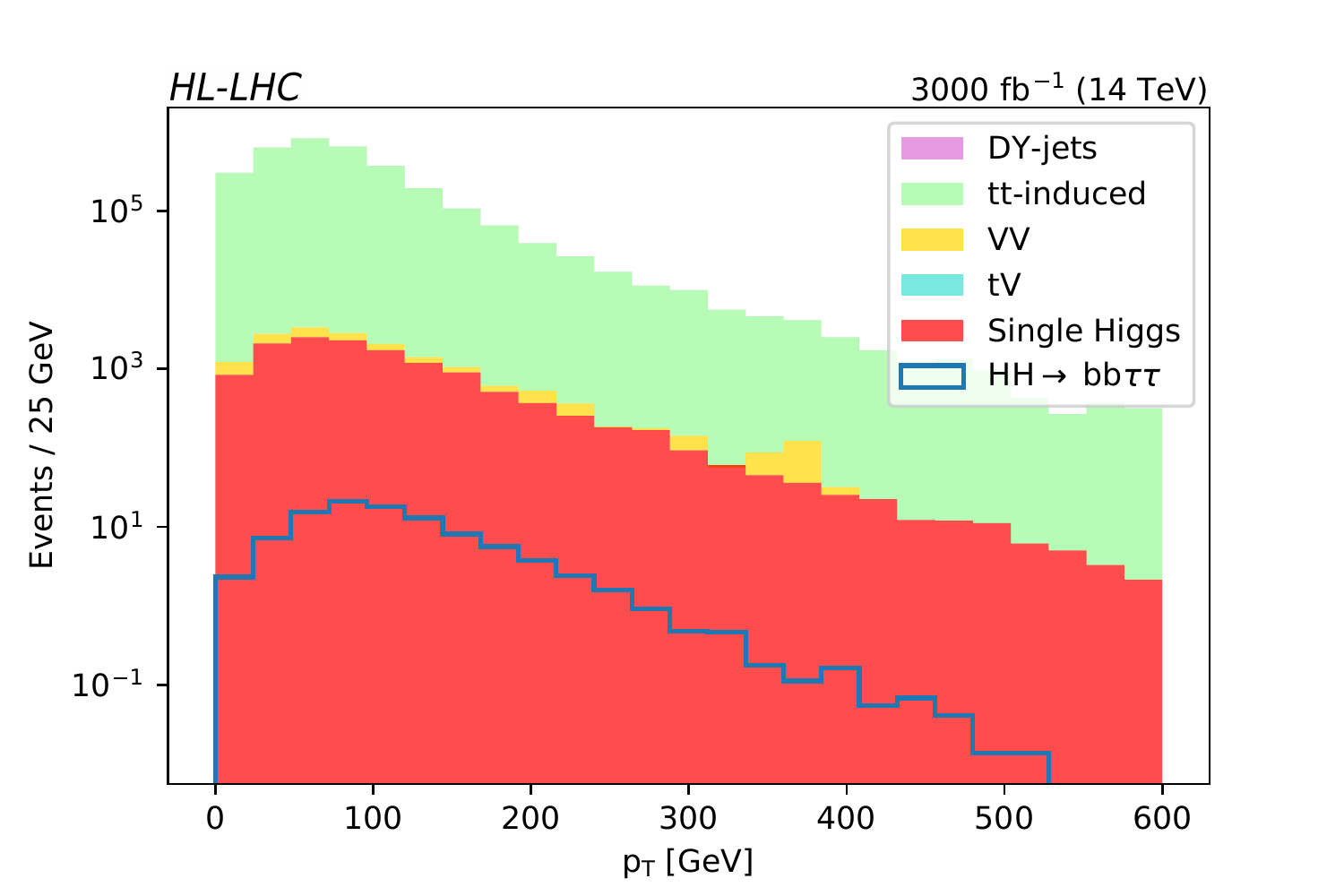}
    \end{subfigure}
    \begin{subfigure}[t]{0.45\textwidth}
    	\centering
        \includegraphics[width=0.99\textwidth]{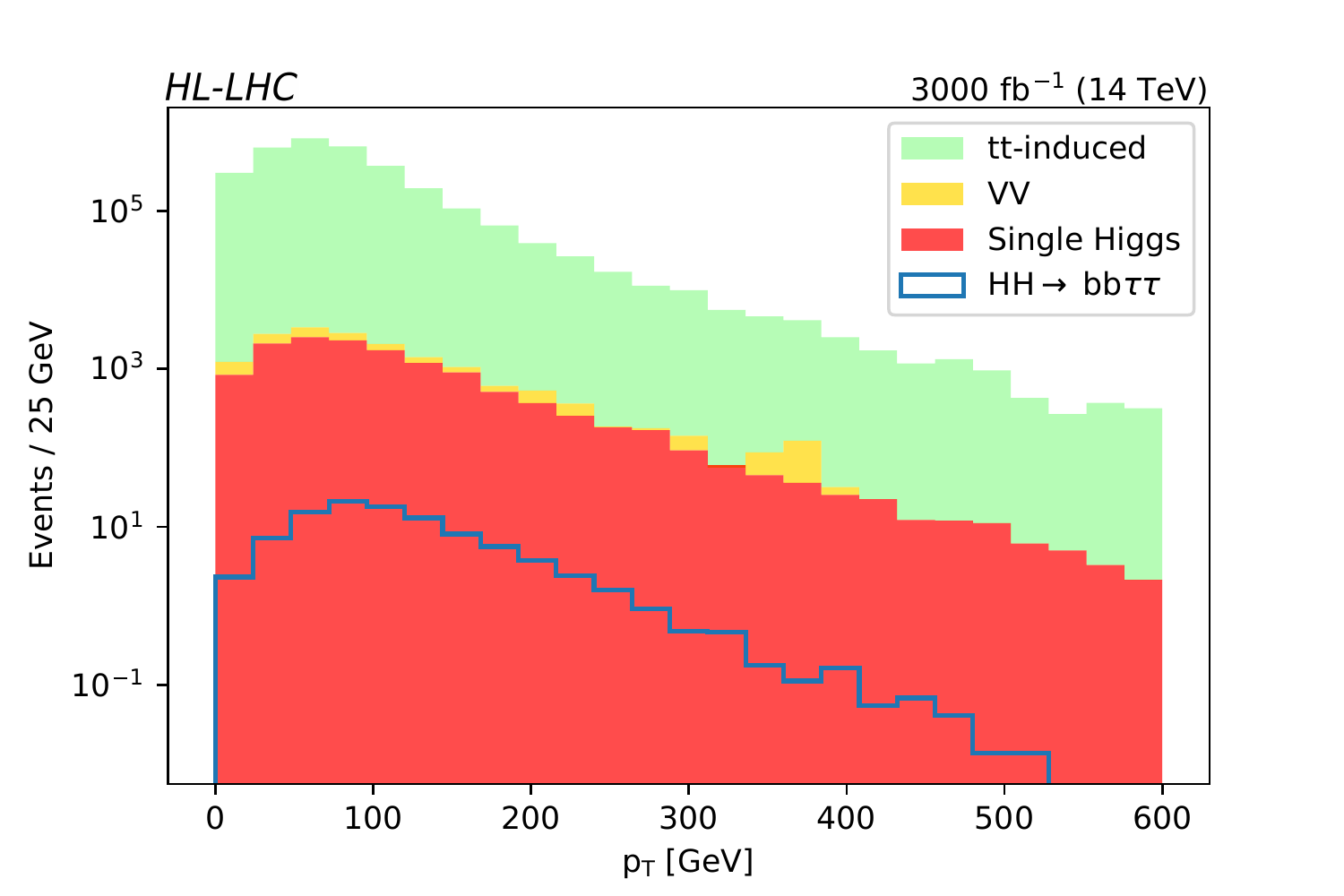}
    \end{subfigure}
   \caption{Transverse momentum distribution of the two selected $\tau$ for the $\mu \tau$ category (up left and right). (bottom) Transverse momentum distribution of the $H \rightarrow \tau\tau$ (left) and $H \rightarrow b\bar{b}$ (right) for the $\mu \tau$ category}
   \label{pic:kin}
\end{figure}

\noindent The two Higgs bosons are reconstructed from the selected final state.\\
For the hadronically decay $\tau$ lepton, the 4-mometum of the $\tau$-tagged jet is used, for the leptonically decay $\tau$ lepton, the 4-mometum of the reconstructed lepton is considered.
The 4-momentum of the Higgs boson is then reconstructed as the vectorial sum of the 4-momentum of the $\tau$ lepton in each of the categories.
Next, $H \rightarrow b\bar{b}$ is reconstructed from the vectorial sum of the two jets with the highest $p_T$ in the event. The di-Higgs object is the vectorial sum of the two Higgs.\\
\noindent Kinematic observables of the system are shown in Figure \ref{pic:kin}. 

\begin{figure}[h!]
    \centering
    \begin{subfigure}[t]{0.45\textwidth}
    	\centering
        \includegraphics[width=0.99\textwidth]{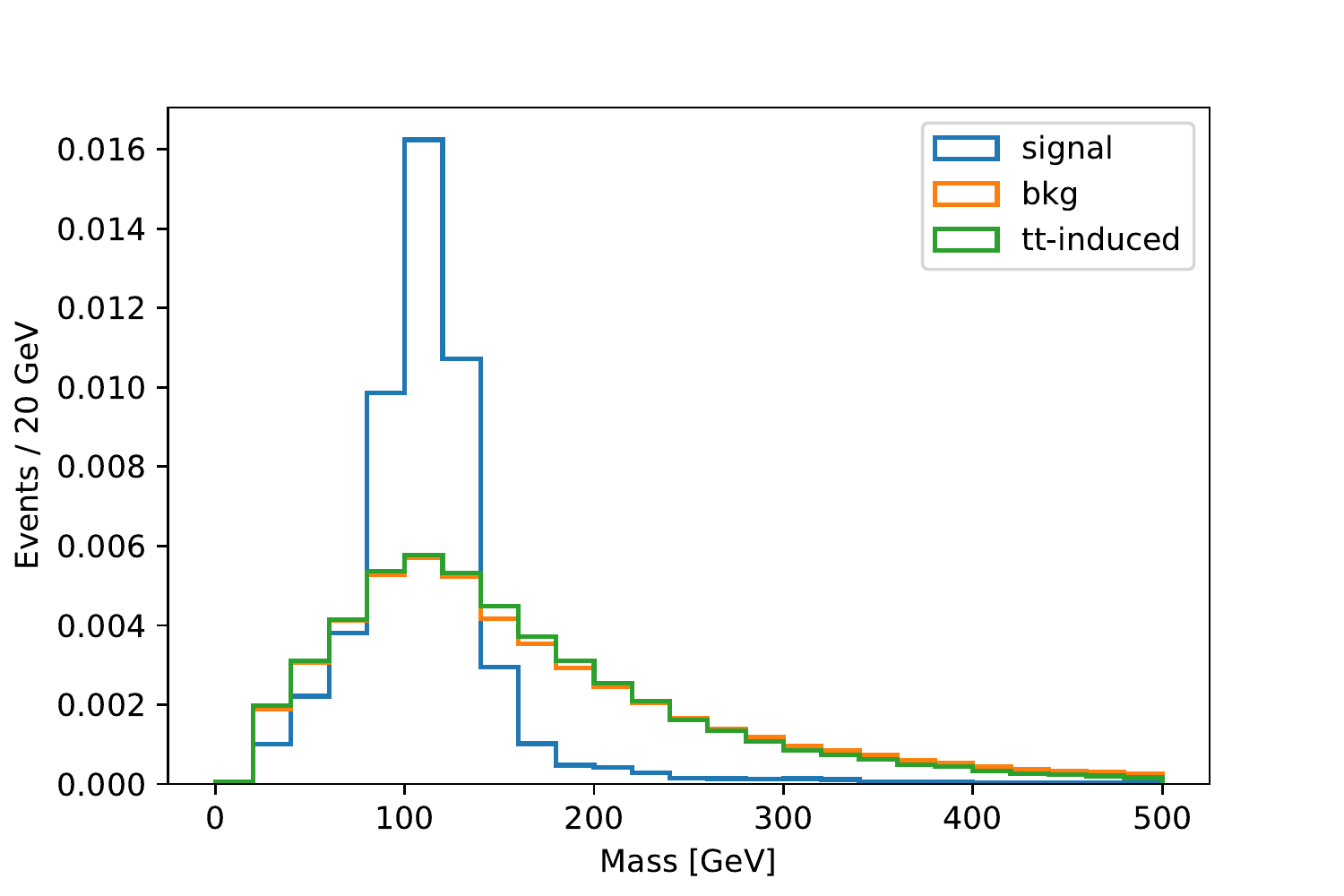}
    \end{subfigure}
    \begin{subfigure}[t]{0.45\textwidth}
    	\centering
        \includegraphics[width=0.99\textwidth]{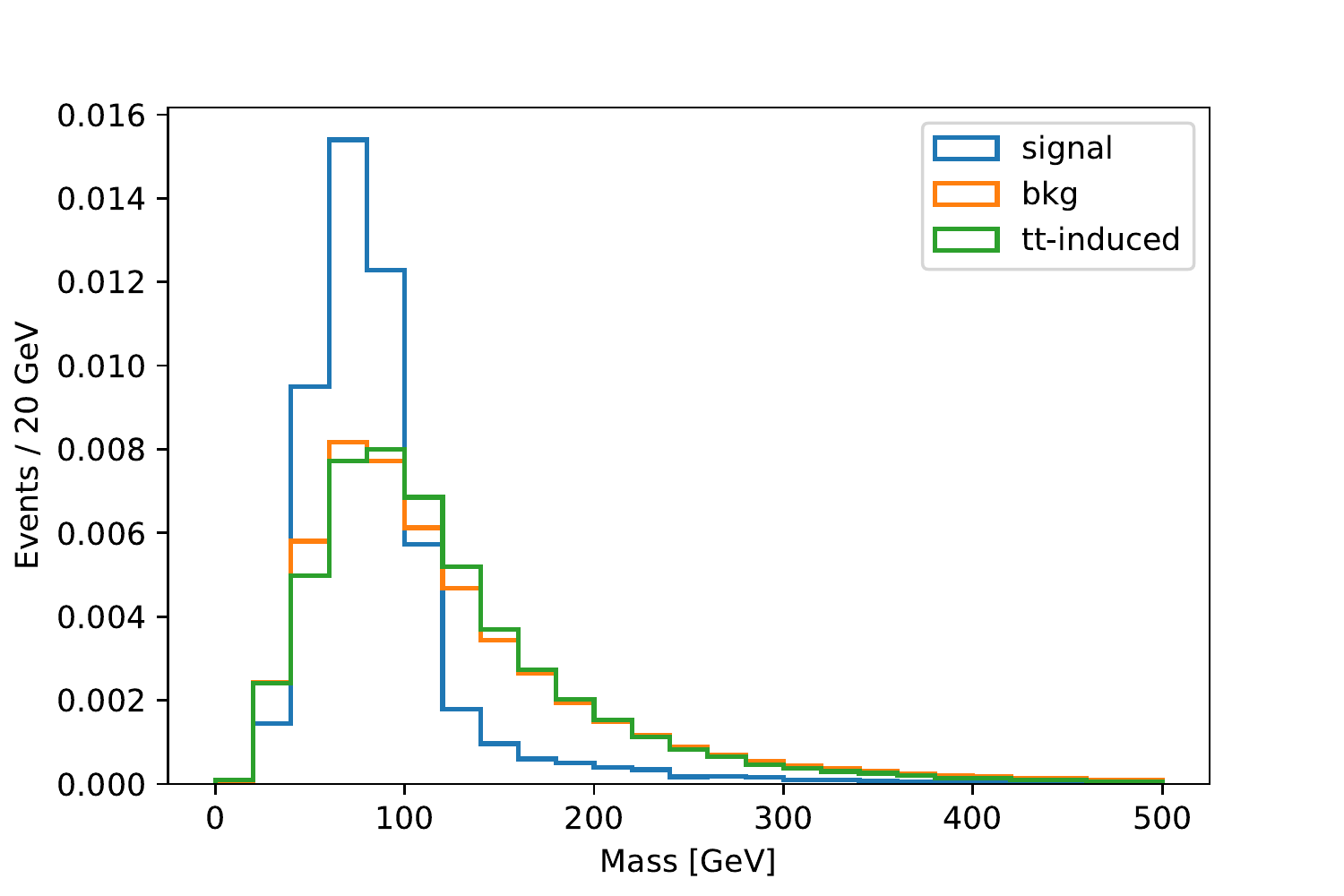}
    \end{subfigure}
    \begin{subfigure}[t]{0.45\textwidth}
    	\centering
        \includegraphics[width=0.99\textwidth]{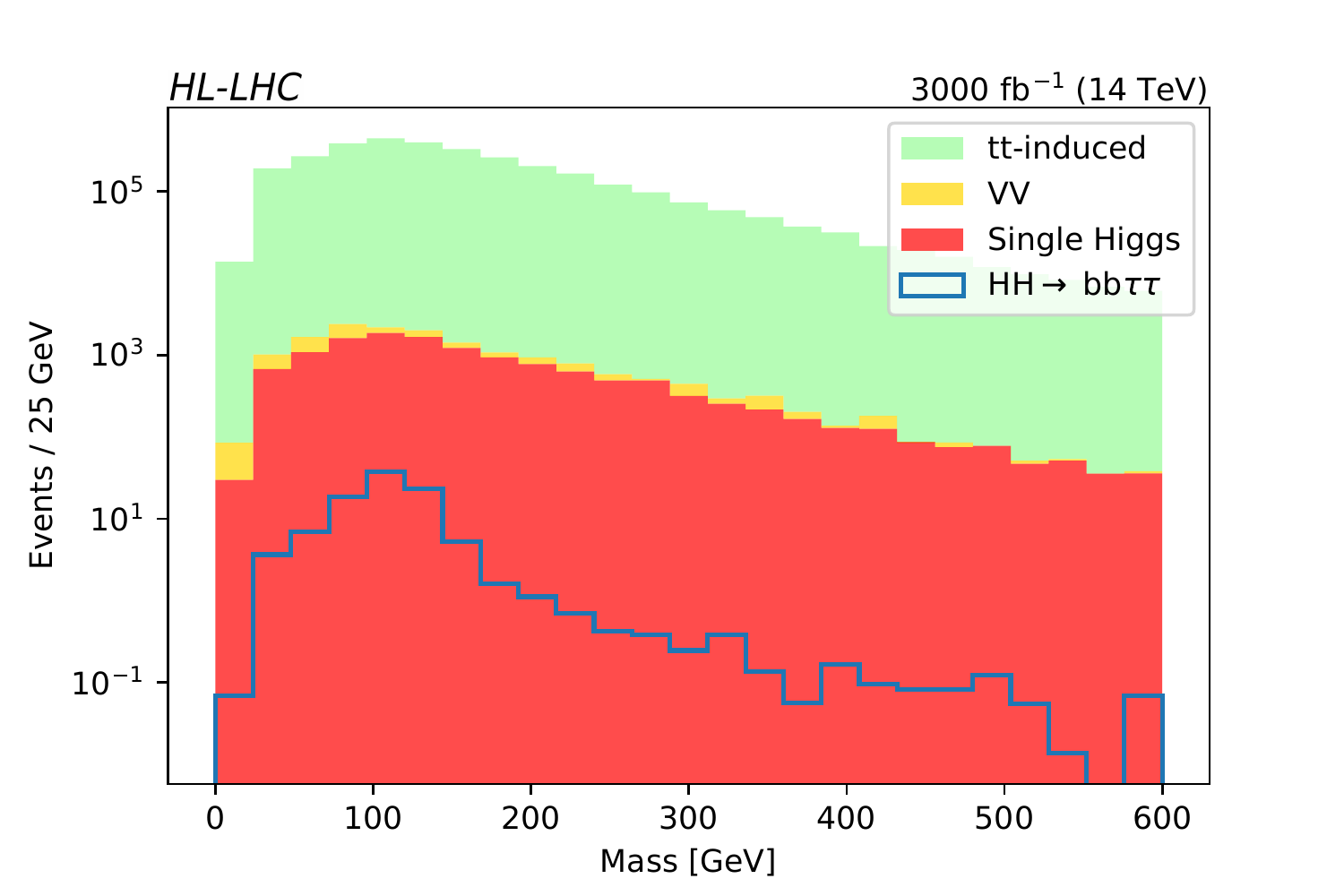}
    \end{subfigure}
    \begin{subfigure}[t]{0.45\textwidth}
    	\centering
        \includegraphics[width=0.99\textwidth]{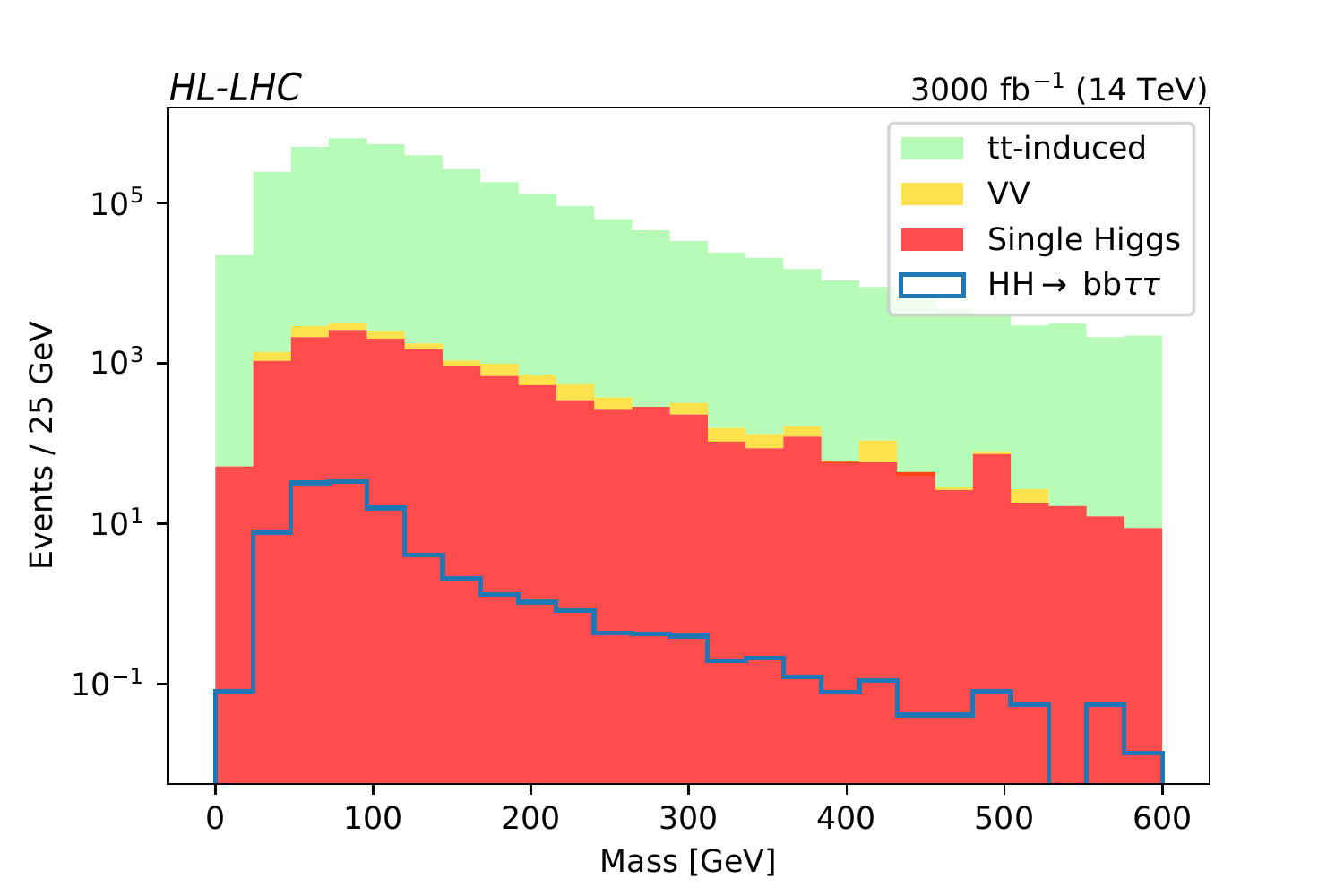}
    \end{subfigure}
    \begin{subfigure}[t]{0.45\textwidth}
    	\centering
        \includegraphics[width=0.99\textwidth]{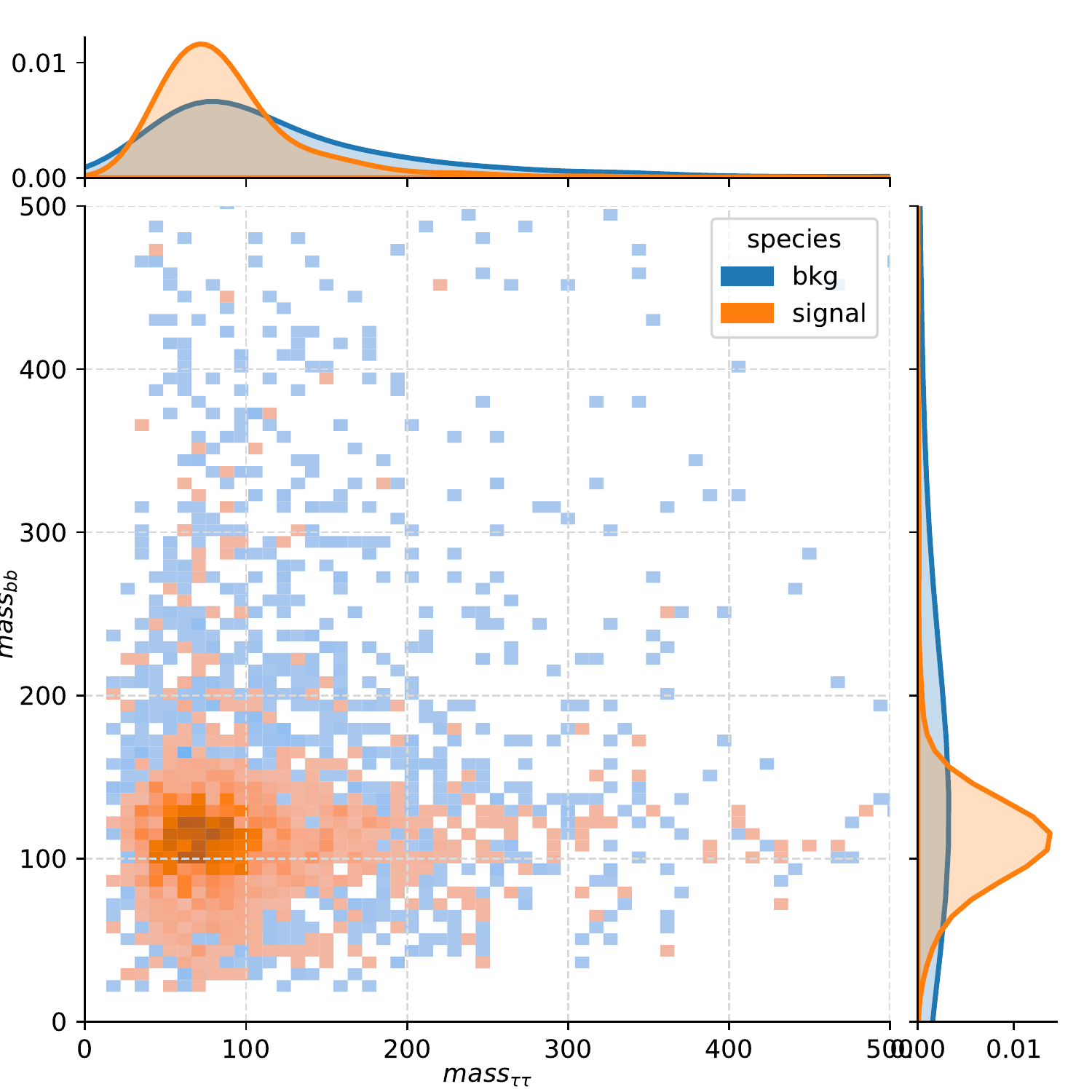}
    \end{subfigure}
   \caption{(Left) bb mass distribution, (Right) $\tau \tau$ mass distribution for the $\mu \tau$ category; distribution are shown with a normalised area and with the expected yields. (Bottom) 2D plot of $m_{\tau \tau}$ and $m_{b\bar{b}}$ for the $\mu \tau$ category}
   \label{pic:mass_obs}
\end{figure}

\noindent At this stage of the analysis, the majority of background comes from tt production and single Higgs processes: both of them completely overwhelmed and mimic the kinematic of the HH process.
It is clear from the plots that a cut and count analysis will never fully isolate the signal from the background contamination, thus the use of machine become the only way to obtain meaningful results for the statistical inference.\\
\noindent Observables of the reconstructed Higgs in $\tau \tau$ are shown in Figure \ref{pic:mass_obs}

\noindent In the invariant $b\bar{b}$ and $\tau \tau$ spectrum, we can notice that the signal is squeezed around the Higgs mass window, as expected.\\

\begin{figure}[h!]
    \centering
    \begin{subfigure}[t]{0.30\textwidth}
    	\centering
        \includegraphics[width=0.99\textwidth]{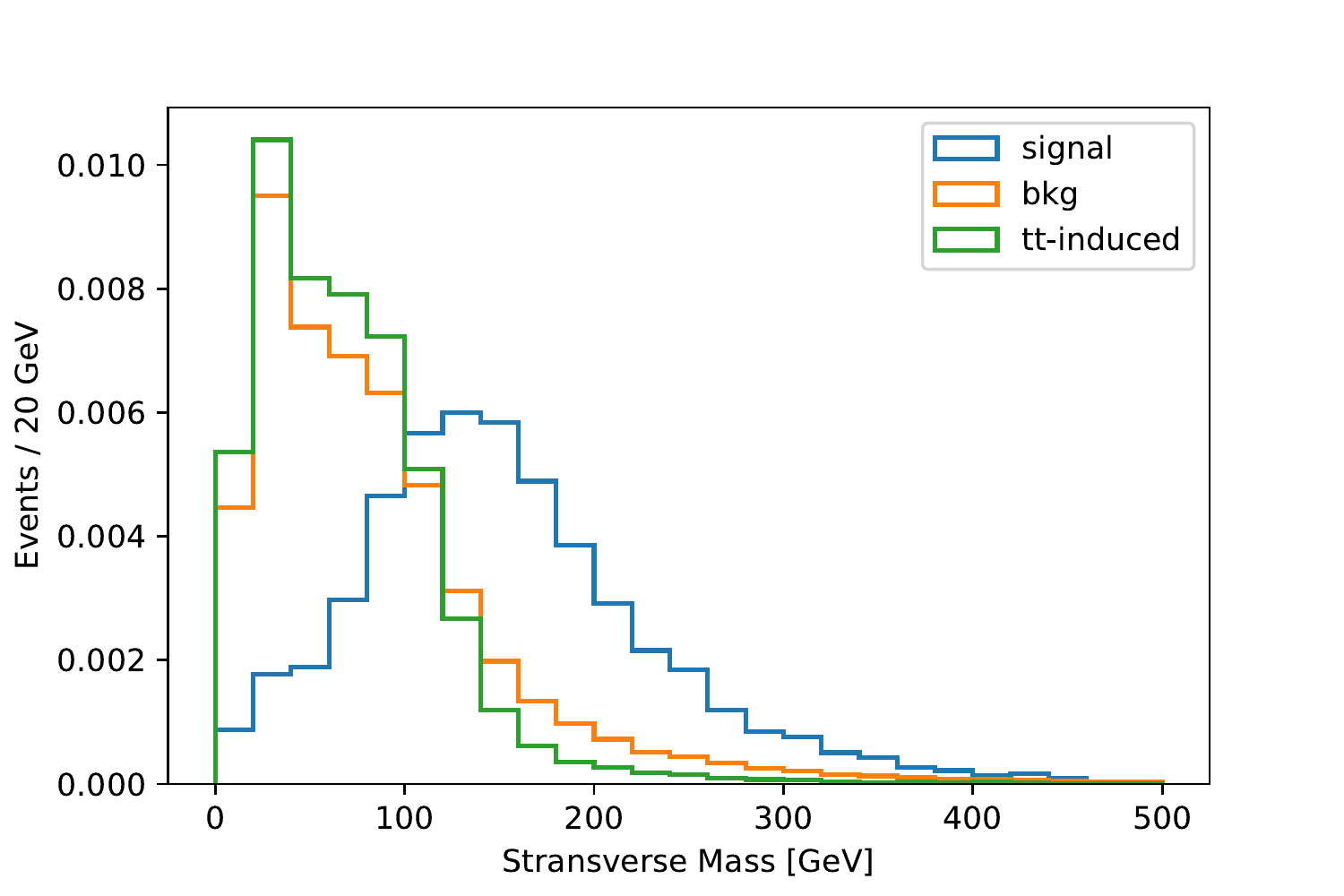}
    \end{subfigure}
    \begin{subfigure}[t]{0.30\textwidth}
    	\centering
        \includegraphics[width=0.99\textwidth]{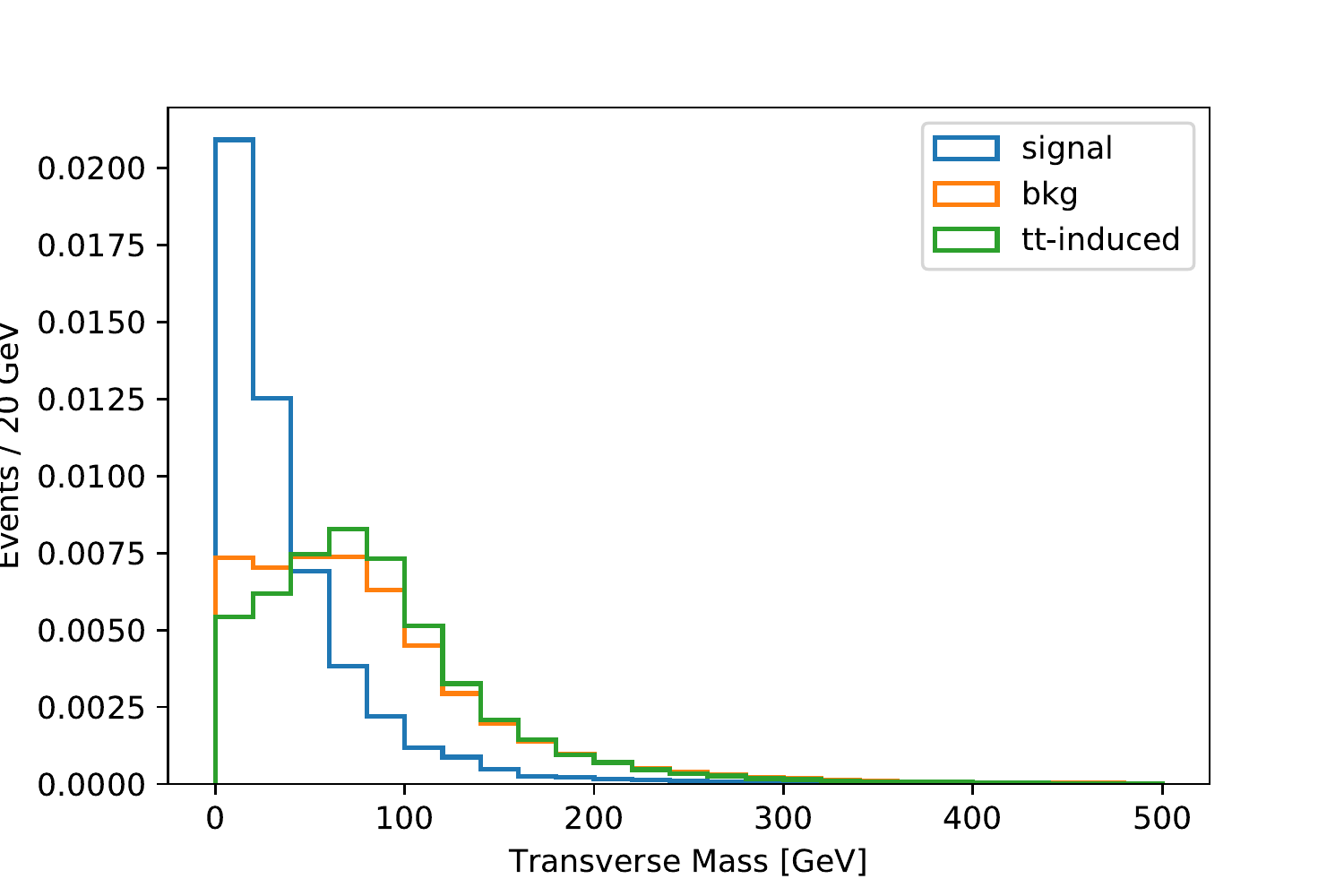}
    \end{subfigure}
    \begin{subfigure}[t]{0.30\textwidth}
    	\centering
        \includegraphics[width=0.99\textwidth]{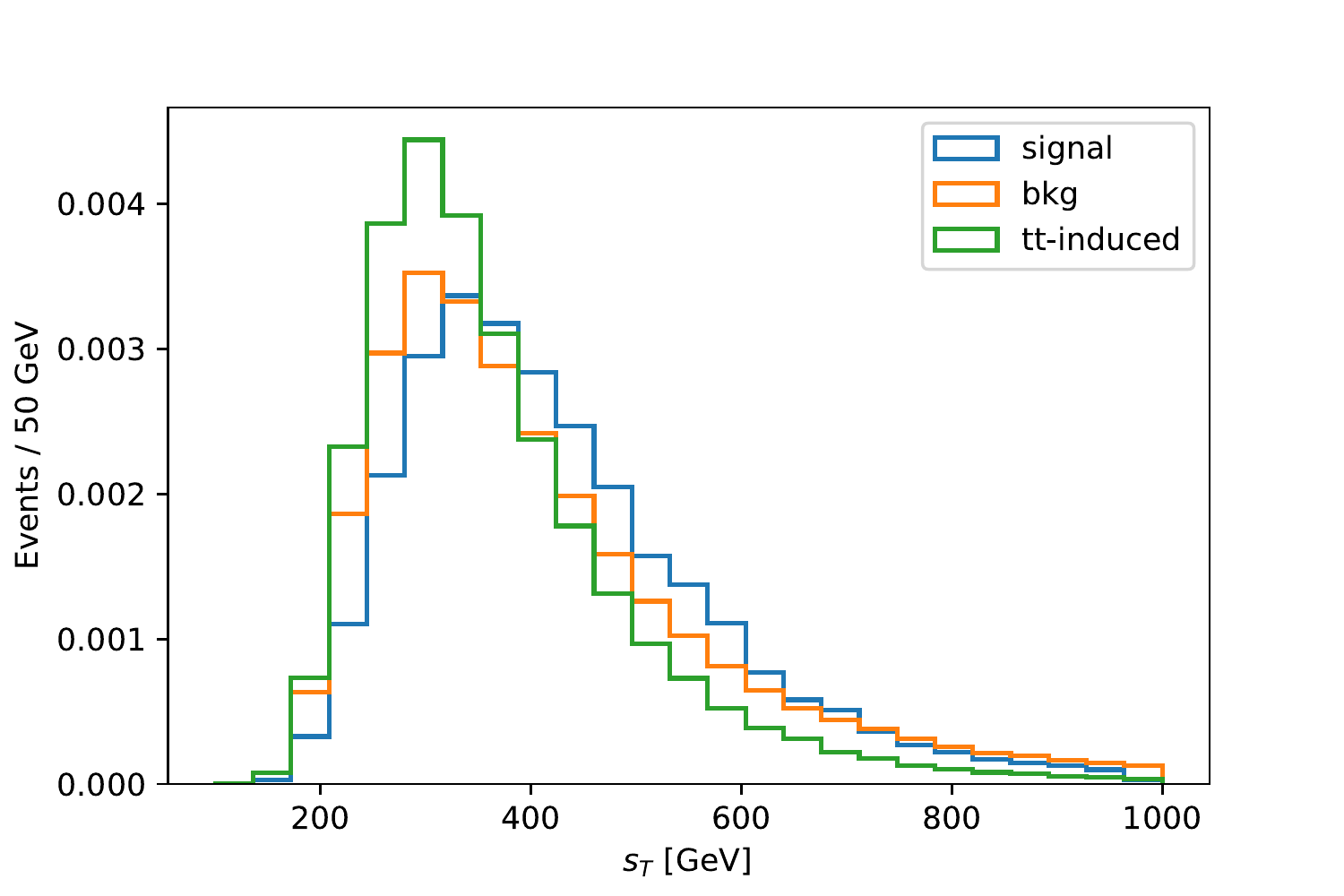}
    \end{subfigure}
    \begin{subfigure}[t]{0.30\textwidth}
    	\centering
        \includegraphics[width=0.99\textwidth]{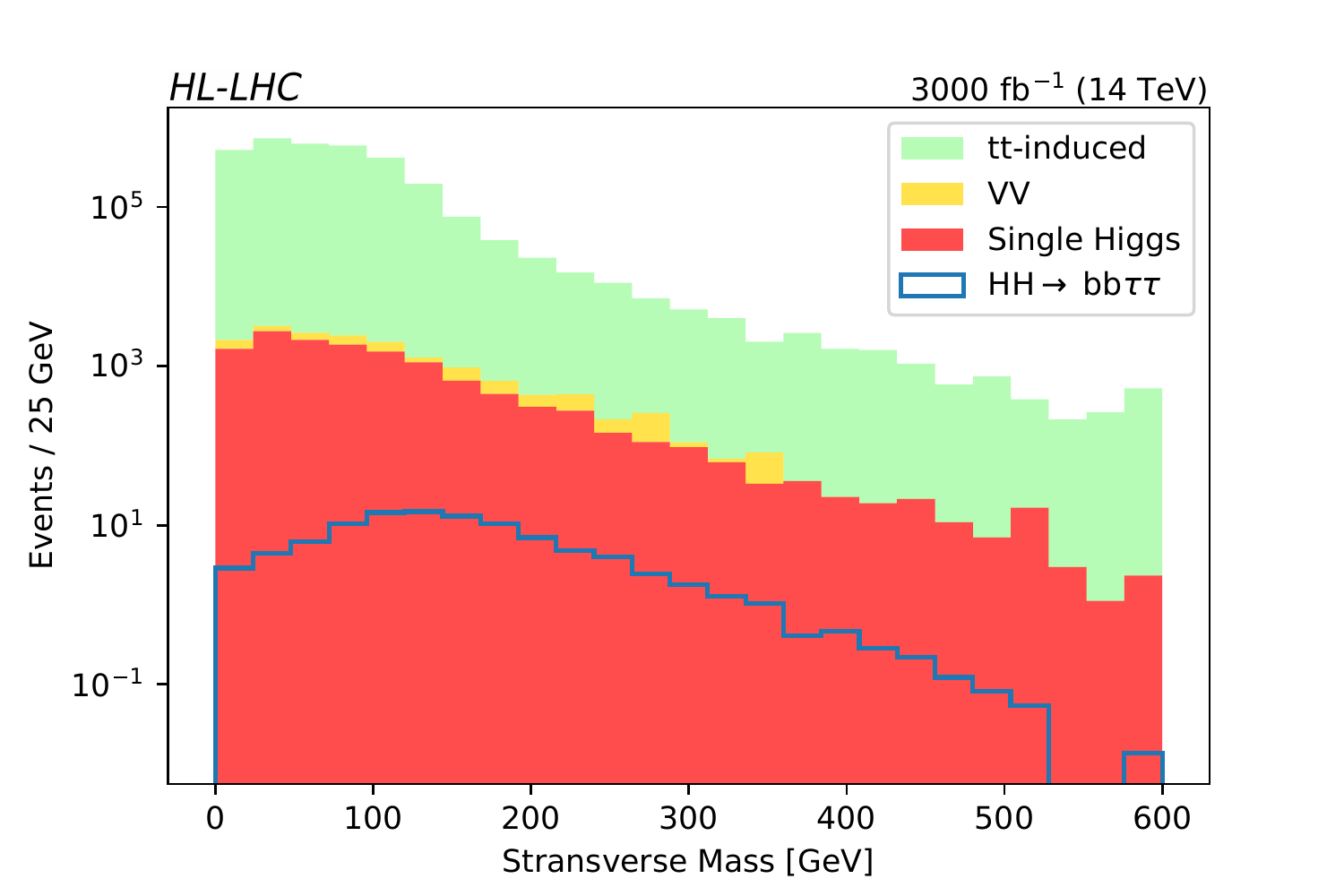}
    \end{subfigure}
    \begin{subfigure}[t]{0.30\textwidth}
    	\centering
        \includegraphics[width=0.99\textwidth]{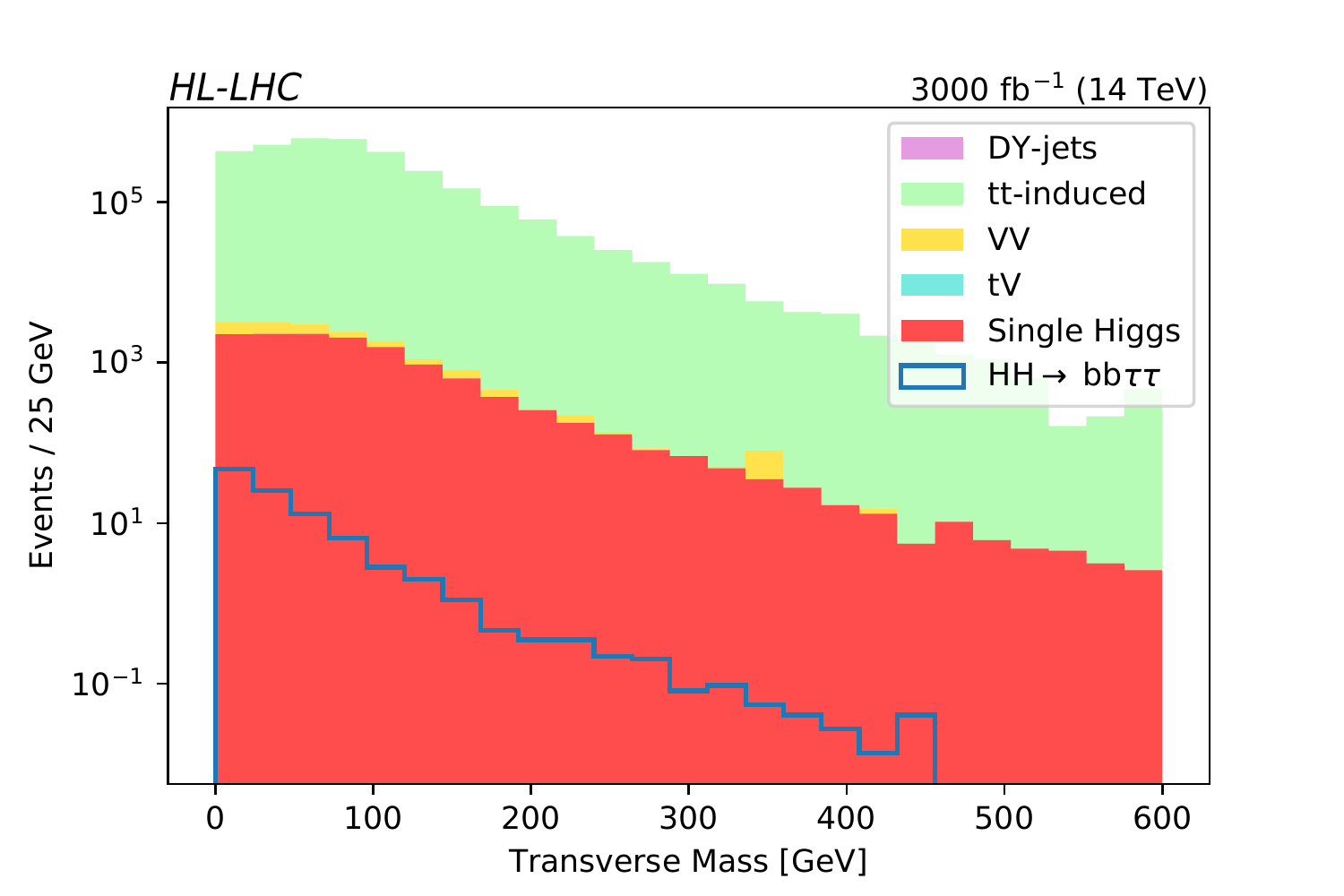}
    \end{subfigure}
    \begin{subfigure}[t]{0.30\textwidth}
    	\centering
        \includegraphics[width=0.99\textwidth]{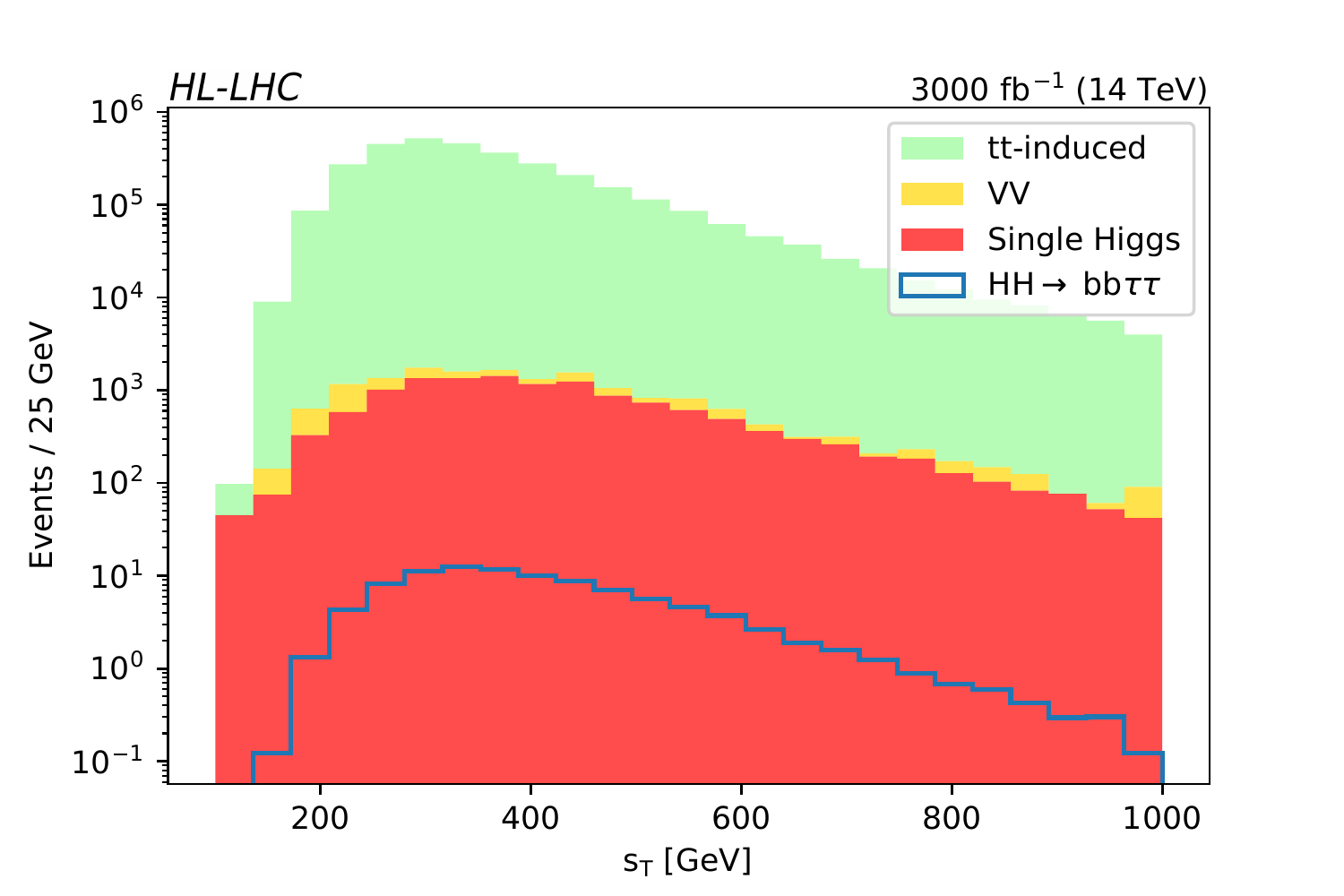}
    \end{subfigure}
   \caption{(Left) Stransverse mass distribution. (Center) Transveres mass distribution (Right) $s_T$ distribution for the $\mu \tau$ category. (Up) Distribution normalised to unit area. (Down) Distribution with expected yields of signal and background}
   \label{pic:str_mass}
\end{figure}

\noindent In the plot of the invariant $\tau \tau$ mass, is possible to see a shift from 125 GeV on the left of the distribution. This happens because some neutrinos in the final state, arising from the hadronic $\tau$ decay, are not properly simulated and considered in the 4-vector sum of the $\tau \tau$ system: the reconstruction of the mass will be lower by construction. This tell us that there is a source of missing energy that is not trivial to simulate but that can be properly taken into account constructing some more sophisticated observables from the HH decay products:

\begin{itemize}
    \item[$\blacksquare$] Stransverse mass ($M_{T2}$), which is able to predict the invisible contribution coming from neutrinos of $\tau$ decay \cite{stranverse_mass}; the distribution is shown in Figure \ref{pic:str_mass} left. As shown in Figure \ref{pic:str_mass} left, the discrimination power of this observable is extremely good: the majority of the background (tt especially) peaks at lower values with respect to the signal. This variables can almost fully capture the kinematics of the HH system, thus it will be used in the statistical inference
    \item[$\blacksquare$] Transverse mass ($m_{T}$), defined as:
    $\sqrt{2p_T(\tau) \times p_T^{miss} \times \Big(1 - cos\Big(\Delta\Phi(\tau, p_T^{miss})\Big)\Big) }$,
    the distribution is shown in Figure \ref{pic:str_mass} center. Here the shape of the signal is complementary with respect to the stransvere mass: the signal peaks at 0 while the background is shifted to the right of the distribution. The good discrimination power of this variable will be used in the DNN training.
    \item[$\blacksquare$] $s_T$, scalar sum of muon, tau, b-jet, and missing energy transverse momentum; the distribution is shown in Figure \ref{pic:str_mass} right.
\end{itemize}

\noindent The general kinematic structure of a typical HH event is hard to fully capture using simple orthogonal cuts on measured quantities; to better discriminate between signal and background events and to improve the sensitivity of the analysis a DNN is trained for each final state.\\
Variables used in the DNNs come mainly from the kinematic of the event, moreover the DNNs are made fully orthogonal to the stransverse mass: the information of both these variables will be used in the signal extraction (Sec \ref{signal_ext}).
Detailed information on the DNNs performances and architecture will be given in the following sections.

\subsection{DNN based approach}

All the obersevables that characterised the event kinemetic are used in as input variables:

\begin{itemize}
    \item[$\blacksquare$] $H \rightarrow \tau\tau$ and $H \rightarrow b\bar{b}$ invariant mass, $\eta$, $\phi$ and $p_{T}$
    \item[$\blacksquare$] Transverse mass ($m_{T}$)
    \item[$\blacksquare$] $s_T$
    \item[$\blacksquare$] missing energy
    \item[$\blacksquare$] $\Delta\Phi$ and $\Delta R$ of the HH system
    \item[$\blacksquare$] Number of jets and number of b jets
\end{itemize}

\begin{figure}[h!]
    \centering
    \begin{subfigure}[t]{0.45\textwidth}
    	\centering
        \includegraphics[width=0.99\textwidth]{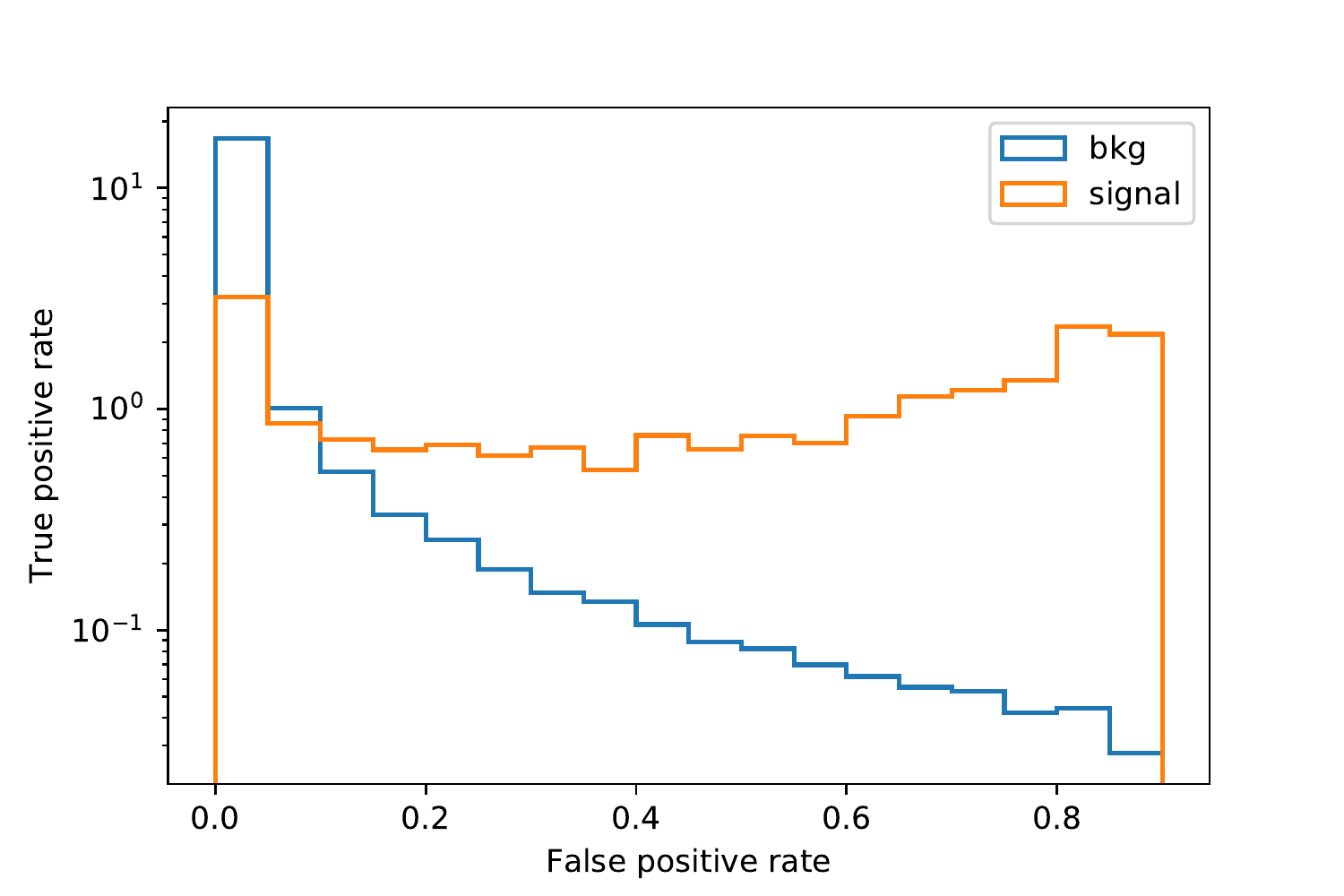}
    \end{subfigure}
    \begin{subfigure}[t]{0.45\textwidth}
    	\centering
        \includegraphics[width=0.99\textwidth]{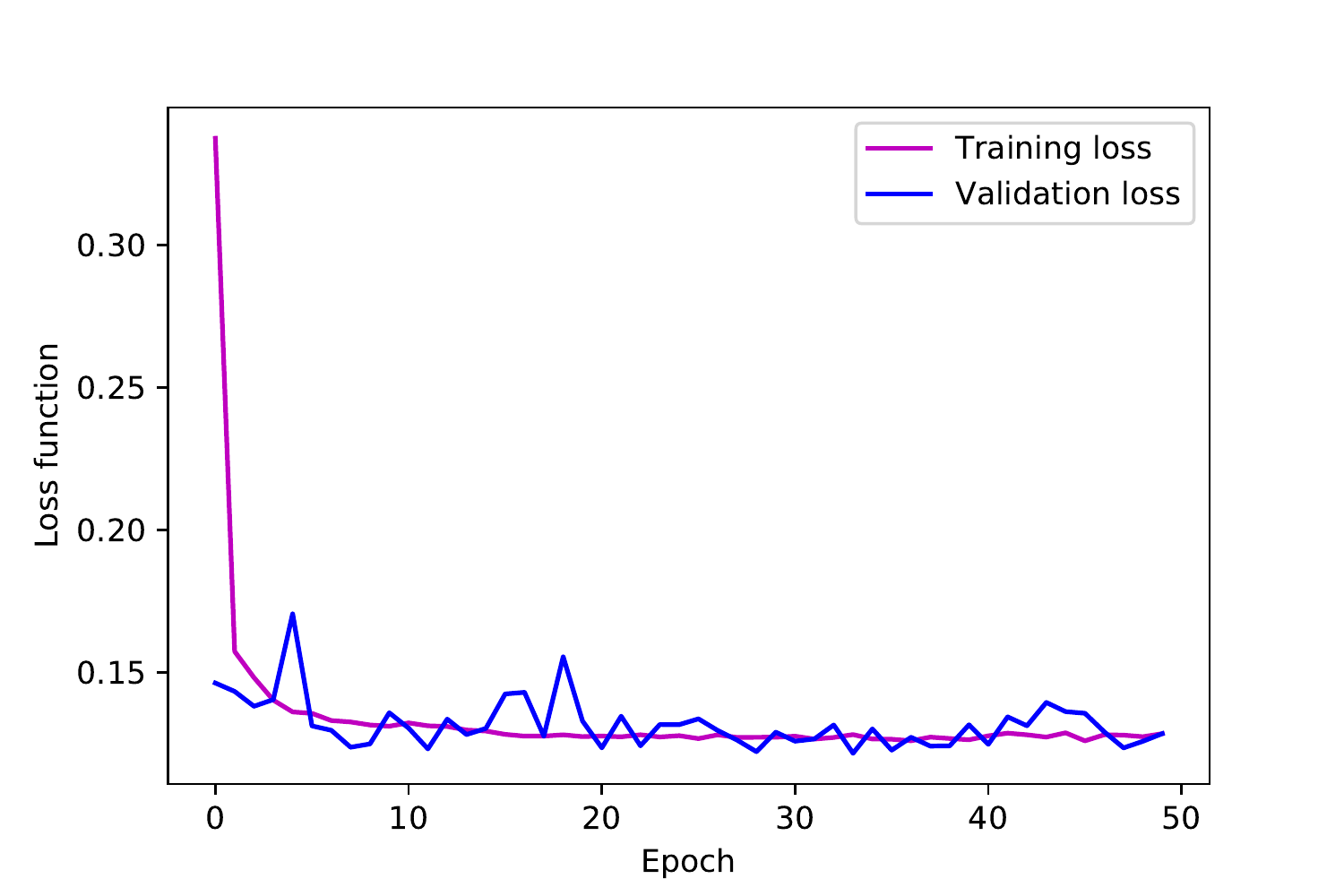}
    \end{subfigure}
   \caption{(Right) DNN score for the $\mu \tau_h$ category (Left) distribution training and validation loss as a function of the epochs}
   \label{pic:dnn_mu}
\end{figure}

\noindent The network is constructed with 3 layers of 132 nodes and with an activation function that varies from channel to channel. The DNN is optimised and trained separately for each category to achieve a AUC of around 90\%.\\
In Figure \ref{pic:dnn_mu} the performance of the DNN in the $\mu \tau_h$ category are shown. Similar performances are found for $e\tau_h$ and $ \tau_h \tau_h$. The learning algorithm has no hints of overtraining or undertraining.

\begin{figure}[h!]
    \centering
    \begin{subfigure}[t]{0.45\textwidth}
    	\centering
        \includegraphics[width=0.99\textwidth]{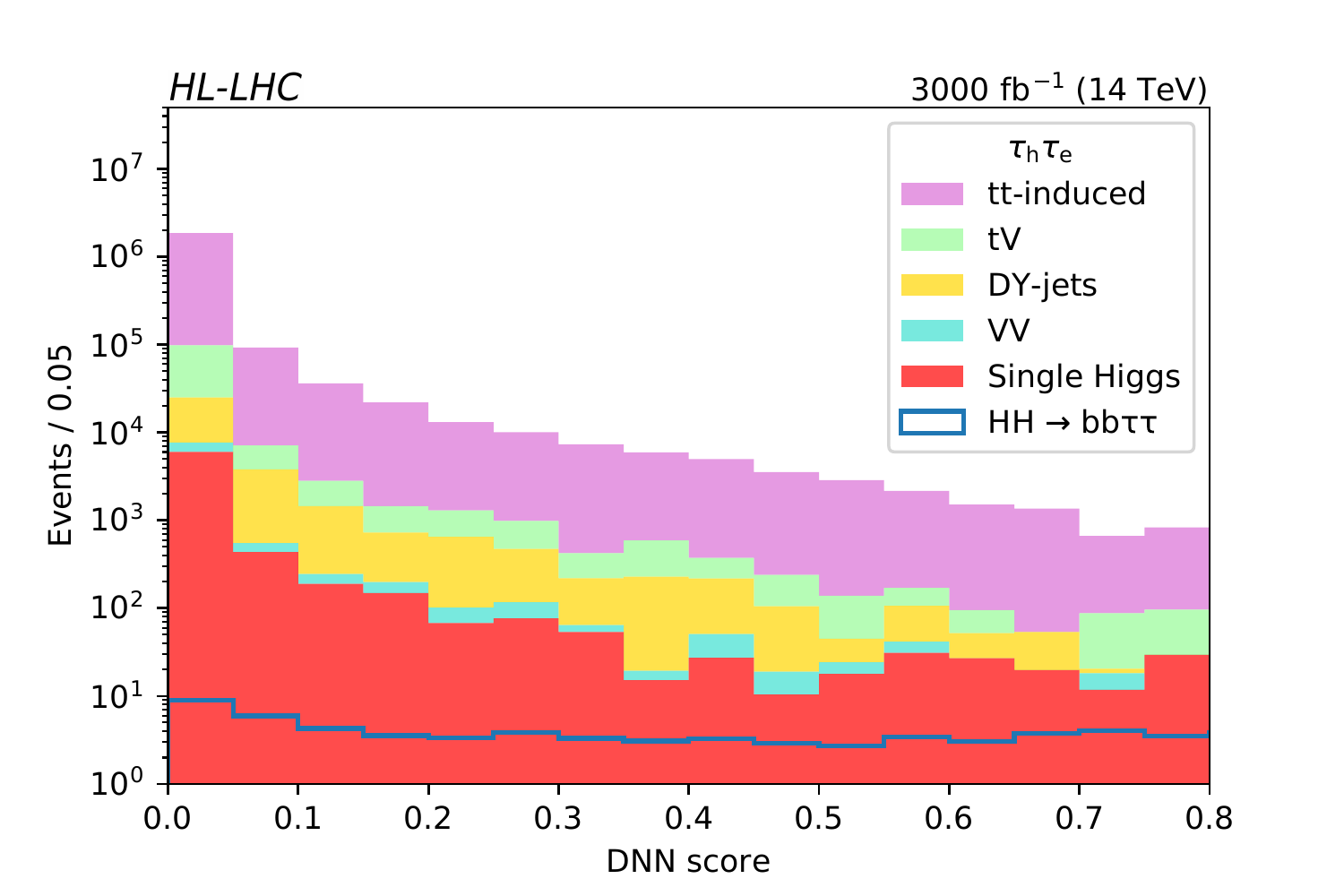}
    \end{subfigure}
    \begin{subfigure}[t]{0.45\textwidth}
       \centering
       \includegraphics[width=0.99\textwidth]{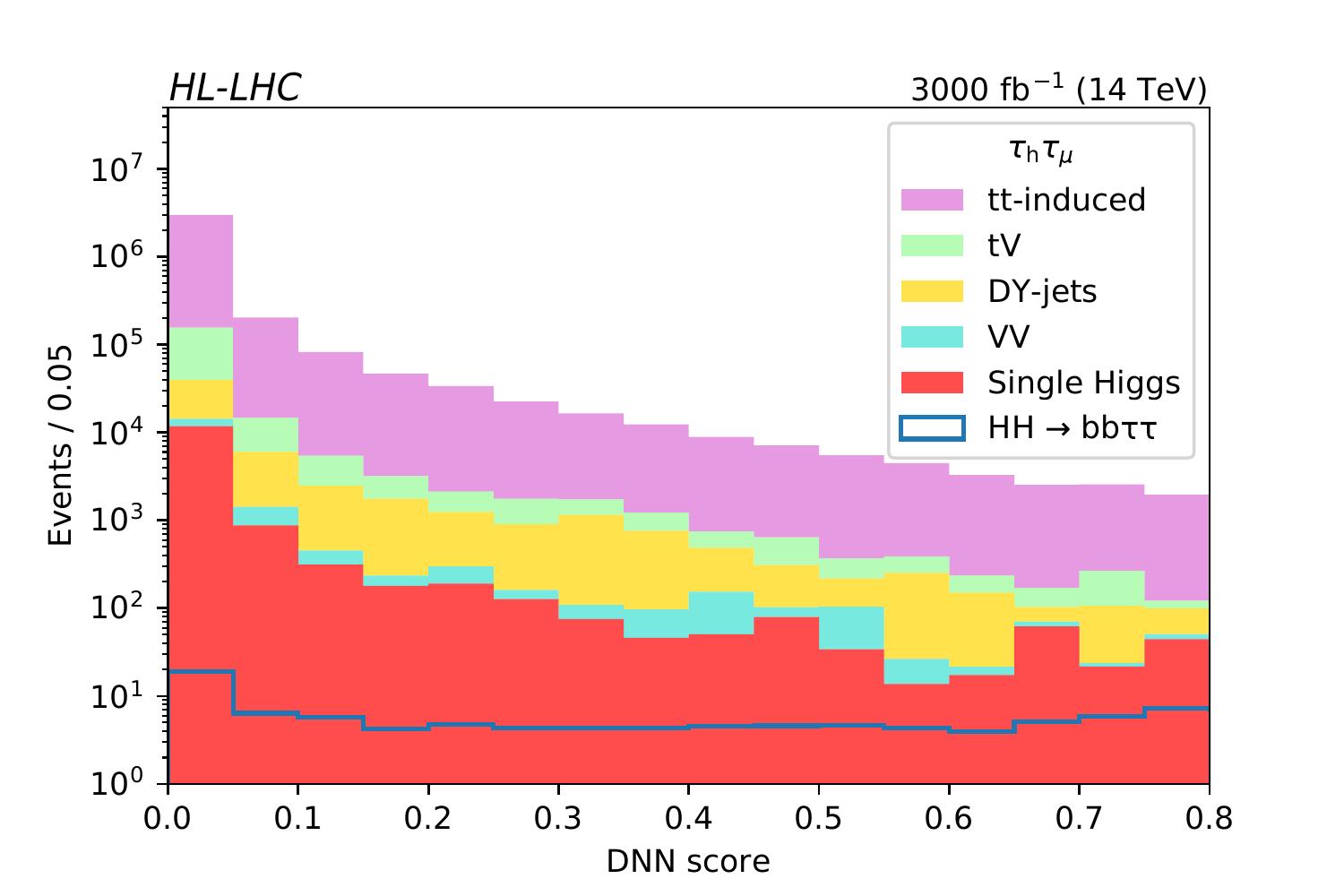}
   \end{subfigure}
     \begin{subfigure}[t]{0.45\textwidth}
       \centering
       \includegraphics[width=0.99\textwidth]{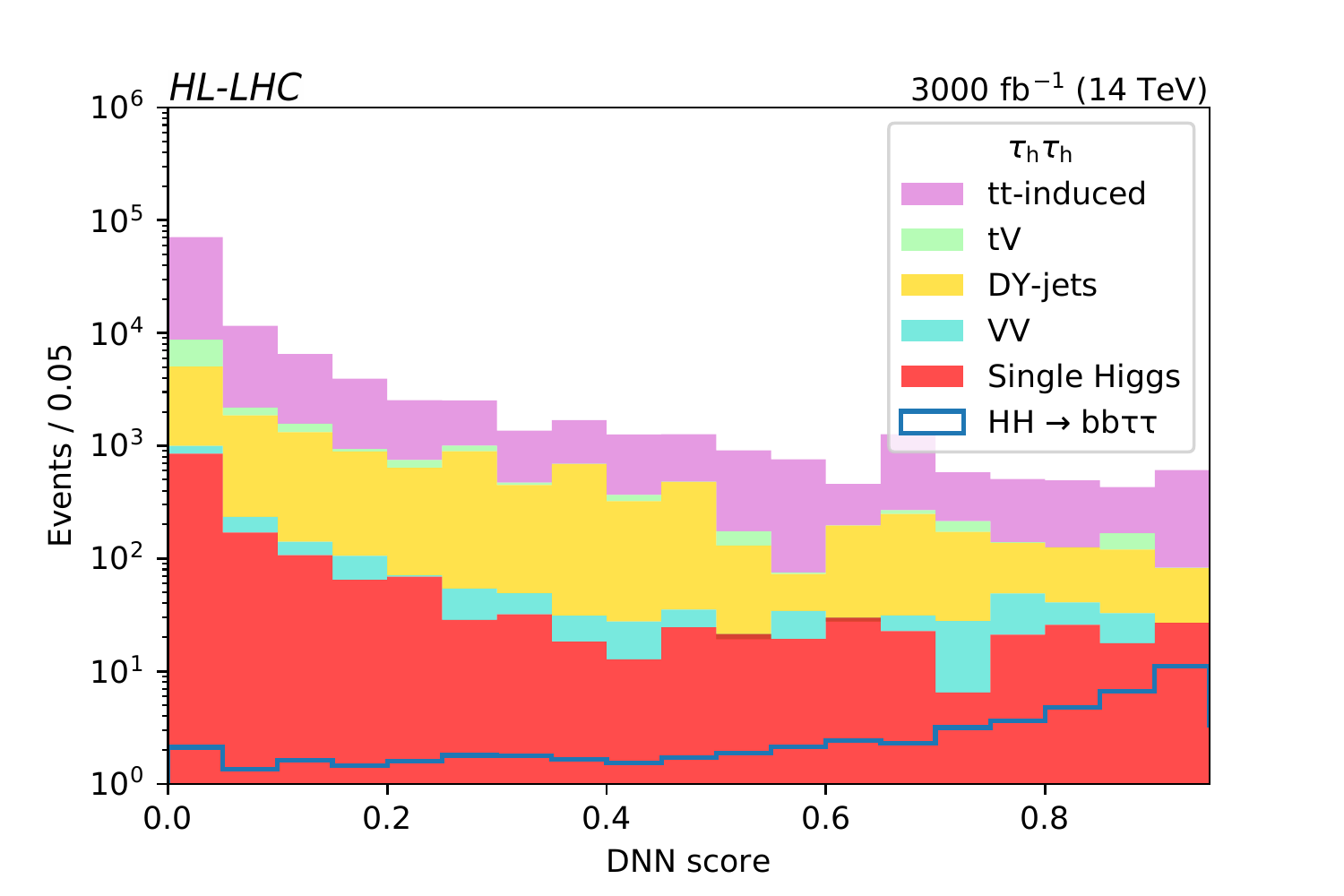}
   \end{subfigure}
 
   \caption{DNN score for  (Right) $e \tau_h$, (Left) $\mu \tau_h$, (Bottom) $\tau_h \tau_h$}
   \label{pic:dnn_score}
\end{figure}

\noindent In Figure \ref{pic:dnn_score} the DNN score for the different categories is shown.

\begin{figure}[h]
    	\centering
        \includegraphics[width=0.5\textwidth]{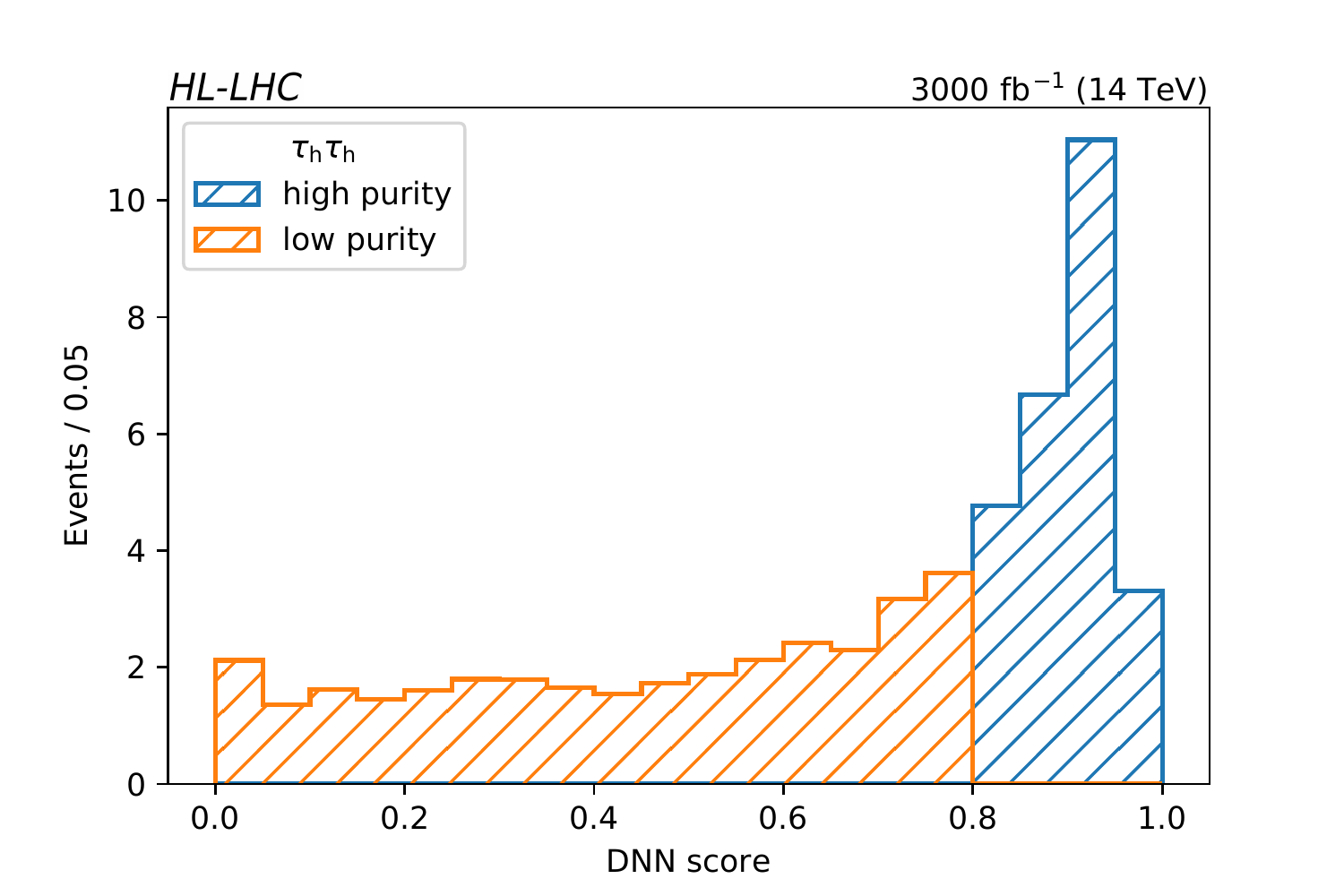}
        \caption{DNN score distrbution for the $\tau_{h}\tau_{h}$ category, showing the division in low and high purity. The histogram is scaled to cross section and luminosity.}
   \label{pic:dnn_divided}
    \end{figure}

\noindent Events are further categorized according to the DNN score. The whole DNN range is divided in two regions: a low purity one characterized by score smaller than a certain threshold $x_{thr}$, a high purity region with score higher than $x_{thr}$, as shown in Figure \ref{pic:dnn_divided} for the most sensible category $\tau_{h}\tau_{h}$. The value of the threshold is chosen with a procedure to maximize the signal to background ratio, that is repeated independently for the three different channels. The resulting values of $x_{thr}$ are reported in Table \ref{tab:x_thr_bbtautau}.

\begin{table}[h!]
	\centering
    \begin{tabular}{c|c } 
    \hline
    
     \textbf{Channel} & $\boldsymbol{x_{thr}}$ \\
    
    \hline
    &\\
    $\tau_{\mu}\tau_{h}$ & 0.6\\
    $\tau_{e}\tau_{h}$ & 0.7\\
    $\tau_{h}\tau_{h}$ & 0.8\\
    &\\
    \hline
	\end{tabular}
	\caption{Chosen values of $x_{thr}$ that defines the high/low purity regions, for the three different channels.}. 
	\label{tab:x_thr_bbtautau}
\end{table}

\begin{figure}[h]
    	\centering
        \includegraphics[width=0.5\textwidth]{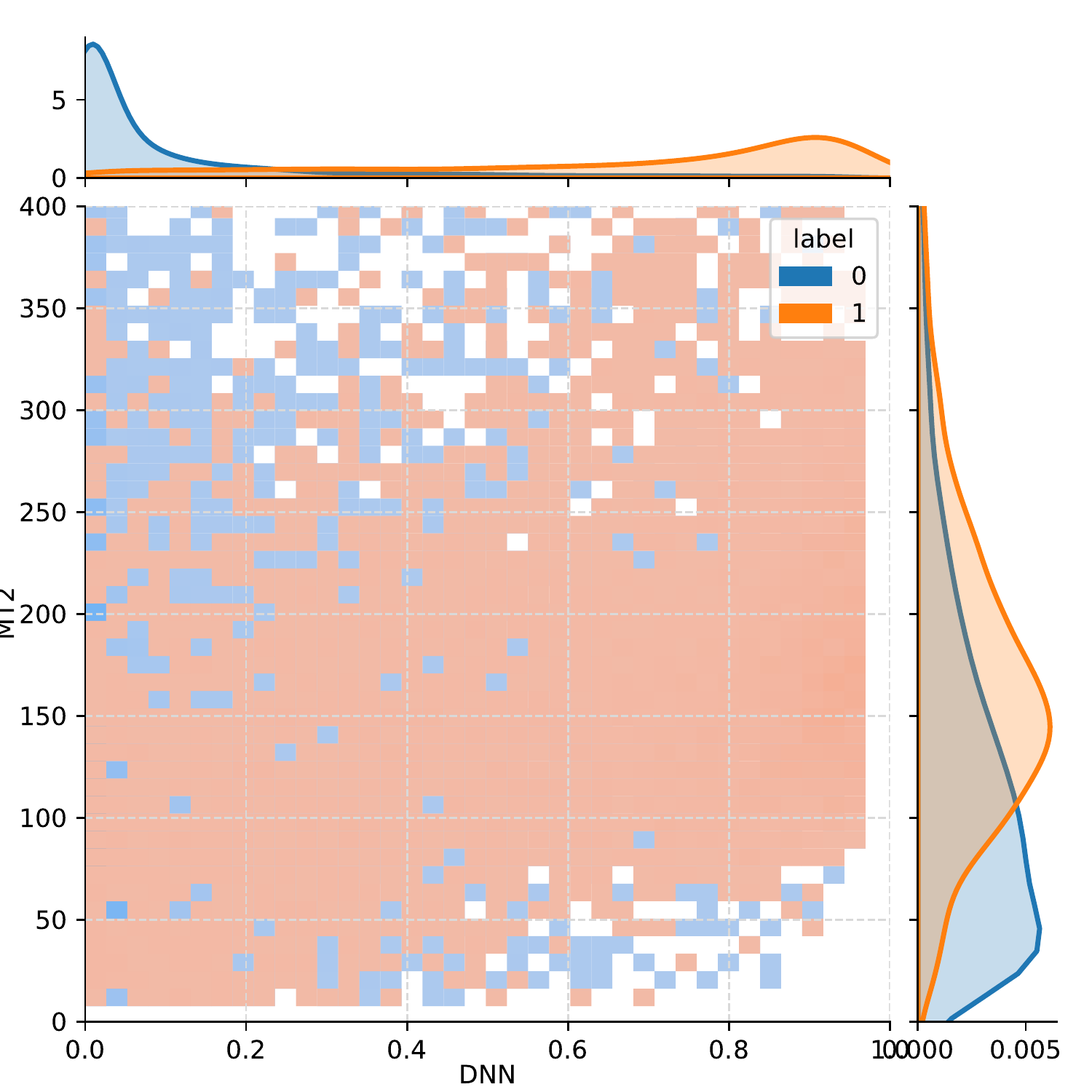}
        \caption{2D distribution of the DNN score and the stransverse mass for the $\tau_h \tau_h$ category, 0 represents the background and 1 represents the signal}
   \label{pic:dnn_str_2d}
    \end{figure}

\noindent For the two regions of the DNN score, the stransverse mass will be used as input for the statistical analysis. 
In Figure \ref{pic:dnn_str_2d} a 2D distribution of the DNN and the stransverse mass is shown.

\subsection{Statistical analysis and results}
\label{signal_ext}

\begin{figure}[h!]
    \centering
    \begin{subfigure}[t]{0.45\textwidth}
    	\centering
        \includegraphics[width=0.99\textwidth]{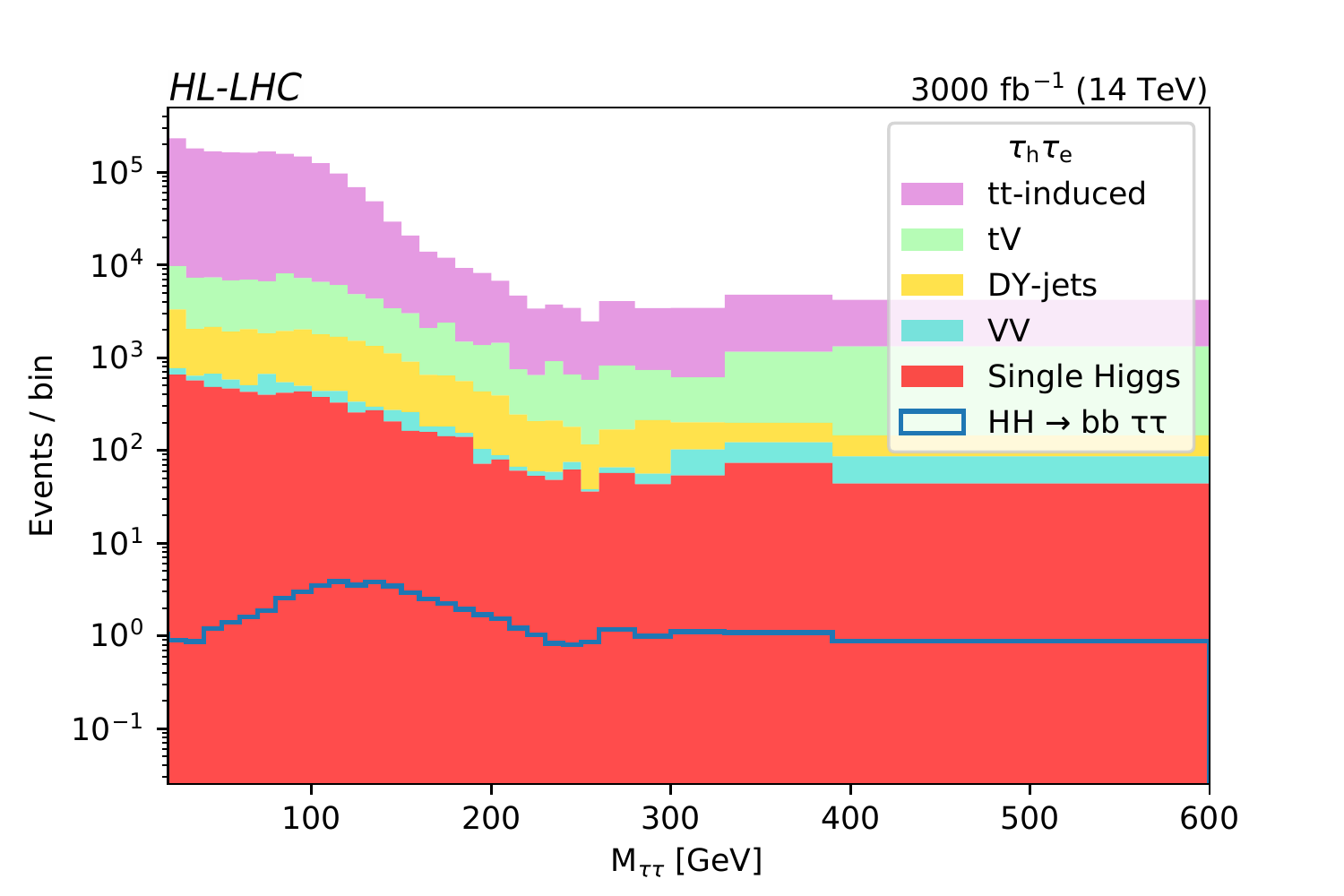}
    \end{subfigure}
    \begin{subfigure}[t]{0.45\textwidth}
       \centering
       \includegraphics[width=0.99\textwidth]{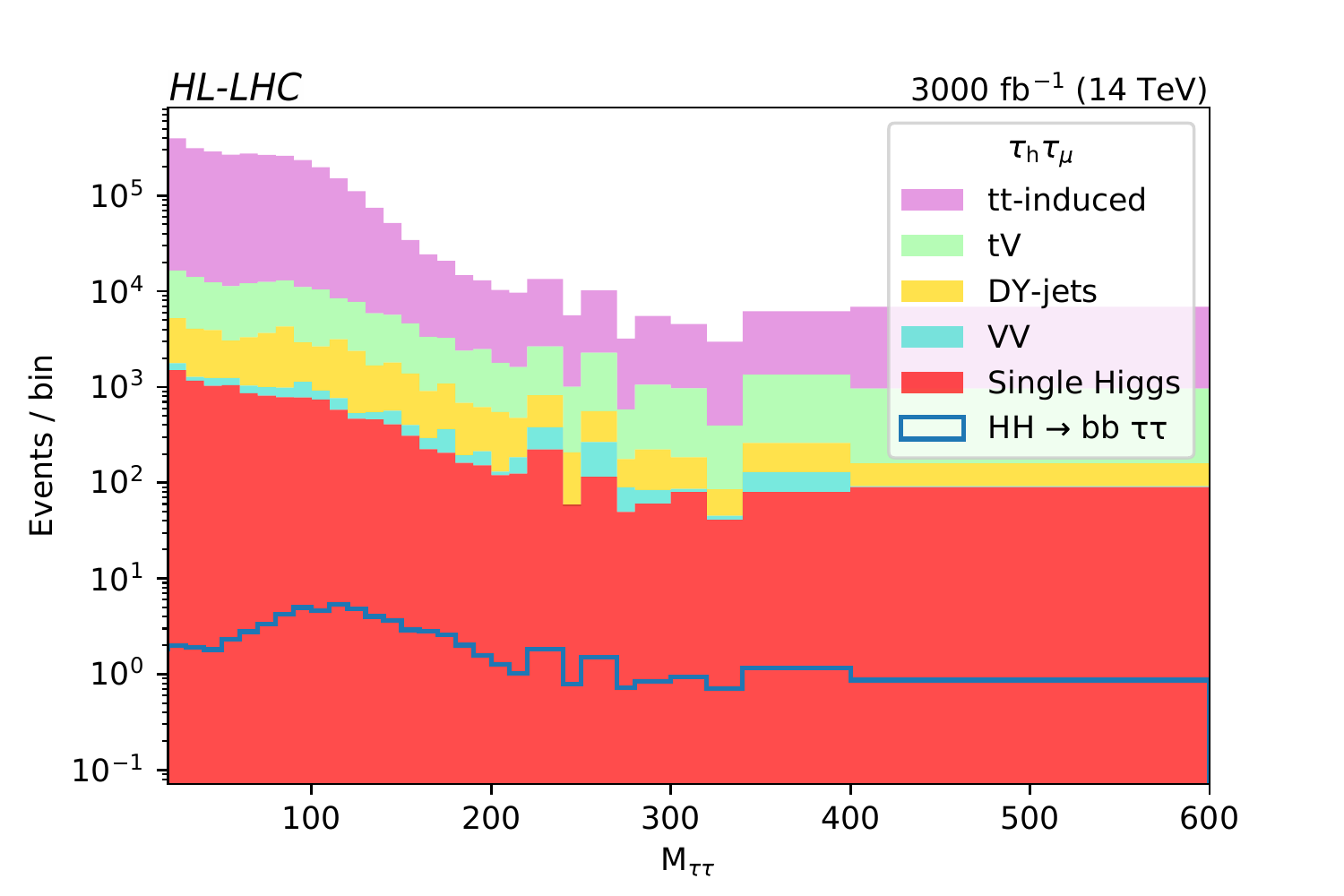}
   \end{subfigure}
     \begin{subfigure}[t]{0.45\textwidth}
       \centering
       \includegraphics[width=0.99\textwidth]{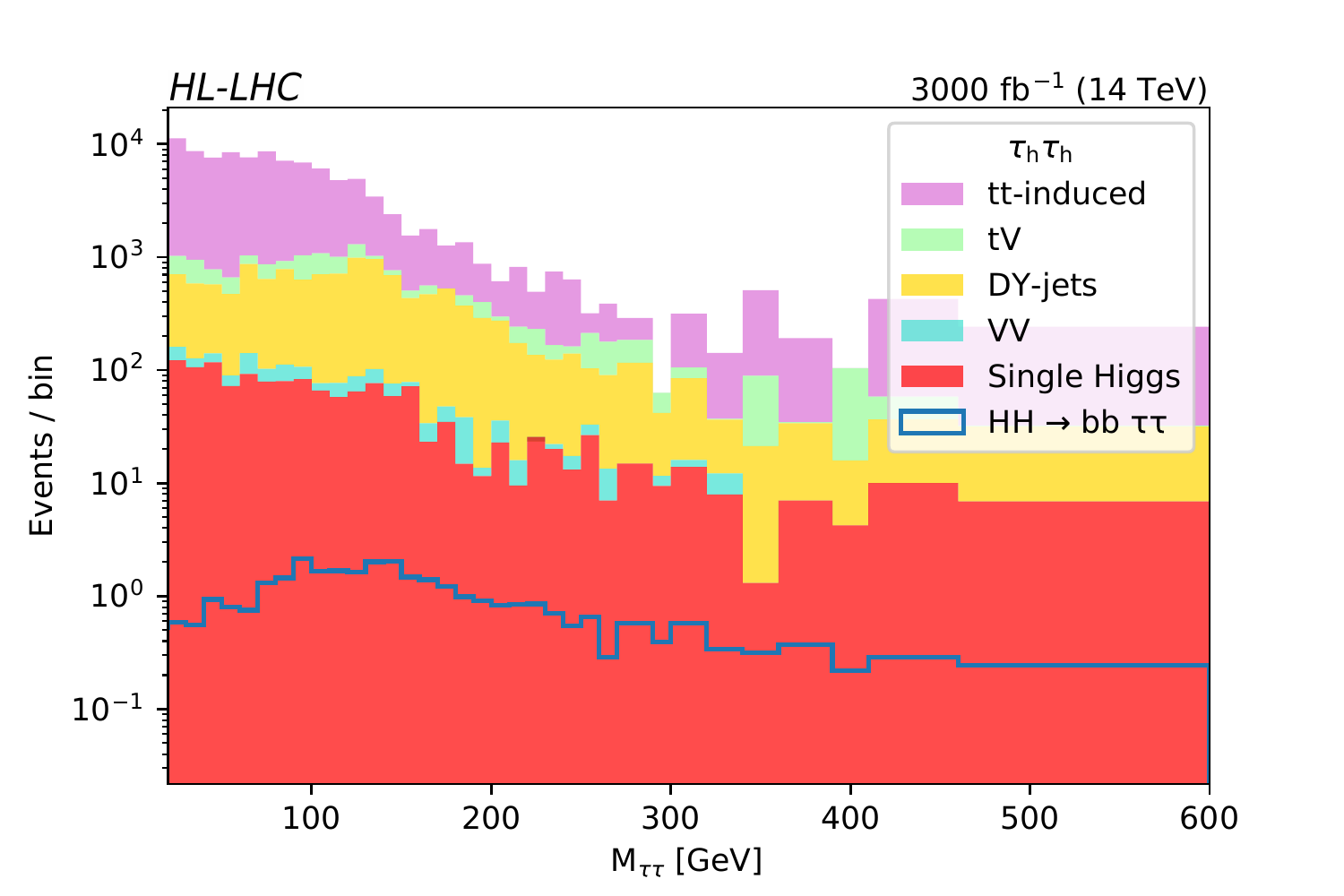}
   \end{subfigure}
 
   \caption{Stransverse mass distribution in the low purity region for  (Right) $e \tau_h$, (Left) $\mu \tau_h$, (Bottom) $\tau_h \tau_h$}
   \label{pic:str_mass_dnn}
\end{figure}

To extract the significance for the SM HH signal and the upper limit at the 95$\%$ CL on the signal strength, $\mu = \sigma_{\mathrm{HH}} / \sigma^{\mathrm{SM}}_{\mathrm{HH}}$, a multi-dimesional fit is performed, using the shape of the stransverse mass, in the low and high purity region. The stransverse mass is binned in a way that ensures the 15\% statistical uncertainties in each bin of the distribution (Figure \ref{pic:str_mass_dnn}). 

\noindent The sources of systematic uncertainties in this analysis come from experimental setup and from theory assumptions, they are summarised in Table \ref{tab:bbtt_unc}. All the uncertainties affected the yields of the final distributions.

    \begin{table}[h!]
	\centering
    \begin{tabular}{c|c } 
    \hline
     \textbf{Systematic uncertainty source} & \textbf{Impact on yields} \\
    \hline
    Luminosity & $\pm$ 1.0 \% \\
    Lepton ID efficiency & $\pm$ 1.0 \% \\
    Tau ID efficiency  & $\pm$ 5.0\% \\
    Photon ID efficiency  & $\pm$ 1.0 \%\\
    Jet Energy Scale  & $\pm$ 1.0\% \\
    B-tag efficiency & $\pm$ 1.0\% \\
	\hline
	QCD scale tt inclusive& +2.4\% /-3.6\% \\
    \hline
    Pdf scale tt inclusive& $\pm$4.2\%  \\
    \hline
    \multirow{3}{*}{{\makecell[c]{Signal theoretical \\ uncertainties}}}
    & +2.1\% /-4.9\% (QCD scale)\\
    & $\pm$3.0\% (pdf scale)\\
    & +4.0\% /-18.0\% (top mass)\\
    \hline
	\end{tabular}
	\caption{Systematic uncertainties for $bb\tau\tau$ channel.}. 
	\label{tab:bbtt_unc}
\end{table}

\begin{figure}[t]
    	\centering
        \includegraphics[width=0.5\textwidth]{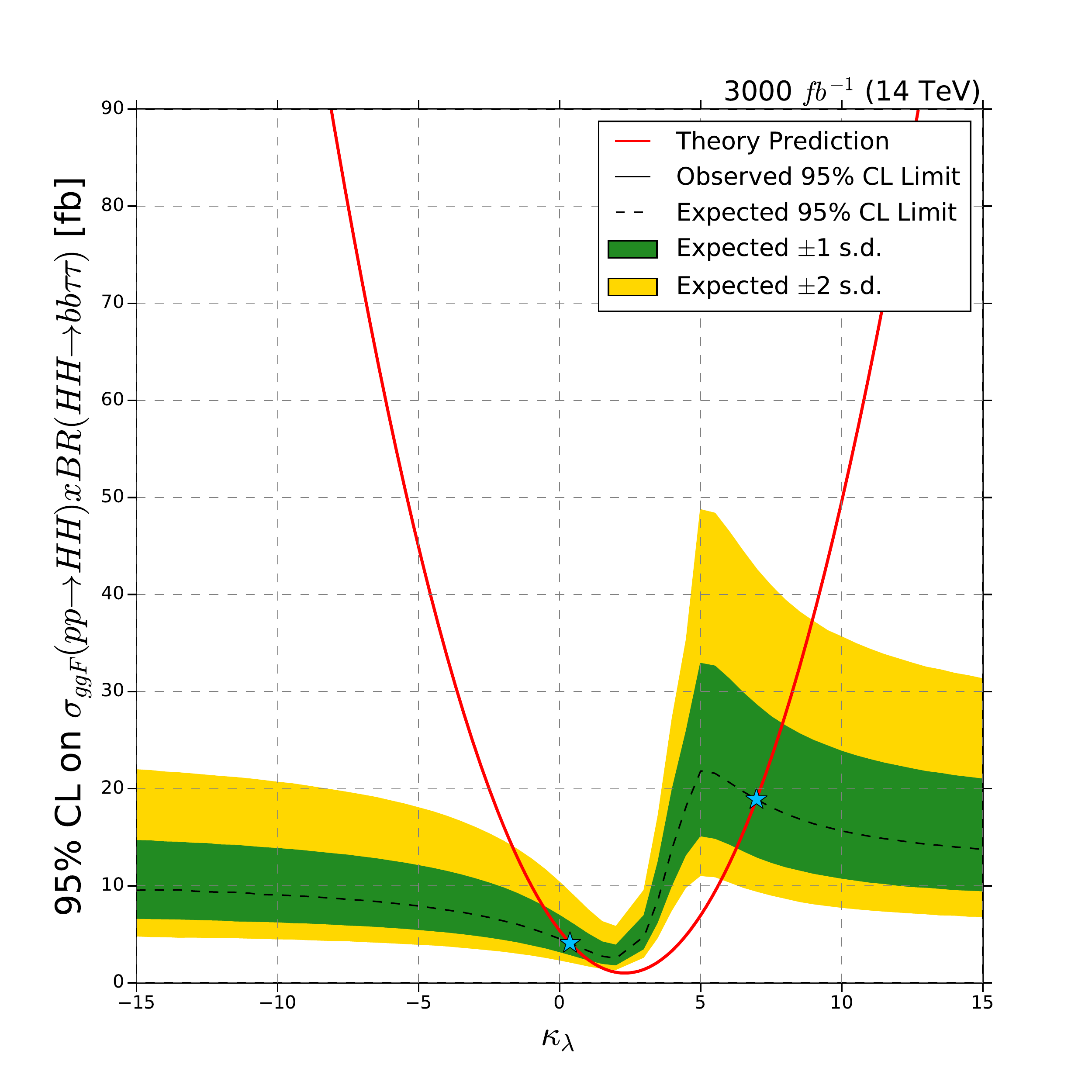}
        \caption{$HH \rightarrow b\bar{b} \tau \tau$ CLs upper limits on the HH production cross section times BR derived as a function of $k_{\lambda} = \lambda_{\text{HHH}} / \lambda_{\text{SM}} $. The red line represents the theoretical value of the cross section times BR.}
   \label{pic:kl_scan_bbtautau}
    \end{figure}

\noindent Expected results in terms of 95\% CL upper limits and HH signal significance are reported in Table \ref{tab:results_bbtautau} with and without systematic uncertainties. An upper limit on the HH cross section times branching fraction of 1.25 times the SM prediction is obtained, corresponding to a significance of 1.72.
\\Prospects for the measurement of the trilinear coupling are also studied. Under the assumption that no HH signal exists, 95\% CL upper limits on the SM HH production cross section are derived as a function of $\kappa_{\lambda}$ as visible in Figure \ref{pic:kl_scan_bbtautau}. A variation of the excluded cross section, directly related to changes in the HH kinematic properties, can be observed as a function of $\kappa_{\lambda}$. The intersection between the expected curve and the theory prediction is used to estimate the constraint on the trilinear coupling. The $\kappa_{\lambda}$ is constrained to be within 0.56 and 6.72.

\begin{table}[h!]
	\centering
    \begin{tabular}{c|c|c|c } 
    \hline
    
     \textbf{Condition} & \textbf{Significance (in $\sigma$)} & \makecell[c]{\textbf{Upper limit on}\\ \textbf{$\mu$ at 95\% CL}} & \textbf{$\kappa_{\lambda}$ constraint}\\
     
    \hline
    
    stat only & 1.72 & 1.25 & [0.56,6.72] \\
    
    stat + sys & 1.70 & 1.37 & [0.37, 6.97] \\
    
    \hline
	\end{tabular}
	\caption{Results for $bb\tau\tau$ channel.}. 
	\label{tab:results_bbtautau}
\end{table}
\clearpage
\section{$\boldsymbol{HH \rightarrow b\bar{b}b\bar{b}}$ analysis}
\label{sec:4b}

Despite the largest branching fraction among the HH decay channels, the $bbbb$ final state suffers from a large contamination from the multijet QCD background that makes it experimentally challenging. 

\subsection{Simulated samples}

Signal samples for SM and BSM hypotheses with different values of the trilinear Higgs boson coupling are simulated at $k_\lambda$= 1,2.45,5 in order to study the expected constraint on the $k_\lambda$.
The backgrounds considered for this analysis consist of single Higgs events and QCD and tt processes. All the relevant samples are listed in Table \ref{tab:ev1}.

 \begin{table}[h!]
	\centering
    \addtolength{\leftskip} {-2.0cm}
    \addtolength{\rightskip}{-2.0cm}
    \begin{tabular}{c|c|c} 
    \hline
    
    &\textbf{Process} &  \makecell[c]{\textbf{Cross section}  (fb)} \\
    
    \hline
    
	\multirow{5}{*}{Signal} & \makecell[c]{$(gg)HH \rightarrow b\Bar{b} b\Bar{b}$ ($\kappa_{\lambda} =1$)} & $2.49 \times 10^{1}$ \\
	& \makecell[c]{$(gg)HH \rightarrow b\Bar{b} b\Bar{b}$  ($\kappa_{\lambda} =2.45$)} &  $1.05 \times 10^{1}$ \\
	& \makecell[c]{$(gg)HH \rightarrow b\Bar{b} b\Bar{b} $  ($\kappa_{\lambda} =5$)}  & $ 7.59 \times 10^{1}$ \\
	
	\hline
	
     \multirow{6}{*}{\makecell[c]{Single \\ Higgs }}& $(gg)H \rightarrow b\bar{b}$ & $3.18 \times 10^{4}$\\
     & $(gg)ZH, Z \rightarrow q\bar{q}, H \rightarrow b\bar{b}$  & $ 5.88  \times 10^{1}$\\
    & $ZH, Z \rightarrow q\bar{q}, H \rightarrow b\bar{b}$  & $ 4.02  \times 10^{2}$\\
    & $W^{+}H, W \rightarrow q\bar{q'}, H \rightarrow b\bar{b}$   & $ 3.62  \times 10^{2}$\\
    & $W^{-}H, W \rightarrow q\bar{q'}, H \rightarrow b\bar{b}$   & $ 2.32  \times 10^{2}$\\
    & $VBFH, H \rightarrow b\bar{b}$   & $ 2.49  \times 10^{3}$\\
    
	\hline
	
	  \multirow{7}{*}{QCD} & HT 200 to 300 & $1.003 \times 10^{8}$\\
	  & HT 300 to 500 &  $2.173 \times 10^{7}$\\
	  & HT 500 to 700 &  $1.945 \times 10^{6}$\\
	  & HT 700 to 1000 &  $3.806 \times 10^{5}$\\
	  & HT 1000 to 1500 & $6.881 \times 10^{4}$\\
	  & HT 1500 to 2000 & $6.053 \times 10^{3}$\\
	  & HT 2000 to Inf & $1.087 \times 10^{3}$\\
    
    \hline
    
    \multirow{2}{*}{ $t\bar{t}$} & $t\bar{t}$ inclusive&
     $8.644\times 10^{5}$\\
    & $t\bar{t}$ extended&
     $8.644\times 10^{5}$\\
    
	\hline
	\end{tabular}
	\caption{List of simulated samples for $bbbb$ channel.}. 
	\label{tab:ev1}
\end{table}

\begin{table}[h!]
	\centering
\begin{tabular}{ c|c}

    \hline
    
    \textbf{Process} & \textbf{Yields}\\
    
    \hline
    
     \makecell[l]{$HH \rightarrow b\bar{b} b\bar{b}$  $\kappa_{\lambda} = 1$} &          $1376 \pm 36$\\
     \makecell[l]{$HH \rightarrow b\bar{b} b\bar{b}$  $\kappa_{\lambda} = 2.45$} &         $2579 \pm 79$ \\
     \makecell[l]{$HH \rightarrow b\bar{b} b\bar{b}$  $\kappa_{\lambda} = 5$} &          $7438 \pm 377$ \\
    
    \hline
    
     $ggH, H \rightarrow b\bar{b}$ &         $1886 \pm 310$ \\

     $(gg)ZH,H \rightarrow b\bar{b}$ &           $686 \pm 19$ \\
     $ZH, H \rightarrow b\bar{b}$ &          $2106 \pm 65$\\
     
     $WH, H \rightarrow b\bar{b}$ &  $209 \pm 12$ \\

    $VBFH, H \rightarrow b\bar{b}$ &           $319 \pm 35$ \\
   
    \hline
    
    QCD &  $18168801 \pm 773221$ \\
    
    \hline
    
    $t \bar{t}$ &  $991631 \pm 25819$ \\

    \hline

\end{tabular}

\caption{Yields and selection efficiencies for $bbbb$ channel.}. 
	\label{tab:ev_bbbb}
	
\end{table}

\subsection{Event selection}

\begin{figure}[h!]
    \centering
     \includegraphics[width=0.5\textwidth]{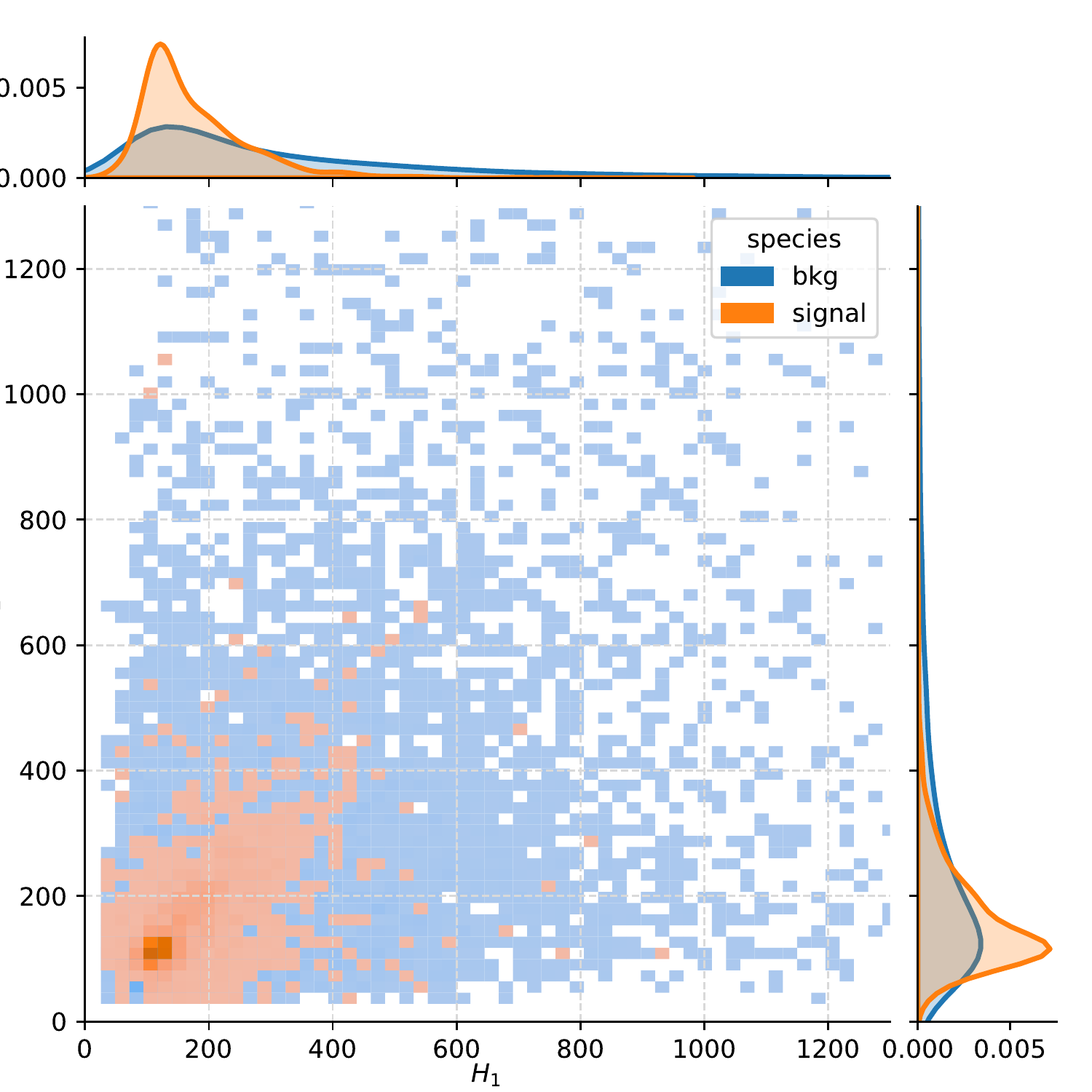}
   \caption{Pairing of the di-jet object to proper identify the $HH \rightarrow 4b$ object. Correct jet pairing is determined as the combination that minimises the difference in the invariant mass of the two jet pairs.}
   \label{pic:bb_pairing}
\end{figure}

Four jets are reconstructed with $p_{T} > 45~GeV$ and $|\eta| < 3.5$ and satisfy the medium b tagging working point. In case more than four jets pass that preselection step, corresponding to less than 7\% of the total signal events, the four highest $p{T}$ candidates are selected to build the double Higgs pair. The jets are paired in order to minimizes the difference in the invariant mass of the two jet pairs; the signature of two resonant $H \rightarrow b\bar{b} $ decays is explored properly in that way  (Figure \ref{pic:bb_pairing}).

\noindent The signal region is defined by considering the events that satisfy the following selection for the invariant mass of the two Higgs boson candidates $H_1$ and $H_2$:

\begin{gather*}
\sqrt {(m_{H_{1}} - 120 \, GeV)^{2} + (m_{H_{2}} - 120 \, GeV)^{2}} < 40 \, GeV
\end{gather*}

\noindent A technique based on the use of a DNN is used to exploit the kinematic differences between the double Higgs signal and the background processes and their correlations.

\subsection{DNN based approach}
\label{sec:4bdnn}

A multivariate discriminant, consisting of a DNN, is built using the following kinematic variables:
\begin{itemize}
\item[$\blacksquare$] the invariant mass of the two Higgs candidates;
\item[$\blacksquare$] the transverse momentum of the two Higgs candidates;
\item[$\blacksquare$] the four-jet invariant mass $m_{HH}$, and the reduced mass $M_{HH} = m_{HH}-(m_{H_1} - 125 \, GeV) - (m_{H_2} - 125 \, GeV)$;
\item[$\blacksquare$] the minimal and max $\Delta \eta$ and $\Delta \phi$ separation of the combinations of the four preselected jets;
\item[$\blacksquare$] the $\Delta \eta$, $\Delta \phi$ and $\Delta R = \sqrt{ (\Delta \eta)^{2} + (\Delta \phi)^2}$ separation of the jets that constitute $H_{1}$ and $H_{2}$;
\item[$\blacksquare$] the cosine of the angle formed by one of the Higgs candidates with respect to the beam line axis in the HH system rest frame.
\end{itemize}

\begin{figure}[h!]
    \centering
    \begin{subfigure}[t]{0.30\textwidth}
    	\centering
        \includegraphics[width=0.99\textwidth]{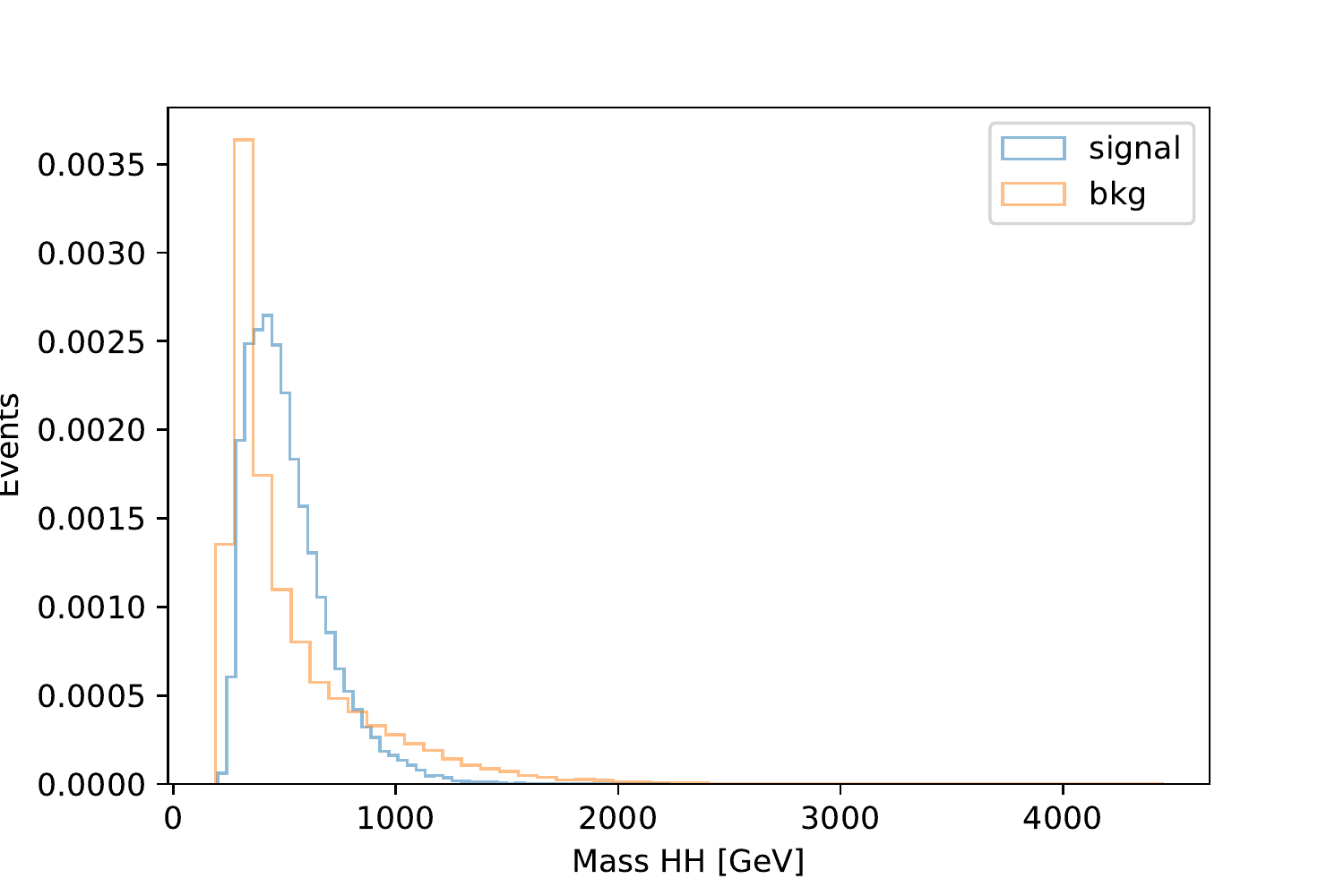}
    \end{subfigure}
    \begin{subfigure}[t]{0.30\textwidth}
       \centering
       \includegraphics[width=0.99\textwidth]{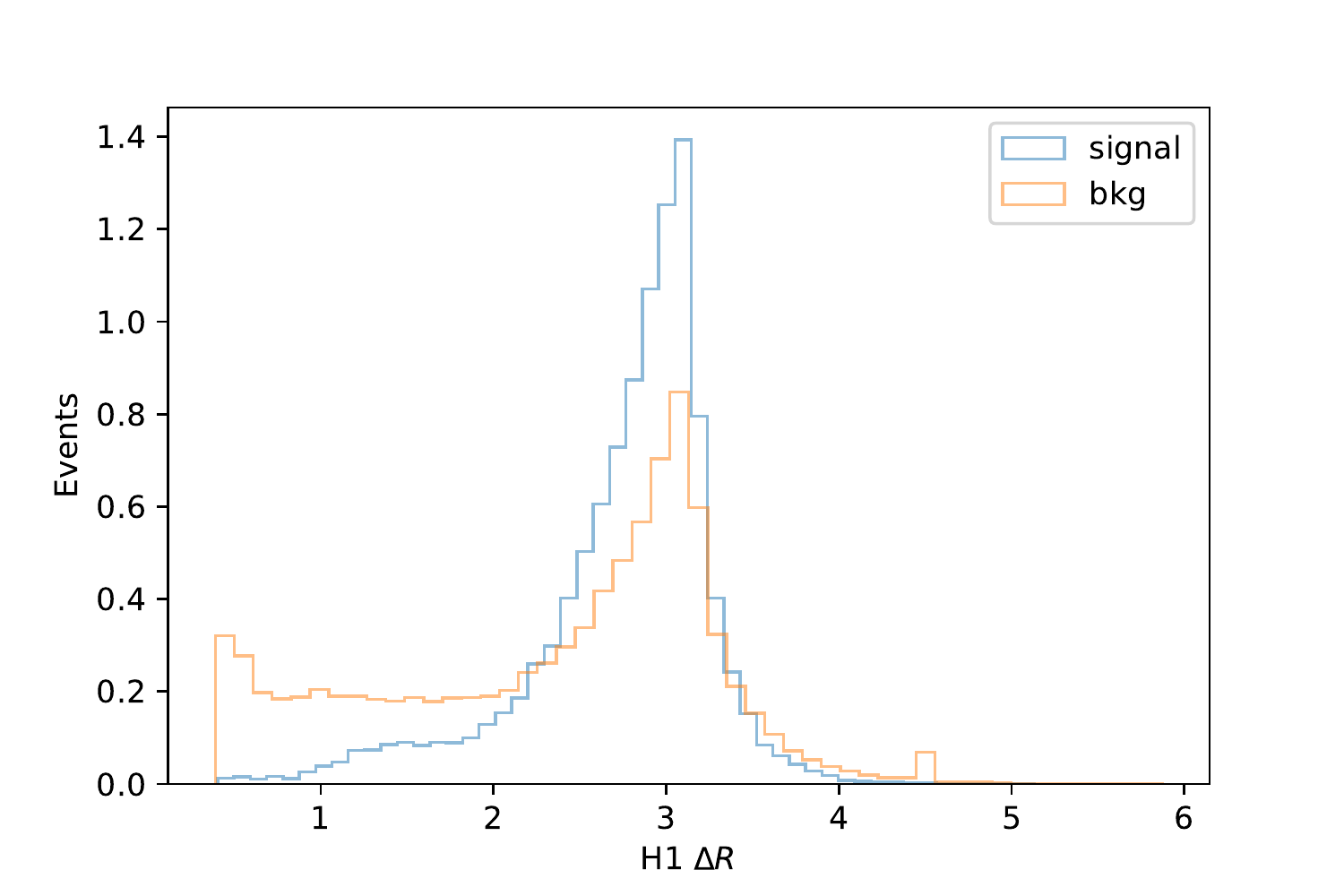}
   \end{subfigure}
   \begin{subfigure}[t]{0.30\textwidth}
       \centering
       \includegraphics[width=0.99\textwidth]{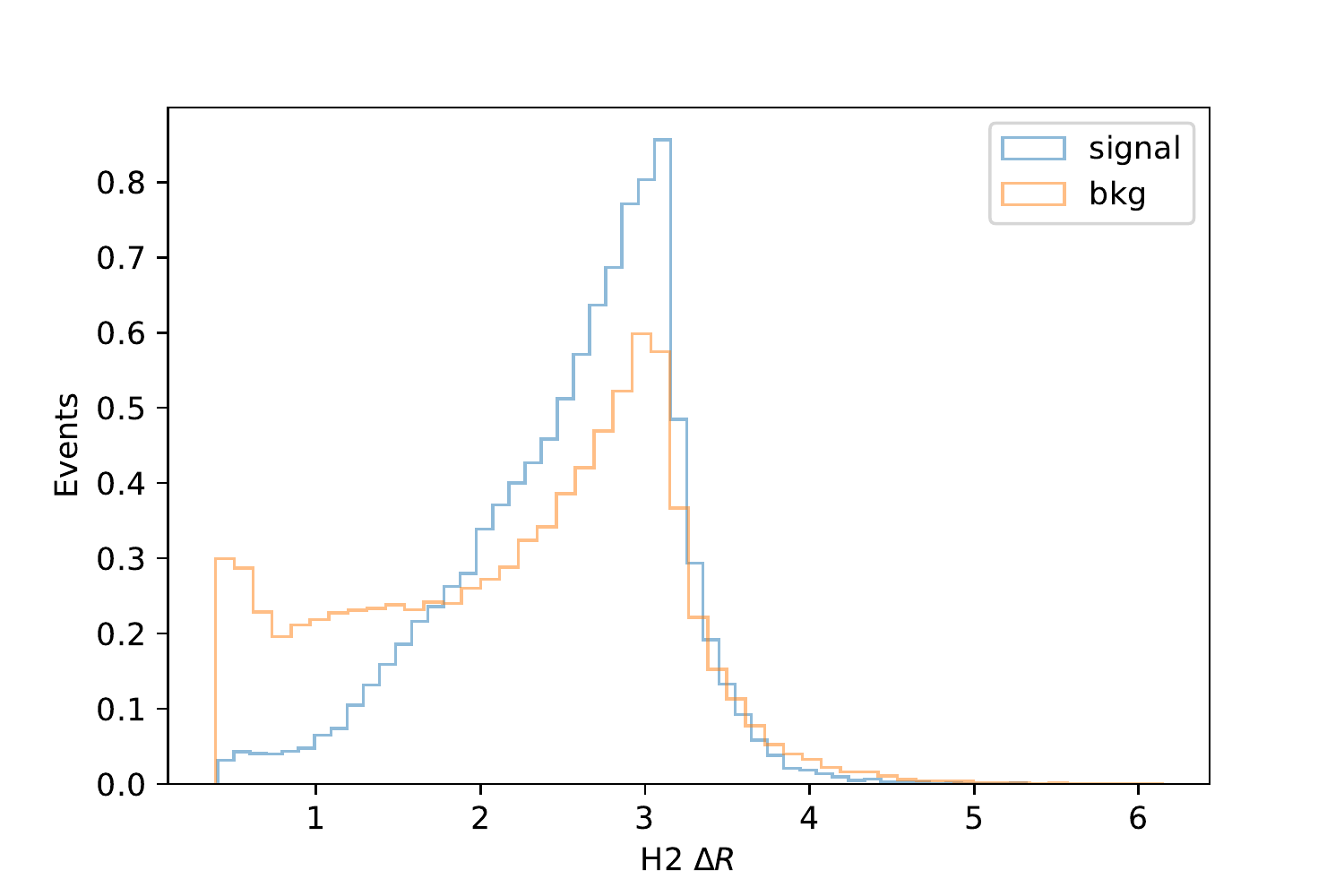}
   \end{subfigure}
    \begin{subfigure}[t]{0.30\textwidth}
       \centering
       \includegraphics[width=0.99\textwidth]{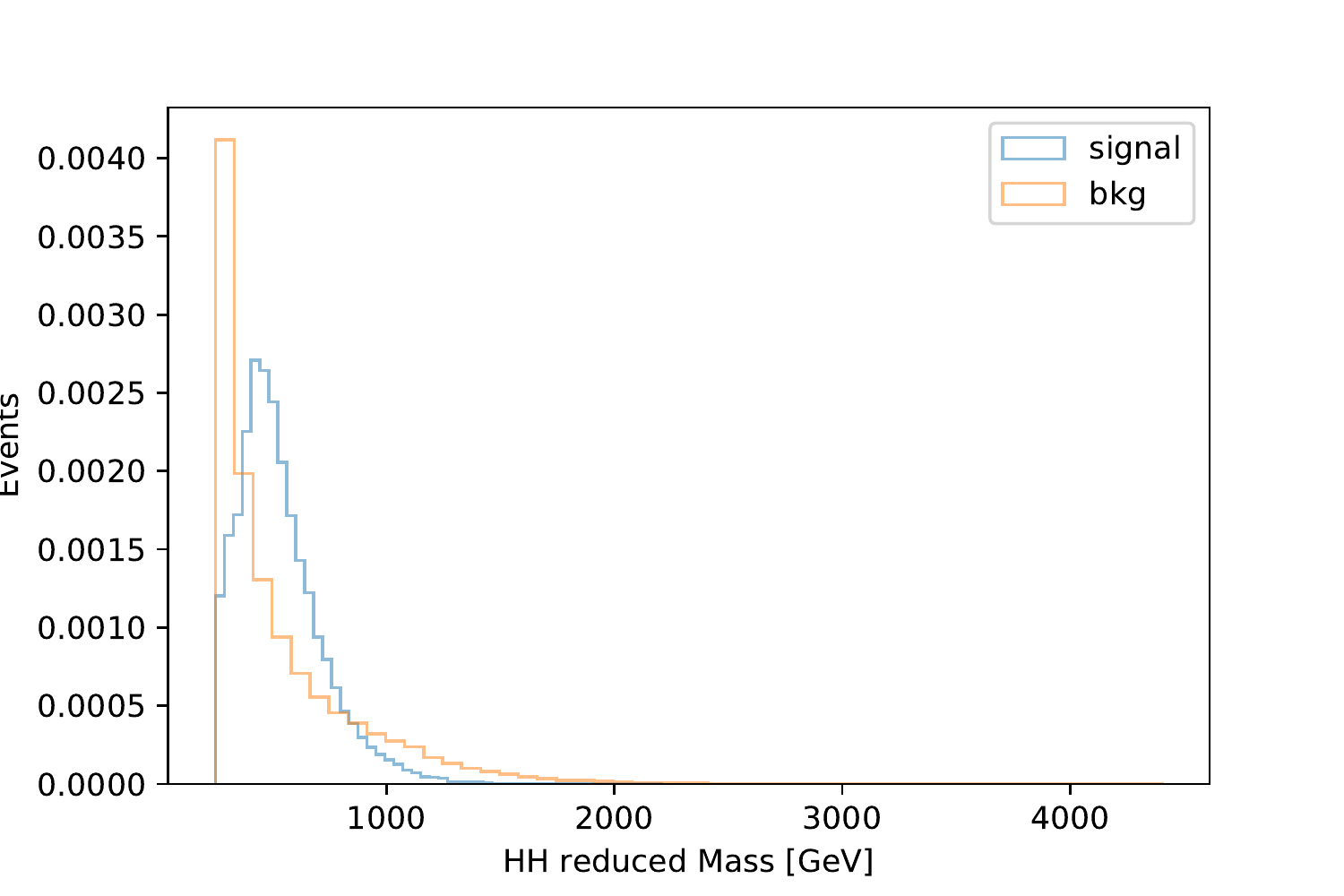}
   \end{subfigure}
    \begin{subfigure}[t]{0.30\textwidth}
       \centering
       \includegraphics[width=0.99\textwidth]{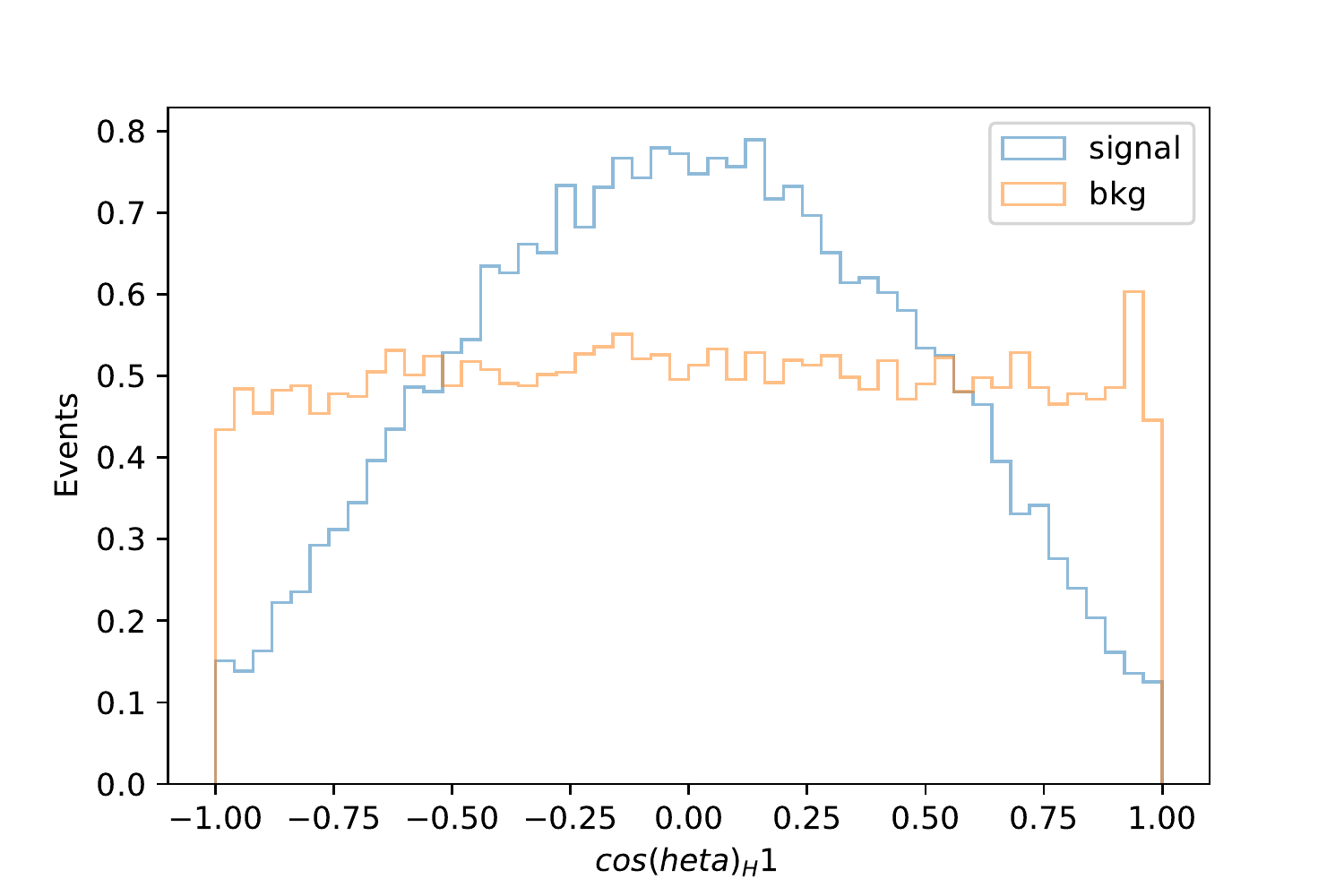}
   \end{subfigure}
   \begin{subfigure}[t]{0.30\textwidth}
       \centering
       \includegraphics[width=0.99\textwidth]{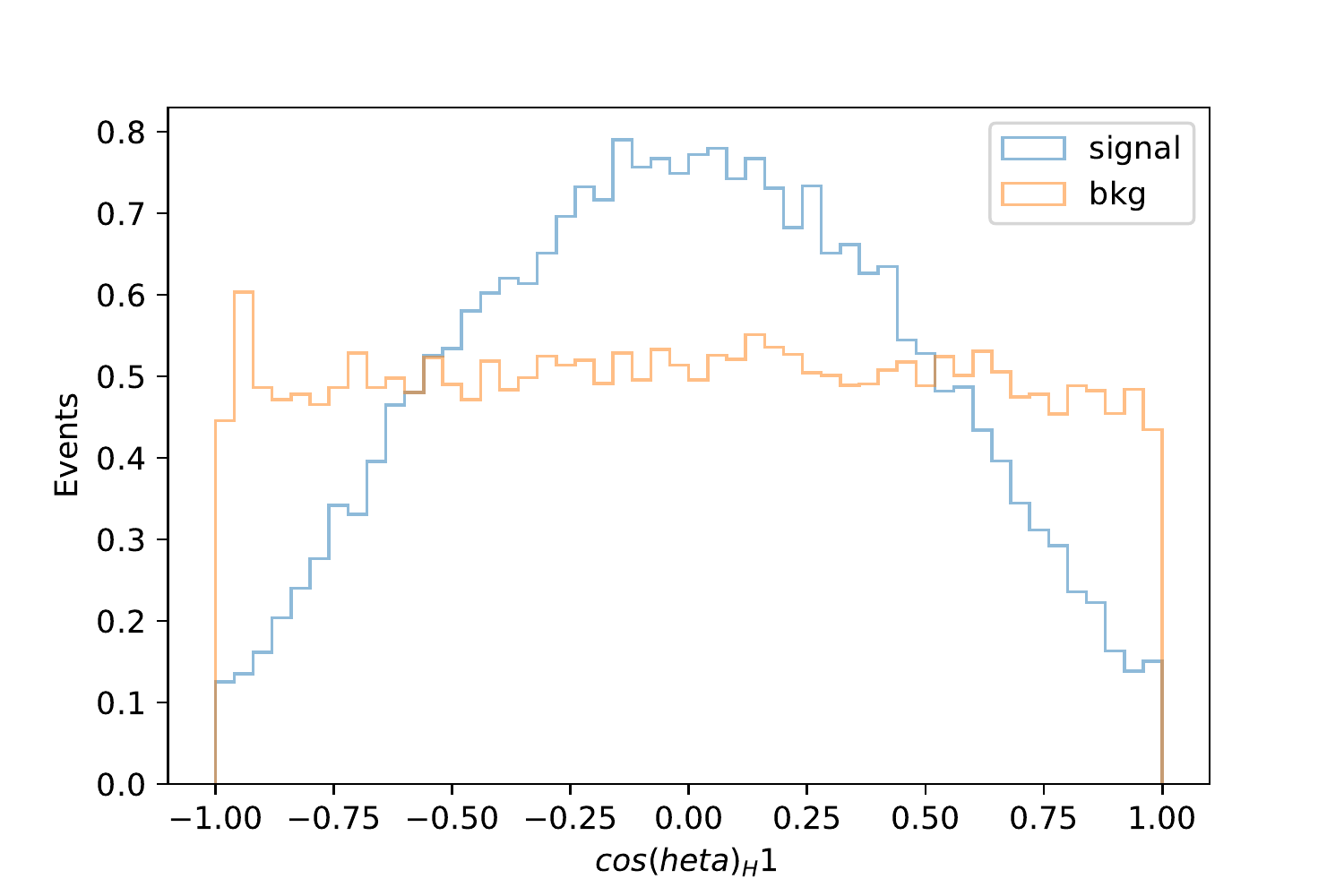}
   \end{subfigure}
   \caption{Most discriminating input variables used in the DNN training}
   \label{pic:DNN4b_input}
\end{figure}

Some of the most discriminating input variables are shown in Figure~\ref{pic:DNN4b_input}.

\begin{figure}[h!]
    \centering
    \begin{subfigure}[t]{0.45\textwidth}
    	\centering
        \includegraphics[width=0.99\textwidth]{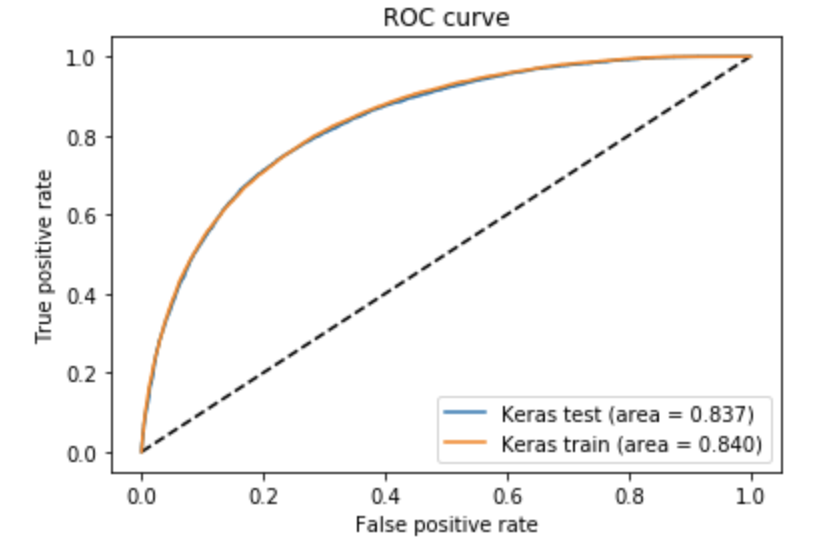}
    \end{subfigure}
    \begin{subfigure}[t]{0.45\textwidth}
    	\centering
        \includegraphics[width=0.99\textwidth]{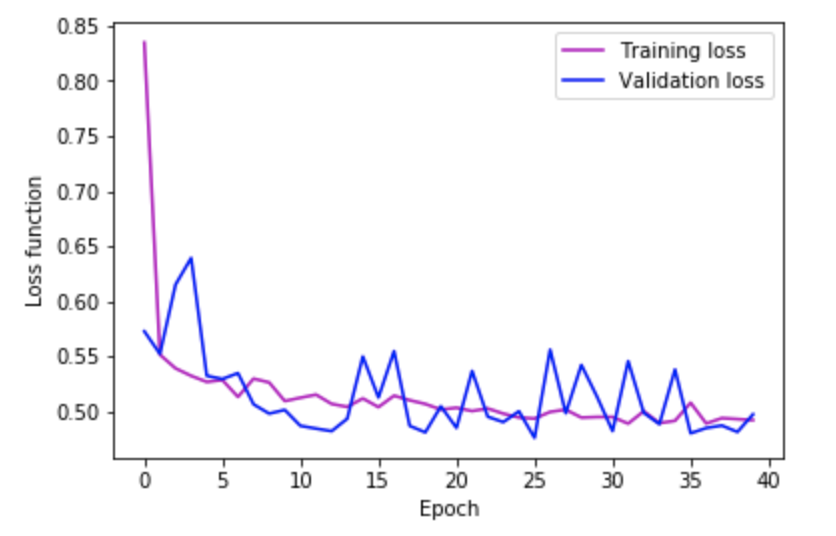}
    \end{subfigure}
   \caption{(Right) AUC for the DNN for the testing and training samples (Left) distribution training and validation loss as a function of the epochs}
   \label{pic:dnn4b_over}
\end{figure}

Moreover Figure \ref{pic:dnn4b_over} shows the reliability of the learning algorithm with no hints of overtraining or undertraining, identified by a training and validation loss curves that decrease to a point of stability with a minimal gap between the two curves.

\begin{figure}[h!]
    \centering
     \includegraphics[width=0.5\textwidth]{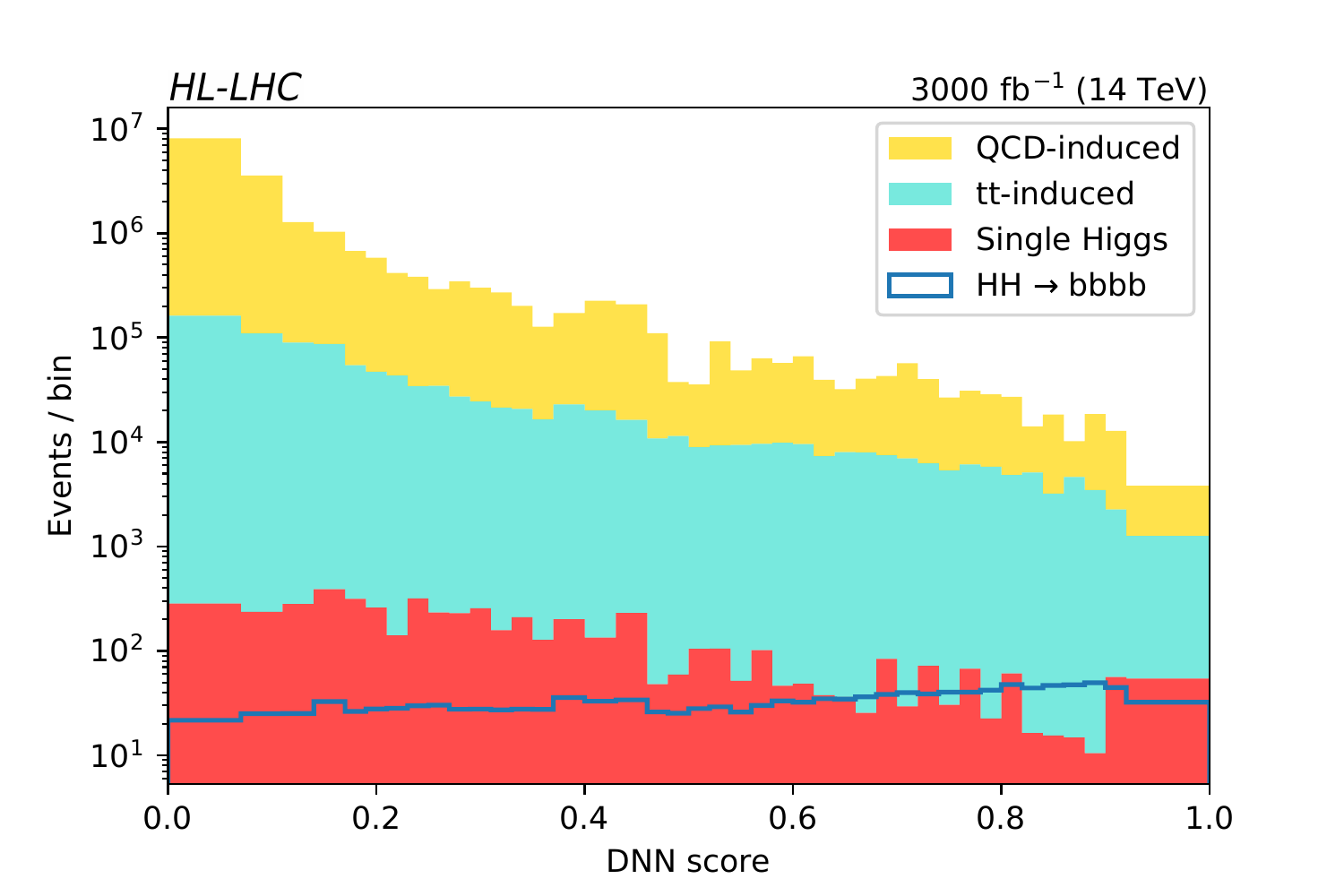}
   \caption{(Left) (Right)}
   \label{pic:DNN4b}
\end{figure}

\noindent The output of the DNN is used as the discriminant variable to look for the presence of a signal as an excess at high output values. The expected distribution of signal and background events is illustrated in Figure~\ref{pic:DNN4b}. The binning of the distribution is optimised to maximise the sensitivity to the SM HH signal and to ensure 5\% stat uncertainties in each bin of the distribution.

\noindent A complete list of the systematic uncertainties used for the bbbb channel and the expected impact on the event yields are reported in Table~\ref{tab:sys4b}.

\begin{table}[h!]
	\centering
    \begin{tabular}{c|c } 
    \hline
     \textbf{Systematic uncertainty source} & \textbf{Impact on yields} \\
    \hline
    Luminosity & $\pm$ 1.0 \% \\
    Jet Energy Scale  & $\pm$ 1.0\% \\
    B-tag efficiency & $\pm$ 1.0\% \\
	\hline
	QCD scale tt inclusive& +2.4\% /-3.6\% \\
    \hline
    Pdf scale tt inclusive& $\pm$4.2\%  \\
    \hline
    \multirow{3}{*}{{\makecell[c]{Signal theoretical \\ uncertainties}}}
    & +2.1\% /-4.9\% (QCD scale)\\
    & $\pm$3.0\% (pdf scale)\\
    & +4.0\% /-18.0\% (top mass)\\
    \hline
	\end{tabular}
	\caption{Systematic uncertainties for $bbbb$ channel.}. 
	\label{tab:sys4b}
\end{table}

\subsection{Statistical analysis and results}

\begin{figure}[h!]
    \centering
     \includegraphics[width=0.5\textwidth]{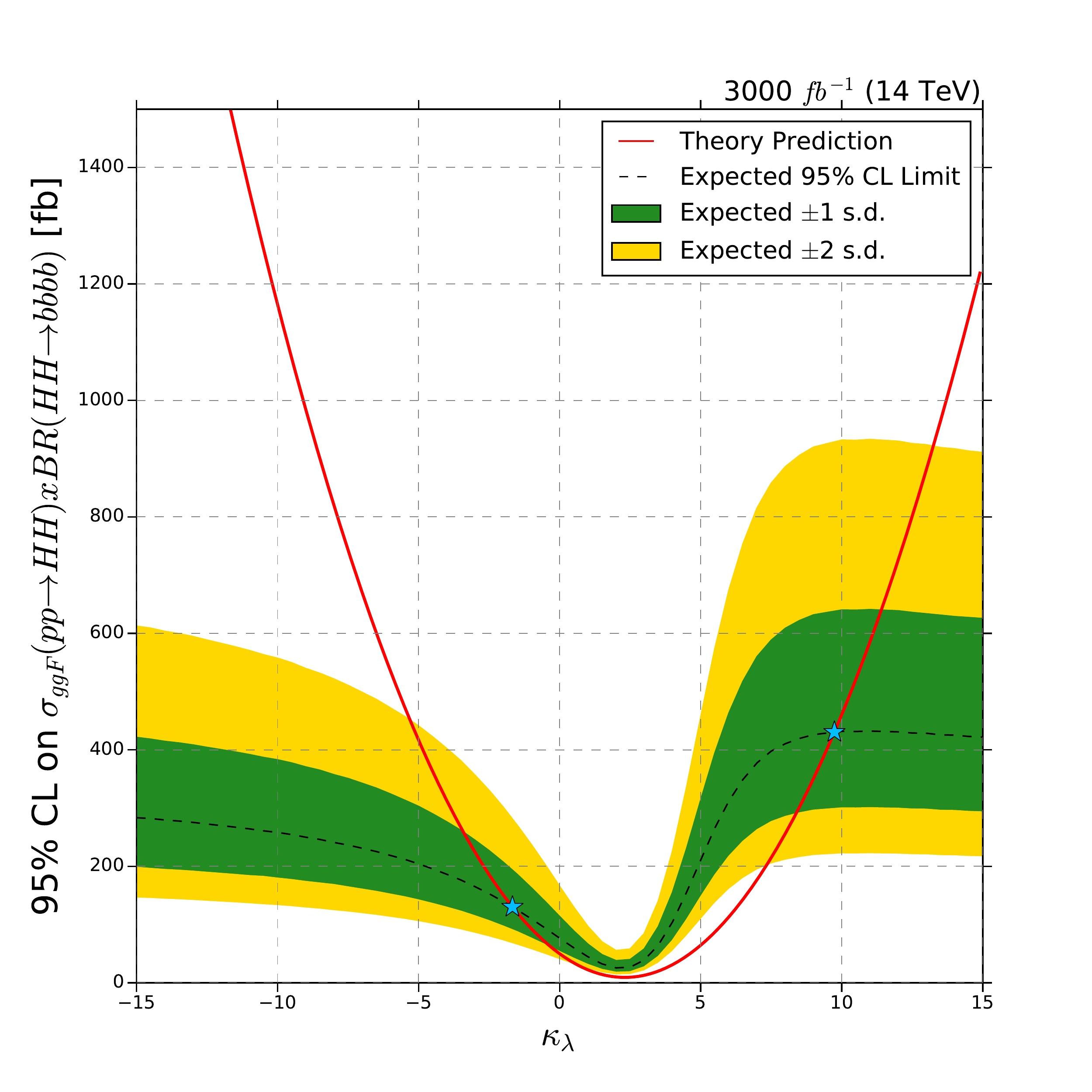}
   \caption{(Left) (Right)}
   \label{pic:klamda}
\end{figure}

The 95\%C.L. on the SM HH signal cross section times the branching fraction for the bbbb decay channel is reported in Table~\ref{tab:results_4b}; the significance for the discovery of the double Higgs for this channel at the HL-LHC is also included while considering the statistical only uncertainty and the combination with the systematic one. 
The 95\%C.L. on the SM HH signal cross section times the branching fraction for the bbbb decay channel is also derived as a function of the $\kappa_{\lambda}$ anomalous coupling ratio to the SM value, as shown in Figure~\ref{pic:klamda}; the expected constraint on the $\kappa_{\lambda}$ is $-1.32<\kappa_{\lambda}< 9.13$ at 68\% C.L, is reported in Table~\ref{tab:results_4b} as well.

\begin{table}[h!]
	\centering
    \begin{tabular}{c|c|c|c } 
    \hline
    
     \textbf{Condition} & \textbf{Significance (in $\sigma$)} & \makecell[c]{\textbf{Upper limit on} \\ \textbf{ $\mu$ at 95\% CL}} & \textbf{$\kappa_{\lambda}$ constraint}\\
     
    \hline
    
    stat only & 1.43 & 1.37 & [0.10,5.60] \\
     
    stat + sys & 1.06 & 2.00 & [-1.67, 9.74] \\
     
    \hline
	\end{tabular}
	\caption{Results for $bbbb$ channel.}. 
	\label{tab:results_4b}
\end{table}

\clearpage
\clearpage
\section{14 TeV HH combination}
 
The results obtained in each of the three decay channels described in this paper are combined together assuming the SM branching fractions for HH decays to the studied final states.\\
The analyses of the three decay channels are designed to be orthogonal thanks to the mutually exclusive object selection used for each channel. Systematic uncertainties on the theoretical assumptions or associated to the same object, such as b tagging efficiency, are treated as correlated, while all the others are left uncorrelated.\\
The upper limit on the signal strength for the HH combination is 0.76 corresponding to a significance of 2.80; those results are improved of about 8\% with respect to the previous projections. Results are summarised in Table~\ref{tab:results_comb}.
 
\begin{table}[h!]
	\centering
    \begin{tabular}{c|c|c|c|c } 
    \hline
     \textbf{Channel} & \textbf{Condition} & \textbf{Significance (in $\sigma$)} & \makecell[c]{\textbf{Upper limit on $\mu$} \\ \textbf{at 95\% CL}} & \textbf{$\kappa_{\lambda}$ constraint}\\
    \hline
    \multirow{2}{*}{$HH\rightarrow b\bar{b}\gamma\gamma$} & stat only & 1.99 & 0.99 & [1.01,4.36] \\
    &stat + sys & 1.94 & 1.09 & [0.87, 4.48] \\
    \hline
    \multirow{2}{*}{$HH\rightarrow b\bar{b}\tau\tau$} & stat only & 1.72 & 1.25 & [0.56,6.72] \\
    & stat + sys & 1.70 & 1.37 & [0.37, 6.97] \\
    \hline
    \multirow{2}{*}{$HH\rightarrow b\bar{b} b\bar{b}$} & stat only & 1.43 & 1.37 & [0.10,5.60] \\
    &stat + sys & 1.06 & 2.00 & [-1.67, 9.74] \\
     \hline
     \hline
    \multirow{2}{*}{\textbf{Combination}} & stat + syst & 2.80 & 0.76 & [1.54, 4.02] \\
   & stat only & 2.99 & 0.66 & [1.86,4.03] \\
    \hline
	\end{tabular}
	\caption{Results for $bb\gamma\gamma$, $bb\tau\tau$ and $bbbb$ channels and their combination.}. 
	\label{tab:results_comb}
\end{table}

\begin{figure}[h!]
    	\centering
        \includegraphics[width=0.5\textwidth]{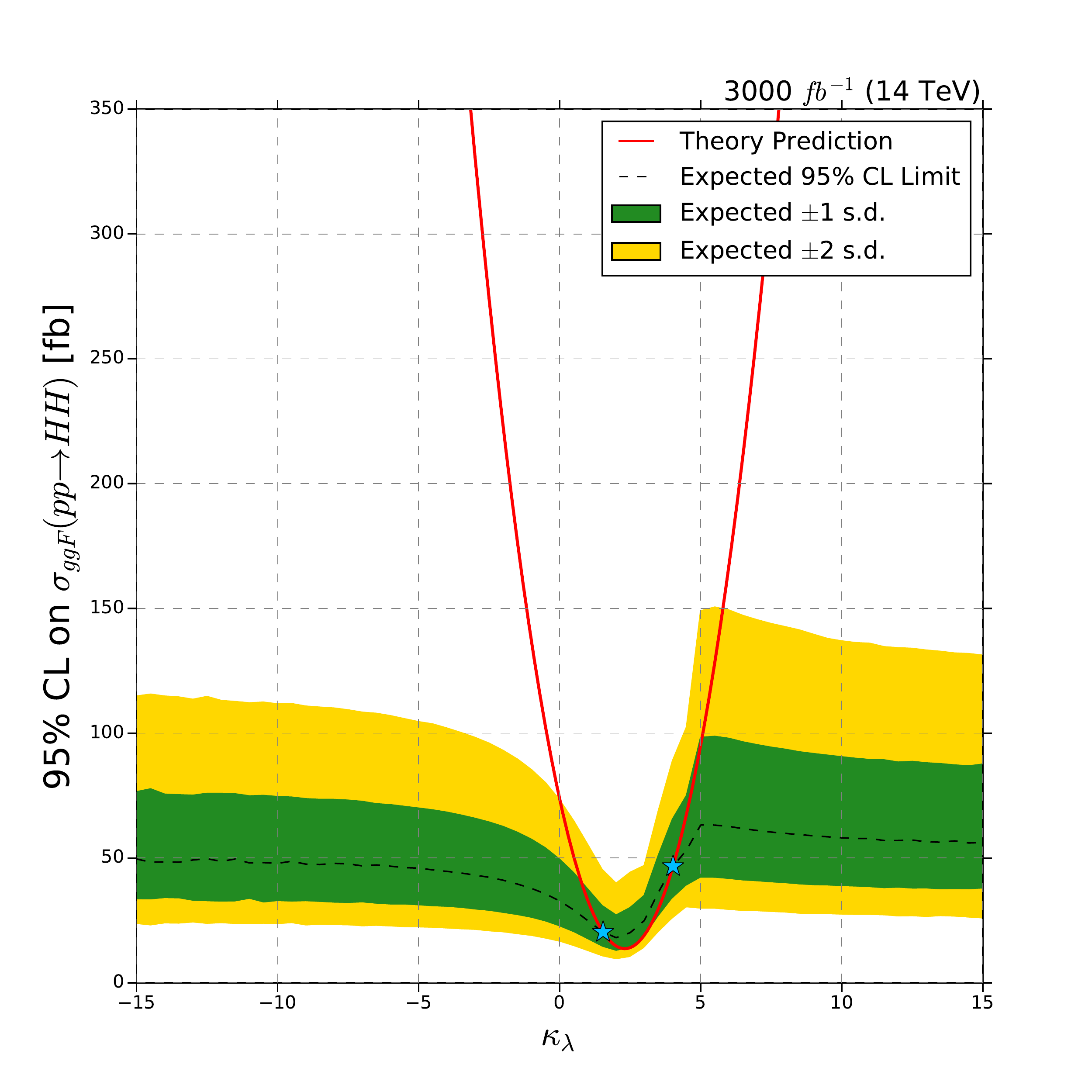}
        \caption{Expected upper limit at the 95\% CL on the HH production cross section as a function of $\kappa_{\lambda}$ with $1\sigma$ and $2\sigma$ bands. The red curve indicates the theoretical prediction.}
   \label{pic:kl_comb}
    \end{figure}

\noindent Under the assumption that no HH signal exists, 95\% CL upper limits on the SM HH production cross section are derived as a function of $\kappa_{\lambda}$ as shown in Figure~\ref{pic:kl_comb}. A variation of the excluded cross section, directly related to changes in the HH kinematic properties, can be observed as a function of $\kappa_{\lambda}$. With the HH combination we are able to constrain the $\kappa_{\lambda}$ to be within 1.54 and 4.02 at 95\% CL.\\
 
\begin{figure}[h!]
    	\centering
        \includegraphics[width=0.7\textwidth]{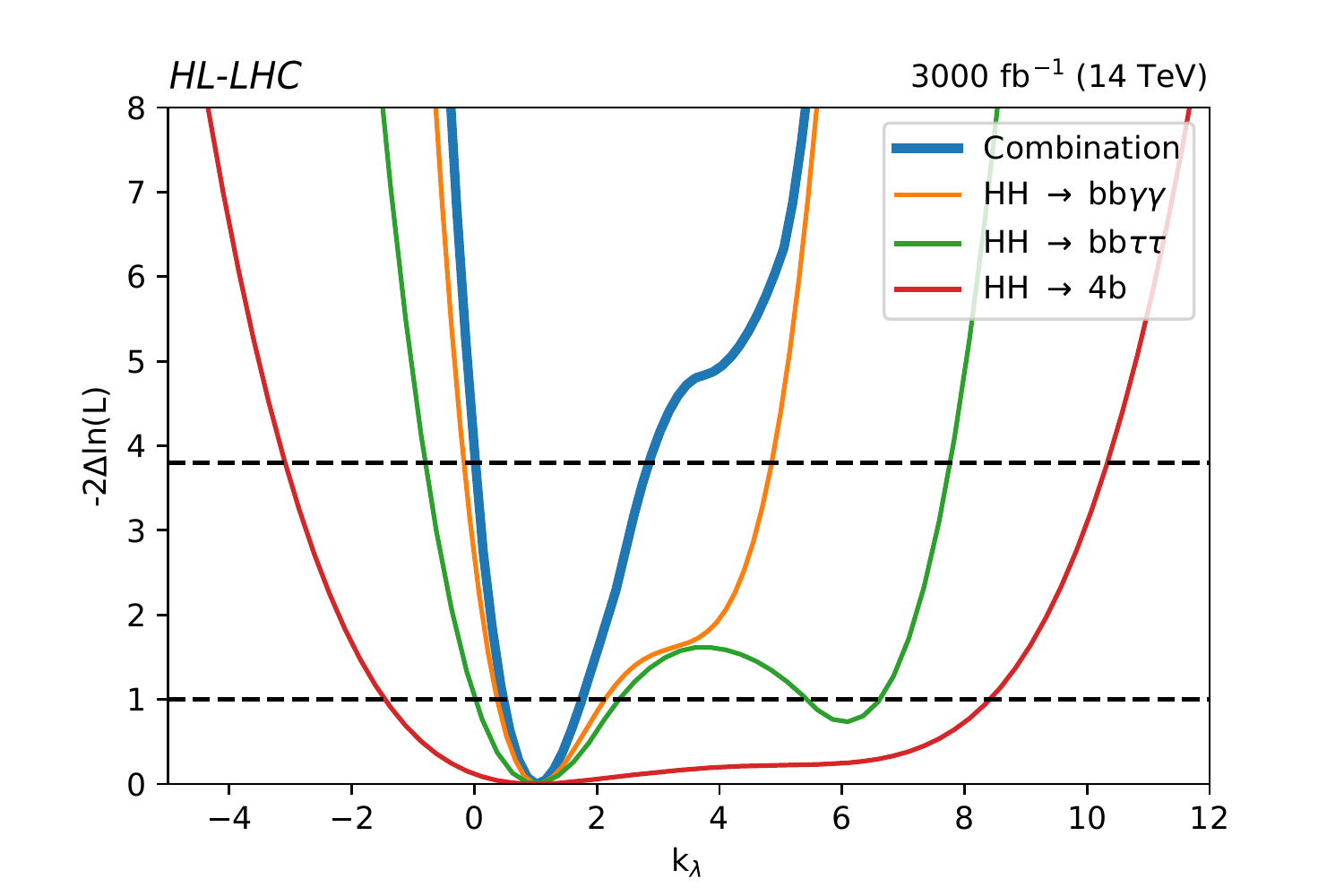}
        \caption{Expected likelihood scan as a function of $\kappa_{\lambda}$. The functions are shown separately for the three decay channels studied and for their combination.}
   \label{pic:log_comb}
    \end{figure}

\noindent Assuming, instead, the presence of a HH signal with the properties predicted by the SM, prospects for the measurement of the $\kappa_{\lambda}$ are given. The scan of likelihood as a function of the $\kappa_{\lambda}$ for each channel and for the combination are shown in Figure~\ref{pic:log_comb}. The expected confidence interval of this coupling for each channel and for the combination are summarised in Table~\ref{tab:const_14}; in particular, for the HH combination, we expect $\kappa_{\lambda}$ in tha ranges [0.47, 1.76] at 68\% CL and [-0.02, 3.05] at 95\% CL.\\

\begin{table}[h!]
\centering
\begin{tabular}{l|l|l}
\hline

                                        & \textbf{$k_{\lambda}$ constraint at 68\% CL}   & \textbf{$k_{\lambda}$ constraint at 95\% CL} \\
                                        
\hline

$HH \rightarrow b\bar{b}\gamma\gamma$     & {[}0.37, 2.13{]}                     & {[}-0.20, 4.9{]}                  \\

$HH \rightarrow b\bar{b}\tau^{+}\tau^{-}$ & {[}0.01, 2.35{]} \& {[}5.41, 6.61{]} & {[}-0.84,7.75{]}                   \\

$HH \rightarrow 4b$
& {[}-1.46, 8.41{]}                    & {[}-3.16, 10,41{]}                 \\

\hline
\hline

\textbf{HH combination}                            & {[}0.46, 1.73{]}                     & {[}-0.02, 3.05{]}                   \\

\hline
\end{tabular}
\caption{$k_{\lambda}$ constraint for $bb\gamma\gamma$, $bb\tau\tau$ and $bbbb$ channels and their combination.}
\label{tab:const_14}
\end{table}

\noindent The peculiar HH likelihood function structure, characterised by a double minimum can be observed. This shape is explained recalling that the HH cross section has a the quadratic dependence on the $k_{\lambda}$ with a minimum at $k_{\lambda} = 2.4$, that corresponds to the maximum interference of the box and triangle Feynman diagram of the HH production. Moreover, the kinematic differences for signals with $\kappa_{\lambda}$ values symmetric around this minimum are relevant in the low region of $m_{HH}$ spectrum.
Consequently, a partial degeneracy can be observed between the $k_{\lambda} = 1$ value and a second $k_{\lambda}$ value. The exact position and the height of this second minimum depends on the sensitivity of the analysis to the $m_{HH}$ spectrum.\\
For $b\bar{b} \tau\tau$ and $b\bar{b}b\bar{b}$ the degeneracy is partially removed thanks to the wide use of machine learning techniques that, having $m_{HH}$ as input, are able to fully capture the $m_{HH}$ vs $k_{\lambda}$ dependency. For $b\bar{b} \gamma \gamma$ that has a dedicated $m_{HH}$ low region categorisation the discrimination on the second minimum is even better.
In the combination, all these effects are enhanced: the double minimum is almost gone and appears for values higher than 3$\sigma$. This is another strong proof of the improvement of the analyses techniques in the three different final states.

\clearpage
\section{Projections for FCC-hh at 100 TeV}

The same channels have been studied also in a future hypothetical scenario at 100 TeV, to investigate the physics potential of the proposed hadronic machine FCC-hh in the double Higgs searches. 
\\For these projections, new signal samples has been simulated at 100 TeV for the three different values of $\kappa_{\lambda}$, while the background processes are scaled from the 14 TeV samples taking into account only the most important ones and considering 30 $ab^{-1}$ of integrated luminosity and 1000 pile up events.\\
For 100 TeV scenario two sets of systematics are considered:

\begin{itemize}
    \item[$\blacksquare$] scenario 14 TeV like: same sets of systematics used at 14 TeV. This scenario is the most conservative one, hereafter referred as scenario 1;
    \item[$\blacksquare$] optimistic 100 TeV scenario (Table \ref{tab:sys_bbgg_scen0}): the set of systematics here has been chosen taking into account a substantial improvement that will happen in the 100 TeV scenario, hereafter referred as scenario 0.
\end{itemize}

\begin{table}[h!]
	\centering
    \begin{tabular}{c|c } 
    \hline
     \textbf{Systematic uncertainty source} & \textbf{Impact on yields} \\
    \hline
    Luminosity & $\pm$ 0.5\% \\
    Photon ID efficiency  & $\pm$ 1.0 \%\\
    B-tag efficiency & $\pm$ 1.0\% \\
    Lepton ID efficiency  & $\pm$ 1.0\% \\
    Tau ID efficiency  & $\pm$ 2.0\% \\
	\hline
    Theoretical uncertainties & $\pm$ 1\% \\
    \hline
	\end{tabular}
	
	\caption{Systematic uncertainties in the optimistic scenario.}. 
	\label{tab:sys_bbgg_scen0}
\end{table}

\noindent At 100 TeV the HH production will be observed in all the channels, thus in this section we will provide the precision as a function of the luminosity with which we can measure both the signal strength and the Higgs self coupling, assuming that the HH signal exist with the properties predicted by the SM.

\subsection{$HH \rightarrow b\bar{b}\gamma\gamma$}

\begin{figure}[h!]
    \centering
    \begin{subfigure}[t]{0.45\textwidth}
    	\centering
        \includegraphics[width=0.99\textwidth]{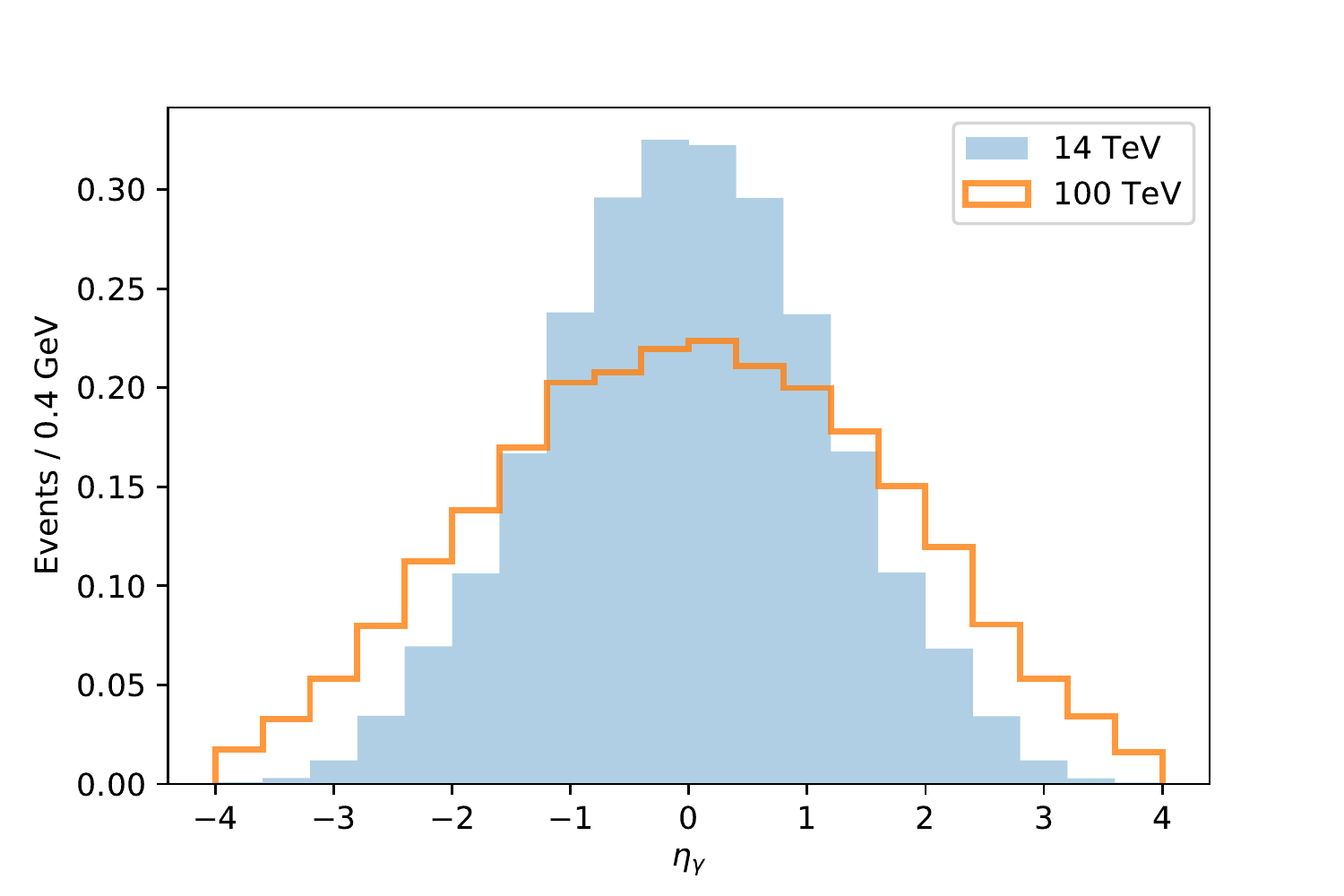}
    \end{subfigure}
    \begin{subfigure}[t]{0.45\textwidth}
       \centering
       \includegraphics[width=0.99\textwidth]{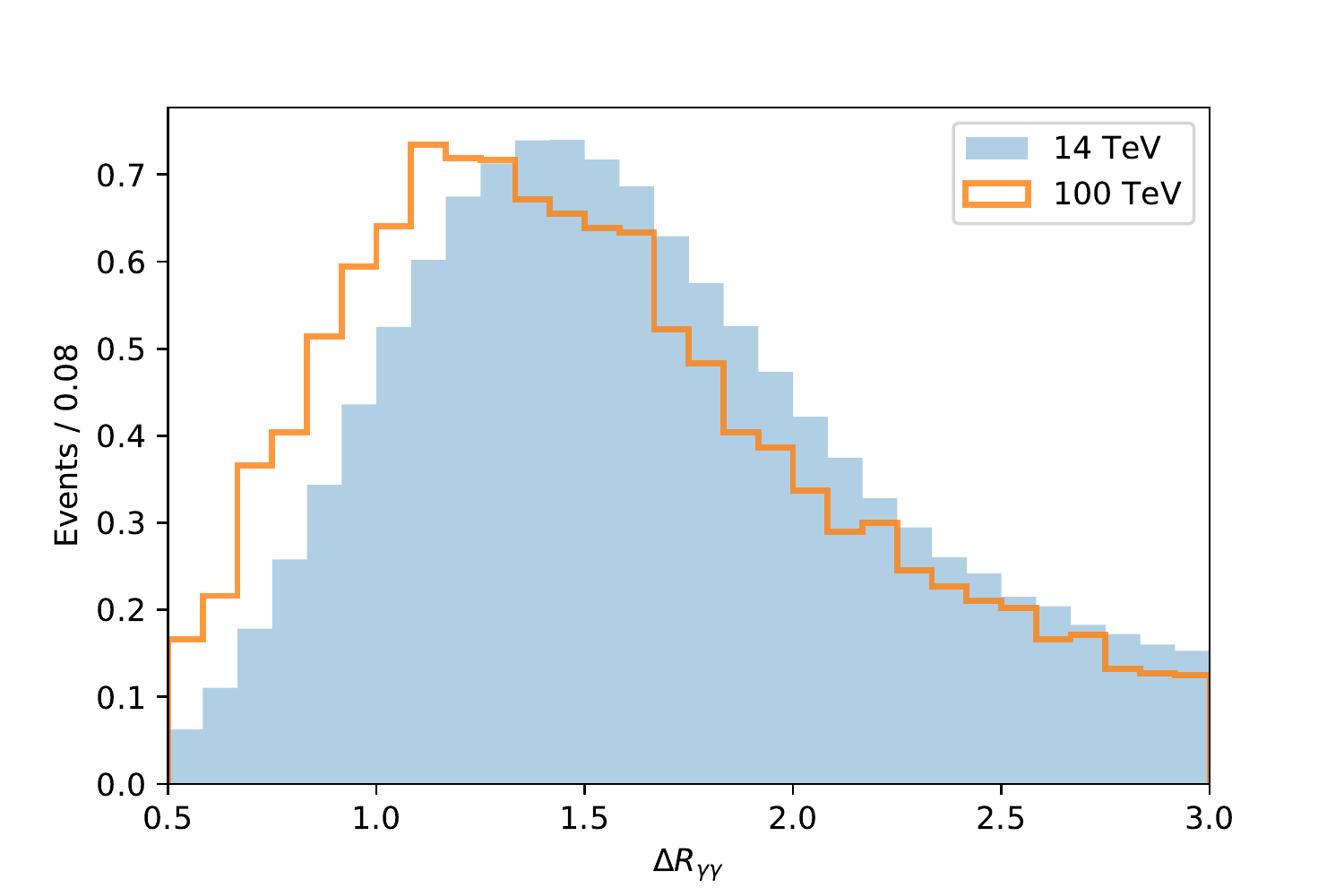}
   \end{subfigure}
   \caption{(Left) Pseudorapidity distribution of the leading photon and (Right) $\Delta R$ between the two selected photons for 100 TeV and 14 TeV scenarios. Histograms are normalized to unity. }
   \label{pic:eta_100_14}
\end{figure}

\begin{figure}[h!]
    \centering
    \begin{subfigure}[t]{0.45\textwidth}
    	\centering
        \includegraphics[width=0.99\textwidth]{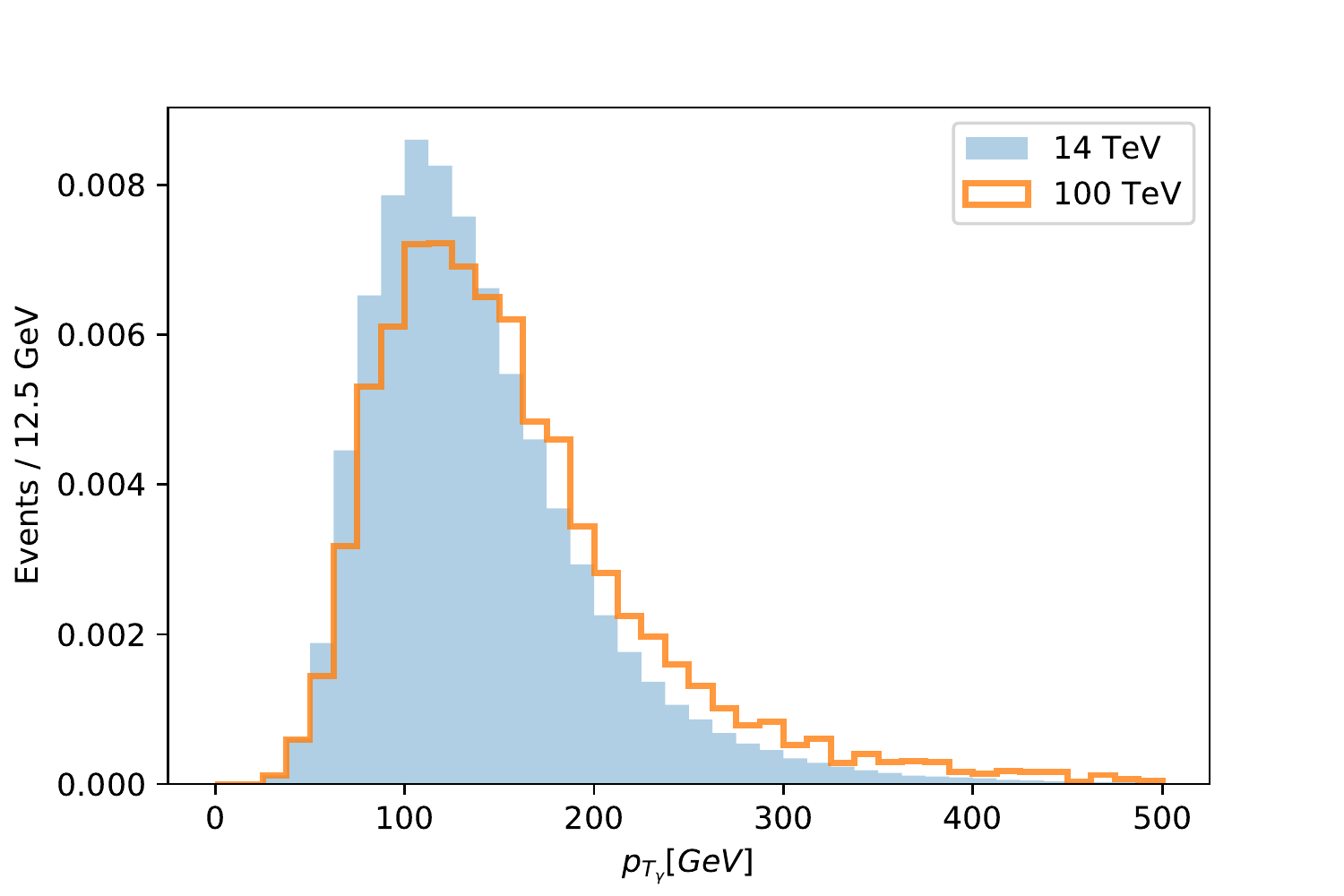}
    \end{subfigure}
    \begin{subfigure}[t]{0.45\textwidth}
       \centering
       \includegraphics[width=0.99\textwidth]{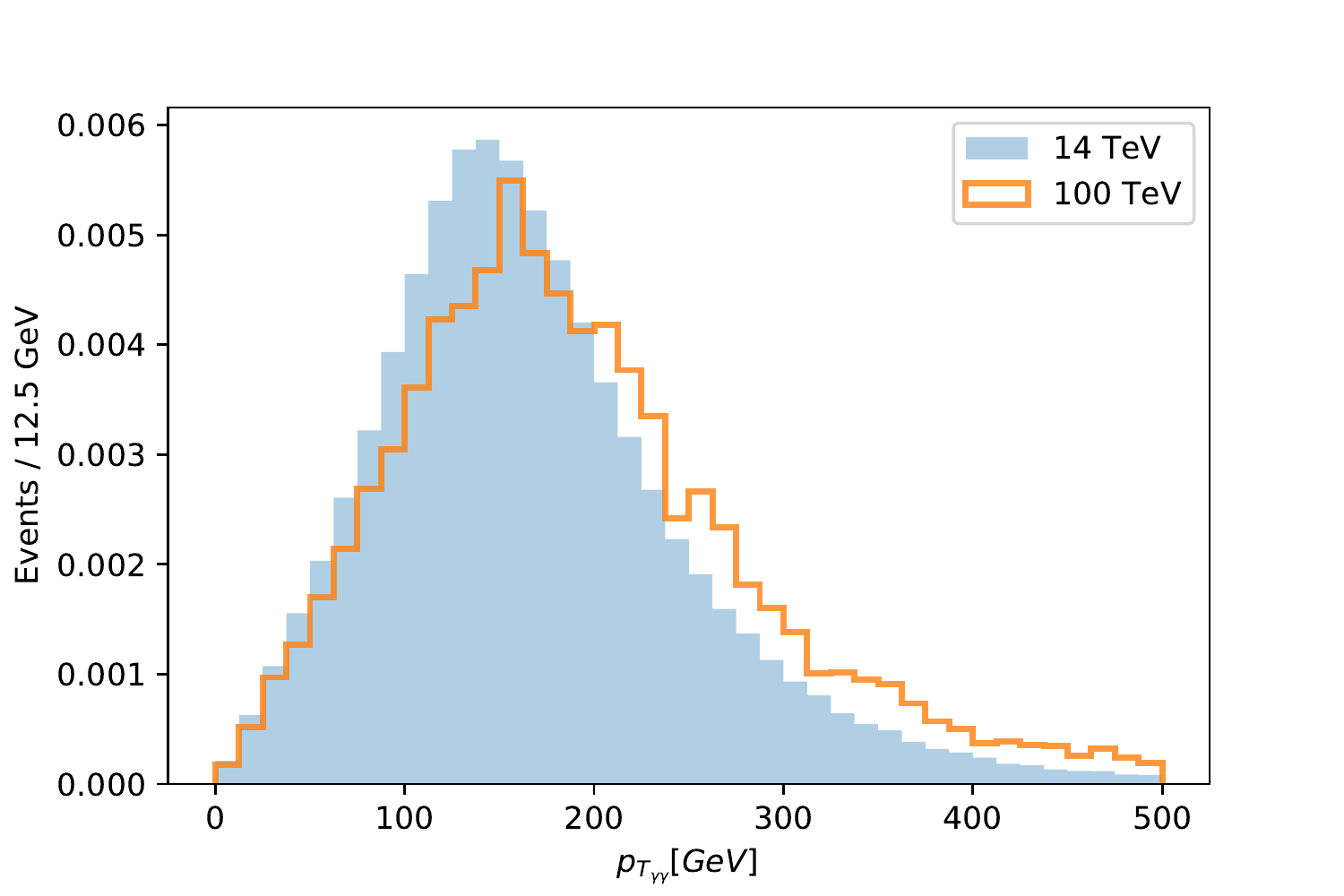}
   \end{subfigure}
   \caption{Transverse momentum distribution of the (Left) leading photon and of the (Right) di-photon candidate for 100 TeV and 14 TeV scenarios. Histograms are normalized to unity.}
   \label{pic:pt_10_14}
\end{figure}

\begin{figure}[h!]
    \centering
    \begin{subfigure}[t]{0.45\textwidth}
    	\centering
        \includegraphics[width=0.99\textwidth]{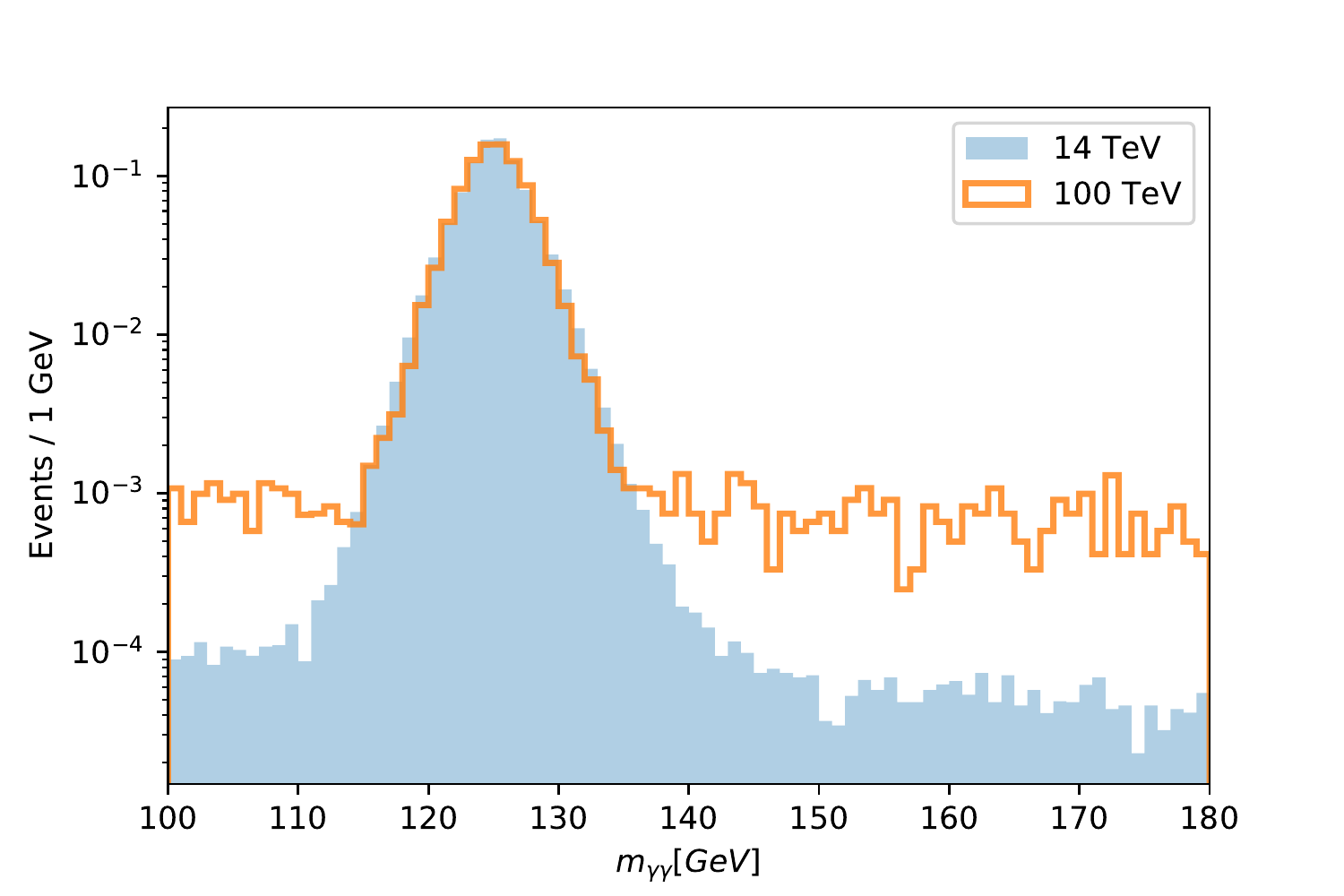}
    \end{subfigure}
    \begin{subfigure}[t]{0.45\textwidth}
       \centering
       \includegraphics[width=0.99\textwidth]{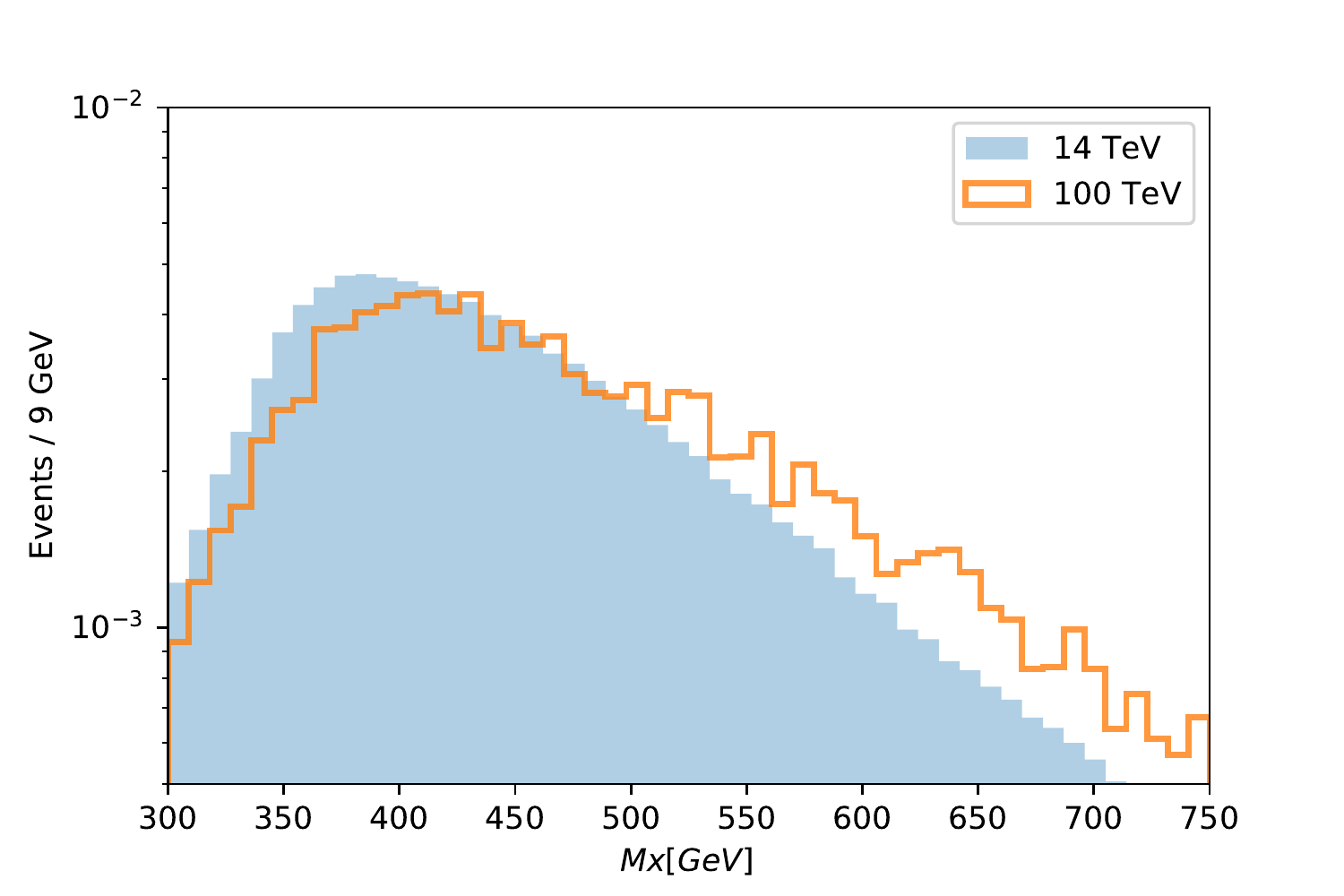}
   \end{subfigure}
   \caption{(Left) Di-photon invariant mass distribution and (Right) Mx distribution for 100 TeV and 14 TeV scenarios. Histograms are normalized to unity.}
   \label{pic:inv_mass_100_14}
\end{figure}

For the $bb\gamma\gamma$ channel, the principal backgrounds are the di-photon and single photon + jets for the non resonant ones and ttH and ggH as the resonant ones. The kinematic of the signal changes with the increase of the center of mass energy: the photon and jet objects are produced less centrally (Figure \ref{pic:eta_100_14} left), are more closed to each other (Figure \ref{pic:eta_100_14} right) and have a harder transverse momentum spectra (Figure \ref{pic:pt_10_14}). Also the invariant mass shapes are slightly changed, with the tails of the $m_{\gamma\gamma}$ distribution more populated and the Mx distribution shifted towards higher values in the 100 TeV case (Figure \ref{pic:inv_mass_100_14}).

\noindent For this reasons, some requirements are relaxed in $bb\gamma\gamma$ channel to retain the $\approx$15\% signal selection efficiency: the pseudorapidity is required to be only less than 4.0 and the restrictions on $p_{T}/m_{\gamma\gamma}$ ratio are removed.\\

\begin{figure}[h!]
    \centering
    \begin{subfigure}[t]{0.45\textwidth}
    	\centering
        \includegraphics[width=0.99\textwidth]{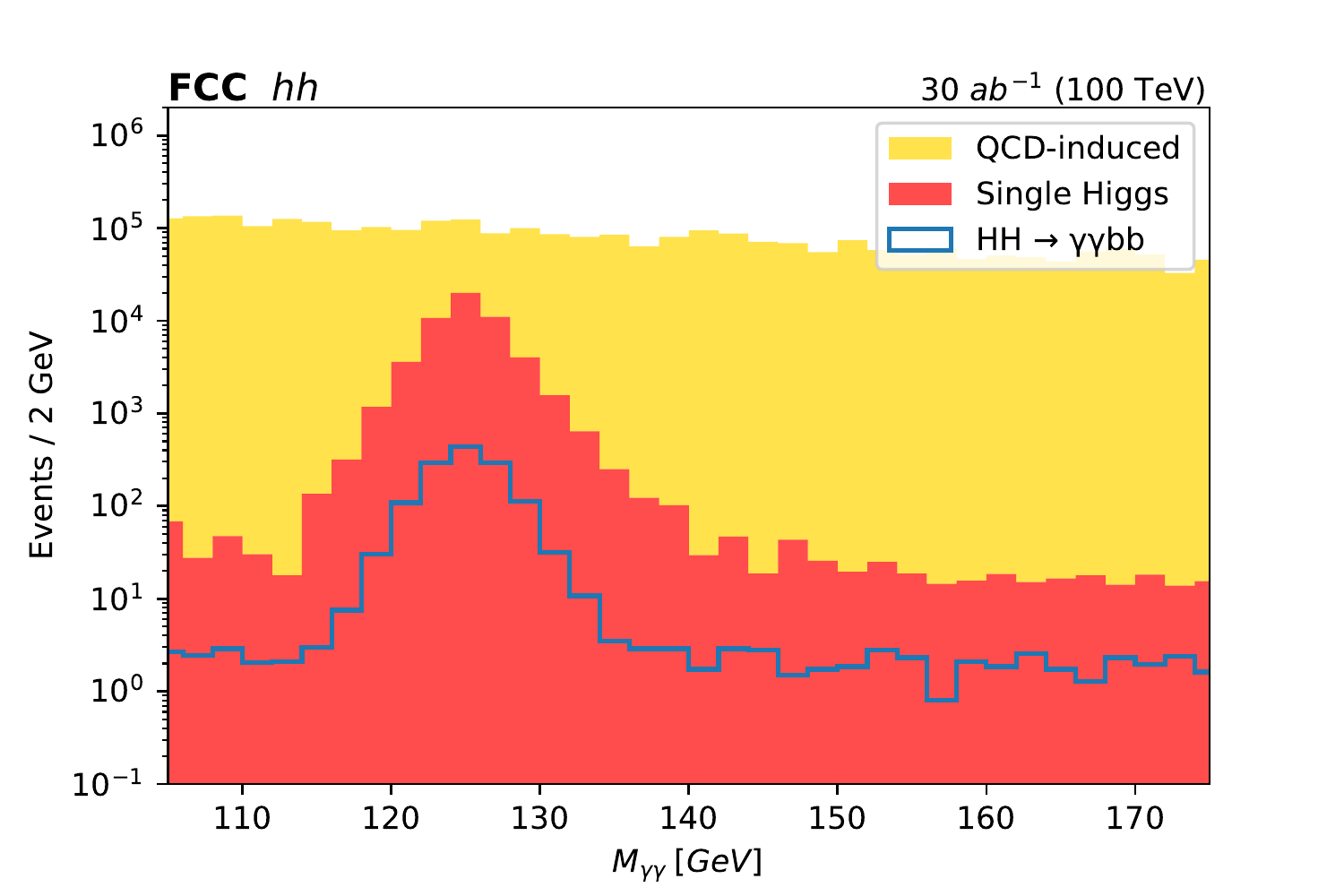}
    \end{subfigure}
    \begin{subfigure}[t]{0.45\textwidth}
       \centering
       \includegraphics[width=0.99\textwidth]{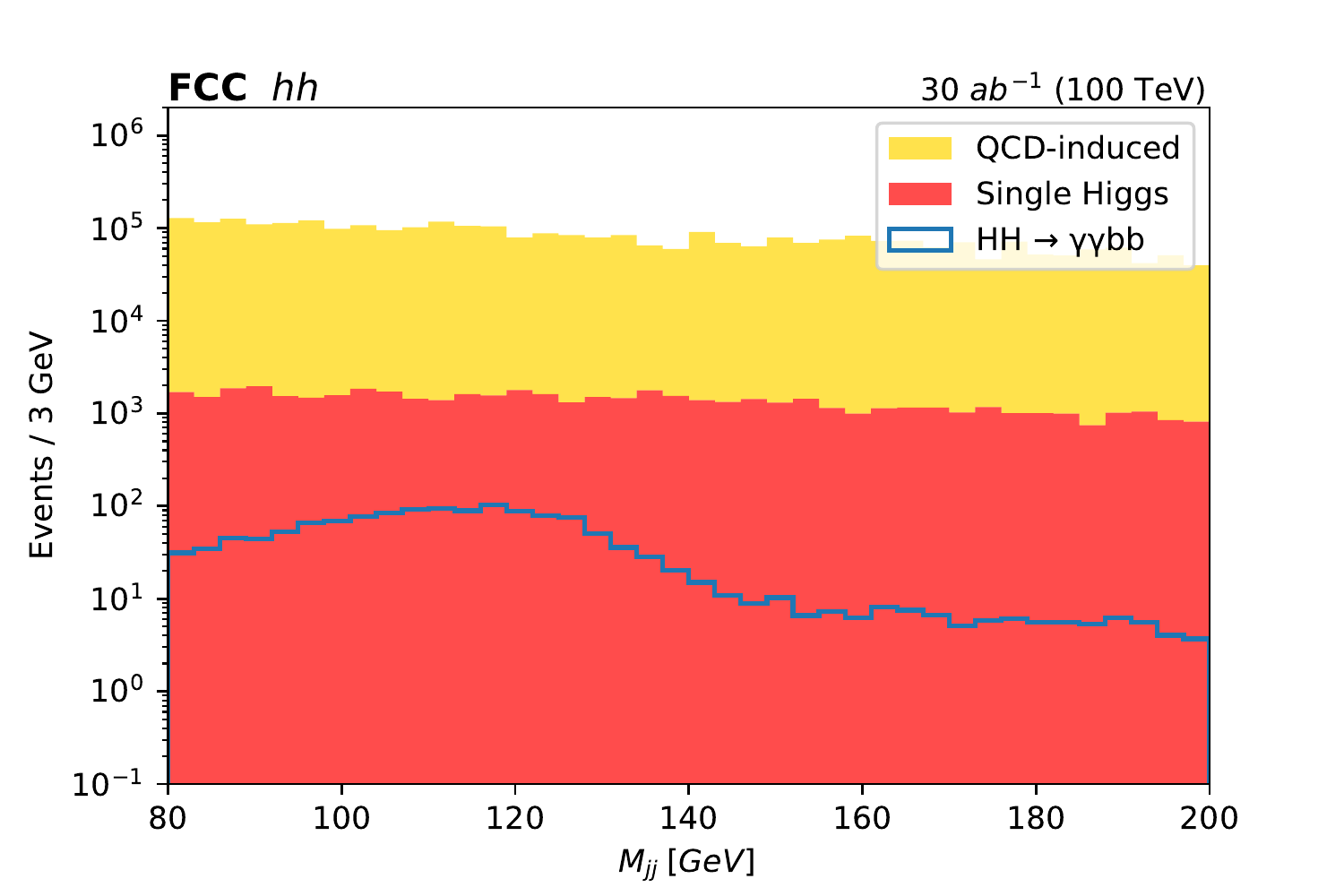}
   \end{subfigure}
   \caption{(Left) Di-photon and (Right) di-jets invariant mass after kinematic selections, for signal and background processes. Histograms are scaled to cross section and luminosity.}
   \label{pic:inv_mass_100}
\end{figure}

\begin{figure}[h!]
    \centering
        \includegraphics[width=0.5\textwidth]{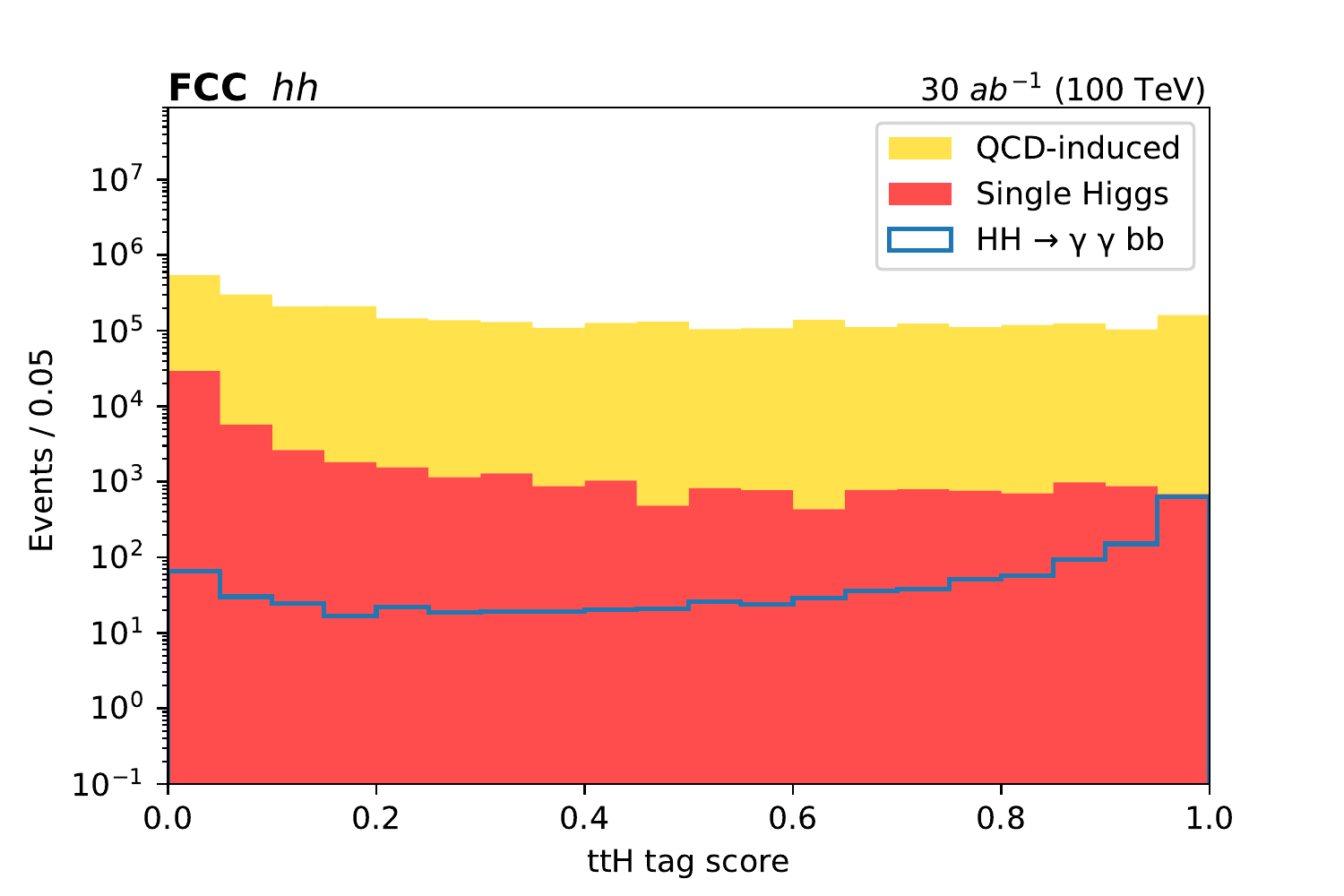}
   \caption{ttH tagger score for HH signal and stacked background scaled to cross section and luminosity.}
   \label{pic:ttH_100}
\end{figure}

\begin{figure}[h!]
    \centering
    \begin{subfigure}[t]{0.45\textwidth}
    	\centering
        \includegraphics[width=0.99\textwidth]{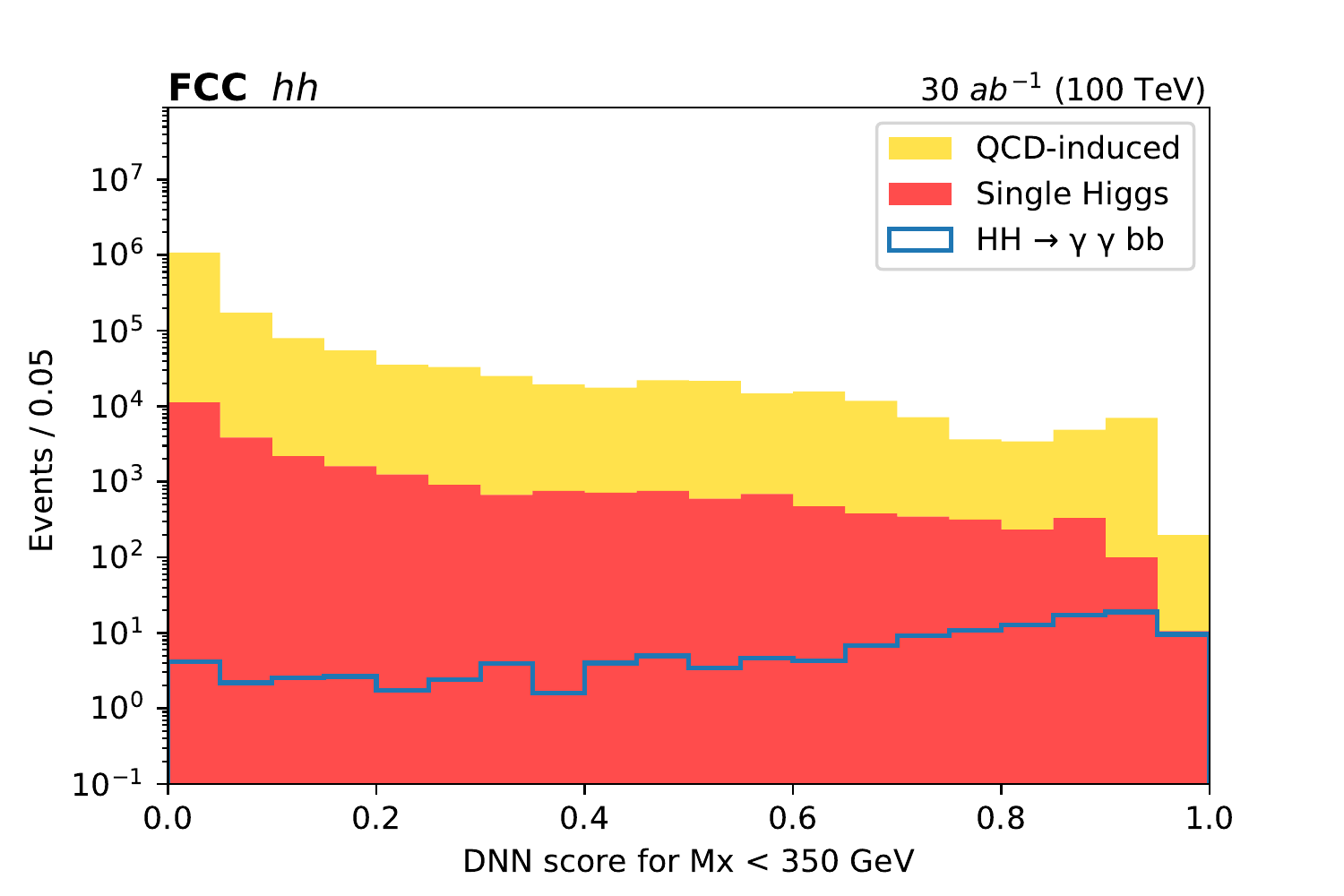}
    \end{subfigure}
    \begin{subfigure}[t]{0.45\textwidth}
       \centering
       \includegraphics[width=0.99\textwidth]{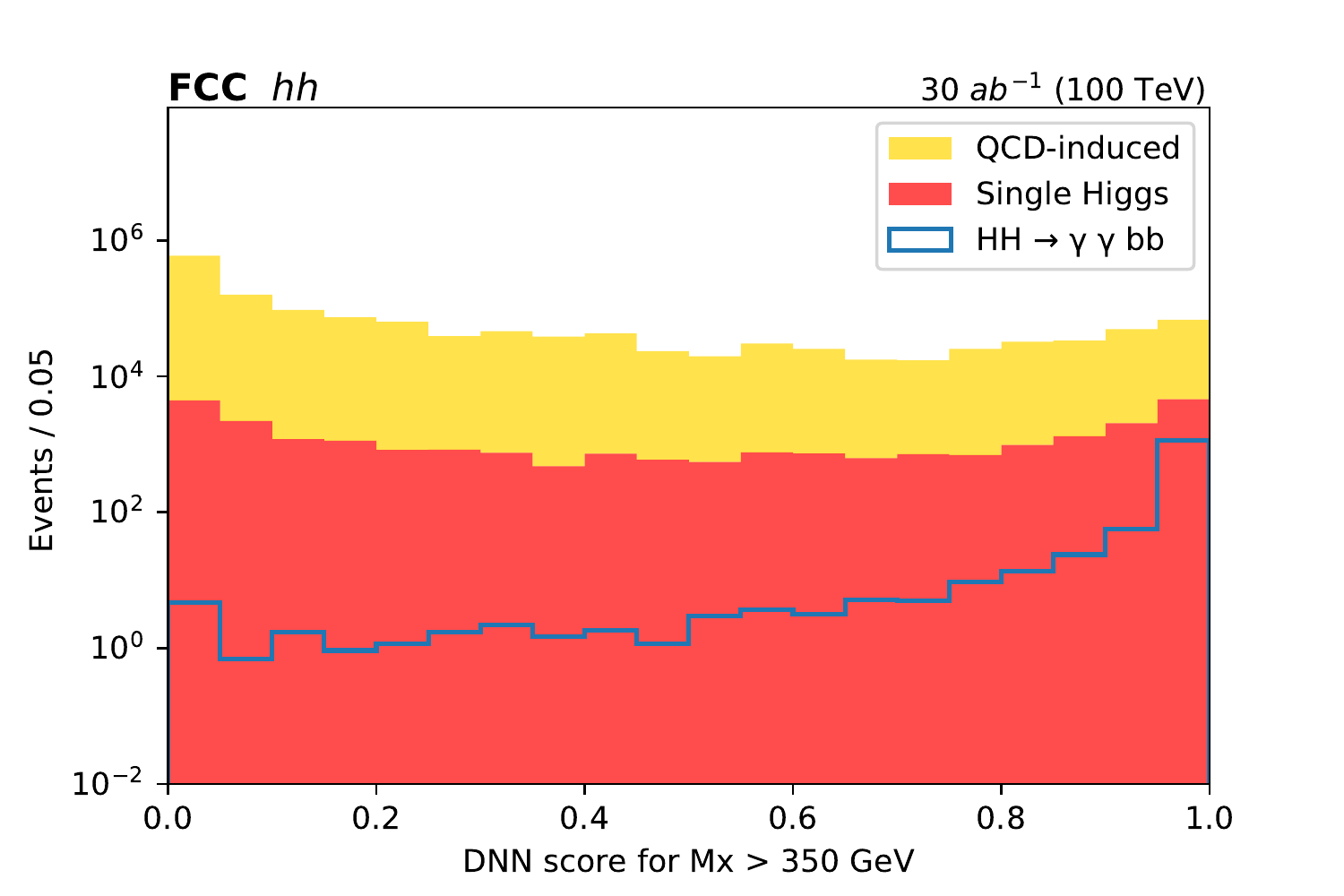}
   \end{subfigure}
   \caption{(Left) DNN score for the M< 350 GeV region.(Right) DNN score for the M> 350 GeV region. Histograms are scaled to cross section and luminosity. The background processes are stacked.}
   \label{pic:dnn_gloabal_100}
\end{figure}

\noindent The same analysis flow described in Sec \ref{sec:bbgg} is followed; to quickly summarise, the steps are:
\begin{itemize}
    \item[$\blacksquare$] Event selection and identification of the two Higgs boson candidates (candidate mass distributions in Figure \ref{pic:inv_mass_100})
    \item[$\blacksquare$] Application of the ttH tagger (distribution in Figure \ref{pic:ttH_100})
    \item[$\blacksquare$] DNN classification (Figure \ref{pic:dnn_gloabal_100}) to separate signal from all other background processes
    \item[$\blacksquare$] Event categorization according to Mx and di-jet mass region, and to purity of the DNN score
    \item[$\blacksquare$] Signal extraction for each category using the di-photon invariant mass as a figure of merit. The mass distribution is binned in such a way that for each bin a relative uncertainty of 30\% is guarantee 
\end{itemize}

\begin{figure}[h!]
    \centering
    \begin{subfigure}[t]{0.45\textwidth}
    	\centering
        \includegraphics[width=0.99\textwidth]{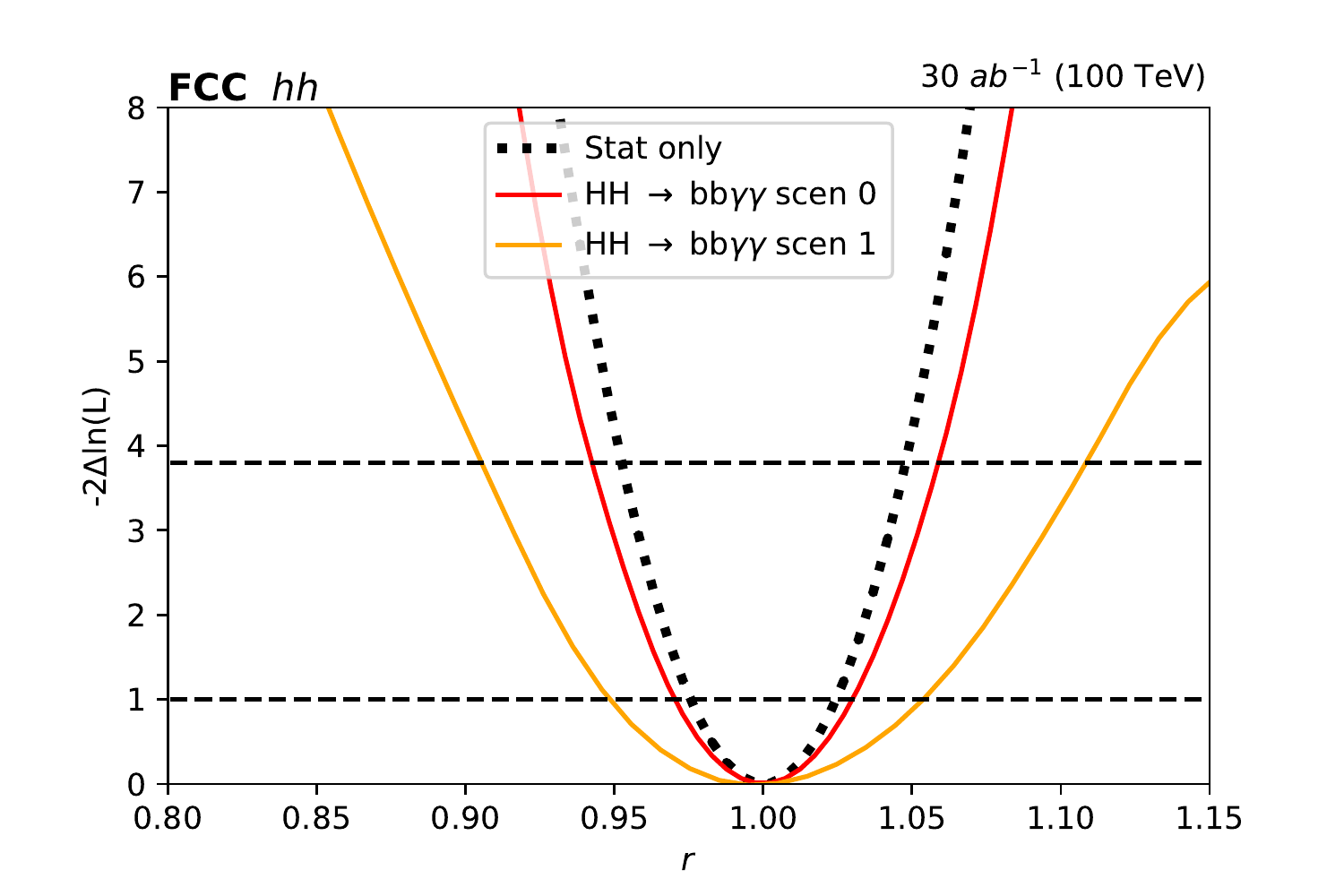}
    \end{subfigure}
    \begin{subfigure}[t]{0.45\textwidth}
       \centering
       \includegraphics[width=0.99\textwidth]{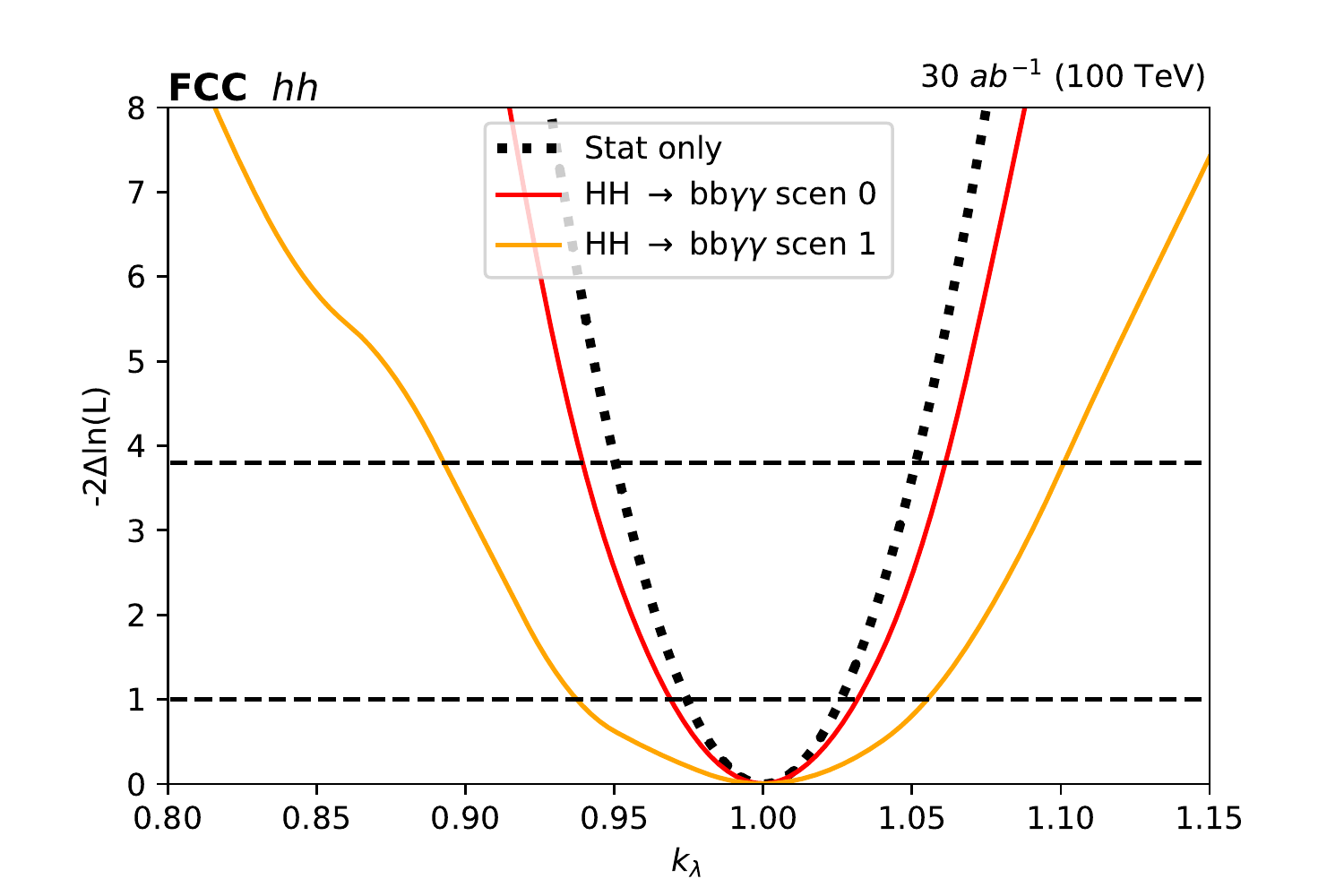}
   \end{subfigure}
  \caption{(Left) Precision on the determination of the signal strength (Right) Precision on the determination of the $\kappa_\lambda$}
   \label{pic:bbgg_res_100}
\end{figure}

\noindent Assuming a luminosity of 30 $ab^{-1}$, we can measure the signal strength of the HH production with a precision that varies between 3-5.5 at 68\% CL and 5.9-10.9 at 95\% CL depending on the systematic scenario considered (Tab \ref{tab:100tev_sum_68}).\\
In the hypothesis of the presence of a HH signal with the same properties of the SM, we can measure the Higgs self coupling with a precision that varies between 3.1-5.6 at 68\% CL and 6.2-10.8 at 95\% CL (Tab \ref{tab:100tev_sum_95}). Both sets of results are summarised in the plot Fig \ref{pic:bbgg_res_100}.

\clearpage
\subsection{$HH \rightarrow b\bar{b} \tau\tau$}

\begin{figure}[h!]
    \centering
    \begin{subfigure}[t]{0.30\textwidth}
    	\centering
        \includegraphics[width=0.99\textwidth]{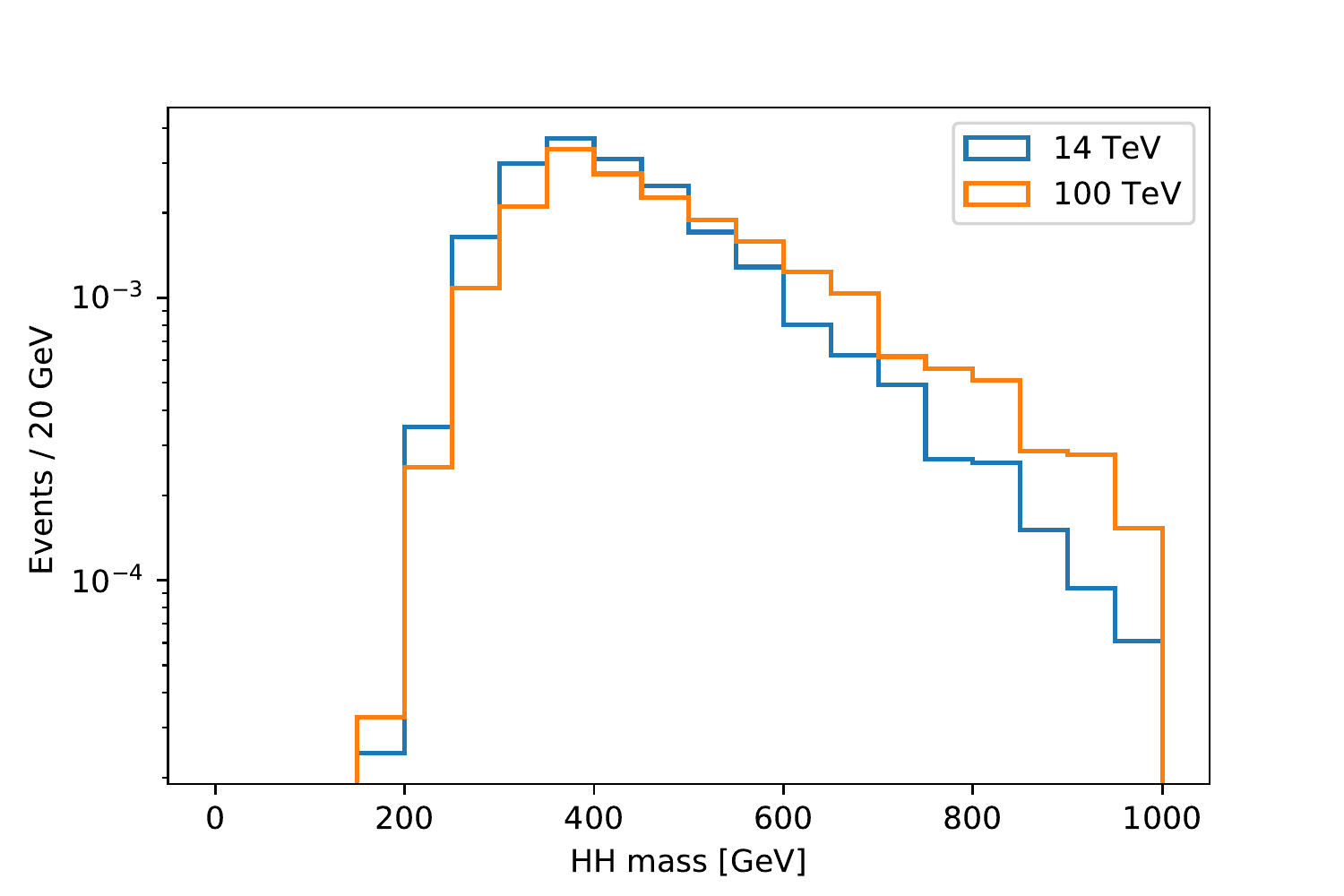}
    \end{subfigure}
    \begin{subfigure}[t]{0.30\textwidth}
       \centering
       \includegraphics[width=0.99\textwidth]{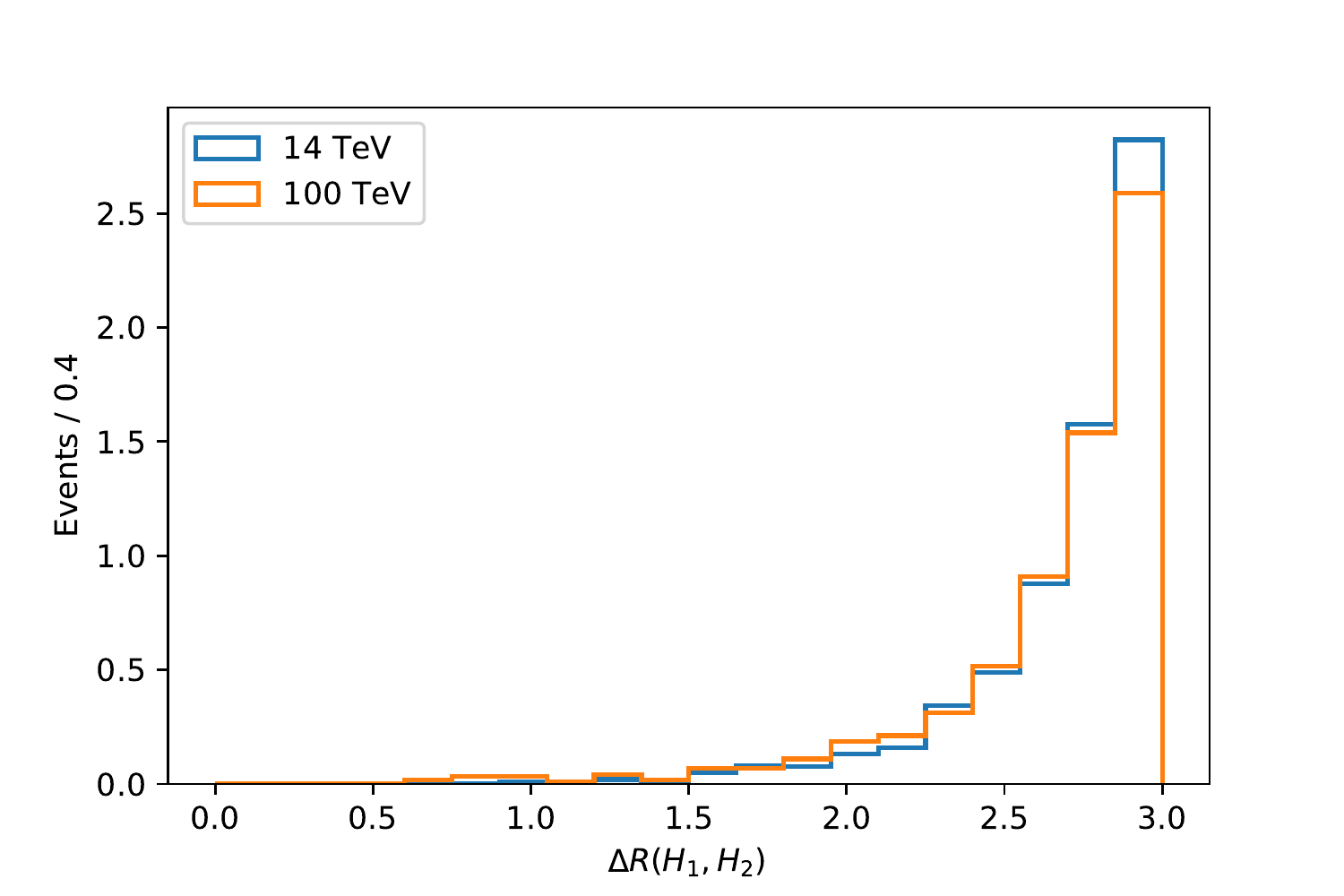}
   \end{subfigure}
   \begin{subfigure}[t]{0.30\textwidth}
       \centering
       \includegraphics[width=0.99\textwidth]{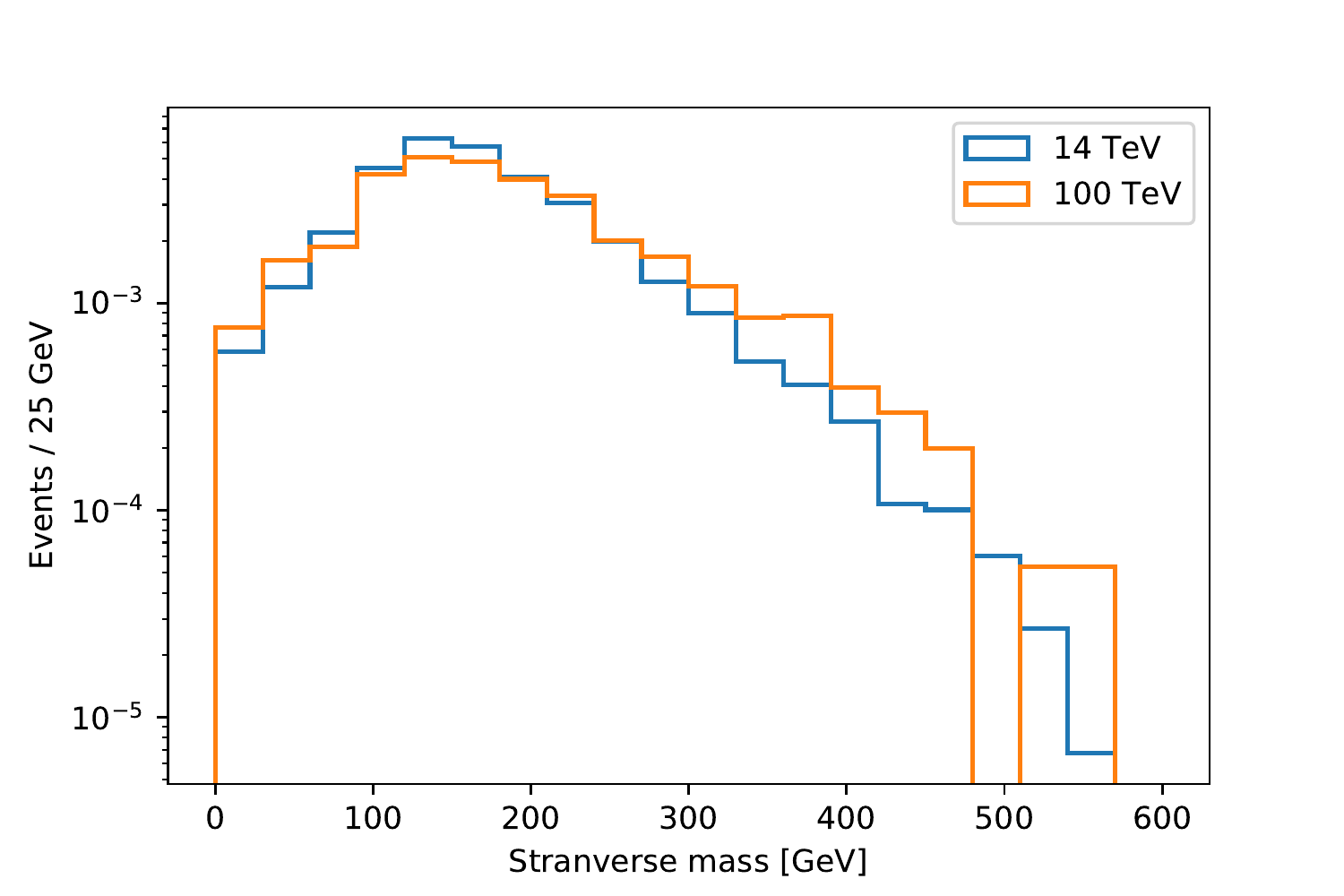}
   \end{subfigure}
    \begin{subfigure}[t]{0.30\textwidth}
       \centering
       \includegraphics[width=0.99\textwidth]{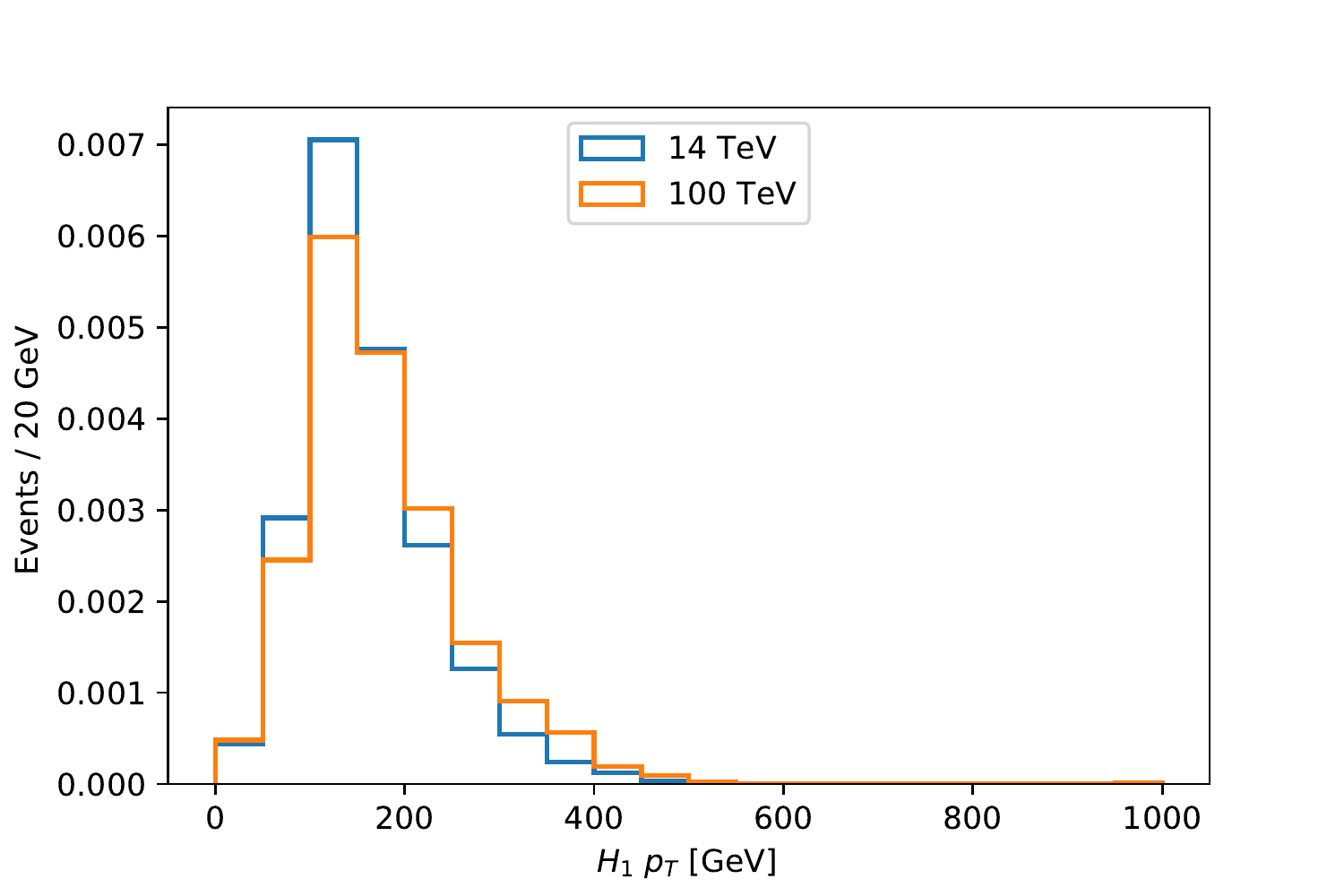}
   \end{subfigure}
    \begin{subfigure}[t]{0.30\textwidth}
       \centering
       \includegraphics[width=0.99\textwidth]{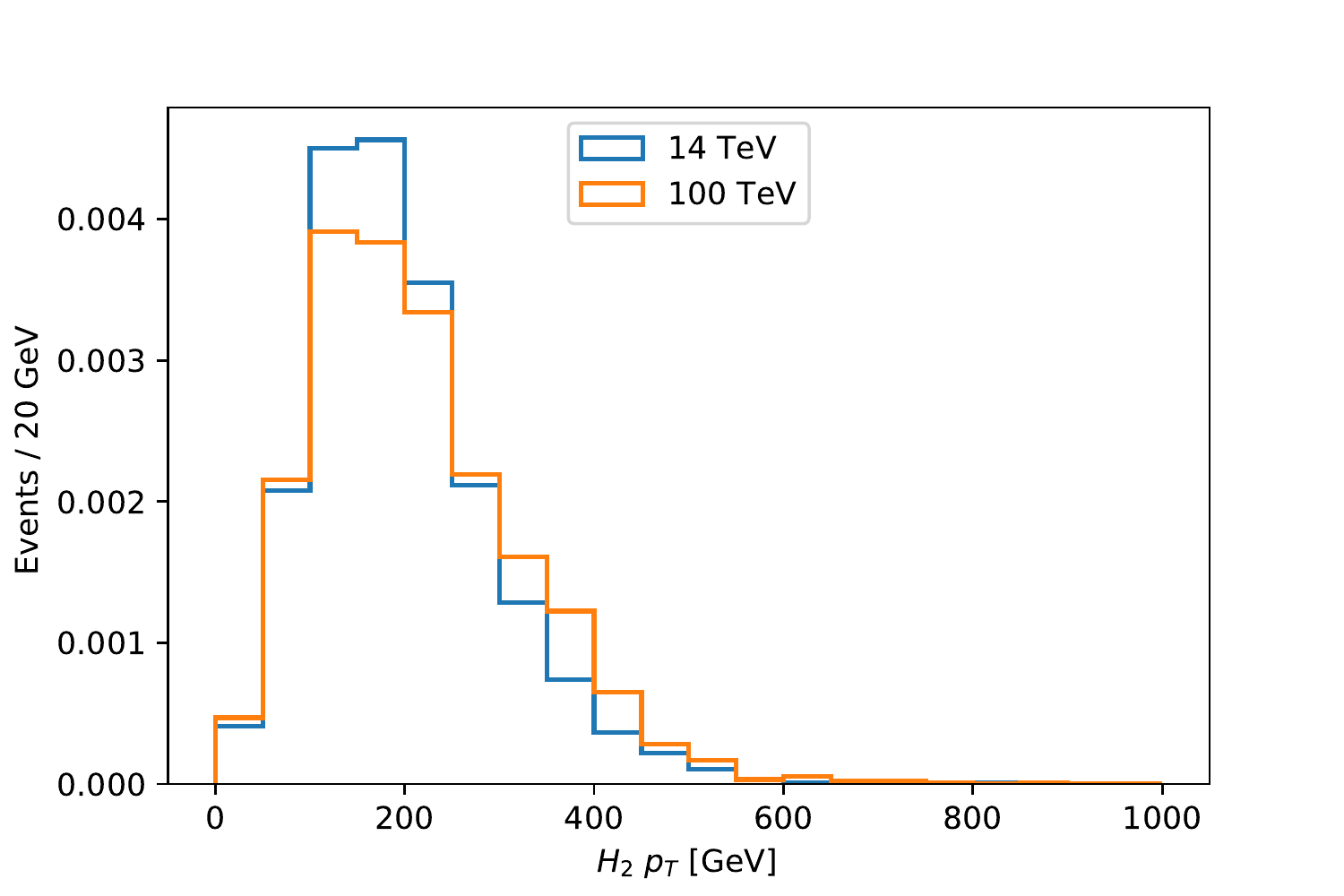}
   \end{subfigure}
   \caption{Comparison of 14 TeV and 100 TeV for signal observables. Histograms are normalized to unity. (Bottom Right) H1 transverse momentum (Bottom left) H2 transverse momentum}
   \label{pic:bbtautau_comp}
\end{figure}

For the $b\bar{b} \tau\tau$ channel, the principal backgrounds are the tt production, Drell-Yan and single Higgs production.\\
In Figure \ref{pic:bbtautau_comp} follows a comparison on the observables.

\noindent After a preliminary study on the acceptance efficiency at 100 TeV, we decided to use the same sets of cuts and categories as described in Section \ref{sec:bbtautau}.
The analysis flow can be quickly summarised into:

\begin{itemize}
    \item[$\blacksquare$] pre-selection quality cuts on the $p_T$ and the acceptance of the objects
    \item[$\blacksquare$] three categories defined depending on the lepton flavor: $\mu \tau_h$, $e \tau_h$ and $\tau_h \tau_h$
    \item[$\blacksquare$] DNN classification, making the classifier orthogonal to the stransverse mass
    \item[$\blacksquare$] construction of the high and low purity region from the DNN classifier
    \item[$\blacksquare$] signal extraction for each category in each purity region, using the stransvere mass as a figure of merit. The stransverse mass is binned in such a way that for each bin a relative uncertainty of 30\% is guarantee 
\end{itemize}

\begin{figure}[h!]
    \centering
    \begin{subfigure}[t]{0.45\textwidth}
    	\centering
        \includegraphics[width=0.99\textwidth]{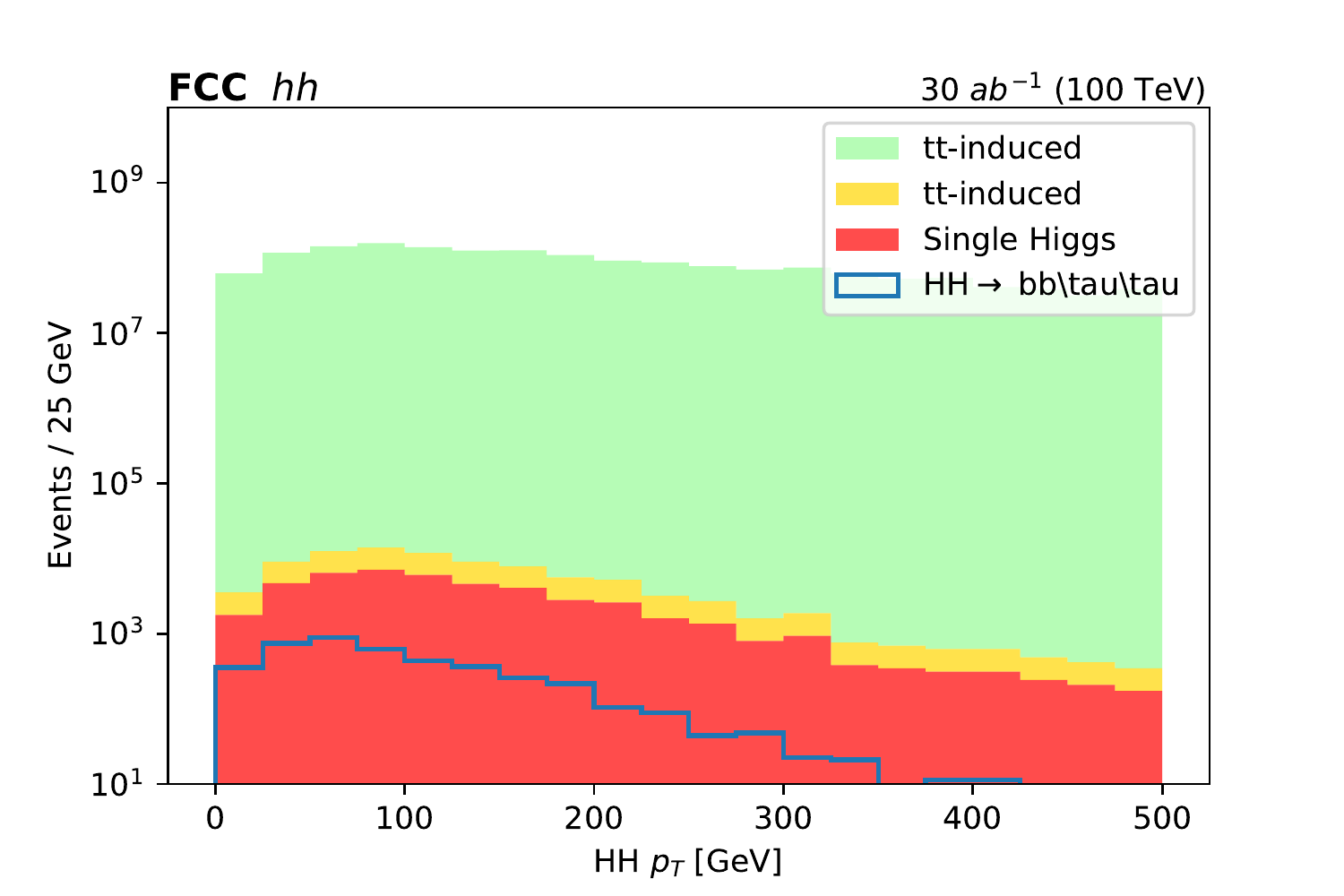}
    \end{subfigure}
    \begin{subfigure}[t]{0.45\textwidth}
       \centering
       \includegraphics[width=0.99\textwidth]{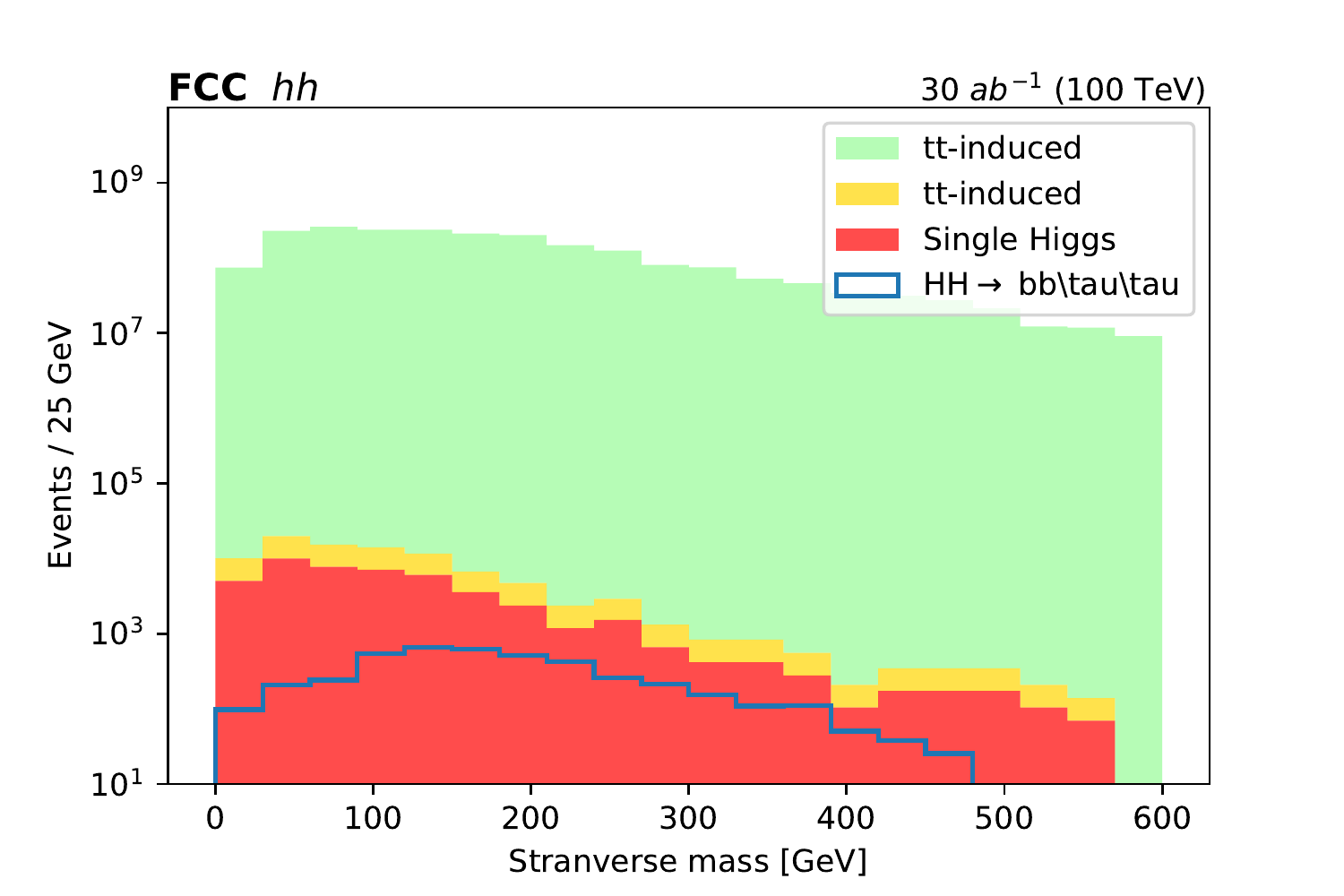}
   \end{subfigure}
  \caption{(Left) HH transverse momentum (Right) Stransverse mass}
   \label{pic:bbtautau_stack}
\end{figure}

\noindent Kinematic variables for the $\mu \tau_h$ category are shown in Fig \ref{pic:bbtautau_stack}.

\noindent The highest background contribution comes from tt production and DY as expected.\\
Assuming a luminosity of 30 $ab^{-1}$, we can measure the signal strength of the HH production with a precision that varies between 3.4-5.3 at 68\% CL and 6.8-11.6 at 95\% CL depending on the systematic scenario considered (Table \ref{tab:100tev_sum_68}).\\
In the hypothesis of the presence of a HH signal with the same properties of the SM, we can measure the Higgs self coupling with a precision that varies between 4-6.6 at 68\% CL and 8.3-13.6 at 95\% CL (Table \ref{tab:100tev_sum_95}). Both sets of results are summarised in the plot Figure \ref{pic:bbgg_res_100}

\begin{figure}[h!]
    \centering
    \begin{subfigure}[t]{0.45\textwidth}
    	\centering
        \includegraphics[width=0.99\textwidth]{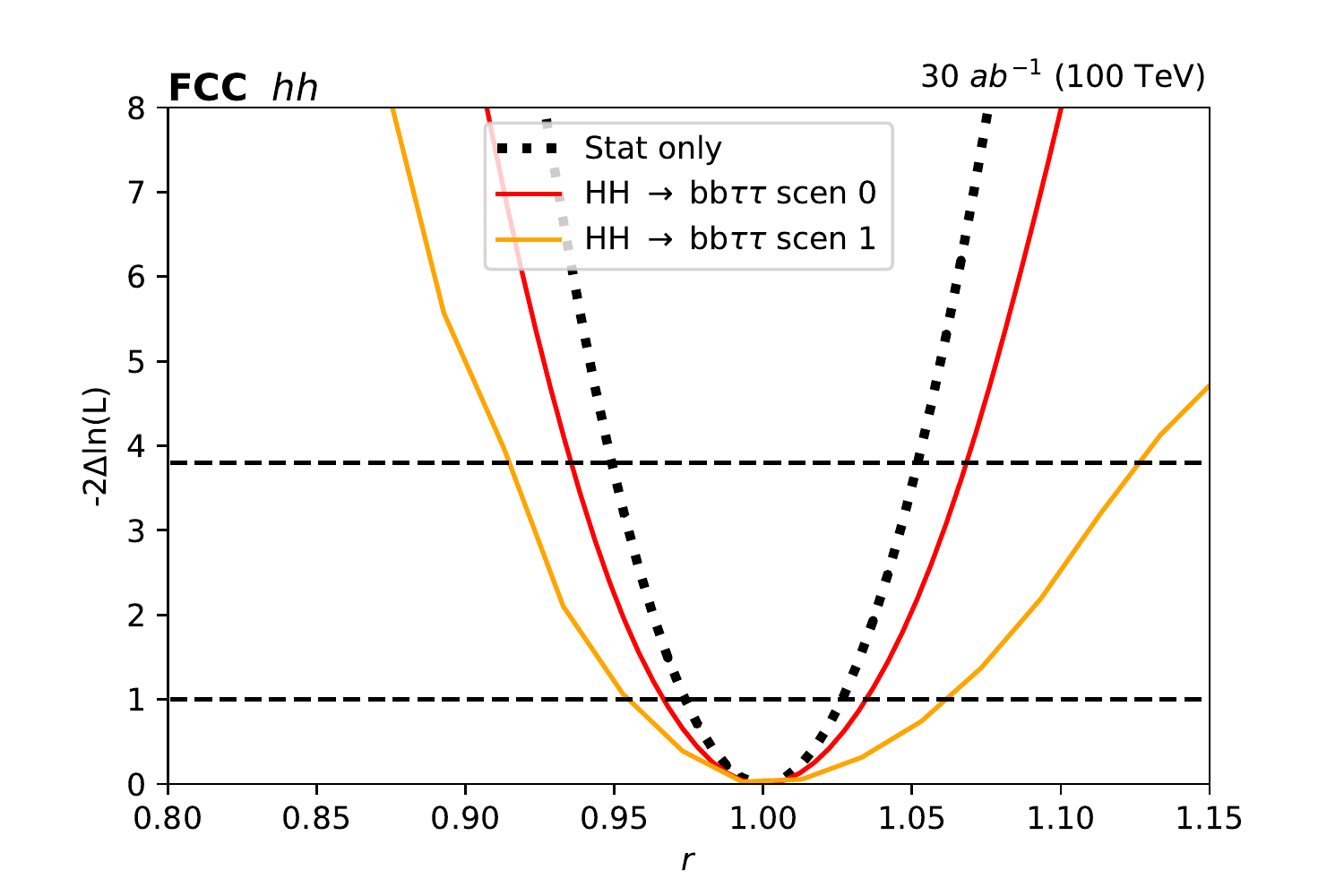}
    \end{subfigure}
    \begin{subfigure}[t]{0.45\textwidth}
       \centering
       \includegraphics[width=0.99\textwidth]{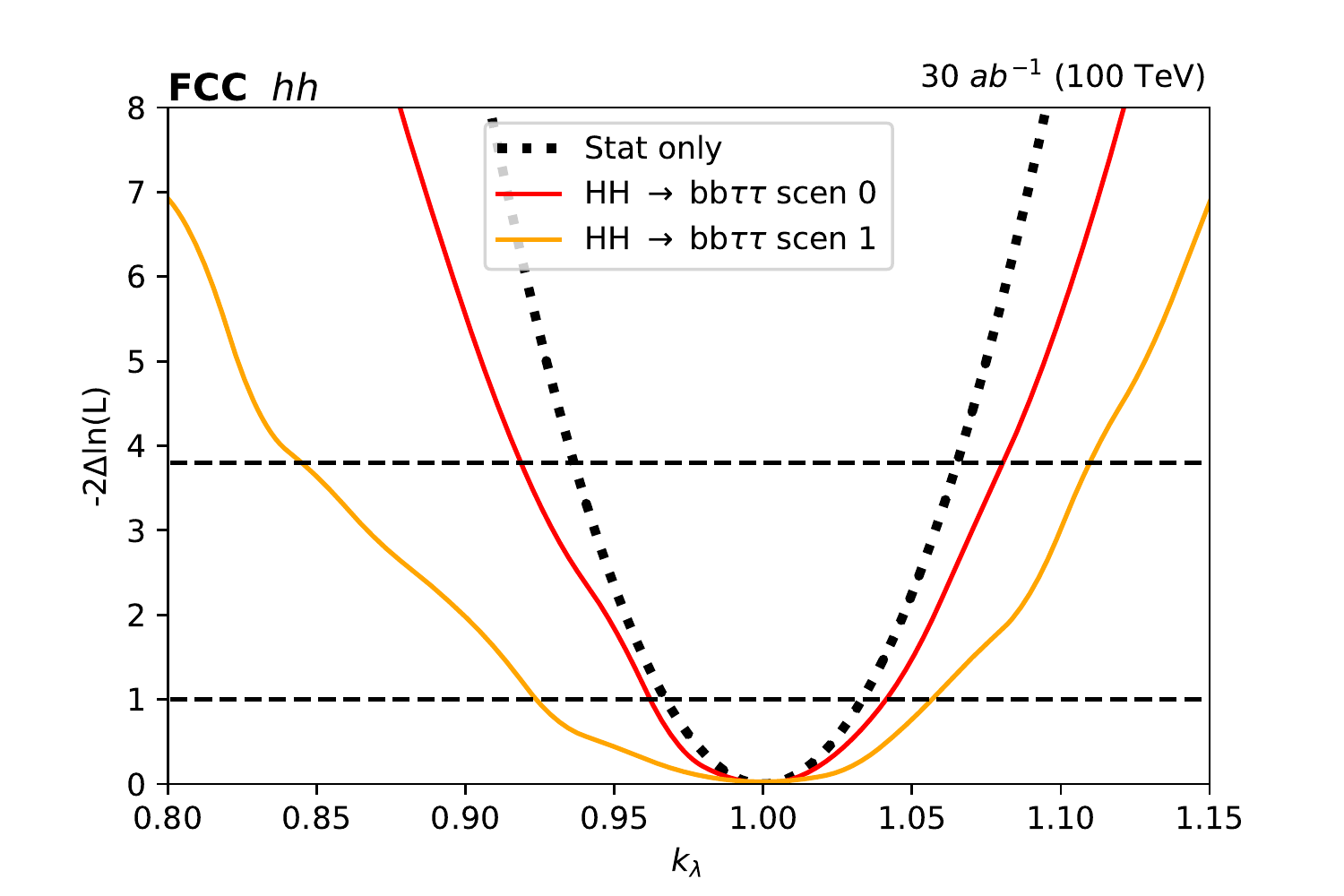}
   \end{subfigure}
  \caption{(Left) Precision on the determination of the signal strength (Right) Precision on the determination of the $\kappa_\lambda$}
   \label{pic:bbtautau_res_100}
\end{figure}

\newpage

\subsection{$HH \rightarrow b\bar{b}b\bar{b}$}

\begin{figure}[h!]
    \centering
    \begin{subfigure}[t]{0.30\textwidth}
    	\centering
        \includegraphics[width=0.99\textwidth]{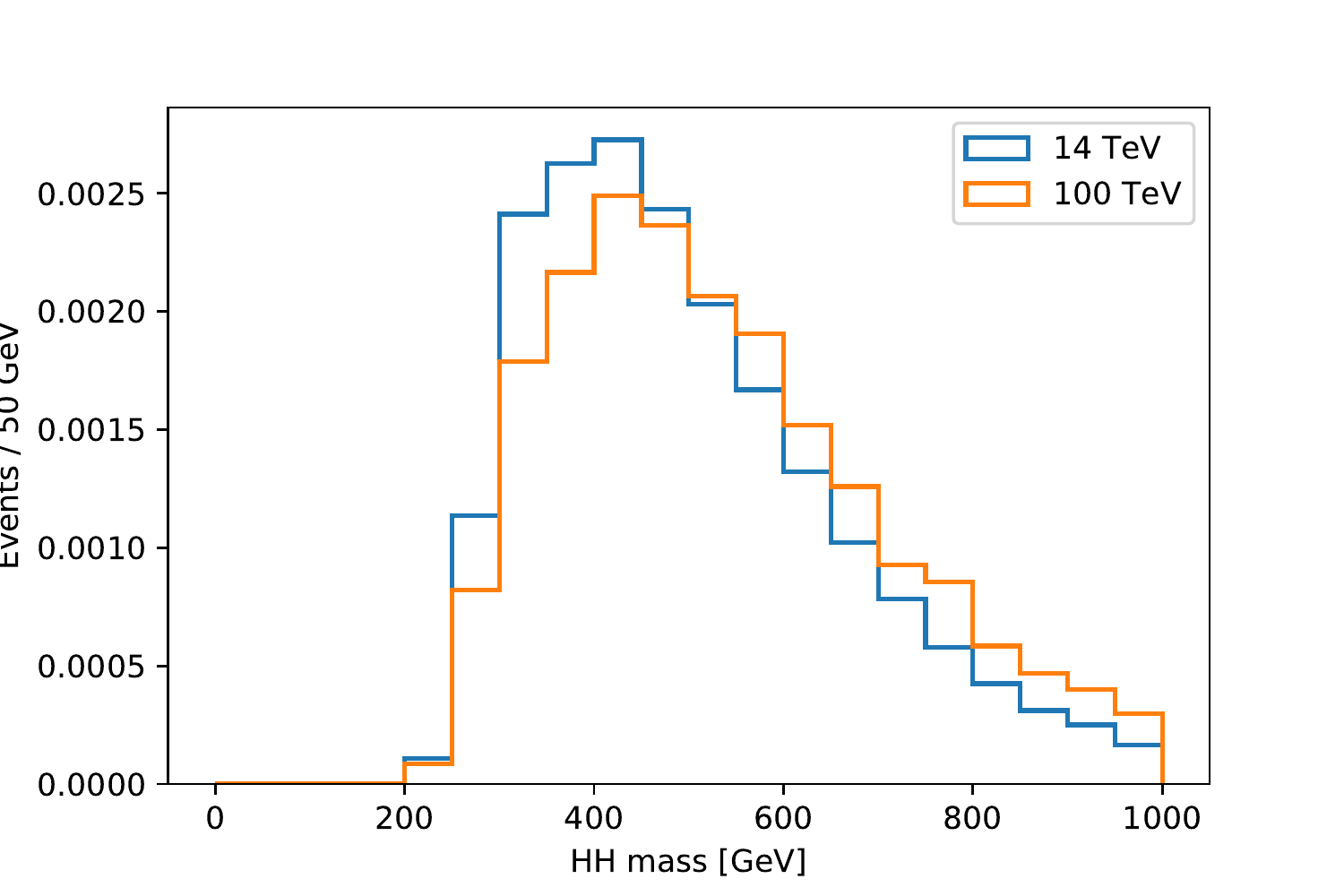}
    \end{subfigure}
    \begin{subfigure}[t]{0.30\textwidth}
       \centering
       \includegraphics[width=0.99\textwidth]{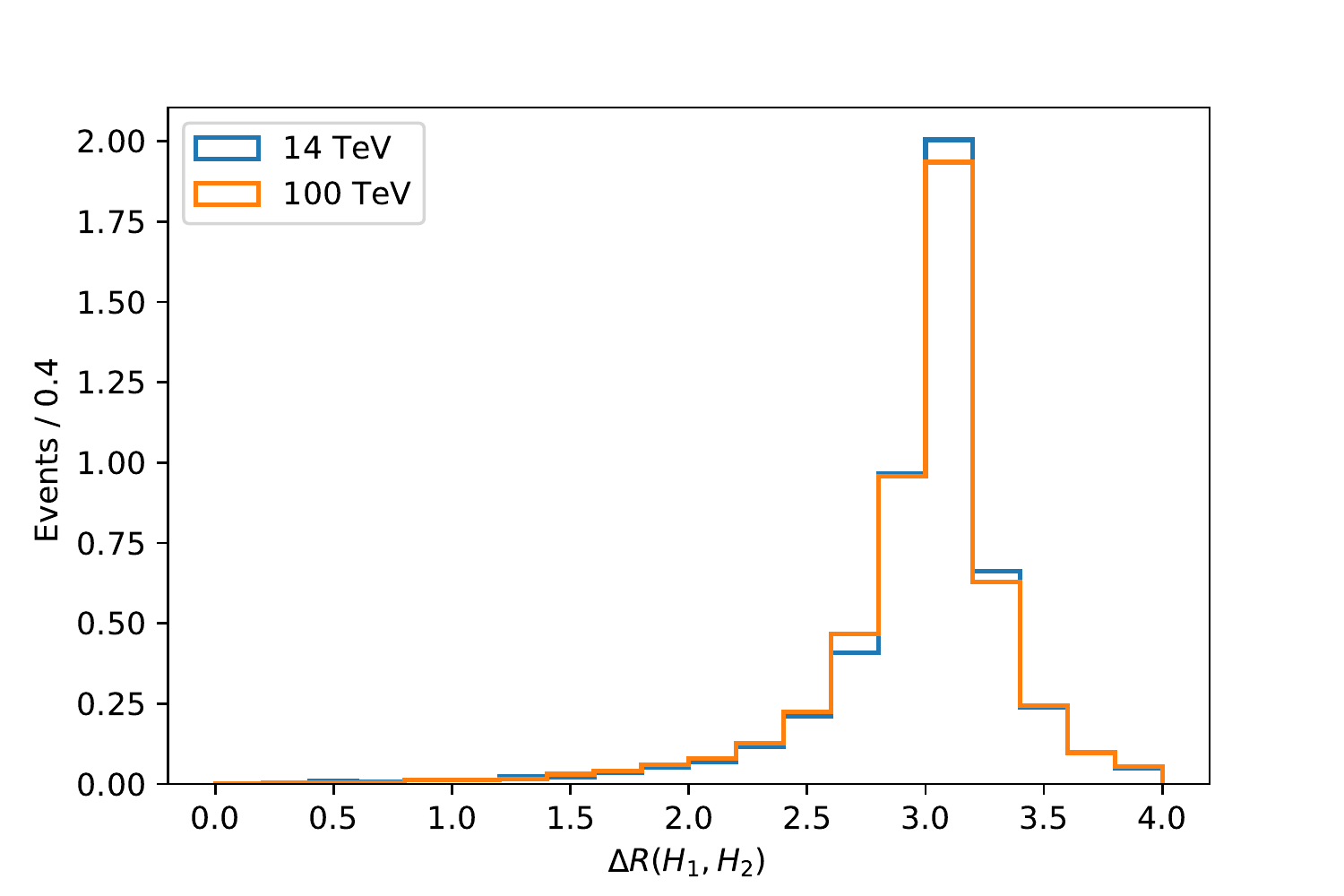}
   \end{subfigure}
   \begin{subfigure}[t]{0.30\textwidth}
       \centering
       \includegraphics[width=0.99\textwidth]{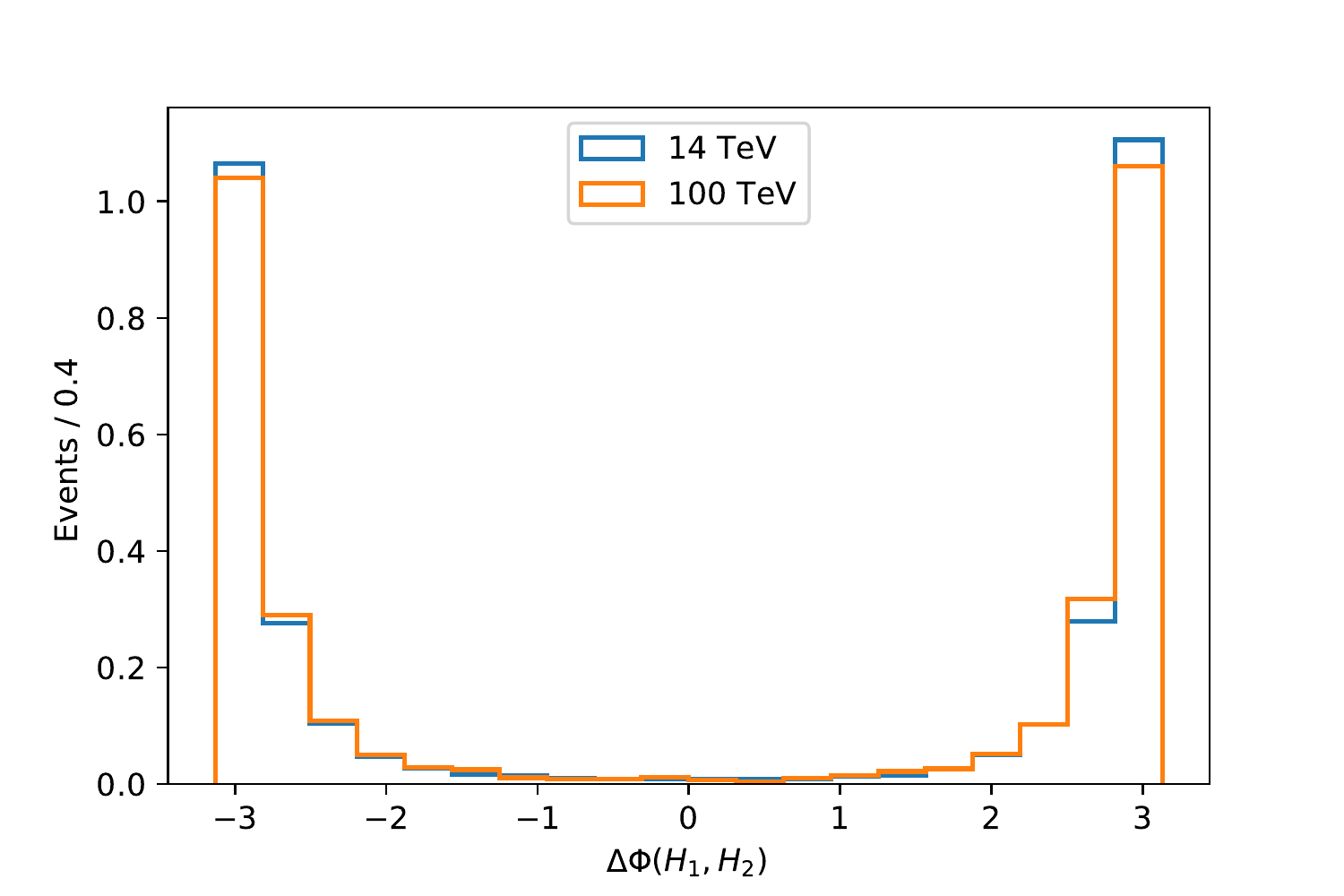}
   \end{subfigure}
    \begin{subfigure}[t]{0.30\textwidth}
       \centering
       \includegraphics[width=0.99\textwidth]{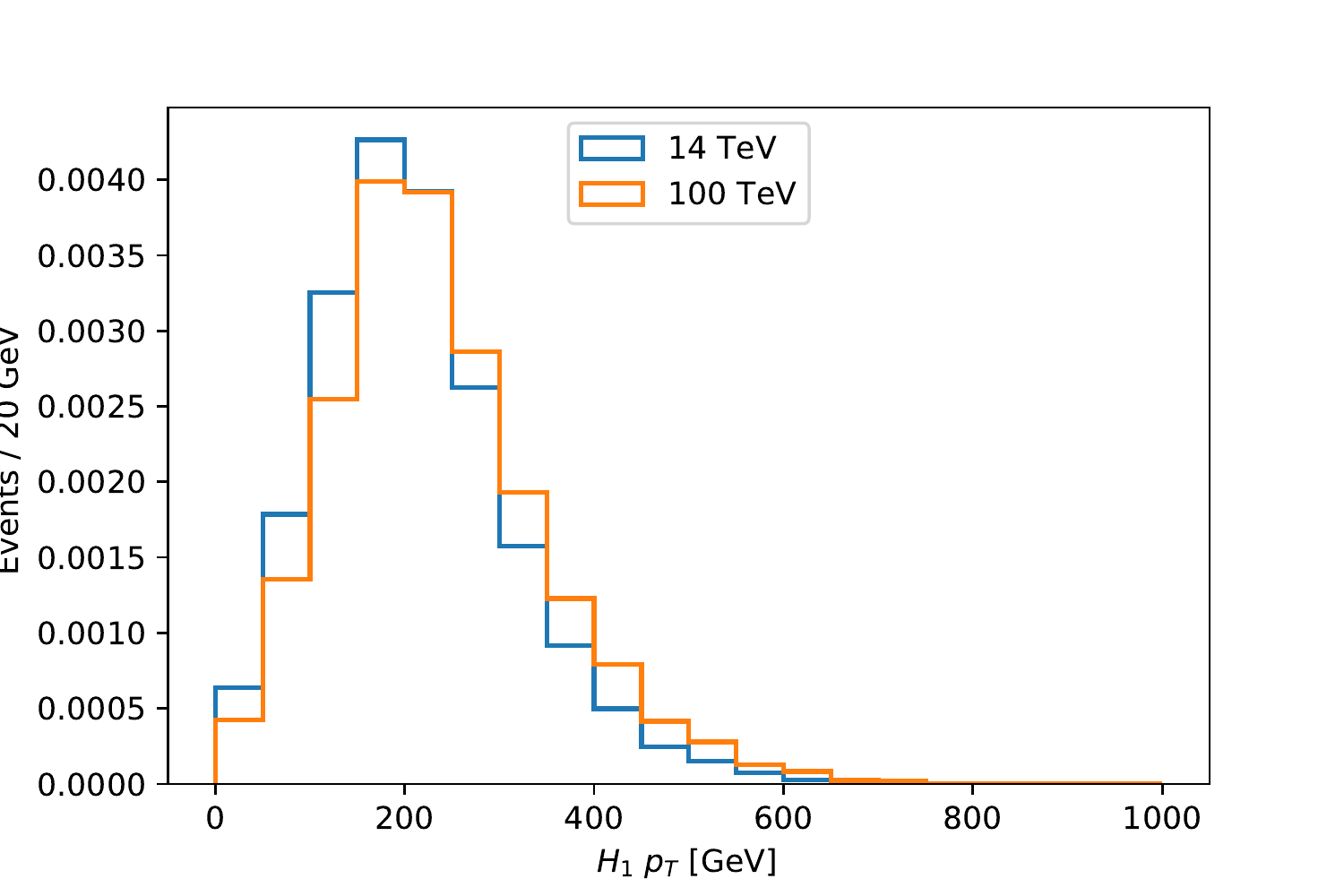}
   \end{subfigure}
    \begin{subfigure}[t]{0.30\textwidth}
       \centering
       \includegraphics[width=0.99\textwidth]{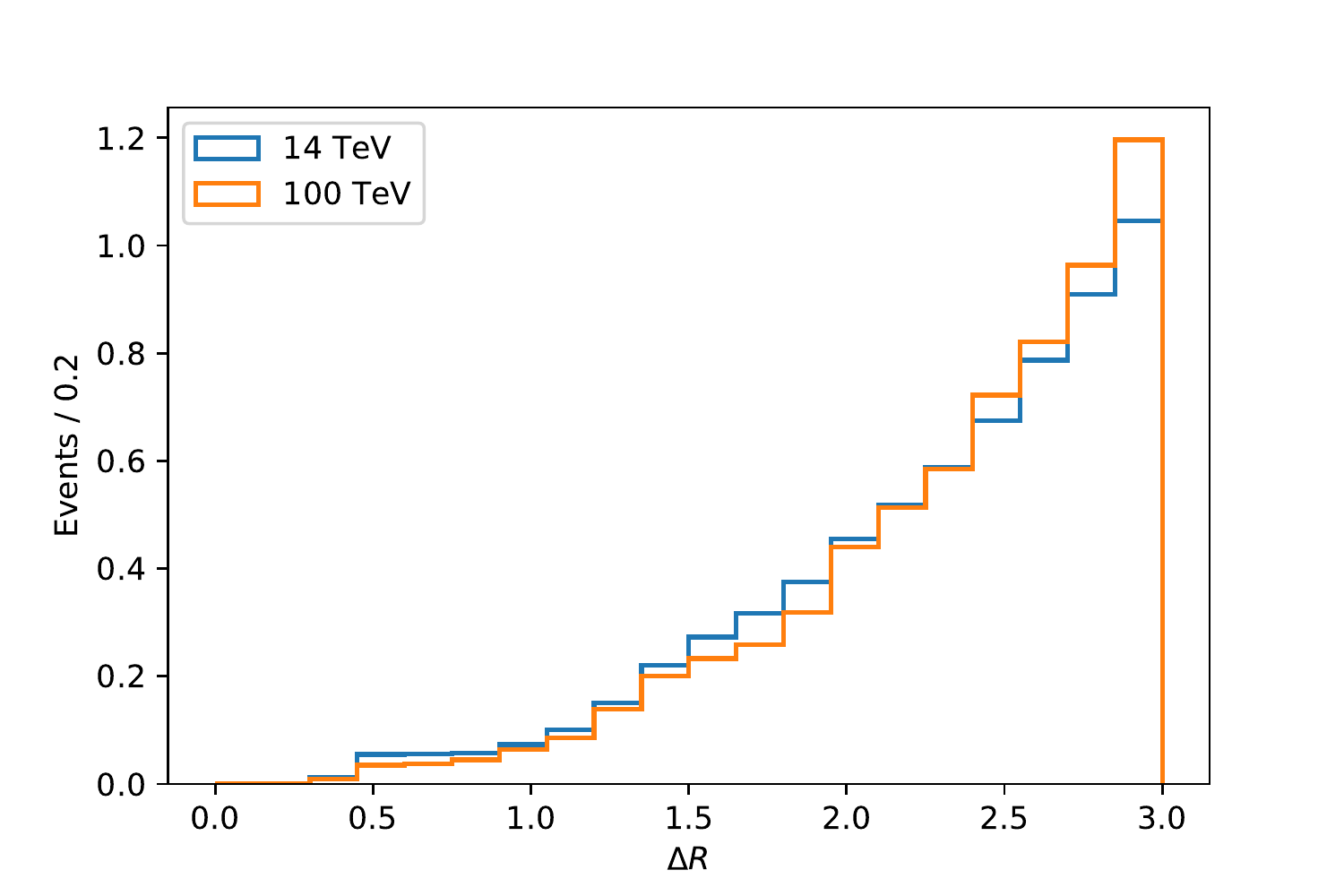}
   \end{subfigure}
   \caption{Comparison of 14 TeV and 100 TeV for signal observables. Histograms are normalized to unity. (Bottom Right) H1 transverse momemtum (Bottom left) H2 transverse momentum}
   \label{pic:4b_comp}
\end{figure}

For the $b\bar{b}b\bar{b}$ final state, the principal backgrounds are the QCD, tt and single Higgs production.\\
In Figure \ref{pic:4b_comp} a comparison of the 14 TeV and 100 TeV follows.

\begin{figure}[h!]
    \centering
    \begin{subfigure}[t]{0.45\textwidth}
    	\centering
        \includegraphics[width=0.99\textwidth]{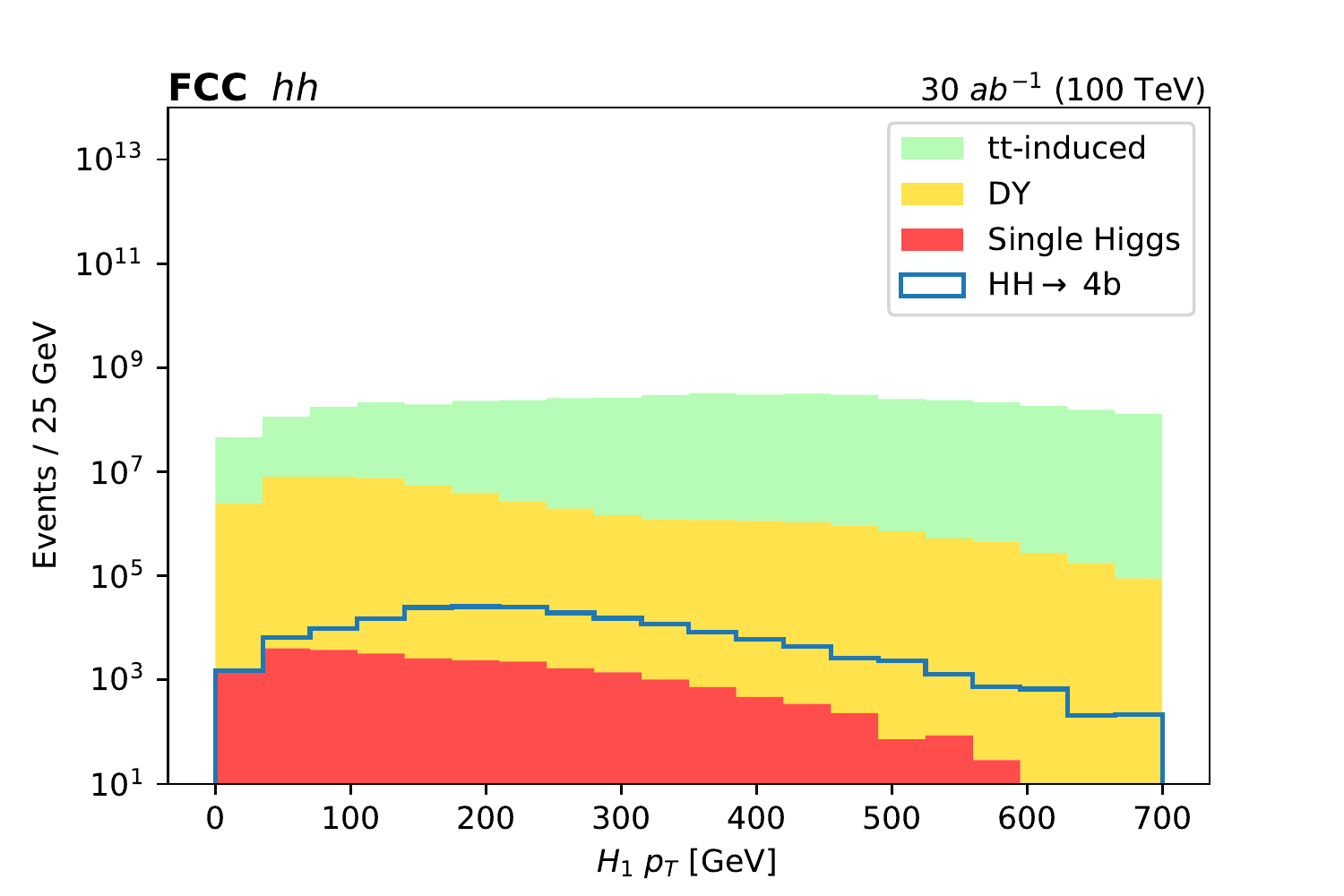}
    \end{subfigure}
    \begin{subfigure}[t]{0.45\textwidth}
       \centering
       \includegraphics[width=0.99\textwidth]{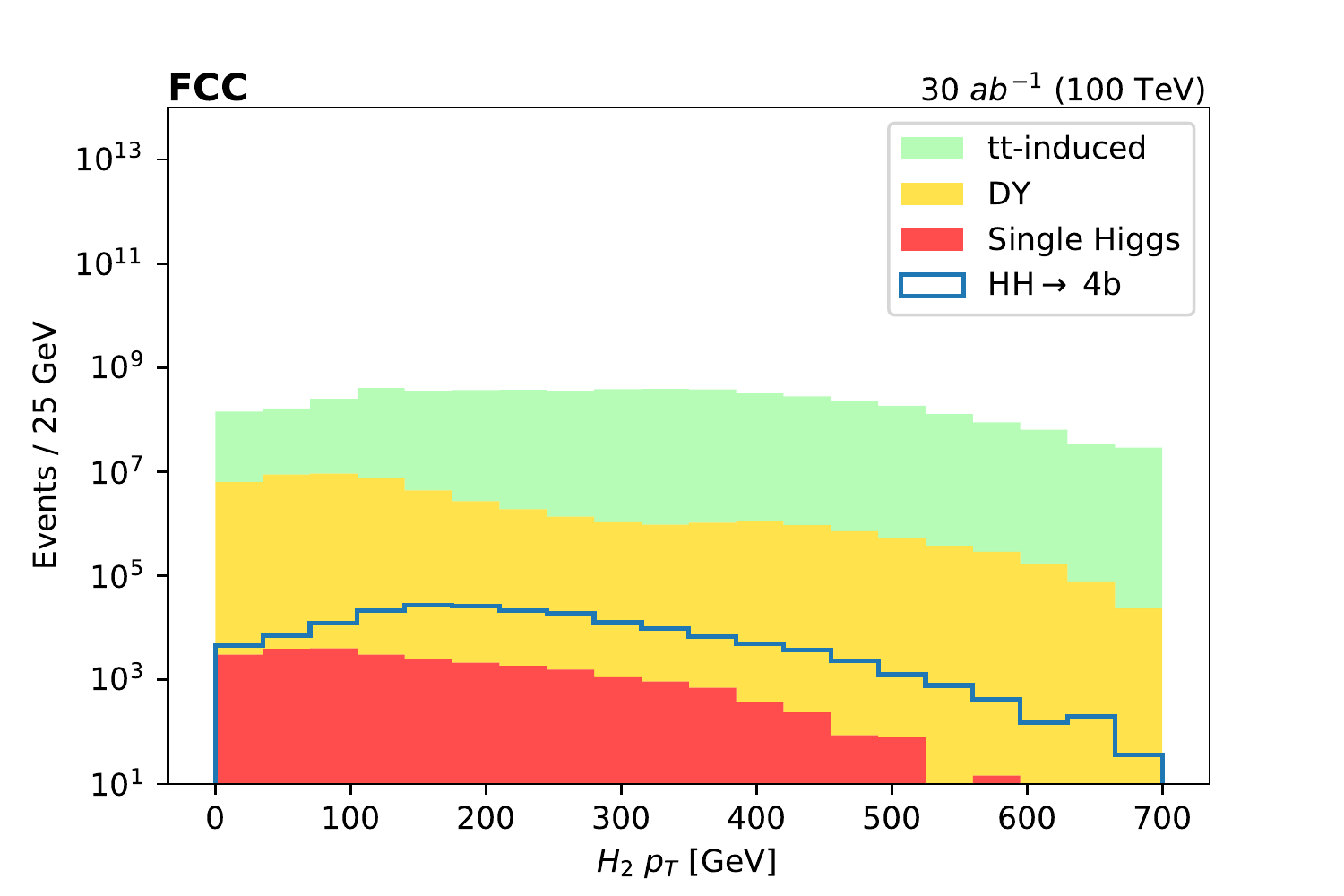}
   \end{subfigure}
  \caption{(Left) H1 transverse momentum (Right) H2 transverse momentum}
   \label{pic:4b_stack}
\end{figure}

\noindent Even if the 100 TeV scenario has harder objects with respect to 14 TeV scenario, after a preliminary study on the acceptance efficiency at 100 TeV, we decided to use the same sets of cuts described in Section \ref{sec:4b}.\\
Quality cuts on $p_T$, acceptance and b tagging of the jets are used to enhance the signal sensitivity; a further requirement on the number of b jets and cut on the $m_{jj}$ is used to suppress the QCD and tt background.\\
In Figure \ref{pic:4b_stack} some kinematic variables of the HH system are shown.

\begin{figure}[h!]
    \centering
    \begin{subfigure}[t]{0.45\textwidth}
    	\centering
        \includegraphics[width=0.99\textwidth]{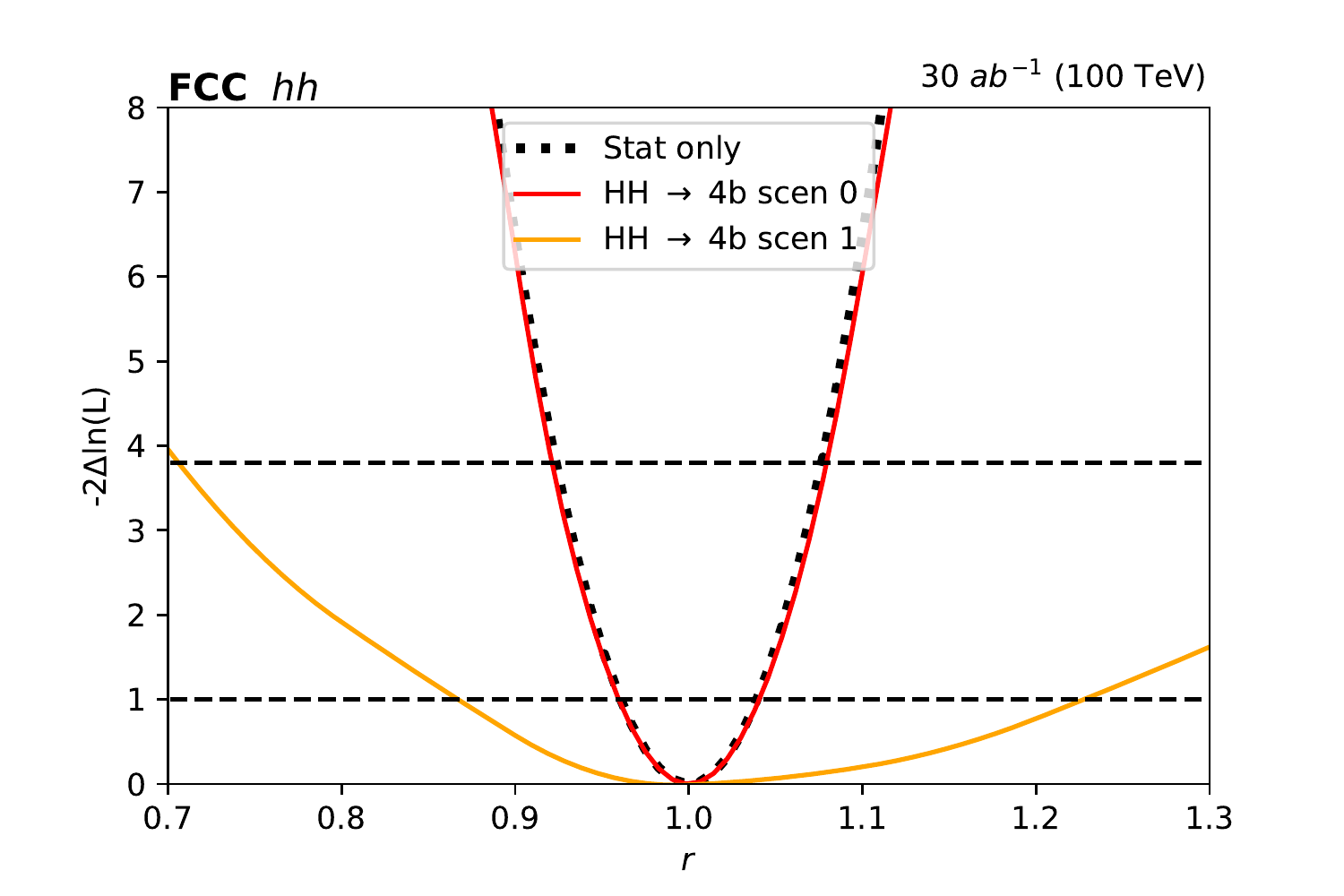}
    \end{subfigure}
    \begin{subfigure}[t]{0.45\textwidth}
       \centering
       \includegraphics[width=0.99\textwidth]{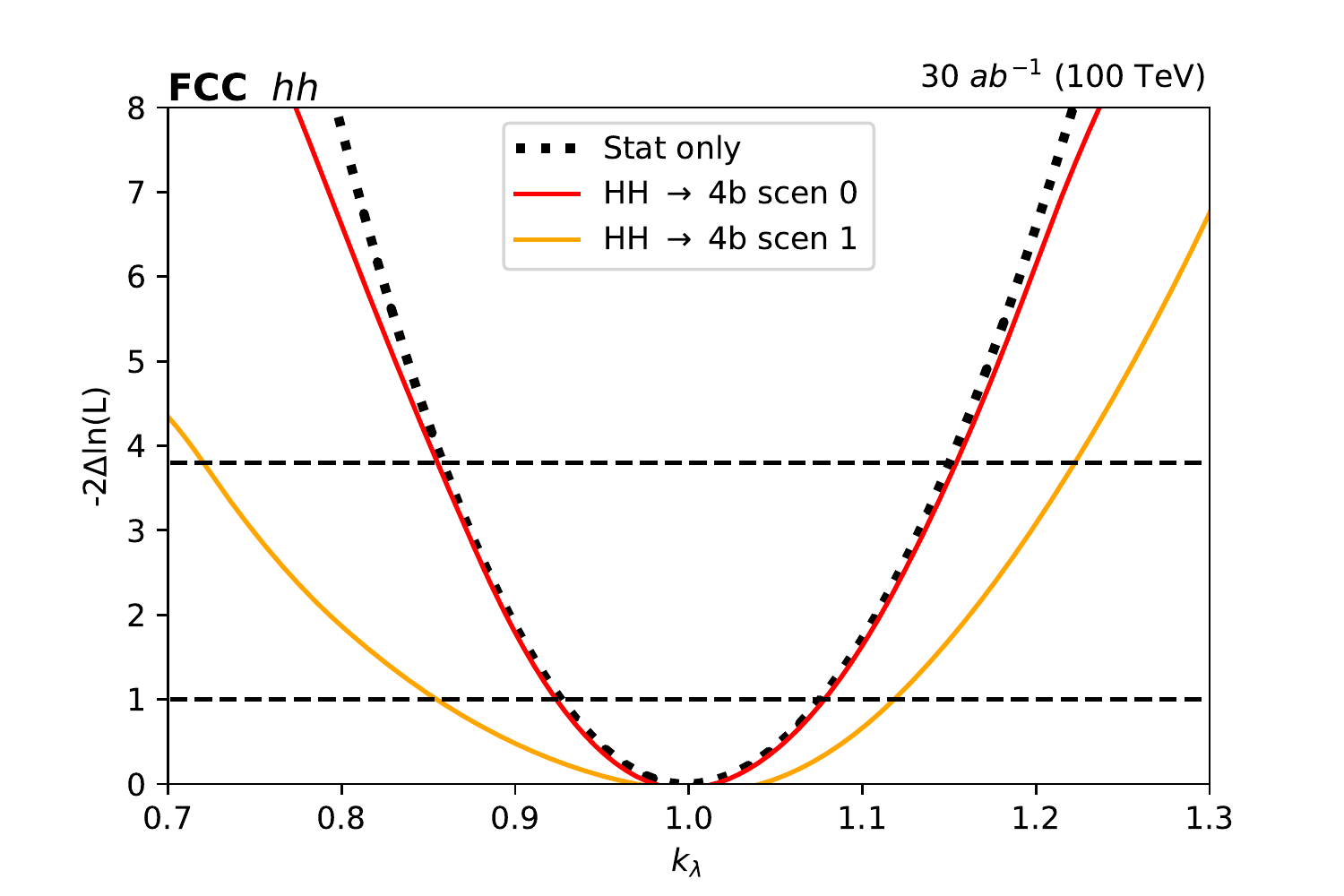}
   \end{subfigure}
  \caption{(Left) Precision on the measurement of the signal strength (Right) Precision on the measurement of the $\kappa_\lambda$}
   \label{pic:4b_res_100}
\end{figure}

\noindent The highest background contribution comes from tt production and DY as expected.\\
As for 14 TeV, a DNN classifier is used for the signal extraction (see Section \ref{sec:4bdnn}).\\
Assuming a luminosity of 30 $ab^{-1}$, we can measure the signal strength of the HH production with a precision that varies between 4-18.2 at 68\% CL and 8.1-37.4 at 95\% CL depending on the systematic scenario considered (Table \ref{tab:100tev_sum_68}).\\
In the hypothesis of the presence of a HH signal with the same properties of the SM, we can measure the Higgs self coupling with a precision that varies between 9.4-13.5 at 68\% CL and 18.9-28 at 95\% CL (Table \ref{tab:100tev_sum_95}). Both sets of results are summarised in the plot Figure \ref{pic:4b_res_100}

\subsection{$HH$ 100 TeV combination}

\begin{figure}[h!]
    \centering
    \begin{subfigure}[t]{0.45\textwidth}
    	\centering
        \includegraphics[width=0.99\textwidth]{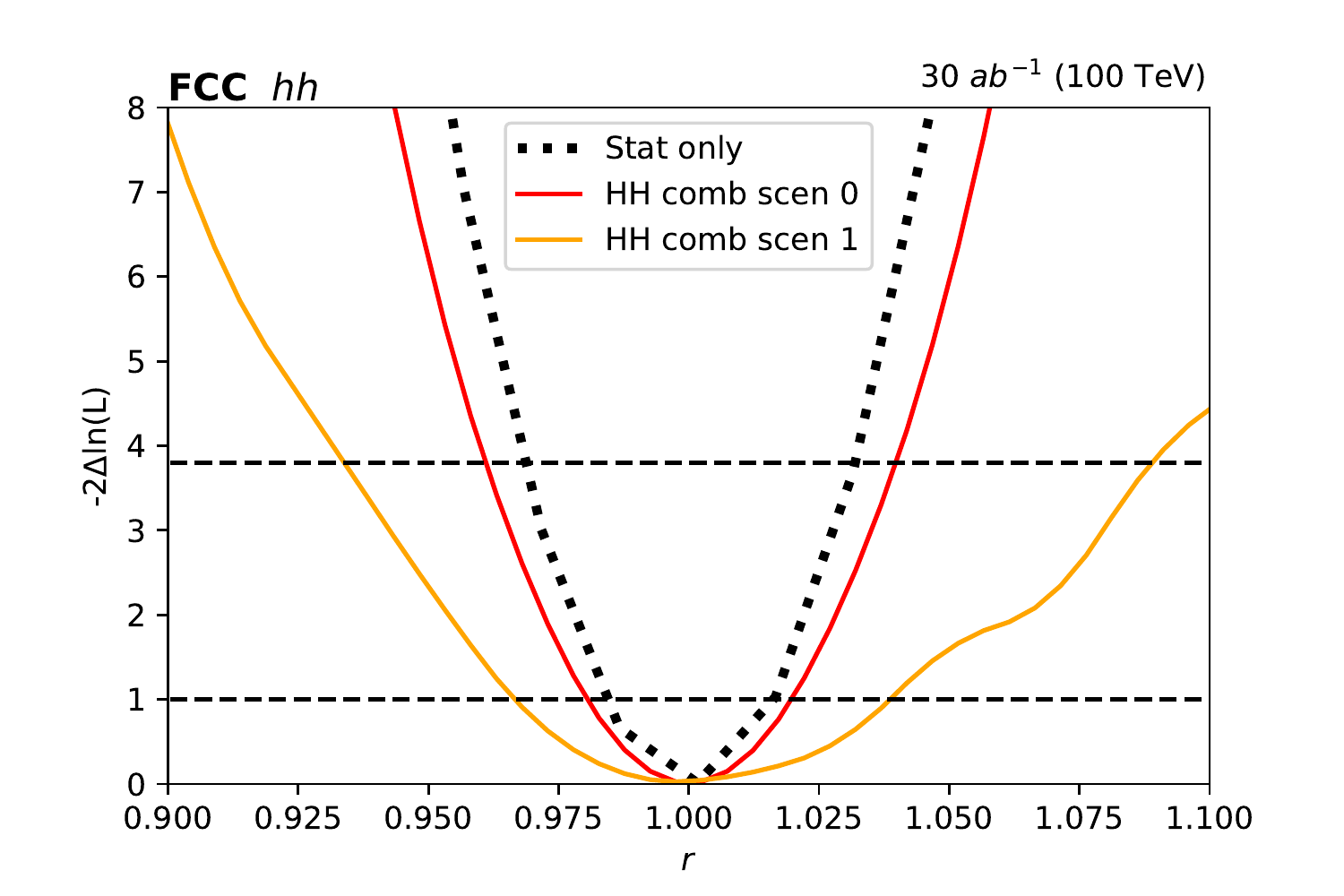}
    \end{subfigure}
    \begin{subfigure}[t]{0.45\textwidth}
       \centering
       \includegraphics[width=0.99\textwidth]{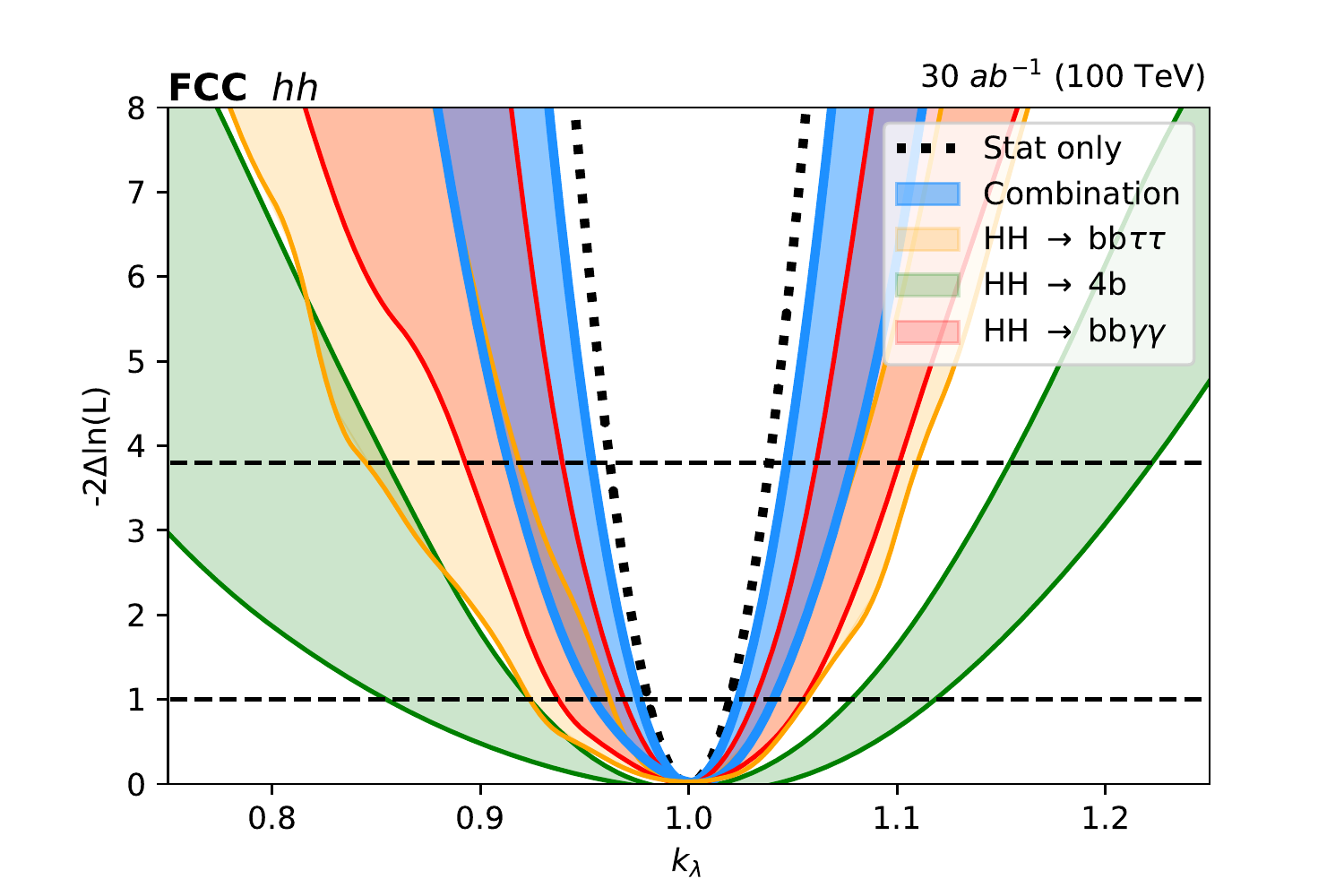}
   \end{subfigure}
  \caption{(Left) Precision on the determination of the signal strength (Right) Precision on the determination of the $\kappa_\lambda$}
   \label{pic:comb_res_100}
\end{figure}

The results obtained in each of the three decay channels are combined together assuming the SM branching fractions for HH decays to the studied final states.\\
The analyses of the three decay channels are designed to be orthogonal thanks to the mutually exclusive object selection used for each channel. Systematic uncertainties on the theoretical assumptions or associated to the same object, such as b tagging efficiency, are treated as correlated, while all the others are left uncorrelated.\\
Combining all the channels together the expected precision on the signal strength at 30 $ab^{-1}$ is 2-3.6 at 68\% and 4-8 at 95\%, depending on the systematic scenario considered (Tab \ref{tab:100tev_sum_68}). The precision on the measurement on the Higgs self coupling assuming the presence of a HH signal with the same properties of the SM, is 2.4-3.9 at 68\% and 4.8-8.5 at 95\% (Fig \ref{pic:comb_res_100}).\\

\begin{figure}[h!]
    \centering
    \begin{subfigure}[t]{0.45\textwidth}
    	\centering
        \includegraphics[width=0.99\textwidth]{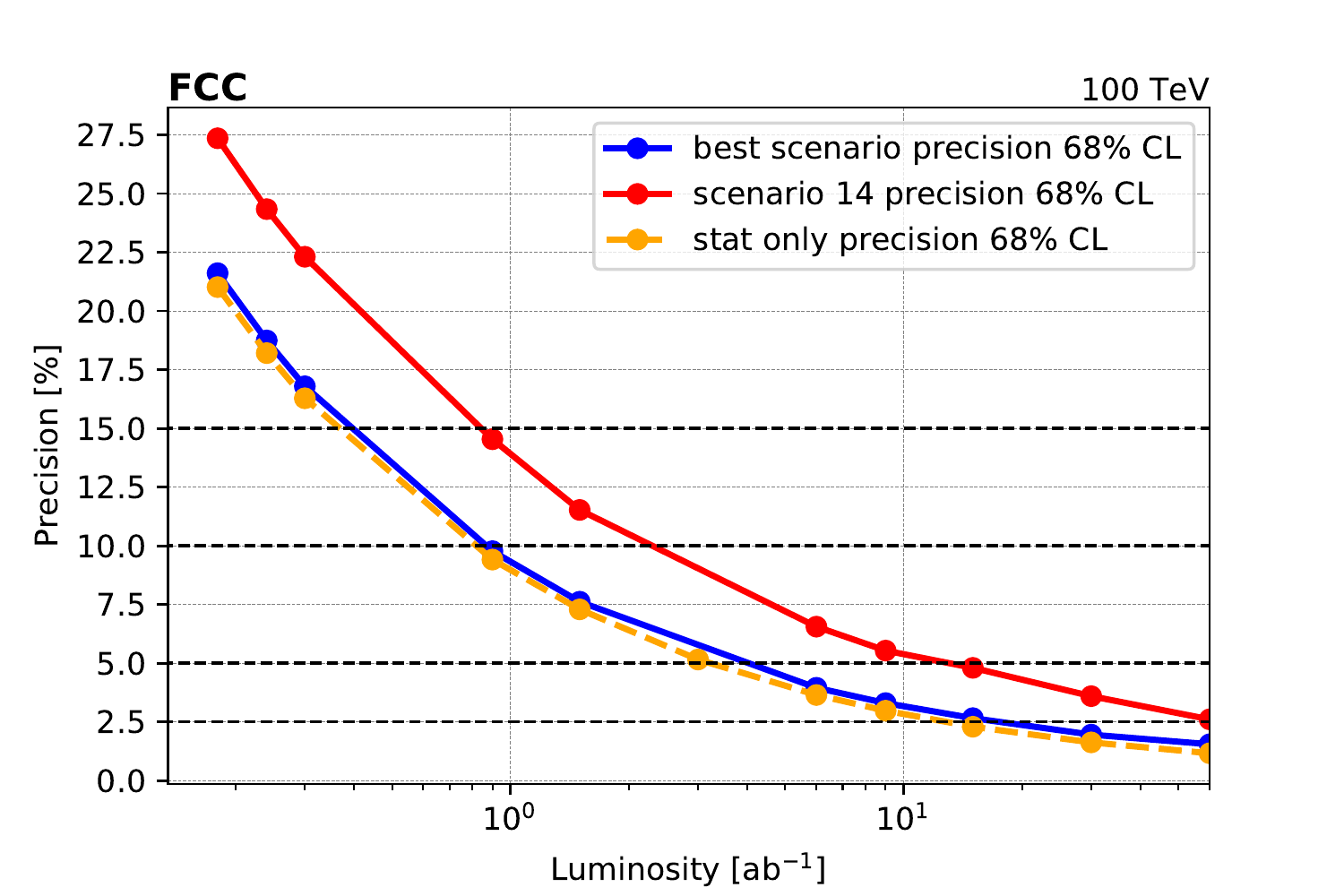}
    \end{subfigure}
    \begin{subfigure}[t]{0.45\textwidth}
       \centering
       \includegraphics[width=0.99\textwidth]{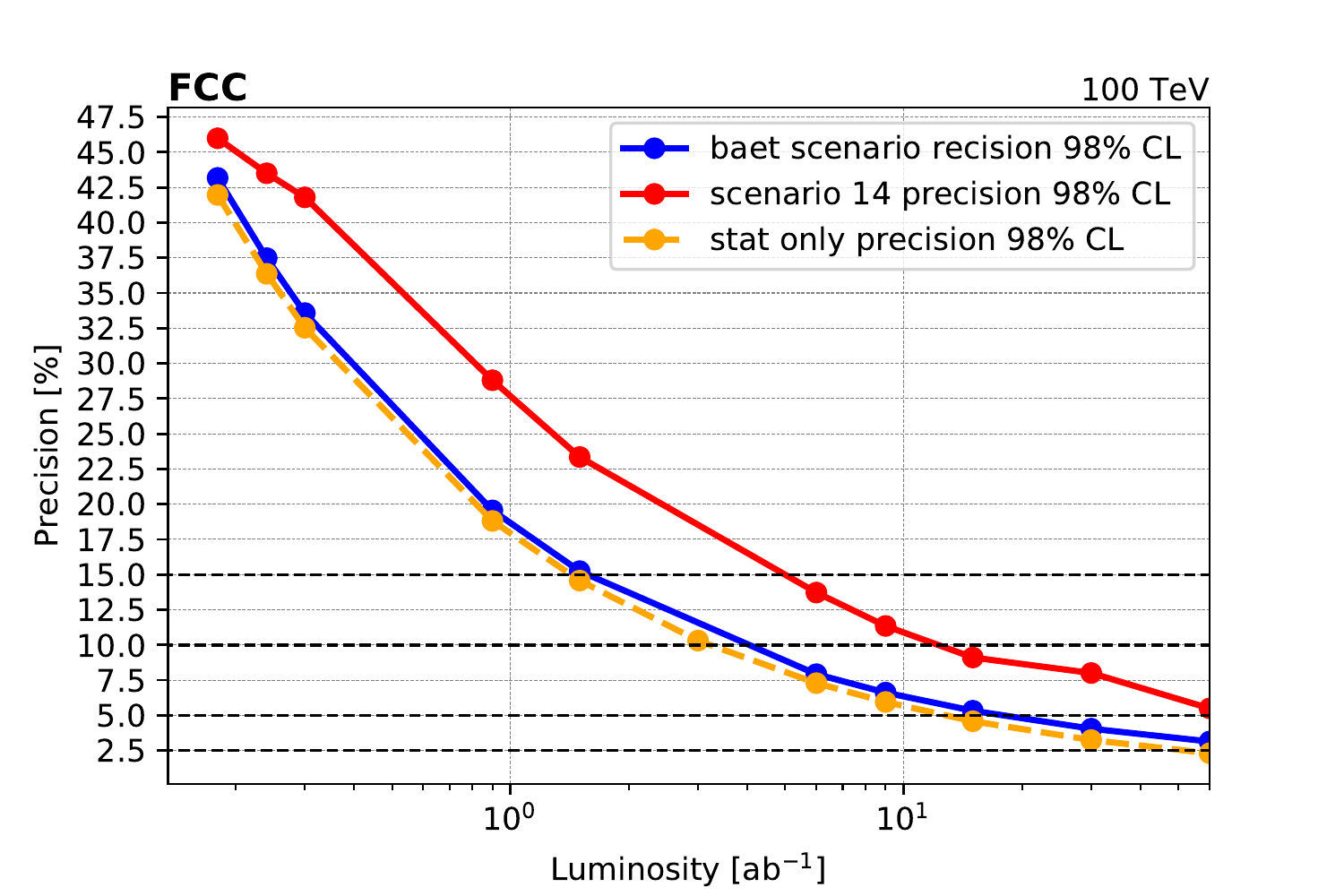}
   \end{subfigure}
  \caption{(Left) Precision on the determination of the signal strength as a function of the luminosity at 68\% CL (Right) Precision on the determination of the signal strength as a function of the luminosity at 95\% CL}
   \label{pic:pr_r}
\end{figure}

\noindent The precision on the signal strength and on the self coupling is also measured as a function of the luminosity, as reported in Fig \ref{pic:pr_r}.

\begin{table}[!h]
    \centering
    \begin{tabular}{l|l|l|l|l}
    \hline
        ~ & $HH \rightarrow  b\bar{b}\gamma\gamma$  & $HH \rightarrow b\bar{b}\tau\tau$  & $HH \rightarrow 4b$  & \textbf{HH combination} \\
    \hline
    \textbf{Precision on the signal strength at 68\% CL} & ~ & ~ & ~ & ~ \\ 
    \hline
    stat only & 2.4 & 2.6 & 3.9 & 1.6 \\ 
    scen 0 & 3 & 3.4 & 4 & 2 \\ 
    scen 1 & 5.5 & 5.3 & 18.2 & 3.6 \\

    \hline
    \textbf{Precision on the $k_{\lambda}$ at 68\% CL\%} & ~ & ~ & ~ & ~ \\ 
    \hline

    stat only & 2.6 & 3.3 & 8 & 2 \\ 
    scen 0 & 3.1 & 4 & 9.4 & 2.4 \\ 
    scen 1 & 5.6 & 6.6 & 13.5 & 3.9 \\ 

    \hline
    \end{tabular}
    \caption{Precision on the measurement at 68\% CL of the signal strength and $k_{\lambda}$ for each channel and for the combination}
    \label{tab:100tev_sum_68}
\end{table}

\begin{table}[!h]
    \centering
    \begin{tabular}{l|l|l|l|l}
    \hline
         & $HH \rightarrow b\bar{b} \gamma\gamma$  & $HH \rightarrow b\bar{b} \tau \tau$  & $HH \rightarrow 4b$  & \textbf{HH combination} \\ \hline
        \textbf{Precision on the signal strength at 95\% CL} &  &  &  &  \\
        \hline
        stat only & 4.9 & 5.3 & 7.9 & 3.3 \\ 
        scen 0 & 5.9 & 6.8 & 8.1 & 4 \\ 
        scen 1 & 10.9 & 11.6 & 37.4 & 8 \\ 
        \hline
        \textbf{Precision on the $k_{\lambda}$ at 95\% CL\%} &  &  &  &  \\
        \hline
        stat only & 5.2 & 6.6 & 16 & 3.9 \\ 
        scen 0 & 6.2 & 8.3 & 18.9 & 4.8 \\ 
        scen 1 & 10.8 & 13.6 & 28 & 8.5 \\
        \hline
    \end{tabular}
        \caption{Precision on the measurement at 95\% CL of the signal strength and $k_{\lambda}$ for each channel and for the combination}
    \label{tab:100tev_sum_95}
\end{table}

\clearpage
\section{Conclusions}

\noindent In this paper, we performed a detailed analysis of double Higgs production (through gluon-gluon fusion process) in the most sensitive decay channels $b\bar{b} \gamma\gamma$, $b\bar{b} \tau\tau$, $b\bar{b}b\bar{b}$ for several future colliders options: the HL-LHC at 14 TeV and FCC-hh at 100 TeV, assuming respectively 3 $ab^{-1}$ and 30 $ab^{-1}$ of integrated luminosity.\\
\\The sensitivity was studied by using a fast simulation tool for the Phase-2 upgraded CMS (FCC-hh) detector assuming 200 (1000) pileup events. \\The analysis benefits significantly from the usage of Deep Neural Networks, trained with the most relevant topological features of the events, to efficiently discriminate the HH signal from the much more abundant background. 
In both the scenarios, the $b\bar{b}\gamma\gamma$ channel is found to be the most sensitive one, favoured by the really high precision of photon reconstruction.
In the end, the three channels were combined to enhanced the overall significance for the di-Higgs production observation. \\

\noindent In the HL-LHC scenario, the combined significance is expected to be $2.8\sigma$, considering both the statistic and the systematic uncertainties. Since in this case the significance of the process is not enough to claim its observation, the results of this study are used to derive an upper limit on the production rate of the process, which we estimate to be 0.76 times the SM prediction, at the 95\% CL. Prospects for the measurement of the trilinear coupling are also studied, leading to a constraint on $\kappa_{\lambda}$ of $[-0.02,3.05]$ at the 95\% CL.\\
\\In the FCC-hh scenario, the significance for a HH signal is expected to lead to an observation. Depending on the assumed detector performance and systematic uncertainties, the Higgs boson trilinear self-coupling and the signal strength will be measured with a precision in the range $4.8-8.5$\% at 95\% CL ($2.4-3.9$ at 68\% CL) and $4-8$\% at 95\% CL ($2-3.6$\% at 68\% CL), respectively. 

\newpage
\bibliographystyle{splncs03_unsrt}                      \bibliography{b}
\end{document}